\documentstyle[12pt,titlepage]{book}
\def\be{\begin{equation}}
\def\ee{\end{equation}}
\def\bea{\begin{eqnarray}}
\def\eea{\end{eqnarray}}
\def\im{{\hbox{\rm Im}}} 
\def\ker{{\hbox{\rm Ker}}}
\def\cok{{\hbox{\rm Coker}}} 
\def\tr{{\hbox{\rm Tr}}}
\def\too{\longrightarrow}
\def\ind{{\hbox{\rm index}}}
\def\di{{\hbox{\rm dim}}}
\def\sp{{\hbox{\rm Spin}}}
\def\cl{{\hbox{\rm Cl}}}
\def\ccl{{\hbox{{\bf \rm C}{\rm l}}}}

\begin{document}
\begin{titlepage}
\begin{flushright} US-FT-2/97\\ hep-th/9701128
\end{flushright}
%\vspace*{20pt}
\bigskip
\begin{center} {\LARGE The Geometry of Supersymmetric Gauge Theories in Four
Dimensions}
\footnote{Ph.D.  Thesis presented in October 1996}
\vskip 0.9truecm

{Marcos Mari\~no}

\vspace{1pc} {\it  Departamento de F\'\i sica de Part\'\i culas,\\
Universidade de Santiago de Compostela,\\ E-15706 Santiago de Compostela,
Spain.}

\vspace{5pc}

{\large \bf Abstract}
\end{center} We review the relations between (twisted) supersymmetric gauge
theories in four dimensions  and moduli problems in four-dimensional
topology, and we study in detail the non-abelian  monopole equations from
this point of view. The relevance of exact results in $N=1$ and 
$N=2$ supersymmetric gauge theories to the computation of topological
invariants is emphasized.  Some background material is provided, including an
introduction to Donaldson theory, the  twisting procedure and the
Mathai-Quillen formalism.
 
\vfill

\end{titlepage}

\newpage
\thispagestyle{empty}

\tableofcontents
\newpage
\thispagestyle{empty}

\chapter{Introduction}

This work has its roots in the recent developments in four-dimensional
geometry and the  interrelations with the understanding of the
non-perturbative behaviour of  supersymmetric gauge theories in four
dimensions. Perhaps it is useful to give a brief  historical overview to
understand the context of this relationship. 

Donaldson theory was born in 1982 as an attempt to provide a new tool for the 
classification of four-manifolds. The usual techniques in Topology were not 
powerful enough and the inspiration was found in the theory of Yang-Mills
fields,  already explored by Atiyah and Bott in their classical works.
Although Donaldson  theory has been extremely successful and solved some of
the longstanding problems  in the field, technical difficulties made this
progress rather hard to pursue. 

On the other hand, Donaldson theory was at the origin of Topological Quantum
Field  Theories, invented by Witten in 1988. Witten formulated  a Quantum
Field Theory starting from $N=2$ supersymmetric Yang-Mills theory in  four
dimensions.  The resulting theory was a kind of exotic coupling to gravity of
the original supersymmetric  theory, and had some astonishing properties: the
physical excitations were projected  out of the Hilbert space and only
remained the vacuum degrees of freedom. The correlation  functions turned out
to be independent of the four-manifold metric and standard path  integral
arguments showed that they were identical to the invariants defined by 
Donaldson. This Topological Quantum Field Theory is now known as
Donaldson-Witten  theory. 

Although this fundamental work established the first deep link between a
supersymmetric  gauge theory and four-manifold topology, it was not clear at
all by that time what  the reasons of this connection were and what new
results could be derived from it.  An important step in the first direction
was made by Atiyah and Jeffrey in 1990. They  showed that Witten's Lagrangian
could be interpreted as the Mathai-Quillen  representative  of the Thom class
of some infinite dimensional vector bundle over the space of  Yang-Mills
connections. 

The situation radically changed in 1994 due to the convergence of some
spectacular  developments both in Donaldson theory and in Quantum Field
Theory: the Kronheimer-Mrowka  structure theorem, the work of Seiberg on
$N=1$ supersymmetry and the rebirth of duality.  The structure theorem of
Kronheimer and Mrowka was a major step towards the understanding  of the
geometric core of Donaldson theory. They were able to show that all the
information  contained in Donaldson invariants, at least for a wide class of
manifolds,
 was encoded in a finite number of two-dimensional cohomology classes, the
basic classes.  Although in their work these classes were conjectured to have
some simple  properties, they were not completely identified by them. This
structure theorem led  Witten to perform the first physical computation of
Donaldson invariants in the K\"ahler  case. Witten showed that on a K\"ahler
manifold it was possible to softly break the 
$N=2$ theory down to an $N=1$ theory in a topologically trivial way from the
point  of view of the twisted theory. This allowed him to compute the
Donaldson invariants  from the non-perturbative behaviour of $N=1$ super
Yang-Mills theory, and to identify  in a precise way the basic classes of
Kronheimer and Mrowka (in the K\"ahler case). 

The crucial step in this story involved the use of electric-magnetic duality
and the  progress in the understanding of $N=1$ supersymmetric gauge theories
made by Seiberg.  Using  these two ingredients, Seiberg and Witten were able
to obtain the exact low-energy  effective action of $N=2$ super Yang-Mills
theory. The alternative description of the infrared  physics of this theory
allowed them to formulate a dual version of Donaldson theory. As  the
effective behaviour of $N=2$ super Yang-Mills can be described at the
relevant  points in the moduli space by an abelian $N=2$ gauge theory coupled
to a magnetic monopole,  the twist of this $N=2$ theory led to a powerful
moduli theory based on  the {\it Seiberg-Witten monopole equations}. This
theory gives a new viewpoint on Donaldson theory and  completely
characterizes the basic classes found by Kronheimer and Mrowka. Moreover, 
the new equations constitute by themselves an extremely useful  tool to
obtain topological information on four-manifolds. 

The possibility of coupling twisted matter hypermultiplets to
Donaldson-Witten  theory was considered  before in the literature, but the
physical an mathematical tools that have been  developed in the last two
years make possible a thorough understanding of this family  of theories. In
this work we make a systematic analysis of the resulting non-abelian 
monopole theory, from many different points of view. Mathematically, these
equations  define a moduli space which is the natural generalization of both
Donaldson theory and  Seiberg-Witten theory, as it contains the moduli spaces
of ASD connections and of  Seiberg-Witten pairs as subspaces. We describe
mathematically this moduli space and  we relate the non-abelian equations to
the twisted $N=2$ Yang-Mills theory coupled to  a matter hypermultiplet. The
relation is carefully established through the Mathai-Quillen  formalism. We
also compute, using both the $N=1$ approach on K\"ahler manifolds and  the
$N=2$ approach, the topological correlation functions of this theory in the
case of  an
$SU(2)$ group. We show that the corresponding topological invariants can be 
expressed in terms of Seiberg-Witten invariants, providing some evidence that
we  should classify the moduli problems associated to four-dimensional
topology in  universality classes. We also develop a generalization of the
Mathai-Quillen formalism  which makes possible to give a clear topological
meaning to theories involving twisted  {\it massive} hypermultiplets. This
generalization also has interesting two-dimensional applications. 

The organization of this work is as follows: the first Chapters are devoted
to a rather  general presentation of the relation between moduli spaces and
Topological  Quantum Field Theories. In Chapter 1 we give an introduction  to
Donaldson theory. Many mathematical tools which will be useful later are
presented  here, as Donaldson theory is still the canonical model of how to
extract topological  information starting from a set of partial differential
equations defined on a  four-manifold. In Chapter 2 we briefly review the
twisting procedure in two and four  dimensions. In Chapter 3 we present the
Mathai-Quillen formalism as well as its  equivariant extension with respect
to vector field actions, and we give an illustrative  example: the
topological sigma model with potentials. After this, we focus on the 
non-abelian monopole theory. In Chapter 4 we study the equations  and the
moduli space they define, from a mathematical point of view. In Chapter 5  we
construct the topological action of the theory using the Mathai-Quillen
formalism and  we show that it comes from the twist of $N=2$ super Yang-Mills
theory  coupled to an $N=2$  hypermutiplet. The equivariant extension gives
precisely the massive theory, providing  in this way the second example of
the construction in Chapter 3. In the remaining Chapters  we analyze the
non-abelian monopole theory from the point of view of the  untwisted physical
theory.  In Chapter 6 we give the $N=1$ point of view, and we first introduce
some techniques  about the non-perturbative behaviour of $N=1$ supersymmetric
gauge theories. As a result,  we compute the topological correlation
functions on a K\"ahler manifold. In Chapter  7 we use the exact solution of
Seiberg and Witten of the $N=2$ theory to compute again  the invariants and
we reanalyze the K\"ahler case from the 
$N=2$ point of view. Finally, we give some Conclusions and prospects  for
future work in the field. An Appendix contains our Euclidean spinor 
conventions as well as some basic facts about 
$\sp$ and $\sp^c$-structures.

\chapter{Donaldson theory}
 A general feature of the topological invariants associated to  Cohomological
Field Theories is that they are defined in terms  of solutions of partial
differential equations on a manifold.  The model for this kind of
construction is Donaldson theory, and in this chapter a short overview of
this  theory is given. This will also be useful to understand moduli
problems  associated to monopole equations. 

The organization of this chapter is as follows: in section 1 we present
general  facts about the geometry of Yang-Mills fields. In section 2 the
anti-self-dual (ASD)
 equations are introduced as well as local characterizations of the moduli
space and  the instanton deformation complex of Atiyah, Hitchin and Singer.
In section  3 we discuss some properties of ASD equations on K\"ahler
manifolds. Finally, in section  4 we briefly introduce the Donaldson
invariants of smooth four-manifolds and  some additional  facts about moduli
spaces.

\section{The geometry of Yang-Mills fields}

Donaldson theory defines differentiable invariants of smooth four-manifolds
starting from  Yang-Mills fields on a vector bundle over the manifold. In
this section we briefly  introduce  the relevant geometrical constructions
associated to gauge theories. Standard references  on connections on bundles
are \cite{kn, gh, chern}. The geometry of gauge theories  is analyzed in
\cite{dv, dk, fu}.

Let $G$ be a Lie group (usually we will take $G=U(1)$ or $SU(2)$).  Let $P
\rightarrow M$ be a principal $G$-bundle over a manifold $M$ with  a
connection $A$, taking values in the Lie algebra of $G$, ${\bf g}$. Given a 
vector space $V$ and a representation $\rho$ of $G$ in $GL(V)$, we can form
an  associated vector bundle
$E=P \times_{G} V$ in the standard way. $G$ acts on $V$  through the
representation
$\rho$. 
 The connection 
$A$ on $P$ induces a connection on the vector bundle $E$ (which we will  also
denote by $A$) and a covariant derivative 
$d_{A}$. Notice that, while the connection $A$ on the principal bundle is an
element  in $\Omega^{1}(P,{\bf g})$, the induced connection on the vector
bundle $E$ is better  understood in terms of a local trivialization
$U_{\alpha}$. On each
$U_{\alpha}$,  the connection 1-form $A_{\alpha}$ is a $gl(V)$ valued one-form
(where $gl(V)$ denotes  the Lie algebra of $GL(V)$) and the transformation
rule which glues together  the different descriptions is given by:
\be A_{\beta}=g_{\alpha \beta}^{-1}A_{\alpha}g_{\alpha \beta}+ g_{\alpha
\beta}^{-1} dg_{\alpha \beta},
\label{trivialgauge}
\ee where $g_{\alpha \beta}$ are the transition functions of $E$. 

Recall that the representation $\rho$ induces a homomorphism of  Lie algebras
$\rho_{*}:{\bf g} \rightarrow gl(V)$ which is a monomorphism if  the
representation is faithful. We will identify $\rho_{*}({\bf g})={\bf g}$,  
and define the adjoint action of $G$ on $\rho_{*}({\bf g})$ through  the
representation $\rho$. On $M$ one can consider the bundle ${\bf g}_E$ defined
by:
\be {\bf g}_E= P \times_{G}{\bf g},
\label{adbundle}
\ee which is a subbundle of ${\rm End}(E)$. For example, for $G=SU(2)$ and $V$
corresponding  to the fundamental representation, ${\bf g}_E$ consists of
(skew) adjoint, trace-free  endomorphisms of $E$. If we look at
(\ref{trivialgauge}) we see that the difference of  two connections is an
element in $\Omega^{1}({\bf g}_E)$ (the one-forms on $M$  with values in the
bundle ${\bf g}_E$). Therefore, we can think about the space of all 
connections ${\cal A}$ as an affine space with tangent space at $A$ given by 
$T_A{\cal A}=\Omega^{1}({\bf g}_E)$.   

The curvature $F_A$ of the vector bundle $E$ associated to the connection $A$
can be also  defined in terms of the local trivialization of $E$. On
$U_{\alpha}$, the curvature 
$F_{\alpha}$ is a $gl(V)$-valued two-form that behaves under a change of 
trivialization as:
\be F_{\beta}=g_{\alpha \beta}^{-1}F_{\alpha}g_{\alpha \beta},
\label{dostrivial}
\ee and this shows that the curvature can be considered as an element in
$\Omega^{2} ({\bf g}_E)$. 

The next geometrical objects we must introduce are gauge transformations. We
come back  momentarily to the principal bundle $P$. A gauge transformation on
$P$ is a  diffeomorphism $f: P \rightarrow P$ verifying two conditions:

1) $f(pg)=f(p)g$, $p \in P$, $g \in G$.

2) $f$ descends, because of (1), to a map ${\hat f}:M \rightarrow M$. The
second  condition is that this map is the identity on $M$. This means that
there is a map 
$\alpha: P \rightarrow G$ such that $f(p)=p \alpha(p)$. Notice that, because
of (1), 
$\alpha(pg)=g^{-1}\alpha(p)g$.

Equivalently, gauge transformations can be considered as sections of the
bundle 
${\rm Aut}(P)=P \times_{G}G$, where $G$ acts on itself by conjugation. If we
return  to the associated vector bundle $E$, it is easy to see that gauge
transformations of
$P$  correspond to automorphisms of $E$, $u: E \rightarrow E$, which
preserve  the fibre structure ({\it i.e.}, they map one fibre onto another)
and descend  to the identity on $M$. For instance, given the gauge
transformation $s$ of $P$,  one obtains the gauge transformation of $E$ given
by $u([p,v])=[s(p),v]$. In the  same way, gauge transformations of $E$ can be
described as sections of the  bundle ${\rm Aut}(E)$.  

Gauge transformations of $P$ (or $E$) form an infinite-dimensional Lie group
${\cal  G}$, where  the group structure is given by pointwise multiplication.
The Lie algebra of 
${\cal G}=\Gamma ({\rm Aut}(E))$ is given by ${\rm Lie}({\cal
G})=\Omega^{0}({\bf g}_E)$.  This can be seen by looking at the local
characterization of the sections
$s: M \rightarrow {\rm Aut}(P)$. On a trivializing open set $U_{\alpha}$ the
gauge  transformation is given by a map $s_{\alpha}:U_{\alpha} \rightarrow
G$.  Using the representation $\rho$, we get in the  same way a local
expression for $u$. It is interesting to notice that ${\rm Aut}(P)$ has a 
trivial subbundle $P \times_{G} C(G)=M \times C(G)$, where $C(G)$ is the
center of  the group $G$. Similarly, the bundle ${\rm Aut}(P)$ is trivial if
$G$ is abelian,  and in this case the group of gauge transformations is given
by ${\rm Map}(M,G)$. 

We can now obtain the action of the group of gauge transformations of $E$ on
the  connection. If we consider the covariant derivative $d_A$ associated to
the connection 
$A$,  and $u \in {\cal G}$, we have
\be d_{u(A)}\sigma =ud_A(u^{-1}\sigma), \,\,\,\,\ \sigma \in \Gamma(E),
\label{gaugecon}
\ee where $u$ is regarded as an automorphism of $E$. As $u$  can be
considered a section of ${\rm End}(E)$, the action of $d_A$ on $u$ is  simply
given by the covariant derivative induced in this bundle by  the connection
on $E$. A local description of this action  can be obtained if we write $d_A
= d+A_{\alpha}$ on
$U_{\alpha}$, and take into account  that $A_{\alpha}$ acts on $u_{\alpha}$
through the adjoint representation,  according to the  previous remark. We
then get the well-known form of gauge transformations:
\be
u^{*}(A_{\alpha})=u_{\alpha}A_{\alpha}u^{-1}_{\alpha}-du_{\alpha}u^{-1}_{\alpha},
\label{ramon}
\ee and on the curvature:
\be u^{*}(F_{\alpha})=u_{\alpha}F_{\alpha}u^{-1}_{\alpha}.
\label{ramondos}
\ee

\section{The moduli space of ASD connections} In this section we will see
that the solutions to a certain equation on a  four-manifold $M$ endowed with
a vector bundle
$E$ form a moduli space of finite  dimension. This is the moduli space of
anti-self-dual connections and is   the main object of Donaldson theory. A
thorough description of this space would take  too much space-time, and then
we content ourselves with a local description which will  be enough for our
purposes. A detailed account can be found in \cite{fu, dk}.  Good
introductory references are
\cite{jicho, nash}.

We suppose now that $M$ is an oriented, compact, Riemannian four-manifold. 
The Riemannian  structure allows us to define the Hodge star operator $*$,
which is related to  the induced metric on the forms by: 
\be
\psi \wedge * \theta= (\psi, \theta) d\mu,
\label{primisima}
\ee where $d\mu$ is the Riemannian volume element. $*$ gives a  splitting of
the two-forms $\Omega^{2}(M)$ in self-dual (SD)  and anti-self-dual (ASD)
forms,  defined as the $\pm 1$ eigenspaces of $*$ and denoted by
$\Omega^{2,+}(M)$ and 
$\Omega^{2,-}(M)$, respectively. This splitting extends in a natural way to 
bundle-valued forms, in particular to the curvature associated to  the
connection
$A$, $F_{A} \in \Omega^{2}({\bf g}_E)$. We call a
 connection ASD if $F_{A}^{+}=0$. It is instructive to consider this
condition  in the case of $M={\bf R}^4$ with the Euclidean metric. If $\{e_0,
e_1, e_2, e_3 \}$  is an oriented orthonormal frame, a basis for SD (ASD)
forms is given by:
\be
\{e_0\wedge e_1 \pm e_2 \wedge e_3, e_0\wedge e_3 \pm e_1 \wedge e_2, 
e_0\wedge e_2
\pm e_3 \wedge e_1 \}.
\label{eformas}
\ee The curvature $F_A$ can then be decomposed as $F_A=F_A^{+} + F_A^{-}$,
where:
\bea F_A^{\pm} &=& {1 \over 2}(F_{01}\pm F_{23})(e_0\wedge e_1 \pm  e_2
\wedge e_3) 
\nonumber\\  &+& {1 \over 2}(F_{03} \pm F_{12})(e_0\wedge e_3 \pm e_1 \wedge
e_2)
\nonumber\\ &+& {1 \over 2}(F_{02}\pm F_{31})(e_0\wedge e_2 \pm e_3 \wedge
e_1) ,
\label{descompone}
\eea and the ASD condition reads:
\bea F_{01}+F_{23}&=&0\nonumber\\  F_{03}+F_{12}&=&0\nonumber\\ 
F_{02}+F_{31}&=&0.
\label{asdeq}
\eea There is a relationship between the ASD condition and the topology of
the  vector bundle $E$. We will restrict ourselves to $SU(2)$ bundles where
$E$ corresponds  to the fundamental representation. Therefore, $E$ will be a
two-dimensional complex  vector bundle. $SU(2)$ bundles over a compact
four-manifold are completely  classified by the second Chern class $c_2(E)$
(for a proof, see
\cite{fu}).  Chern-Weil theory gives a representative of the cohomology class
of
$c_2(E)$ in terms  of the curvature of the connection:
\be c_2(E)=\big[ {1 \over 8 \pi^2} {\rm Tr} F_A^2 \big],
\label{chern}
\ee where $F_A$ is a skew-adjoint, trace-free matrix valued two-form. When
dealing with  ASD connections, the {\it instanton number} $k$
 is better defined as the integral of $c_2(E)$ over $M$. The next step is to 
define Riemannian metrics on the bundles $\Omega^{*}({\bf g}_E)$. If we take
as  the Lie algebra of $SU(2)$ the skew-adjoint, trace-free matrices, which
have the  form:
\be
\xi=\left(\begin{array}{cc}ia& b+ic\\
               -b+ic&-ia \end{array} \right), \,\,\,\,\ a,b, c \in {\bf R},
\label{sudos}
\ee the trace is a negative definite form: 
\be {\rm Tr} \xi^2=-2(a^2+b^2+c^2)= -2|\xi|^2, \,\,\,\,\,\,\ \xi \in {\bf
su}(2).
\label{mimetra}
\ee Therefore we define the Riemannian metric on 
$\Omega^{*}T^{*}M \otimes {\bf g}_E$ using the Riemannian metric on  the forms
(\ref{primisima}) and (\ref{mimetra}):
\be
\langle \psi, \theta \rangle=-{\rm Tr}(\psi,\theta),
\label{primametra}
\ee which is positive definite. It should be noticed that in Field Theory one
usually  defines ${\bf su}(2)$ as the self-adjoint, trace-free matrices, and
therefore  the trace is  a positive definite form. 

Using (\ref{primametra}), we get
\be {\rm Tr}(F_A^2)=-\{|F^+_A|^2-|F^{-}_A|^2\}d\mu,
\label{dona}
\ee and we see that if $A$ is an ASD connection ($F^{+}_A=0$) the instanton
number is  positive. This gives a topological constraint on the existence of
ASD connections. 

The first step to define the moduli space of ASD connections is well known in
Physics.  If $A$ is ASD, then the action of any gauge transformation on $A$
gives another  ASD connection, because of (\ref{ramondos}). One must then
``divide by
${\cal G}$"  in order to obtain a finite dimensional moduli space. We then
consider the map
\be {\cal G} \times {\cal A} \rightarrow {\cal A}
\label{mapa}
\ee and the associated quotient space ${\cal A}/{\cal G}$. The  equivalence
class of connections in this quotient space are denoted by $[A]$.  There are
some requirements on the action of ${\cal G}$ in order to have a
 manifold structure on ${\cal A}/{\cal G}$. The  first obstruction is the
possibility of a non-free action of ${\cal G}$.  We then define the isotropy
group of a connection $A$, $\Gamma_A$, as
\be
\Gamma_A=\{u \in {\cal G}| u(A)=A\}.
\label{isotropia}
\ee Notice from the definition that constant sections of the  trivial
subbundle associated to the center of $G$, $C(G)$, are always in $\Gamma_A$. 
From the description of $u$ as a section of ${\rm Aut}(E)$ and  the action on
$A$ given in  (\ref{ramon}), we see that
\be
\Gamma_A=\{u \in \Gamma({\rm Aut}(E))|d_A u=0\},
\label{otraforma}
\ee {\it i.e.} the isotropy group at $A$ is given by the covariant constant
sections of the  bundle ${\rm Aut}(E)$. It follows that 
$\Gamma_A$ is a Lie group, and its Lie algebra is given by
\be {\rm Lie}(\Gamma_A)=\{f \in \Omega^{0}({\bf g}_E) | d_A f =0\}.
\label{lios}
\ee If $C(G)$ is discrete, as it happens for $SU(2)$, a useful way to detect
if
$\Gamma_A$  is bigger than $C(G)$ (and has positive dimension) is to study the
kernel of $d_A$ in 
$\Omega^0({\bf g}_E)$. 

It is then natural to consider the subset ${\cal A}^{*}$ of connections whose
isotropy  group is minimal:
\be {\cal A}^{*}=\{ A \in {\cal A}|\Gamma_A=C(G) \}.
\label{irre}
\ee Such connections are called {\it irreducible}. From the above
descriptions, and for the  group $G=SU(2)$, there are alternative
descriptions of this definition which turn  out to be very useful (see
\cite{fu}). Reducible connections  correspond to a non-zero kernel of 
\be -d_A: \Omega^{0}({\bf g}_E) \rightarrow \Omega^{1}({\bf g}_E),
\label{conda}
\ee  and they have an isotropy group $\Gamma_A/C(G)=U(1)$. This means,
topologically, that  the bundle $E$ splits as:
\be E \simeq L \oplus L^{-1},
\label{split}
\ee with $L$ a complex line bundle. There is also a topological constraint to
have such a  splitting, because it implies that $c_2(E)=-c_1(L)^2$. As
$SU(2)$ bundles over  compact four-manifolds are completely characterized by
the second Chern class, and in  the same way the first Chern class classifies
line bundles over any manifold, we  see that, for $M$ compact, reductions of
$E$ are in one-to-one correspondence  with cohomology classes $\alpha \in
H^{2}(M;{\bf Z})$ such that
$\alpha^2=-c_2(E)$. 

A natural consequence of all this is that the reduced group of gauge
transformations 
${\hat {\cal G}}={\cal G}/{\bf Z}_2$ acts freely on ${\cal A}^{*}$. We will
restrict  ourselves to this framework when studying the local structure of
the moduli space. 

The most useful way to divide by the action of the gauge group is to consider
slices  of the action of the reduced gauge group ${\hat {\cal G}}$. The
procedure  is simply to consider the derivative of  the map (\ref{mapa}) in
the ${\cal G}$ variable at a point $A \in {\cal A}^{*}$  to obtain
\be  C:{\rm Lie}({\cal G})  \longrightarrow T_A{\cal A},
\label{ope}
\ee which is nothing but (\ref{conda}) (notice the minus sign in $d_A$, which
comes  from the definition of the action in (\ref{ramon})). We have a
principal  bundle structure ${\cal A}^{*} \rightarrow {\cal A}^{*}/{\hat
{\cal G}}$, and  (\ref{ope}) is the map defining fundamental vector fields on
${\cal A}^{*}$.  Using (\ref{primametra}) we can define an $L^2$ metric on
the spaces 
$\Omega^{*}({\bf g}_E)$  by integrating on $M$, and this in turn gives a
formal adjoint operator:
\be d_A^{*}:\Omega^{1}({\bf g}_E) \longrightarrow  \Omega^{0}({\bf g}_E).
\label{ajunto}
\ee Then we can orthogonally decompose the tangent space at $A$ into the
gauge orbit 
${\rm Im} \,\ d_A$ and its complement:
\be
\Omega^{1}({\bf g}_E)={\rm Im} \,\ d_A \oplus {\rm Ker} \,\ d_A^{*}.
\label{descompongo}
\ee  This is precisely the slice of the action we were looking for. Locally,
this means that  the neighbourhood of $[A]$ in ${\cal A}^{*}/{\cal G}$ can
be  modelled by the subspace of $T_A{\cal A}$ given by ${\rm Ker}\,\
d_A^{*}$.  Furthermore, the isotropy  group $\Gamma_A$ has a natural action
on $\Omega^{1}({\bf g}_E)$ given by the adjoint  multiplication, as in
(\ref{ramondos}). If the connection is reducible, the moduli space  is
locally modelled on $({\rm Ker}\,\ d_A^{*})/\Gamma_A$ (see
\cite{dk, fu}). 

We have obtained a local model for the orbit space, but we need to enforce
the  ASD condition to obtain a local model for the moduli space of ASD
connections modulo  gauge transformations, ${\cal  M}_{\rm ASD}$. This model
was introduced by Atiyah, Hitchin and Singer together with  the instanton
deformation complex, and we will closely follow their description 
\cite{ahs}. Let $A$ be an irreducible ASD connection, verifying $F^{+}_A=0$,
and let 
$A+a$ be another ASD connection, where $a \in \Omega^{1}({\bf g}_E)$.  The
condition we get on $a$ starting from $F^{+}_{A+a}=0$ is 
$p^{+}(d_Aa +a\wedge a)=0$, where $p^{+}$ is the projector on the  SD part of
a two-form. To obtain this expression, notice that the curvature  can be
defined as
$F_A=d_Ad_A$,  and the covariant derivative corresponding to $A+a$ is
$d_A+a$.  Thus we get a map:
\be p^{+}F: {\ker}\,\ d_A^{*} \longrightarrow \Omega^{2,+}({\bf g}_E)
\label{pcasi}
\ee given by $p^{+}F(a)=p^{+}(d_Aa +a\wedge a)$. The zero set of this map
describes  the neigbourhood of $[A]$ in ${\cal M}_{\rm ASD}$. To study this
map, one can apply  an infinite-dimensional version of transversality theory.
The first thing to do is  to linearize the map above, to get:
\be p^{+}d_A: \ker \,\ d_A^{*} \longrightarrow \Omega^{2,+}({\bf g}_E).
\label{lineas}
\ee A key point to prove that the moduli space ${\cal M}_{\rm ASD}$ is
finite-dimensional  is to show that the above operator is Fredholm, because by
definition a Fredholm  operator has finite dimensional kernel and cokernel. A
useful way to see this is to  consider the following {\it instanton
deformation complex} or {\it Atiyah-Hitchin-Singer
 (AHS) complex} \cite{ahs}:
\be 0 \too \Omega^{0}({\bf g}_E)
\buildrel d_A \over \too \Omega^{1}({\bf g}_E) 
\buildrel p^{+}d_A \over \too
 \Omega^{2,+} ({\bf g}_E)  \too 0. 
\label{asd}
\ee      This is in fact a complex, because the ASD condition on $A$
guarantees that 
\be p^{+}d_A d_A \phi=[F_A^+,\phi]=0,\,\,\,\,\,\,\,\ \phi \in \Omega^{0}({\bf
g}_E).
\label{complejo}
\ee To see that it is elliptic, we can form the single operator:
\be
\delta_A=d_A^{*} \oplus p^{+}d_A:\Omega^{1}({\bf g}_E) \too \Omega^{0}({\bf
g}_E) 
\oplus \Omega^{2,+} ({\bf g}_E),
\label{elipsis}
\ee which is elliptic. Therefore $\delta_A$ is Fredholm, and we have that 
$\ker \,\ \delta_A=
\ker\,\ d_A^{*} \cap \ker \,\ p^{+}d_A$ is finite-dimensional, as well  as
$\cok \,\
\delta_A = \cok \,\ d_A^{*} \oplus \cok \,\ p^{+}d_A$. But the  first kernel
is just the kernel  of (\ref{lineas}), and from the decomposition
(\ref{descompongo}) and (\ref{complejo})  we see that $\cok \,\ p^{+}d_A
|_{{\rm Ker} \,\ d^{*}_A} =\cok \,\ p^{+}d_A$.  Therefore, (\ref{lineas})  is
also Fredholm, with index:
\be
\ind \,\  p^{+}d_A |_{{\rm Ker} \,\  d^{*}_A}= \ind \,\ \delta_A + \di \,\
\ker \,\ d_A.
\label{relacion}
\ee  Now, if $p^{+}d_A |_{{\rm ker} \,\ d^{*}_A}$ is surjective ({\it i.e}, 
$\cok \,\  p^{+}d_A=0$), and 
$A$ is irreducible,  we can apply the implicit function theorem and get as a
local model for the  moduli space the middle cohomology group of the complex
(\ref{asd}):
\be H^1_A= {\ker \,\ p^{+}d_A \over \im \,\ d_A}, 
\label{primacoho}
\ee which is naturally isomorphic to $ \ker \,\ d_A^{*} \cap \ker  \,\
p^{+}d_A$. This is in  fact the tangent space $T_{[A]}{\cal M}_{\rm ASD}$ to
the moduli space at the  point $[A]$. Irreducible  connections $A$ such that
$\cok \,\ p^{+}d_A=H^2_A=0$ are called  {\it regular} \cite{dk}.  For these
connections the dimension of the moduli space is given by $\ind \,\
\delta_A$,  which is the index of the instanton deformation complex and is
usually called the  {\it virtual} dimension of the moduli space. 

In favourable situations, then, $\ind \,\ \delta_A$ gives $\di \,\ {\cal
M}_{\rm  ASD}$, and can be computed for any gauge group $G$ using the
Atiyah-Singer index  theorem. This is done in \cite{ahs}, and the result for
$SU(N)$ is:
\be
\di \,\ {\cal M}_{\rm  ASD}=4N c_2(E)-{N^2-1 \over 2}(\chi +\sigma),
\label{bunito}
\ee where $\chi$ is the Euler characteristic of $M$ and $\sigma$ its
signature. Recall  that, for a four-manifold, $\chi=2-2b_1(M)+b_2(M)$,
$\sigma=b_2^{+}(M)-b_2^{-}(M)$,  where $b_i(M)$ are the Betti numbers of $M$
and
$b_2^{\pm}$ are the dimensions of 
$H^{2, \pm}(M;{\bf R})$. It follows that $\chi + \sigma =
2-2b_1(M)+2b_2^{+}(M)$. 

The conclusion of this analysis is that, if $A$ is an irreducible,  regular,
ASD connection,  the moduli space in a neighbourhood of this point can be
modelled by the linearization  associated to the instanton deformation
complex and is represented by (\ref{primacoho}).  But one should know what
are the conditions for the moduli space not to contain  reducible,
non-regular ASD connections. Notice that reducibility and regularity depend 
on the metric chosen on the four-manifold $M$, and usually one looks for 
conditions guaranteeing the smoothness of the moduli space for a {\it
generic} metric  on $M$. The meaning of ``generic" can be made more precise
in the context of transversality  theory and the Sard-Smale theorem. It
roughly refers to a dense subset in some space  parametrizing the possible
metrics on $M$. When we say, then, ``for a generic metric",  we mean ``for a
dense subset in the space of metrics". Again, we refer to \cite{dk, fu}  for
more details.           

Concerning reducibility, for $G=SU(2)$, and using the above results on the
classification  of reductions, one sees that reducible $SU(2)$ ASD connections
correspond to  line bundles  whose first Chern class can be represented by an
ASD two-form. It follows, using  transversality theory, that if $b_2^{+}>r$,
ASD $U(1)$ connections do not occur in  generic $r$-dimensional families of
metrics. These guarantees that for $b_2^{+}>0$  there are no reducible
$SU(2)$ ASD connections on a dense subset of metrics. Actually  one needs
more than that, because metric-independence of the Donaldson invariants  can
be guaranteed if one does not meet reducible solutions in a one-parameter
family  of metrics. This is why one requires $b_2^{+}>1$. 

One can also prove with the same techniques that generically $H^2_A=0$
\cite{dk, fu}, and  therefore the moduli spaces are generically smooth when
$b_2^+>1$, because in this case  the two sources of singularities (reductions
and non-regular points) are not present.  However there is an important case
in which one does not deal with generic metrics: this  is the case of $M$
being a K\"ahler manifold. K\"ahler metrics are far  from being generic,  and
then the general results above do not necessarily apply. For these 
manifolds, as we will see, there is a very precise algebro-geometric
description  of the moduli spaces of ASD connections in terms of stable
bundles. Regularity  conditions  do not always hold, but we can still have
smooth moduli spaces (the regularity  condition is  only sufficient) which
will have in general a dimension larger than the virtual one given in
(\ref{bunito}).  This fact has some important consequences for the
computation of the topological  invariants. A possible issue is to perturb
the ASD equations in an appropriate way  in order to obtain a regular,
perturbed moduli space. We will comment on this in section  4. 
\section{The ASD equations on K\"ahler manifolds}

K\"ahler manifolds are an interesting arena in Donaldson theory. They lead to
a  picture of the moduli space of ASD connections which can be formulated in
terms of  algebraic geometry, and explicit computations are often possible.
From the point  of view of Topological Quantum Field Theory, K\"ahler
geometry is also of special  relevance, as we will see in Chapter 6. In this
section we will review some properties  of holomorphic vector bundles over
complex manifolds and state the relation between  ASD connections on compact
K\"ahler manifolds and stable bundles. We will briefly review  the
algebro-geometric construction of local models for the moduli  space and its
relation with the Atiyah-Hitchin-Singer model sketched in the previous 
section. We will also explain the relevant features of symplectic  geometry
which provide a general setting for the equivalence of moduli  spaces that we
will find. The obvious references for this  section are \cite{dk,kobi,fm}.

\subsection{ASD connections and stable holomorphic vector bundles}

We will begin by considering a general situation: the case of $M$ being a 
complex manifold. A complex vector bundle ${\cal E}$ over $M$ is {\it
holomorphic}  if we can find a local trivialization such that the transition
functions are holomorphic.  In that case, one can define a linear operator
for differential forms with values  in the bundle ${\cal E}$:
\be {\overline \partial}_{\cal E}: \Omega^{0, p}({\cal E})
 \too \Omega^{0, p+1}({\cal E}),
\label{miope}
\ee which satisfies the Leibniz rule and gives zero acting on holomorphic
sections of ${\cal  E}$. It is also nilpotent, and this allows us to define a
Dolbeault cohomology:
\be H^{p}({\cal E})= {\ker \,\ {\overline \partial}_{\cal E}
 \over \im \,\ {\overline \partial}_{\cal E} }.
\label{volvo}
\ee A complex vector bundle $E$ is not neccesarily holomorphic. The
introduction of the  operator ${\overline \partial}_{\cal E}$ allows one to
formulate in  differential-geometric terms a necessary and  sufficient
condition for $E$ to be holomorphic. To accomplish this it is useful to 
introduce the concept of {\it partial connection} \cite{dk}. A partial
connection  on a complex vector bundle $E$ is an operator 
\be {\overline \partial}_{\alpha}: \Omega^{0}(E) \too \Omega^{0,1}(E)
\label{parcial}
\ee satisfying the Leibniz rule. In a local trivialization, the partial
connection can be  expressed as ${\overline \partial}_{\alpha}={\overline
\partial}+ \alpha$,  where $\alpha$ is a $(0,1)$-form. It is easy to check
that, if one can find a local  trivialization of $E$ in which $\alpha=0$ on
every trivializing patch, then the  corresponding transition functions are
holomorphic functions and  the bundle is holomorphic. If this is the case,
the partial connection is  called {\it integrable}.  Notice that this notion
of integrability is a holomorphic analogue of flatness.  In the same way, the
necessary and sufficient condition  for ${\overline
\partial}_{\alpha}$ to be integrable is 
${\overline \partial}_{\alpha}^2=0$. 

It is clear that a connection $A$ on a complex vector bundle $E$ defines a
partial  connection: due to the bigrading of differential forms on a complex
manifold, we  can decompose the covariant derivative as $d_A=\partial_A
+{\overline \partial}_A$,  and the antiholomorphic part gives the partial
connection associated  to $A$. The integrability condition reads in this case:
\be {\overline \partial}_A^2=F_A^{0,2}=0.
\label{otia}
\ee We will be interested in unitary connections, ${\it i.e.}$, we will work
with Hermitian  vector bundles $(E,h)$ ($h$ is the Hermitian metric)  and
connections compatible with the Hermitian structure. If a  connection is
unitary, the connection and curvature matrix are skew-adjoint. The main 
result we get is the following: a unitary connection on a  Hermitian complex
vector bundle $(E,h)$ defines a holomorphic structure if and only if  its
curvature is of type $(1,1)$ (because
$F_A^{0,2}=-(F_A^{2,0})^{\dagger}$=0).  Conversely, given a holomorphic,
Hermitian vector bundle $({\cal E},h)$,  there is a unique unitary connection
compatible with both the holomorphic and  the Hermitian  structure. 

It seems that the moduli space of holomorphic structures on a complex,
Hermitian  vector bundle $(E,h)$ would be parametrized by the space of 
unitary connections with curvature of type $(1,1)$, which will be denoted  by
${\cal A}^{1,1}$. However, two different connections $A_1$, $A_2 \in  {\cal
A}^{1,1}$ can give {\it isomorphic} holomorphic structures. Thus we  should
define some equivalence relation in ${\cal A}^{1,1}$ in order to recover in a
 proper way the moduli space of holomorphic structures of $E$. To this end we
must  consider the complex gauge group of general linear automorphisms of
$E$, 
${\cal G}^{\bf C}$. This group is the complexification of the group of gauge 
transformations introduced in section 1. Notice that, if $g \in {\cal G}$
takes values  (in a local trivialization) in $SU(2)$, the elements in the
complexified gauge group  take values in $Sl(2, {\bf C})$. The equivalence
relation in ${\cal A}^{1,1}$ is  associated to the action of ${\cal G}^{\bf
C}$ on the unitary connections, which is  defined as follows. Set ${\hat
g}=(g^{\dagger})^{-1}$. Then, if $g \in {\cal G}^{\bf C}$,  the covariant
derivative transforms under $g$ as:
\bea {\overline \partial}_{g(A)}&=&g{\overline \partial}_{A}g^{-1},
\nonumber\\ {\partial}_{g(A)}&=&{\hat g}{\partial}_{A}{\hat g}^{-1}. 
\label{comple}
\eea If $g \in {\cal G}$, then $g={\hat g}$, and we recover precisely
(\ref{gaugecon}). Notice  that the action of ${\cal G}^{\bf C}$ preserves the
space
${\cal A}^{1,1}$. We say that 
$A_1$, $A_2 \in  {\cal A}^{1,1}$ are equivalent (in the sense that they give
isomorphic holomorphic  structures) if there exists a $g\in {\cal G}^{\bf C}$
such that 
\be  A_2=g(A_1).
\label{equix}
\ee Our final conclusion is that the moduli space of holomorphic structures
on a complex,  Hermitian   vector bundle $(E,h)$, can be identified with the
quotient
\be {\cal A}^{1,1} / {\cal G}^{\bf C}.
\label{modi}
\ee

What is the relevance of this discussion to the study of ASD  connections on
four-manifolds? Suppose that our four-manifold $M$ is a complex surface 
endowed with a Hermitian metric $g$. Associated to this metric, there is a
real 
$(1,1)$ form 
$\omega$, called the {\it K\"ahler form}, with the following expression in
local  coordinates:
\be
\omega = i\sum_{ij}g_{i {\overline j}}dz^i d{\overline z}^j.
\label{oformilla}
\ee
$M$ is K\"ahler if $\omega$ is closed. The interesting thing is that SD
forms  on
$M$ (more precisely, the complexified SD forms) can be related to the
bigraded  decomposition of differential forms associated to the complex
structure. We have the  following equality:
\be
\Omega_{\bf C}^{2,+}(M)=\Omega^{2,0}(M) \oplus \Omega^{0}(M)\cdot \omega 
\oplus \Omega^{0,2}(M),
\label{desco}
\ee where $\Omega^{0}(M)\cdot \omega$ denotes multiples of the $(1,1)$
K\"ahler form.  Correspondingly, ASD forms are given by $(1,1)$-forms
pointwise orthogonal to 
$\omega$ in the  induced metric. If we consider the operator $\Lambda$, the
adjoint to the operator  given by wedge product by $\omega$, we can write the
$(1,1)$ part  of the curvature as
\be F^{1,1}_A={1\over 2}\Lambda F_A \omega + D,
\label{toma}
\ee where $D$ is pointwise orthogonal to $\omega$ (the $1/2$ factor comes
from the  fact that $(\omega, \omega)=n$ in a complex manifold of dimension
$n$).  

We can now marry the decomposition given in (\ref{desco}) with the above
results for  the existence of holomorphic structures on $E$. It is clear that
if $A$ is an ASD  connection  on $(E,h)$, over a Hermitian complex surface
$M$, then $F_A^{0,2}=0$ and there is a  holomorphic structure on $E$
associated to $A$. Conversely, suppose that 
$({\cal E},h)$ is a  holomorphic, Hermitian vector bundle over $M$. A unitary
connection $A$  compatible with the  holomorphic  structure ({\it i.e}, 
${\overline \partial}_A={\overline \partial}_{\cal E}$) is ASD if and only if 
$\Lambda F_A=0$. 

The previous results give a new point of view on the solutions to the ASD
equation:  a solution will be given by a holomorphic structure on the
Hermitian vector bundle
$E$, 
$({\cal E},h)$, such that the associated unitary connection verifies $\Lambda
F_A=0$.  More precisely, recall that a single holomorphic structure is
associated to different  connections in ${\cal A}^{1,1}$ related by complex
gauge transformations, {\it i.e},  to a ${\cal G}^{\bf C}$ orbit. It would be
extremely useful to have a criterium for  the holomorphic vector bundle 
$({\cal E}, h)$ to admit a solution to this equation for (at least) one of 
the connections in the orbit. Remarkably, when $M$ is a compact  K\"ahler
surface, such a criterium exists and is of an  algebro-geometric nature: it
is equivalent to the {\it stability} of the holomorphic  vector bundle ${\cal
E}$. The concept of stability has played a major r\^ole in the  study of
moduli problems over K\"ahler manifolds (see for instance \cite{abone,
jichi,  brad}). We will then precise what a stable holomorphic bundle is and
state the final  result in relation to Donaldson theory. 

Let $M$ be a compact K\"ahler manifold of dimension $n$, and let ${\cal E}$
be a  holomorphic vector bundle over $M$. The ${\it degree}$ of ${\cal E}$ is
defined as:
\be {\rm deg}({\cal E})=\int_M c_1(E) \wedge {\omega ^{n-1} \over (n-1)!},
\label{degri}
\ee and the {\it slope} of ${\cal E}$ as:
\be
\mu({\cal E})={{\rm  deg}({\cal E}) \over {\rm rk}({\cal E})}.
\label{pende}
\ee A holomorphic $Sl(2,{\bf C})$ bundle ${\cal E}$ is ${\it stable}$ if, 
for each holomorphic line bundle ${\cal L}$ for which  there is a non-trivial
holomorphic map
${\cal E} \rightarrow {\cal L}$, we have:
\be {\rm deg}({\cal L}) > 0.
\label{estable}
\ee This is the definition of stability given in \cite{dk}, and  is
sufficient to our expository purposes, but we point out  that in the general
definition one has to consider subsheaves of ${\cal E}$.  See \cite{kobi} for
more details. 

The main results on the existence of solutions to the ASD equations on
K\"ahler
 surfaces and the relation to the stability condition were obtained by 
Donaldson in
\cite{donsta}.  Before we asked for a criterium for a holomorphic bundle
${\cal E}$ to admit a solution  of the equation $\Lambda F_{A }=0$ in its
orbit. Donaldson's theorem gives a complete  answer to this question, in the
case of $SU(2)$ connections. First of all,  any ${\cal G}^{\bf C}$ orbit
contains at most  one
${\cal G}$ orbit of solutions (${\it i.e.}$, one solution modulo  gauge
transformations).  This will be the case if the holomorphic bundle is stable 
(notice that the stability  condition is of algebro-geometric nature and is
preserved by the action of 
${\cal G}^{\bf C}$) or if it splits holomorphically as ${\cal E}=L \oplus 
L^{-1}$, where $L$ is a holomorphic line bundle of ${\rm deg}(L)=0$. In the
first case,  the solution corresponds to an irreducible ASD connection, and
in the second case to  a reducible one. In this way, for $E$ an $SU(2)$
bundle over a K\"ahler  compact surface $M$, we can identify the moduli space
of irreducible ASD  connections with the set of equivalence classes of stable
holomorphic bundles ${\cal E}$  which are topologically equivalent to $E$. 

The above equivalence is just a equivalence at the level of sets. We would
like now  to precise these results by considering the local model for
deformation of holomorphic  structures on vector bundles and comparing the
resulting picture with the one provided  by the AHS deformation complex
\cite{dk, fm}. Let $E$ be a complex vector bundle and  let ${\cal E}$ be a
holomorphic structure on $E$, characterized by the operator 
${\overline \partial}_{\cal E}$. Let us consider a one-parameter family  of
partial connections,   given by
\be {\overline \partial}_t={\overline \partial}_ {\cal E}+\alpha_t,
\,\,\,\,\,\,\,\,\ 
\alpha_t \in \Omega^{0,1}({\rm End}_0({\cal E})),
\label{compledef}
\ee where $\alpha_0=0$ and ${\rm End}_0({\cal E})$ denotes the trace-free
endomorphisms
 of ${\cal E}$. The condition for this new connection to be in ${\cal
A}^{1,1}$ is  easily computed:
\be {\overline \partial}_ {\cal E}\alpha_t+\alpha_t \wedge \alpha_t=0.
\label{ferni}
\ee At first order in $t$, {\it i.e.}, in the linear approximation,
(\ref{ferni}) reads
\be {\overline \partial}_ {\cal E}\alpha=0, \,\,\,\,\,\,\
\alpha=\Big({\partial
\alpha_t 
\over \partial t}\Big)_{t=0}.
\label{primodef}
\ee In this case we must quotient by the group ${\cal G}^{\bf C}$,  and we
must get rid of the linearized  deformations coming from one-parameter
families of complex gauge transformations. The  gauge orbits correspond in
this case to the gauge orbits induced by elements in the  Lie algebra of
${\cal G}^{\bf C}$, $\Omega^0({\rm End}_0({\cal E}))$:
\be
\partial_t=g_t{\overline \partial}_ {\cal E}g_t^{-1},
\label{orbital}
\ee and the linearized deformations induced by the complex gauge
transformations are given  by:
\be
\beta=-{\overline \partial}_ {\cal E}g, \,\,\,\,\,\,\,\,\,\ g= \Big({\partial
g_t 
\over \partial t}\Big)_{t=0} \in \Omega^0({\rm End}_0({\cal E})).
\label{ultimo}
\ee We have then obtained the {\it Kodaira-Spencer complex}:
\be 0 \too \Omega^0({\rm End}_0({\cal E}))\buildrel {\overline \partial}_
{\cal E}
 \over \too \Omega^{0,1}({\rm End}_0({\cal E})) 
\buildrel {\overline \partial}_ {\cal E} \over \too
 \Omega^{0,2} ({\rm End}_0({\cal E}))  \too 0.
\label{kscomplex}
\ee In this case, the cohomology of the complex is given by the Dolbeault or
sheaf  cohomology groups, and the tangent space is then modelled by
\be  H^{0,1}({\rm End}_0({\cal E}))={ \ker \,\  {\overline \partial}_ {\cal
E} \over 
\im \,\ {\overline \partial}_ {\cal E}}.
\label{dolbo}
\ee If we want to compare (\ref{kscomplex}) with  (\ref{asd}), we first
implement the
 natural  isomorphisms 
\bea
\Omega^{1}({\bf g}_E) &\simeq & \Omega^{0,1}({\rm End}_0({\cal E})),
\nonumber \\
\Omega^{0}({\bf g}_E)\oplus \Omega^{2,+}({\bf g}_E) &\simeq& 
\Omega^{0}({\rm End}_0({\cal E})) \oplus \Omega^{0,2}({\rm End}_0({\cal E})).
\eea The holomorphic structure ${\overline \partial}_ {\cal E}$ can be
identified with a  connection $A \in {\cal A}^{1,1}$ such that ${\overline
\partial}_ {\cal E}= {\overline \partial}_ A$, and (\ref{kscomplex}) is the
written as 
\be {\overline \partial}^{\dagger}_ A\oplus {\overline \partial}_ A:
\Omega^{0,1}({\rm End}_0({\cal E})) \too \Omega^{0}({\rm End}_0({\cal E})) 
\oplus \Omega^{0,2} ({\rm End}_0({\cal E})).
\label{otraelipsis}
\ee Using the K\"ahler identities for the Dolbeault operators  \cite{gh},
\be [\Lambda, {\overline \partial}]=-i\partial^*, \,\,\,\,\,\,\,\  [\Lambda,
\partial ] = i{\overline \partial}^*,
\label{kalidem}
\ee
 we see that (\ref{otraelipsis}) can be identified to  (\ref{elipsis}). In
particular we can identify the resulting local  models for the moduli space,
(\ref{dolbo}) and (\ref{primacoho}). For more details  on this equivalence,
see
\cite{dk} and specially \cite{fm}.

\subsection{ASD connections and symplectic geometry} The equivalence above
can be analyzed in a general setting using techniques from  symplectic 
geometry. The main point is to realize the original moduli space as a
symplectic  quotient, and then one can obtain the desired identification on
general grounds. This is  also useful to introduce symplectic and K\"ahler
structures on the moduli space, through  the Marsden-Weinstein theorem. This
elegant framework  was introduced by Atiyah and Bott in their study of the
Yang-Mills equations over  Riemann surfaces \cite{abone} (which can be seen
as a two-dimensional analogue  of Donaldson theory) and also applies to other
moduli problems on K\"ahler manifolds,  as we will see later.  

The basic ingredient of symplectic geometry that arises in this setting is
the {\it  moment map} (for a detailed discussion, see \cite{gs, wei}).
Suppose that 
$(M, \Omega)$ is a symplectic manifold and that there is a  Lie group action
on $M$ preserving the symplectic form. This action defines a map  from the
Lie algebra of
$G$, ${\bf g}$ to the symplectic vector fields on $M$:
\be C: {\bf g} \too {\rm Sym}(M), \,\,\,\,\,\ {\cal L}(C(\xi))\Omega =0,\,\,\ 
\xi \in {\bf g}
\label{symple}
\ee where ${\cal L}(X)$ denotes the Lie derivative with respect to a vector
field 
$X$. From the basic homotopy identity for the Lie derivative and the fact
that 
$\Omega$ is closed one obtains that the one-form
\be
\iota (C(\xi))\Omega
\label{miforma}
\ee is closed, where $\iota(X)$ denotes as usual the inner product with the
vector field $X$.  One can wonder if the one-form (\ref{miforma}) is exact,
and this  leads to the introduction of  Hamiltonian structures and more
generally to the introduction of a moment map. A moment  map is a map
\be
\mu: M \too {\bf g}^{*}
\label{momento}
\ee such that
\be 
\langle d \mu _{x}(v), \xi \rangle =\Omega(v, C(\xi)), \,\,\,\,\,\  v \in
T_{x}M,\,\,\ \xi \in {\bf g},
\label{mome}
\ee where $\langle \,\ , \,\ \rangle $ denotes here the dual pairing between
${\bf g}$ and 
${\bf g}^{*}$. If a moment map exists, then $\iota (C(\xi))\Omega$ is
obviously exact.  The case of interest for us is when the moment map not only
exists, but it is also  equivariant with respect to the action of the group
$G$. The action of 
$G$ on ${\bf g}^{*}$ is  given by the coadjoint action:
\bea  G \times {\bf g}^{*} &  \rightarrow &  {\bf g}^{*}\nonumber \\ (g,
\lambda) & 
\mapsto & ({\rm ad }_g^{*} \lambda)(\xi)=\langle \lambda, {\rm ad }_g(\xi)
 \rangle 
\,\,\,\,\,\,\,\,\ \xi \in {\bf g},
\label{coaju}
\eea and equivariance means 
\be
\mu(gx)=({\rm ad }_g^{*}) \mu(x), \,\,\,\,\,\ x \in M,\,\,\,\ g \in G.
\label{twins}
\ee If (\ref{twins}) holds, $\mu^{-1}(0)$ is an invariant subset for the $G$
action and one  can consider the {\it symplectic quotient} or reduced phase
space
\be W=\mu^{-1}(0)/G.
\label{wein}
\ee If $0 \in {\bf g}^{*}$ is a regular value for $\mu$ and $G$ acts freely 
on
$\mu^{-1}(0)$, then $W$ is a  smooth manifold, and the Mardsen-Weinstein
reduction endows it with  a symplectic structure  inherited from $M$. 

In the Yang-Mills framework, as usual, we will deal with  infinite-dimensional
generalizations of this situation. As an important example,  let $(M,
\omega)$ be a compact symplectic manifold of dimension $2n$. In the  space of
connections ${\cal A}$ on a unitary bundle $E$  we can define an induced
symplectic form  given by
\be
\Omega_{\cal A} (\psi, \theta)=\int_M \tr (\psi \wedge \theta) \wedge
\omega^{n-1},
\label{rayo}
\ee where $\psi$, $\theta \in \Omega^{1}({\bf g}_E)$ are tangent vectors to
${\cal A}$. The  group action corresponds to the gauge transformations ${\cal
G}$, and preserves  (\ref{rayo}). We use the natural inner product on
Lie$({\cal G})$ to identify 
${\rm Lie}({\cal G})^{*}= \Omega^{2n}({\bf g}_E)$. We can construct a moment
map  for this action given by 
\be
\mu(A)=F_A \wedge \omega^{n-1}.
\label{mediolimon}
\ee This is easily checked as follows. The differential of (\ref{mediolimon})
is 
\be d\mu_A(\psi)=d_A\psi \wedge \omega^{n-1}
\label{difer}
\ee and, after integrating by parts, one obtains
\be
\langle d\mu_A(\psi), \phi \rangle=-\int_M \tr(\psi d_A \phi)\wedge
\omega^{n-1}= 
\Omega(\psi, C(\phi)),
\label{cal}
\ee which is precisely (\ref{mome}). Equivariance of $\mu$ can also be easily
checked. 

Finally we consider the relevant situation to Donaldson theory. Suppose $(M,
\omega)$ is  now a K\"ahler manifold, therefore symplectic with symplectic
form
$\omega$. As before,  assume that a group $G$ acts on $M$ isometrically and
then preserving the symplectic  form. As $M$ is complex, there is a natural
extension of the action of $G$ to the  complexification $G^{\bf C}$. In this
context there is a natural identification between  the symplectic quotient of
$M$ by $G$ and the space of orbits of $G^{\bf C}$ in $M$. An  orbit of the
complexified group is called {\it stable} if it contains a  zero of the
moment map and its points have no continuous isotropy groups. Stable points 
are those lying on stable orbits, and this subset is denoted as $M_S$. Then, 
the complex quotient $M_S/ G^{\bf C}$ can be identified with the symplectic 
quotient $W^{*}$ of $M^{*}$ by $G$, where $M^{*}$ denotes the subset of
points  with no continuous isotropy groups under the $G$ action. 

We will see now that this general result encompass the conclusions about the
moduli  space of ASD connections in the case of a K\"ahler manifold. We have
seen that the  space of connections on a unitary bundle $E$ over a symplectic
manifold $M$ inherits a  natural symplectic structure. The same phenomenon
occurs when $M$ is K\"ahler.  The complex  structure on ${\cal A}$ is
obtained by identifying the holomophic tangent vectors as  the elements in
$\Omega^{0,1}({\bf g}_E)$. The Hermitian metric on $M$  also gives ${\cal A}$
a Hermitian structure in such a way that ${\cal A}$  becomes a flat K\"ahler
manifold, with the K\"ahler form given by the  expression in (\ref{rayo}).
The action of the gauge group is by isometries, and  of course its
complexification with respect to the complex structure introduced in 
${\cal A}$ is the complexified gauge group ${\cal G}^{\bf C}$ (see
\cite{kobi}  for details). 

In the study of holomorphic bundles, we should restrict ourselves to ${\cal
A}^{1,1}$,  and the above remarks also hold. The moment map in this case has
the form:
\be 
\mu(A)=F_A\wedge \omega^{n-1}={1 \over  n}\Lambda F_A \omega^n,
\label{moka}
\ee which is the $n$-dimensional version of (\ref{toma}). We see that, for
four-dimensional  manifolds, the zeros of the  moment map are the ASD
connections, and the symplectic quotient $W^{*}$  is just the moduli space 
of irreducible ASD connections. The identification obtained with  symplectic
techniques is precisely  the one we discussed in the previous subsection. In
fact Donaldson's theorem 
\cite{donsta} also identifies the stability concept arising in algebraic
geometry  and stability in the sense of gradient flows, as it appears in the
framework of  the orbits of the complexified action. These equivalences,
exploited in the seminal  work of Atiyah and Bott \cite{abone}, also appear
in  other moduli problems, in particular in the context of vortex or 
monopole equations \cite{gpdos, mfm, lmna, okone, okotwo}. Another example 
will be studied in Chapter 4.

\section{Donaldson invariants}

One of the main goals of Donaldson theory is to define topological invariants
starting  from the solutions of a certain differential equation on a
manifold. This is a  well-known problem in topology, and the most familiar
situation of this kind is the  case of a vector field $X$ on a manifold $M$.
The zero locus of this vector  field is generically  a set of discrete
points, and to each point one can associate  (if the manifold is orientable)
the index of the vector field at this point, 
$\ind (X,p)=
\pm 1$. The sum of the indices is, by the Poincar\'e-Hopf theorem, a
topological  invariant of $M$, namely the Euler characteristic $\chi(M)$.
This example is actually  a special case of a general situation which one can
generalize to an infinite-dimensional  setting and provides a construction of
Donaldson invariants. It is precisely this special   construction the natural
one arising in Topological Quantum Field Theory, through  the Mathai-Quillen
construction, as we will see in chapter 3. 

We will reformulate Donaldson theory in terms of Fredholm differential
topology. Let 
${\cal A}^{*}$ the space of irreducible connections. With the action of 
${\hat {\cal G}}={\cal G}/C(G)$, 
${\cal A}^{*}$ becomes a principal bundle over the space of irreducible
connections  modulo gauge transformations, ${\cal B}^*$:
\be
\begin{array}{ccc}{\hat {\cal G}}& \too & {\cal A}^{*} \\
                   {}&{}& \downarrow \\
                    {}&{}&{\cal B}^* \end{array}
\label{irrefibra}
\ee The group ${\hat {\cal G}}$ also acts on the vector space 
$\Omega^{2,+}({\bf g}_E)$, and we can construct an associated vector bundle
\be {\cal E}={\cal A}^{*} \times_{{\hat {\cal G}}}\Omega^{2,+}({\bf g}_E).
\label{mivector}
\ee over ${\cal B}^*$. The ASD connection can be regarded as a 
${\cal G}/C(G)$-equivariant map 
\bea s: {\cal A}^{*} &\too &\Omega^{2,+}({\bf g}_E)\nonumber\\ A & \mapsto&
s(A)=F_A^{+},
\label{secciona}
\eea and this induces a section ${\hat s}: {\cal B}^* \rightarrow {\cal E}$
of the  associated vector bundle given in (\ref{mivector}). Notice that the
zero locus  of
${\hat s}$ is precisely the moduli space of irreducible ASD connections,
${\cal M}_ {\rm ASD}$. Most of our previous remarks on the construction of
the moduli space  translate  to this framework. In particular, ${\hat s}$ is
Fredholm, and for a generic metric  on the four-manifold $M$, the zero locus
of ${\hat s}$ is regular, its dimension is  given by the index of
(\ref{elipsis}) and equals (\ref{bunito}). 

Before pursuing this line of thought, let us consider a different question. 
The Donaldson  invariants are  roughly defined in terms of integrals of
differential forms in the  moduli space of irreducible ASD connections. These
differential forms come from the  rational cohomology ring of ${\cal B}^*$,
and it is necessary to have an explicit  description of this ring. The
construction involves the {\it universal bundle} or  {\it universal
instanton} associated to this moduli problem. Let $P$ be the principal 
$G$-bundle  over $M$ associated to $E$, and consider the space ${\cal A}^{*}
\times P$. This  space can be consider as a pullback bundle:
\be
\begin{array}{ccc} \pi_2^*(P)={\cal A}^{*} \times P& {} & P\\
\downarrow &{}& \downarrow \\ {\cal A}^{*} \times M& \buildrel \pi_2 \over
\longrightarrow & M \end{array}
\label{familia}
\ee  In this construction, the space ${\cal A}^{*} \times P$ is called a  {\it
family of tautological connections}. In fact, the natural conection on 
${\cal A}^{*} \times P$ is tautological in the $P$ direction and trivial  in
the
${\cal A}^*$ direction: at the point $(A,p)$, the connection is  given by
$A(p)$. 

On the other hand, the group  of gauge transformations ${\hat {\cal G}}$ acts
on both factors, and the quotient 
\be {\bf P}={\cal A}^* \times_{{\hat {\cal G}}} P
\label{univ}
\ee is a $G/C(G)$-bundle over ${\cal B}^*\times M$. This is the {\it universal
bundle}  associated to $P$. 

The next step is to construct a  connection on the universal bundle and
compute its curvature. This was done in 
\cite{as} in relation to anomalies in Quantum Field Theory. To define the
connection  on the universal bundle, we need a connection on the principal
bundle  ${\cal A}^{*} 
\rightarrow {\cal B}^*$. We define this connection through a horizontal
distribution  on ${\cal A}^*$. The procedure is very general and has a key
r\^ole in the  construction of Topological Gauge Theories in the
Mathai-Quillen  formalism
\cite{aj, cmr}.  We will  describe it here in general. Let $P$ be a principal
$G$-bundle  endowed with a Riemannian metric $g$ on $P$ which is
$G$-invariant. We use this  metric to define the connection
 on $P$, by declaring the horizontal subspace to be the orthogonal complement
to the vertical
 one. More explicitly, one starts from the map defining fundamental vector
fields on
$P$:
 \be C_p= R_{p{*}}:{\bf g} \rightarrow T_pP.
 \label{funda}
 \ee
 Consider now the following differential form on $P$ with  values in ${\bf
g}^{*}$:
 \be {\tilde \nu}_p(Y_p, A)=g_p(R_{p{*}}A, Y_p),\,\,\,\,\ Y_p \in T_pP,\,\,\
A \in {\bf g}.
\label{bautizo}
 \ee We can use the Killing form on ${\bf g}$ to obtain a one-form on $P$
with  values in ${\bf g}$, denoted by $\nu$. Notice that
$\nu_p=C^{\dagger}_{p}$, the adjoint  of $C_p$, which is defined by the
metric on $P$ together with the Killing form  on ${\bf g}$. If
$R=C^{\dagger}C$, the connection one-form is defined by:
\be
 \theta=R^{-1}\nu.
\label{conecta}
\ee The curvature of this connection can be easily computed on horizontal
vectors.  In this case, it is simply given by:
\be K=d\theta=R^{-1}d\nu,
\label{curvatriz}
\ee as the other terms vanish on the horizontal subspace \cite{aj}.  
  
In the case we are considering, $P={\cal A}^*$, we have seen that the
Riemannian metric  on $M$ endows the space of differential forms with values
in ${\bf g}_E$ with the  metric (\ref{primametra}). The fundamental vector
fields are the tangents to the gauge  orbits, and are given at $A\in {\cal
A}^{*}$ by $\im \,\ d_A$, as it follows from  (\ref{ope}) and (\ref{conda}).
The horizontal vectors are obtained according to the  othogonal decomposition
in (\ref{descompongo}), {\it i.e.}, they are the one-forms  with values in
${\bf g}_E$ in the kernel of $d^{*}_A$. In this case, the operator 
$C^{\dagger}$ is given by $-d^*_A$, $R$ is the laplacian $d_A^*d_A$, and its 
inverse is the  Green function $G_A$ associated to this laplacian. The
connection one-form is then, at  the point $A$:
\be
\theta_A =-G_Ad^*_A.
\label{doncon}
\ee The curvature can be obtained explicitly evaluating $d\nu$ on two
horizontal vectors 
\cite{dk}, but we don't need that expression. 

Once we have obtained a connection on the principal bundle ${\cal A}^*$, we
can obtain   a connection on the universal bundle ${\bf P}$. Its curvature
has two pieces. One  of them comes from the tautological connection on ${\cal
A}^*\times P$, and the other  comes from the connection that we have just
defined on ${\cal A}^*$. The total curvature  is a form in $\Omega^2({\cal
A}^* \times M,{\bf g}_P)$, and splits according to the  bigrading of
$\Omega^*({\cal A}^* \times M)$. Adding both pieces,  one gets at $([A],x)$
\cite{as,dk}:
\bea K_{\bf P}(X,Y)&=&F_A(X,Y), \nonumber\\ K_{\bf P}(a,X)&=&a(X), \nonumber\\
K_{\bf P}(a,b)&=&G_A d\nu(a,b),
\label{todacurva}
\eea where $X$, $Y \in T_xM$, and $a$, $b  \in \Omega^1({\bf g}_P)$ are
 horizontal with respect to the distribution in ${\cal A}^*$. The  first and
last components come from the tautological connection and  the horizontal
distribution, respectively. The second component is the  horizontal
projection of the ``mixed" contribution in the tautological connection. 

In the case of $G=SU(2)$, the universal bundle ${\bf P}$ is a $SU(2)/{\bf
Z}_2=SO(3)$  bundle due to the non-triviality of the center. Using Chern-Weil
theory, we can  represent  the Pontriagin class $-p_1({\bf P})/4$ by the
four-form
\be {1 \over 8 \pi^2}\tr(K_{\bf P} \wedge K_{\bf P}).
\label{ponti}
\ee In this Pontriagin class, there is a piece in $H^4({\cal B}^*)$ (taking
only the last  component in (\ref{todacurva})). This gives a four-form on the
space of irreducible  connections modulo gauge transformations:
\be {\cal O}={1 \over 8 \pi^2}\tr (G_A d\nu \wedge G_A d\nu).
\label{clasefour}
\ee This form is originally defined on ${\cal A}^*$, but it is gauge 
invariant and its horizontal  part projects to a form on ${\cal B}^*$. There
is also a piece in (\ref{ponti})  which is in $H^2({\cal B}^*)\otimes
H^2(M)$. We can then take the slant product of  this piece with a homology
class in $H_{2}(M)$, $[\Sigma]$. This gives a two-form on 
${\cal B}^*$ for every $[\Sigma]$, with the explicit expression:
\be I(\Sigma)_{[A]}(a,b) ={1 \over 4 \pi^2}\int_{\Sigma}\tr \Big(G_A
d\nu(a,b)F_A+ {1\over 2}(a\wedge b) \Big),
\label{dosclase}
\ee where again $a$, $b \in \Omega^1({\bf g}_P)$ are horizontal. With this
construction  we have solved the problem of how to obtain the cohomology ring
of
${\cal B}^*$, because  one can prove \cite{dk} that, for $SU(2)$ connections
over a simply-connected  four-manifold, this ring is generated by the
cohomology classes  given in (\ref{clasefour}) and (\ref{dosclase}) for every
$[\Sigma]$ in $H_2(M)$.  This mathematical construction appears in
Topological Gauge Theories as the  {\it descent  procedure}. This is because
the BRST cohomology of these theories is essentially the  equivariant
cohomology (with respect to the gauge group ${\cal G}$) of the space of 
connections. The cohomology classes which are found (and called there {\it
observables})
 are  essentially the generators of the cohomology of ${\cal B}^*$, and in
this way one  finally obtains (\ref{clasefour}) and (\ref{dosclase}).
 
We come back to the question of the definition of invariants. In
finite-dimensional  situations, the zero locus of a transversal section of a
vector  bundle $E$ is a smooth submanifold of the  base space $M$. This
submanifold represents the Poincar\'e dual of the Euler  class of $E$.  As
such, it has a topological meaning and it is independent of the Riemannian
structures  involved in the problem. When the zero locus of the section
consists of single zeros,  {\it i.e.}, $\di (M) = {\rm rk} (E)$, and $E$ and
$M$ are orientable,  the Euler number of 
$E$ can be obtained by counting the points in the zero locus with appropriate
signs.  This is roughly the way in which the first Donaldson invariant is
defined, when  the moduli space of ASD connections has zero-dimension. To
define higher dimensional  invariants, we consider the product of an adequate
number of generating cohomology  classes (\ref{clasefour}), (\ref{dosclase})
with the homology class of 
${\cal M}_{\rm ASD}$ in ${\cal B}^*$. Of course, this definition is not
precise from the  mathematical point of view, and one can consult
\cite{donpi, dk, fm} for a  rigorous construction of the invariants. Two
important requirements for this definition  to make sense are the
orientability and the compactness of the moduli space. The moduli  space
${\cal M}_{\rm ASD}$ is not compact, but there is a standard
compactification  which we won't consider here, see \cite{dk, fu, fm}. On the
other hand, to prove the  orientability of the moduli space, one must
trivialize the highest  exterior power of the tangent bundle. For regular,
irreducible connections, 
$H^0_A=H^{2}_A=0$, and this line bundle is given by
\be
\Lambda^{\rm max}T_A {\cal M}_{\rm ASD}=\Lambda^{\rm max}\ker \,\
\delta_A,
\label{lineao}
\ee where $\delta_A$ is defined in (\ref{elipsis}). Therefore, under the
above  conditions, ${\rm coker}\,\ \delta_A=0$ and one must prove the
triviality of the determinant line bundle of the operator
$\delta_A=d_A^{*}\oplus p^{+}d_A$, 
${\rm det}\,\ {\rm ind} \,\ \delta_A$. This has been done by Donaldson 
\cite{donpaper, donor}. Actually, Donaldson proves  a stronger  result, that
the determinant line bundle of $\delta_A$ over ${\cal B}^*$ is trivial,  and
that a canonical orientation is obtained from an orientation of the  space
$H^1(M)\oplus H^{2,+}(M)$. In the case of zero-dimensional moduli spaces, 
the orientation is required to associate a well-defined sign to every point.
This sign  can  be understood as the sign of ${\rm det} \,\ \delta_A$. 

With the above requirements about compactness and orientability, one can
define  the invariants  along the lines we sketched before. A non-rigorous
formulation of the  above remarks can be given, which is nonetheless useful
to make contact with Topological  Quantum Field Theory \cite{aj, cmr}. Let
$\Phi({\cal E})$ be the Thom class of the  vector bundle (\ref{mivector}). It
is well-known that the Euler class of ${\cal E}$, 
$\chi({\cal E})$, can  be represented by the pullback of $\Phi({\cal E})$ by a
transversal section of 
${\cal E}$, for instance the section ${\hat s}$ induced by (\ref{secciona}).
We will  represent the Donaldson invariants through a generating function. 
We choose a basis for the two-dimensional homology of $M$, $[\Sigma_a]$,
where 
$a=1, \cdots, \di (H_2(M))$. The Donaldson invariants are then formally
defined as
\be 
\langle {\rm exp}\Big( \sum_a \alpha_aI(\Sigma_a) + \mu {\cal O} \Big)
\rangle = 
\int_{{\cal B}^*} \chi({\cal E}) \wedge {\rm exp}\Big( \sum_a
\alpha_aI(\Sigma_a) +
\mu {\cal O} \Big),
\label{corre}
\ee where $\chi({\cal E})={\hat s}^*(\Phi ({\cal E}))$, and a sum over all
instanton numbers  of $E$ is understood. If we denote
 by $i: {\cal M}_{\rm ASD} \hookrightarrow {\cal B}^*$ the inclusion of the 
moduli space of irreducible ASD connections in ${\cal B}^*$, we can take into
account  that
$\chi({\cal E})$ is the Poincar\'e dual of the zero locus ${\hat s}^{-1}(0)=
{\cal M}_{\rm ASD}$, and write the Donaldson invariants as:
\be
\langle {\rm exp}\Big( \sum_a \alpha_aI(\Sigma_a) + \mu {\cal O} \Big)
\rangle = 
\int_{{\cal M}_{\rm ASD}}{\rm exp}\Big( \sum_a \alpha_a i^*(I(\Sigma_a))
 + \mu i^*({\cal O}) \Big).
\label{otrodon}
\ee As we are summing over instanton numbers, the dimension of ${\cal M}_{\rm
ASD}$  will vary. It is clear that, from the expansion of the exponent in
(\ref{otrodon}),  only the differential forms whose degree equals the
dimension of ${\cal M}_{\rm  ASD}$ will give a non zero invariant. This has a
nice interpretation in the  framework of Field Theory, as we will see in
Chapter 4. On the other hand,  the Mathai-Quillen formalism,  which will be
studied in chapter 3, gives a representative  for the Thom form of a vector
bundle $\Phi({\cal E})$ which can be formally  generalized to  the
infinite-dimensional setting. The Euler class of
${\cal E}$, 
${\hat s}^*(\Phi ({\cal E}))$, in this representation, can be seen to
coincide with 
${\rm e}^{-S}$, where $S$ is the action of a Topological Quantum Field
Theory.  According to (\ref{corre}), the Donaldson invariants can be regarded
as  the correlation functions of certain operators which are precisely
(\ref{clasefour}),  (\ref{dosclase}), as we have already mentioned. This
gives the underlying geometrical  reason of this fact, discovered by Witten
using ordinary path-integral techniques  in \cite{tqft}.

To define the invariants, we have assumed in our discussion regularity of the
moduli  space. We have already mentioned that for non-generic metrics, for
instance K\"ahler  metrics, the moduli spaces can still be smooth manifolds
with an actual dimension  greater than the virtual one. In some cases we can
compute  the invariants in this situation, taking into account the
non-vanishing  of the obstruction cohomology (which in the case of Donaldson
theory is the  cohomology $H^2_A$). We consider the fibre bundle over the
moduli space whose  fibre at each point is given by the obstruction
cohomology, and then we include  in the right hand side of (\ref{otrodon})
(or its equivalent expression for other kind  of invariants) the Euler class
of this finite-dimensional bundle. This procedure to  compute invariants in
some non-generic cases, specially in the realm of  algebro-geometric 
descriptions of the moduli spaces, was introduced in \cite{wzw} and applied
in some  two-dimensional situations \cite{am}. It has also led to the
definition of  the so-called Euler character theories \cite{blaueu, cmrdos,
vw}.

As a general conclusion concerning this presentation of Donaldson theory, we
would like  to summarize the main steps that make possible to define
topological invariants  starting from a system of partial differential
equations \cite{don},  and that we have illustrated  in the case of the
equations for ASD connections.

1) The maps involved must be Fredholm, {\it i.e.}, the linearisation of the
equations  should give an elliptic complex whose index is the virtual
dimension of the moduli space. 

2) The resulting moduli space must be orientable.

3) One must not have reducible solutions in a one-parameter family of
solutions. 

4) The moduli space  must be compact, or should be compactified in some 
appropriate way. 
  
\chapter{Twisted $N=2$ SUSY}
 
In this Chapter we will review some basic facts about the twisting  procedure
and its relation with the construction of Topological Quantum Field 
Theories. The twist of $N=2$ supersymmetry was introduced by Witten
\cite{tqft, tsm} as
 a general  way to construct theories in which the correlation functions do
not  depend (at least formally) on the underlying metric of the manifold. The
starting  point are $N=2$ supersymmetric theories in two and four dimensions,
although (as it  should be clear from the procedure) it can also be applied
to $N=4$ supersymmetric  theories \cite{yamron,vw, marcus, high}. There are
excellent reviews on the  subject, and we refer to them for a more complete
survey \cite{blth, cmr}. First of all,  we will describe the twisting
procedure in four dimensions, but we will also  describe some of the relevant
features of the two-dimensional case.  

\section{The twist in four dimensions}

The standard way to introduce the twisting procedure is as follows: in ${\bf
R}^4$  the global symmetry group of $N=2$ supersymmetry is
\be {\cal H}=SU(2)_L\times SU(2)_R\times SU(2)_I\times U(1)_{\cal R}
\label{susygrupo}
\ee where
${\cal K}=SU(2)_L\times SU(2)_R$ is the rotation group, and $SU(2)_I$ and
$U(1)_{\cal R}$ are internal symmetry groups. The supercharges $Q_{i\alpha}$
and
${\bar Q}^{i\dot\alpha}$ of $N=2$ supersymmetry transform under ${\cal H}$ as
$(1/2,0,1/2)^1$ and $(0,1/2,1/2)^{-1}$, respectively, and satisfy:
\bea
\{ Q_{i\alpha},{\bar Q}^j_{\dot\beta } \}& =  & \delta_i^j
P_{\alpha\dot\beta},
\nonumber\\
\{ Q_{i\alpha},Q_{j\beta} \} &=  & \epsilon_{ij} C_{\alpha\beta} Z, 
\label{apple}
\eea where $\epsilon_{ij}$ and $C_{\alpha\beta}$ are $SU(2)$ invariant
tensors, and
$Z$ is the central charge generator. The twist consists of considering as the
rotation group the group ${\cal K}'=SU(2)_L'\times SU(2)_R$ where 
$SU(2)_L'$ is the diagonal subgroup of $SU(2)_L\times SU(2)_I$. Under the new
global symmetry group 
\be {\cal H}'= SU(2)_L'\times SU(2)_R\times U(1)_{\cal R}
\label{nuevasim}
\ee the supercharges transform as $(1/2,1/2)^{-1}\oplus (1,0)^1 \oplus
(0,0)^1$. The twisting is achieved replacing any isospin index $i$ by a
spinor index
$\alpha$ so that $Q_{i\alpha} \rightarrow Q_{\beta \alpha}$ and
$\bar Q_{i\dot\beta} \rightarrow G_{\alpha \dot\beta}$. The $(0,0)^1 $
rotation  invariant operator is $Q=Q_{\alpha}{}^\alpha$  and satisfies the
twisted version of the $N=2$ supersymmetric algebra (\ref{apple}), often
called {\it topological algebra}:
\bea
\{ Q,G_{\alpha\dot\beta} \} &=  & P_{\alpha\dot\beta}, \nonumber\\
\{ Q,Q\}& =  & Z. \label{orange}
\eea In a theory with trivial central charge the right hand side of the last
of these relations effectively vanishes and one has the ordinary situation in
which
$Q^2=0$. The first of these relations is at the heart of the standard
argument to conclude that the resulting twisted theory will be topological.
With the momentum tensor being $Q$-exact it is likely that the whole
energy-momentum tensor is
$Q$-exact. This would imply that the vacuum expectation values of
$Q$-invariant  operators which do not involve the metric  are metric
independent, {\it i.e.}, that the theory is topological. To our knowledge,
all the twisted $N=2$ theories which have been studied satisfy this property. 

The operator $Q$ does not have to be nilpotent to define a Topological
Quantum Field  Theory. For instance, supersymmetric theories involving
Yang-Mills fields close the supersymmetric algebra up to a gauge
transformation. This implies that in a twisted theory one does not have that
$Q^2$ vanishes but that it is a gauge transformation. This is the case for
Donaldson-Witten theory in which the gauge parameter on the right hand side
of the equation for $Q^2$ is one of the scalar fields of the theory. In
Chapter 4 we will construct a Topological Quantum Field Theory (a
generalization of Donaldson-Witten  theory \cite{tqft}) with this behaviour.
In this case, the observables of the theory  ({\it i.e.}, the operators in
the cohomology of $Q$) must be gauge invariant. Therefore,  the relevant
cohomology in these cases is the equivariant cohomology with respect to the 
gauge group. Similarly, in the presence of a non-trivial central charge, the
first relation in (\ref{orange}) holds and therefore one has the same
expectations to obtain a Topological Quantum Field Theory as in the ordinary
case. One can regard the second relation in (\ref{orange}) as a situation
similar to the case of Donaldson-Witten theory where the gauge symmetry is a
global $U(1)$ symmetry. In addition, this analogy implies that the  correct
mathematical framework to formulate these theories must involve an
equivariant extension. We will see  in Chapters 3 and 4 that some  theories
with non-trivial central charge have an interesting interpretation in terms 
of equivariant cohomology with respect to a vector field action.

An alternative (and equivalent) point of view on the twisting  procedure is
obtained when it is regarded as a gauging of an internal symmetry  group, in
which a  global symmetry of the underlying supersymmetric model is promoted
to a  space-time symmetry. In many cases, the gauging is performed by adding
to  the Lagrangian of  the original theory a new term, involving the coupling
of the internal current  to the Spin connection of the underlying manifold
\cite{egu,cv,cmr}. We will now  discuss this approach to the twisting
procedure in some detail, in the case of $N=2$  Yang-Mills theory, the model
originally considered by Witten in the seminal  paper
\cite{tqft}. The resulting Topological Quantun Field Theory is equivalent, as
it  is well  known, to Donaldson theory: it describes the moduli space of ASD
connections and the  topological  correlation functions are the Donaldson
polynomials.  

The field content of the minimal $N=2$ supersymmetric Yang-Mills theory with
gauge group $G$ on ${\bf R}^4$ is the following: a gauge field $A_{\mu}$,  
two Majorana spinors $\lambda_{i\alpha}$, $i=1,2$, and their conjugates 
$\overline\lambda^{i\dot\alpha }$, a complex scalar $B$, and an auxiliary
field
$D_{ij}$ (symmetric in $i$ and $j$). The indices $i$, $j$ denote  the isospin
indices of the  internal symmetry group $SU(2)_I$ of $N=2$ supersymmetry. The
two Majorana spinors 
$\lambda_1$, $\lambda_2$ form a doublet of $SU(2)_I$. All these fields are
considered in the adjoint representation of the gauge group $G$. The action
and field content of this model have a useful  description in terms of $N=1$
superspace. In $N=1$ superspace only one of the supersymmetries is manifest,
and therefore the
$N=1$  superfields do not have well defined quantum numbers with  respect to
the internal $SU(2)_I$ symmetry. The $N=2$ supersymmetric multiplet contains
an $N=1$ vector multiplet and an $N=1$ chiral multiplet. These multiplets are
described in
$N=1$ superspace in terms of $N=1$ superfields $W_\alpha$ and $\Phi$
satisfying the constraints $\overline D_{\dot\alpha} W_\alpha =0$,
$D^\alpha W_\alpha + \overline D^{\dot\alpha} \overline  W_{\dot\alpha}=0$ and
$\overline D_{\dot\alpha} \Phi =0$, where $D_\alpha$ and $\overline
D_{\dot\alpha}$ are 
$N=1$ superspace covariant derivatives (we use the conventions in 
\cite{wb}).  The
$N=1$ superfields $W_\alpha$ and $\Phi$ have $U(1)_{\cal R}$ charges $-1$ and
$-2$ respectively.  The component fields of the $N=1$ superfields $W_\alpha$ 
and $\Phi$ are:
\bea W_\alpha, \;\; \overline W_{\dot\alpha}
 \;\; & \longrightarrow &\;\;  A_{\alpha\dot\alpha}, \;\;  \lambda_{1\alpha},
\;\; 
\overline\lambda^{1\dot\alpha }, \;\; D_{12}, \nonumber\\
\Phi, \;\; \Phi^\dagger \;\;& \longrightarrow &\;\;  B, \;\; 
\lambda_{2 \alpha},  \;\; D_{22},  \;\; B^\dagger, \;\;
\overline\lambda^{2\dot\alpha }, \;\; D_{11}. 
\label{trapecio}
\eea The $U(1)_{\cal R}$ transformations of the $N=1$ superfields are:
\be W_\alpha \rightarrow {\rm e}^{-i\phi} W_{\alpha}({\rm e}^{i\phi}\theta),
\;\;\; {\rm and} \;\;\;
\Phi \rightarrow {\rm e}^{-2i\phi} \Phi({\rm e}^{i\phi}\theta).
\label{salto}
\ee

In $N=1$ superspace the action of $N=2$ supersymmetric Yang-Mills theory
takes the form:
\be S_{\rm YM}=\int d^4x \, d^2\theta d^2\overline\theta\,\Phi^\dagger {\rm
e}^V
\Phi+
\int d^4x \, d^2\theta \, W^\alpha W_\alpha +
\int d^4x \, d^2\overline\theta \,\overline  W^{\dot\alpha} 
\overline W_{\dot\alpha},
\label{superespacio}
\ee where $V$ is the vector superpotential. An important feature of this
action is that due to the constraint $D^\alpha W_\alpha + \overline
D^{\dot\alpha} \overline  W_{\dot\alpha}=0$ the last two terms in
(\ref{superespacio}) differ by a term which is proportional to the second
Chern class. The $SU(2)_I$ current of this  model is given by:
\be j^{\mu}_a={\overline \lambda}{\sigma}^{\mu}{\sigma_a}\lambda,
\label{corriente}
\ee where ${\sigma}^{\mu}$ are the matrices given in the Appendix,  after
(A.2.40) (in Euclidean space).   

If we try to formulate the theory on a general four-manifold $X$,  we use the
prescription  of minimal coupling to gravity in the Lagrangian, and we couple
the spinor fields to the  Spin connection in the usual way (using the
expression given in (\ref{cova})).  This construction can always  be done 
locally, but globally we have a topological obstruction associated  to the
second Stiefel-Whitney class,
$w_2(X)$. If $X$ is not Spin, we cannot  consistently  couple the original
$N=2$ theory to gravity. The twist of this theory involves the  gauging of
the $SU(2)_I$ group as the spacetime symmetry group $SU(2)_{L}$, the
structure  group of the positive-chirality spinor bundle. The  only fields
charged with respect to $SU(2)_I$ are spinors, therefore the gauging is 
achieved  after adding to the Lagrangian the term
\be -\omega_{\mu}^a j^{\mu}_a=-{\overline \lambda}^{i}_{\dot
\alpha}(\sigma^{\mu})^{\alpha 
\dot \alpha}(\sigma_a)_i{}^j\omega_{\mu}^a \lambda_{j \alpha}.
\label{acoplo}
\ee The only change in the Lagrangian is in the fermion kinetic term, which
becomes
\be -{\overline \lambda}^{i}_{\dot \alpha}(\sigma^{\mu})^{\alpha 
\dot \alpha}\{ \delta_{\alpha}{}^{\beta}\delta_{i}{}^{j}-i\omega_{\mu}^a 
\big((\sigma_a)_{\alpha}{}^{\beta}\delta_{i}{}^{j} +
(\sigma_a)_i{}^{j}\delta_{\alpha}{}^{\beta}\big)\}\lambda_{j\beta}.
\label{kin}
\ee The connection appearing here is the tensor product connection on the
bundle 
$S^{+}\otimes S^{+}$, which is isomorphic to $\Omega^{0}_{\bf C}\oplus 
\Omega^{2,+}_{\bf C}$. The scalar part corresponds to the symmetric part of 
$\lambda_{w\beta}$, and the self-dual two-form corresponds to the
antisymmetric one.  We can write (\ref{kin}) in terms of space-time fields,
and to do this we introduce a  scalar $\eta$, a one-form $\psi_{\mu}$, and a
self-dual two-form $\chi_{\mu \nu}$ as:
$$
\eta=-\lambda_{\beta}^{\beta},\,\,\,\,\,\,\ 
\psi_{\mu}=i(\sigma_{\mu})^{\dot \alpha \alpha}  {\overline \lambda}_{\alpha
\dot
\alpha},
$$
\be
\lambda_{(w\beta)}=-{1 \over {2 {\sqrt 2}}} C^{\dot \alpha \dot \beta}  (\bar
\sigma_{\mu})_{w \dot \alpha}(\bar \sigma_{\nu})_{\beta \dot \beta}
\chi^{\mu \nu}.
\label{formas}
\ee    In terms of these fields, the fermion kinetic terms can be written,
after a lengthy  computation, as
\be -{i \over 2}\psi^{\mu}D_{\mu} \eta - {\sqrt 2}\psi_{\mu}D_{\nu}\chi^{\nu
\mu}.
\label{kindos}
\ee which are the standard fermion kinetic terms of Donaldson-Witten theory.
As a  byproduct,  notice that the gauging of the $SU(2)_I$ global symmetry
allows one to define the  theory on any smooth four-manifold, as the fields
of the resulting, twisted theory  are differential forms. They have the 
geometrical structure prescribed by the new symmetry group (\ref{nuevasim}),
and they  are redefined as follows
\bea A_{\mu} \;\; (1/2,1/2,0)^0 \;\; & \longrightarrow & 
 A_{\mu} \;\; (1/2,1/2)^0, \nonumber\\
\lambda_{i\alpha } \;\; (1/2,0,1/2)^{-1} \;\;&\longrightarrow & 
\eta  \;\; (0,0)^{-1}, \;\;\; \chi_{\alpha\beta} \;\; (1,0)^{-1},\nonumber\\
\overline\lambda_{i\dot\alpha } \;\; (0,1/2,1/2)^1 \;\;& \longrightarrow &
\psi_{\alpha \dot\alpha} \;\; (1/2,1/2)^1, \nonumber\\ B \;\; (0,0,0)^{-2}
\;\;&
\longrightarrow &  i {\sqrt 2} \lambda \;\; (0,0)^{-2},
\nonumber\\ B^\dagger \;\; (0,0,0) \;\;&\longrightarrow& {\phi\over 2 {\sqrt
2}}
 \;\; (0,0)^2,\nonumber\\ D_{ij} \;\; (0,0,1)^0 \;\;& \longrightarrow& 
H_{\alpha\beta} \;\; (1,0)^0,
\label{twist}
\eea where we have indicated the quantum numbers carried out by the fields
relative to the group ${\cal H}$ (\ref{susygrupo}) before the twisting, and
to the group
${\cal H}'$ (\ref{nuevasim}) after the twisting. Notice  that the $U(1)_{\cal
R}$ assignment in (\ref{twist}) is consistent  with the transformations in
(\ref{salto}). The numerical coefficients are chosen in order  that the
twisted action coincides with the the topological action to be constructed 
following  the  Mathai-Quillen formalism in Chapter 5. 
 
The definitions of the twisted fields in terms of the untwisted ones are the
obvious ones from (\ref{formas}) and (\ref{twist}). For $\chi_{\alpha \beta}$
we take 
$\chi_{\alpha \beta}= {\sqrt 2}\lambda_{(\alpha \beta)}$. We also have:
\be
\lambda_{{\bar 1} 2}={i\over 2} \eta +{ 1 \over {\sqrt 2}}\chi_{12},
\;\;\;\;\;\;\;
\lambda_{{\bar 2} 1}=-{i\over 2} \eta + { 1 \over {\sqrt 2}}\chi_{12}.
\label{debbie}
\ee The bar denotes old isospin indices. In these definitions we mainly
follow the  conventions of \cite{wb} (see also \cite{sym,high}).

Although the differential forms in  (\ref{formas}) are complex, they have of
course a natural underlying real structure.  One must restrict the resulting 
fields to be {\it real} differential forms, in order to have the same number
of  degrees of freedom in the untwisted and twisted theory. In general, this 
counting of degrees of freedom must be taken into account if one is
interested in  extracting some information from the dynamics of the
untwisted, physical theory. 

In general, the gauging of a global symmetry can generate anomalies in the
resulting  theory. In the case of $N=2$ Yang-Mills theory, the possible 
anomalies are related to  the global $SU(2)$ anomaly discovered by Witten in
\cite{sudosano}, and it is easy  to see \cite{tqft} that they only appear
when the corresponding moduli space is not  orientable. As we saw in Chapter
1, this is not the case for the moduli space  of ASD connections, then the
twisted theory is anomaly-free. 

\section{The twist in two dimensions}   In ${\bf R}^2$ the global symmetry
group of
$N=2$ supersymmetry is 
\be {\cal H}=SO(2)\times U(1)_L\times U(1)_R
\label{sinti}
\ee
 where ${\cal K}=SO(2)$ is the rotation group, and $U(1)_L$ and $U(1)_R$ are
left and right moving chiral symmetries. There are four supercharges
$Q_{\alpha a}$  transforming under ${\cal H}$ as $(-1/2,1,0)$, $(-1/2,-1,0)$,
$(1/2,0,1)$ and $(1/2,0,-1)$. They  satisfy: 
\bea 
 \{ Q_{\alpha +},Q_{\beta -} \} &=  &  P_{\alpha\beta}, \nonumber\\ 
\{ Q_{\alpha +},Q_{\beta +} \} &= &
\{ Q_{\alpha -},Q_{\beta -} \} = \epsilon_{\alpha\beta} Z, 
\label{appledos}
\eea where  $\epsilon_{\alpha\beta}$ is an antisymmetric $SO(2)$ invariant
tensor, and $Z$ is the central charge generator. The twist consists of
considering as the rotation group the diagonal subgroup of $SO(2)\times
SO(2)'$, where $SO(2)'$ has as generator $(U_L-U_R)/2$ being $U_L$ and $U_R$
the generators of $U(1)_L$ and
$U(1)_R$ respectively. Under the new global symmetry group ${\cal H}'= SO(2)
\times U(1)_{F}$, where $U(1)_{F}$ has as generator the combination
$U_L+U_R$, the supercharges transform as 
$(0,1)\oplus (-1,-1) \oplus (0,1) \oplus (1,-1) $. The twisting is achieved
thinking of the second index of $Q_{\alpha a}$ as an $SO(2)$ isospin index
and, as in the four dimensional case, replacing any isospin index $a$ by a
spinor index $\beta$ so that $Q_{\alpha a} \rightarrow Q_{\alpha\beta}$.  One
of the two  rotation  invariant operators is $Q=Q_{\alpha}{}^\alpha$. It
satisfies the twisted version of the $N=2$ supersymmetric algebra
(\ref{appledos}) or  topological algebra:  
\bea
\{ Q,G_{\alpha\beta} \} &=  & P_{\alpha\beta}, \nonumber\\
\{ Q,Q\}& = & Z, 
\label{orangedos}
\eea where $G_{\alpha\beta}$ is the symmetric part of $Q_{\alpha\beta}$.
Notice that one could have taken  the  combination $(U_L+U_R)/2$ instead of
$(U_L-U_R)/2$ in order to carry out the twisting. We will consider in some
detail the two-dimensional twisting in the  case of the $N=2$ non-linear
sigma model, as another illustration of the twisting as  a gauging of global
currents. The fact that we have two different global symmetries  gives rise
to the two possible topological sigma models considered in \cite{tmtwo,mm}.
This  example also sheds light on the possible anomalies associated to the
gauging procedure.  

First of all we recall a few standard  facts on non-linear sigma models in two
dimensions. Non-linear sigma models involve mappings from a two-dimensional
compact  Riemann surface $\Sigma$ to an $n$-dimensional target manifold $M$.
The local coordinates of this mapping can be regarded as bosonic
two-dimensional fields which might be part of different types of
supersymmetric multiplets. In $N=2$ supersymmetry there are two types of
multiplets, chiral multiplets and twisted chiral multiplets. The possible
geometries of the target manifold $M$ are severely restricted by the
different choices of multiplets taking part of a given model. In models
involving only chiral multiplets $N=2$ supersymmetry requires that $M$ is a
K\"ahler manifold 
\cite{zumo,ag}. In the situations where both multiplets are allowed, $M$ can
be a Hermitian locally product space 
\cite{gates}. Twistings of models involving both types of multiplets have been
considered in \cite{tsm,vafa,tmtwo,mm}. 

From the point of view of the gauging procedure, the two different
possibilities  to obtain topological sigma models are better understood
starting from a $N=2$ model  involving chiral multiplets and choosing two
different global currents. The target manifold 
$M$ is then K\"ahler and we have a bosonic field $\phi: \Sigma
\longrightarrow M$  and a Dirac spinor $\psi^{I}_{\pm} \in 
\Gamma(\Sigma, S^{\pm}\otimes \phi^{*}(TM))$, where $TM$ is the holomorphic
tangent bundle  to $M$. The kinetic fermion term in the action is:
\be S_{\rm f}=\int_{\Sigma}d^{2}z ig_{{\bar I}J}(\psi^{\bar I}_{+}D_{\bar
z}\psi^{J}_{+}+
\psi^{\bar I}_{-}D_{ z}\psi^{J}_{-}),
\label{fermiondos}
\ee where the covariant derivatives are given in local coordinates by
\bea D_{\bar z} \psi^{J}_{+}&=&\partial_{\bar z}\psi^{J}_{+}+ {i \over
2}\omega_{\bar z}\psi^{J}_{+}+
\Gamma^{J}_{KL}
\partial_{\bar z} \phi^{K} \psi^{L}_{+},
\nonumber\\ D_{z} \psi^{J}_{-}&=&\partial_{z}\psi^{J}_{-}-{i \over
2}\omega_{z}\psi^{J}_{-}+
\Gamma^{J}_{KL}
\partial_{z} \phi^{K} \psi^{L}_{-}.
\label{covados}
\eea This theory has a conserved, non anomalous vector current
$j^{\mu}_{V}=g_{{\bar I}J} {\overline \psi}^{\bar I} {\gamma}^{\mu}
\psi^{J}$, with components
\be j^{z}_{V}=2g_{{\bar I}J}\psi^{\bar I}_{-} \psi^{J}_{-},\,\,\,\,\,\ j^{\bar
z}_{V}= 2g_{{\bar I}J}\psi^{\bar I}_{+} \psi^{J}_{+}
\label{jvector}
\ee and an anomalous axial current $j^{\mu}_{5}=g_{{\bar I}J}{\overline
\psi}^{\bar I} {\gamma}^{\mu} \gamma_5 \psi^{J}$ with components
\be j^{z}_{5}=2g_{{\bar I}J}\psi^{\bar I}_{-} \psi^{J}_{-},\,\,\,\,\,\ j^{\bar
z}_{5}= -2g_{{\bar I}J}\psi^{\bar I}_{+} \psi^{J}_{+}.
\label{axial}
\ee The anomaly is given by the index of the Dirac operator and reads
\be
\int_{\Sigma} \phi^{*}(c_1(M))
\label{anomalia}
\ee  To twist the model we can gauge the $U(1)_V$ or the $U(1)_A$ symmetries.
The first choice  leads to the {\it A model} and the second one to the {\it B
model}. As in Donaldson- Witten theory, we promote the abelian global
symmetry to a worldsheet spacetime  symmetry, and in this case this amounts
to add to the Lagrangian the coupling of the  corresponding currents to the
worldsheet Spin connection. For the A model we have:
\bea S_{\rm f}-{i \over 4} \int_{\Sigma} d^2 z {\sqrt
h}\omega_{\mu}j^{\mu}_{V}  &=&
\int_{\Sigma} d^2 z {\sqrt h} i g_{{\bar I}J}\{ \psi^{\bar I}_{+}  ({\overline
\partial} \psi^{J}_{+}+\Gamma^{J}_{KL} {\overline \partial} \phi^{K}
\psi^{L}_{+})\nonumber\\
 & &\,\,\,\,\,\,\,\,\,\,\,\,\,\, +\psi^{\bar I}_{-}(\partial \psi^{J}_{-}
-i\omega_{z}\psi^{J}_{-}+ \Gamma^{J}_{KL}
\partial \phi^{K} \psi^{L}_{-}) \},
\label{tipoa}
\eea and for the B model
\bea
 S_{\rm f}-{i \over 4} \int_{\Sigma} d^2 z {\sqrt h}\omega_{\mu}j^{\mu}_{A}
 &=&\int_{\Sigma} d^2 z{\sqrt h}ig_{{\bar I}J}\{ \psi^{\bar I}_{+} 
({\overline
\partial}\psi^{J}_{+}+i\omega_{\bar z}\psi^{J}_{+}+\Gamma^{J}_{KL}
\partial_{\bar z} \phi^{K} \psi^{L}_{-})
\nonumber\\
 & & \,\,\,\,\,\,\,\,\,\,\,\,\,\, +\psi^{\bar I}_{-}(\partial\psi^{J}_{-}-
i\omega_{ z}\psi^{J}_{-} +\Gamma^{J}_{KL}
\partial \phi^{K} \psi^{L}_{+}) \}.
\label{tipob}
\eea      We see that, in the twisted models, the fermion fields have changed
their spin content. 
$\psi^{J} _{-}$ is a $(0,1)$-form $\rho_{\bar z}^{J}$, while $\psi^{J} _{+}$
is an scalar 
$\chi^{J}$ in the type A model,  and a $(1,0)$-form $\rho^{J}_{z}$ in the
type B model. 

It turns out \cite{tsm} that the type A model can be formulated on any almost
Hermitian  target manifold. However, the type B model was obtained through the
gauging of an  anomalous current, and this can give ill-defined models: the
anomaly in  the global current, given in (\ref{anomalia}), is inherited in
the twisted model as a  global anomaly in the  fermion determinant. This
leads to additional restrictions on the geometry of the  target space, as
pointed out in \cite{mm}, because the fact that the $U(1)$ current is 
chiral  leads to a non-linear sigma model anomaly. We will present here an
computation of  this  global anomaly using the strategy of \cite{aetio},
based on the index theorem  for families.   The fermionic kinetic term of the
type B model is:
\be S_{\rm B}= \int_{\Sigma} d^2 z{\sqrt h}ig_{{\bar I}J} \{ \chi^{\bar I}D_z 
\rho^{J}_{\bar z}  + \theta^{\bar I}D_{\bar z}\rho^{J}_{ z}\},
\label{kinb}
\ee  where $\theta^{\bar I}=\psi_{+}^{\bar I}$ is an scalar in the twisted
theory. The effective  action is then a section of the line bundle
\be {\cal L}={\cal L}_1\otimes {\cal L}_2=({\rm det}\,\ D_z) \otimes  ({\rm
det}\,\ D_{\bar z}),
\label{ef}
\ee and the global, topological anomaly is measured by $c_1({\cal
L})=c_1({\cal L}_1) +  c_1({\cal L}_2)$. This can be computed  using the
index theorem for families as in \cite{aetio}. Consider first the  evaluation
map 
\bea {\hat \phi}: {\rm Map}(\Sigma, M) \times \Sigma &\rightarrow& M
\nonumber\\ (\phi, \sigma) &\mapsto& \phi(\sigma).
\label{sigevalu}
\eea By pulling back  differential forms on $M$ through ${\hat \phi}^*$, we
get differential forms on 
${\rm Map}(\Sigma, M) \times \Sigma$, with the natural bigrading given by the 
product structure:
\be {\hat \phi}^*(\omega)={\cal O}^{(0)}_{\omega}+{\cal O}^{(1)}_{\omega}+
{\cal O}^{(2)}_{\omega},
\label{bigrado}
\ee where the ${\cal O}^{(i)}_{\omega}$ are of degree $i$ with respect to
$\Sigma$. 
 This is precisely the descent procedure of \cite{tsm}.  The topological
obstructions are given by:
\be c_1({\cal L}_{1,2})=\int_{\Sigma} {\rm ch}({\hat \phi}^*({\overline
{TM}}))
\{\pm 1 - {1 \over 2}c_1(\Sigma)\},
\label{cherndet}
\ee where we only keep the degree two forms on ${\rm Map}(\Sigma, M)$. In
$c_1({\cal L})$
 only  the first descendant of ${\hat \phi}^*({\rm ch}({\overline {TM}}))$
contributes,  and we  finally  get
\be c_1({\cal L})=\int_{\Sigma}c_1(\Sigma){\cal O}^{(0)}_{c_1(M)},
\label{anomala}
\ee where ${\cal O}^{(0)}_{c_1(M)}={\hat \phi}_{\sigma}^{*}(c_1(M))$ is the 
descendant of zero degree with respect to $\Sigma$,  and ${\hat
\phi}_{\sigma}$ is the map obtained from ${\hat \phi}$ by fixing a point 
$\sigma\in \Sigma$ (the cohomology class of the pulled-back form does not
depend  on the $\sigma$ chosen). This result says then that the anomaly in
the global  current
$U(1)_A$ is inherited in the  twisted B model as a sigma model anomaly. The B
model has no topological anomalies  if the target is  a Calabi-Yau manifold
or if the worldsheet is a torus ($c_1(\Sigma)=0$). The last  possibility is
natural,  as in this case  the twist does nothing (the torus is a
hyperk\"ahler manifold) and the original $N=2$  supersymmetric model is
anomaly-free.

The models we have just considered have no central charges, but, as in the
four-dimensional
 case, the presence of a central charge does not  invalidate the topological
character of the twisted theory. In the next Chapter  we will consider a
two-dimensional model where the central charge generator of $N=2$
supersymmetry acts as a Lie derivative with respect to a Killing vector
field, and the  resulting structure has a nice interpretation in terms of the
equivariant cohomology  with respect to a vector field action on the moduli
space. 

\chapter{The Mathai-Quillen formalism} The Mathai-Quillen formalism is of
paramount importance in the study of Cohomological  Field Theories, because
it provides the fundamental link between topological Lagrangians  and
geometrical constructions associated to moduli spaces. This formalism, 
introduced in \cite{mq}, provides a particular representative of the Thom
class of  an oriented vector bundle, using equivariant differential forms.
The relevance of this  representative to Cohomological Field Theories was
realized by Atiyah and Jeffrey 
\cite{aj}  in a fundamental work. They showed that the topological Lagrangian
obtained by Witten  as a twisted $N=2$ supersymmetric Yang-Mills theory was
precisely the natural  generalization  of the Mathai-Quillen representative in
Donaldson theory. As we discussed in  Chapter I, the moduli space of
irreducible ASD connections can be considered as the  zero locus of a section
of an infinite dimensional vector bundle, and  the zero-dimensional 
Donaldson  invariant can be understood in terms of an integral of the
pullback of the  corresponding Thom class. The work of Atiyah and Jeffrey
therefore explains from first  principles why topological correlation
functions should be formally identified  with Donaldson  invariants, and also
gives a topological interpretation of the BRST cohomology of  the twisted
model in terms of equivariant cohomology. 

In this chapter we first give a brief review of the Mathai-Quillen formalism
following  the original approaches. We don't intend to be exhaustive, because
there are by now  excellent and comprehensive expositions of this topic
\cite{cmr, blau, bt}.  We will mainly focus on the generalization of this
formalism to take into  account vector field actions. This generalization was
constructed in \cite{lmt}  motivated by the twisting of $N=2$ supersymmetric
theories with central charges  in the supersymmetry algebra. One of the
advantages of the formalism developed in 
\cite{lmt} is that it makes more apparent the geometry of the Mathai-Quillen 
formalism, in the sense that it doesn't involve algebraic models of the  de
Rham cohomology, but rather local descriptions of the relevant differential 
forms.  

The chapter is organized as follows: in section 1 we give an overview of
equivariant  cohomology in the Weil and Cartan models. In section 2 we
present the Mathai-Quillen  formalism following the original work of Mathai
and Quillen in \cite{mq}. In section 3  we present the equivariant extension
of \cite{lmt} and we also develop the BRST  cohomology, the Atiyah-Jeffrey
formulation and the basic connections with the  Field Theory framework. In
order to  illustrate the construction, we analyze a two-dimensional example, 
the topological sigma model with potentials. 

\section{Equivariant cohomology}

Equivariant cohomology appears when studying a topological space $M$ with the
action  of a  group $G$. It can be realized in many different ways, but the
most useful  ones to us involve the differentiable category and appropriate
extensions of  the de Rham cohomology. From a more general
 point of view,  equivariant cohomology is defined as the ordinary cohomology
of the space 
\be M_{G}=EG \times _G M,
\label{uni}
\ee where $EG$ is, as usual, the universal $G$-bundle. We will instead work
with algebraic  models, the Weil and the Cartan model. Although they are
equivalent, the Cartan model  is simpler and more natural in Topological
Gauge Theories, and we will also use it  to study  vector field actions.
However, when dealing with group actions associated to principal  bundles,
the Cartan model must be supplemented with a horizontal projection. In these 
cases it is important to keep in mind the underlying Weil model, because, as
we will  see, it is specially suited for the principal bundle setting. 

In the first subsection we will consider the Weil model, and in the second
one  the Cartan model. Finally, in subsection 3 we analyze the equivariant
cohomology  associated to vector field actions. Good references on
equivariant  cohomology are
\cite{ab, cmr, bgv, kal}.  
 
\subsection{Weil Model}

Let $G$ be a Lie group with Lie algebra ${\bf g}$. First we define the  {\it
Weil algebra} ${\cal W}({\bf g})$ as
\be {\cal W}({\bf g})=\Lambda {\bf g}^{*} \otimes S {\bf g}^{*},
\label{weilal}
\ee given by the tensor product of the exterior algebra and the symmetric
algebra of the  dual ${\bf g}^*$. We take generators of $\Lambda {\bf g}^{*}$,
$\{\theta^a\}_{a=1, 
\cdots ,{\rm dim}(G)}$ with degree $1$, and of $S {\bf g}^{*}$, $\{ u^a
\}_{a=1, 
\cdots ,{\rm dim}(G)}$ with degree $2$. In this way ${\cal W}({\bf g})$
becomes a  graded  algebra. Let $c^a_{bc}$ be the structure constants of
${\bf g}$ associated to  the generators $\theta^a$. We define a differential
operator $d_{\cal W}$ from their  action on the generators:
\bea d_{\cal W}\theta^a &=& -{1 \over 2}c^a_{bc}\theta^b \theta^c + u^a,
\nonumber
\\  d_{\cal W}u^a &=&c^a_{bc}u^b \theta^c .
\label{weilco}
\eea We also define an inner product operator $i_a$ as follows:
\be i_a\theta^b=\delta_{ab}, \,\,\,\,\,\,\,\ i_a u^b=0.
\label{inner}
\ee We can also define a Lie derivative operator using the basic identity 
${\cal L}_a=i_ad_{\cal W}+ d_{\cal W}i_a$. The motivation to define such a
complex comes from the following: if 
$P$ is a principal $G$-bundle with connection $\theta \in \Omega^{1}(P, {\bf
g})$
 and curvature $K \in \Omega^2(P, {\bf g})$, we can expand $\theta$ and $K$ 
as follows:
\be
\theta =\theta^a T_a,\,\,\,\,\,\, K=u^a T_a, 
\label{lilith}
\ee where $\{T_a\}_{a=1,  \cdots ,{\rm dim} (G)} $ is a basis of ${\bf g}$,
and 
$\theta^a \in \Omega^{1}(P)$, $u^a \in \Omega^{2}(P)$. Consider the inner 
product
$\iota (C(T_a))$, where $C(T_a)$ is the fundamental vector  field asociated
to the generator of the Lie algebra ${\bf g}$. The components  defined in
(\ref{lilith}) verify precisely the equations (\ref{weilco}) and 
(\ref{inner}),  with the geometrical realizations of the inner product $i_a$
as $\iota (C(T_a))$ and  the differential operator $d_{\cal W}$ as the usual
de Rham operator on $P$. We then see  that the Weil model is essentially an
algebraic or universal realization of the  basic relations defining
connections and curvatures on principal bundles. 

Consider now a $G$-manifold $M$ and the complex 
\be {\cal W}({\bf g}) \otimes \Omega^{*}(M), 
\label{haciati}
\ee where $\Omega^{*}(M)$ is the complex of differential forms on $M$. On this
complex we  can define a differential operator, an inner product and a Lie 
derivative operator by taking the tensor product of the corresponding
operators in  the two complexes. An element of this complex is called {\it
basic} if it is in the  kernel of the inner product operators (it is
horizontal) and in the kernel of the  Lie derivative operators (it is
invariant). The subalgebra consisting of these elements  is denoted  by
$\Omega_{G}(M)$, and its cohomology is called the {\it algebraic equivariant 
cohomology  of $M$ in the Weil model}:
\be H_{G}^{*}=H^{*}(\Omega_{G}(M)).
\label{equis}
\ee As we will see in section 2, the Weil model is specially useful to study
the cohomology  of associated vector bundles and for this reason has an
important r\^ole in the  Mathai-Quillen formalism.

\subsection{Cartan model}
 
The fact that the construction of the Weil model for equivariant cohomology
involves  only a subcomplex of ${\cal W}({\bf g}) \otimes \Omega^{*}(M)$
suggests that a smaller  complex can be chosen from the very beginning. In
the Cartan model, one starts with 
\be S{\bf g}^* \otimes \Omega^{*}(M)
\label{complejillo}
\ee and defines an operator 
\be d_{\cal C} u^a=0, \,\,\,\,\,\,\,\,\ d_{\cal C}\omega = d\omega
-u^a\iota(C(T_a))\omega, 
\label{carta}
\ee where $\omega \in \Omega^{*}(M)$ and summation over $a$ is understood.  In
general, $d_{\cal C}$ is not nilpotent. Rather we  have
\be d_{\cal C}^2=-u^a{\cal L}(C(T_a)),
\label{nonil}
\ee but we can restrict ourselves to the invariant  subcomplex  
\be
\Omega_{G}(M)=(S{\bf g}^* \otimes \Omega^{*}(M))^{G},
\label{invi}
\ee where $d_{\cal C}^2=0$. The elements in this complex are called  {\it
equivariant differential
 forms}. We used the same notation for the basic subcomplex in the Weil
model,  because in fact they  are isomorphic \cite{mq}. As $d_{\cal C}^2=0$ on
$\Omega_{G}(M)$, we  can define the cohomology of $d_{\cal C}$, which is
precisely the  {\it algebraic equivariant cohomology  of $M$ in the Cartan
model}:
\be H^{*}_{G}(M)=H^{*}((S{\bf g}^* \otimes \Omega^{*}(M))^{G}).
\label{mascoho}
\ee Of course, the Weil and the Cartan models give isomorphic cohomology
theories. 

\subsection{Equivariant cohomology and vector field actions}

An special example of the Cartan model of equivariant cohomology, which will 
be useful later, is the case of the equivariant cohomology associated to a
vector field action. As we will explain below,  this can be regarded as a
particularly tractable realization of the construction  explained in
subsection 2. 

Let $X$ be a vector  field acting on a manifold $M$. Recall that every vector
field is associated to a locally defined one-parameter group of
transformations of $M$,
$\phi: I\times M \rightarrow  M$, with $I \subset {\bf R}$ being an open
interval containing $t=0$.  If we put $\phi_m(t)=
\phi_t(m)=\phi(t,m)$,  the vector field corresponding to $\phi$ is given by:
\be X(m)=\phi_{m * 0} \Big({d \over dt}\Big)_{t=0},
\label{vector}
\ee  where $*$ denotes as usual the differential map between tangent spaces.
We denote  by $\Omega^{*}_{X}(M)$ the kernel of  ${\cal L}(X)$ in
$\Omega^{*}(M)$. We consider now the polynomial ring generated by a generator
$u$ of degree $2$ over  $\Omega^{*}(M)$, denoted by 
$\Omega^{*}(M)[u]$. On this ring we define the equivariant exterior
derivative with respect to $X$ as follows: 
\be d_{X} \omega =d\omega - u \iota (X) \omega,\,\,\,\,\,\ \omega \in
\Omega^{*}[u], \label{ext}
\ee  Notice that
\be d_X^2 =- u{\cal L}(X),
\label{nilp}
\ee and therefore $d_{X}$ is nilpotent on 
 $\Omega^{*}_{X}(M)[u]$. The elements of $\Omega^{*}_{X}(M)[u]$ are  the
equivariant differential forms with respect to the vector field action 
associated to $X$. An equivariant differential form $\omega$ verifying $d_X
\omega =0$ is said to be {\it  equivariantly closed} with  respect to $X$.
Notice that, if $\omega
\in \Omega^{*}(M)[u]$ and $d_X \omega =0$, necessarily $\omega \in 
\Omega^{*}_{X}(M)[u]$ because of (\ref{nilp}).

When $X$ corresponds to a circle ($U(1)$) action on $M$, the above
construction  recovers precisely the Cartan model of the previous subsection.
The reason is that,  if $G=U(1)$, the symmetric exterior algebra has a single
generator:
\be S {\bf g}^* ={\bf R}[u]
\label{solito}
\ee associated to the vector field $X$ on $M$. The invariant subcomplex
(\ref{invi}) has  a very simple description in the abelian case because
$S{\bf g}^*$ is already invariant: 
\be
\Omega^{*}_{U(1)}(M) = \Omega^{*}_{X}(M)[u].
\label{guau}
\ee

Given a closed invariant differential form, {\it i.e.}, a form $\omega \in
\Omega^{*}_{X}(M)$ with $d\omega=0$,  we don't get an equivariantly closed
differential form unless $\iota (X) \omega =0$. But it might be possible to
find some polynomial $p$ in the ideal generated by $u$ in 
$\Omega^{*}_{X}(M)[u]$ such that the resulting form $\omega '=\omega +p$ is
equivariantly closed.  The form $\omega '$ is called an {\it equivariant
extension} of $\omega$. In section 3 we will see some interesting examples
of  equivariant extensions with a deep geometrical interpretation.

\section{The Mathai-Quillen formalism}

The Mathai-Quillen formalism \cite{mq} provides an explicit representative of
the Thom form of a vector bundle $E$. Usually this form is introduced in the
following way: consider an oriented vector bundle $\pi : E \rightarrow M$
with fibre $V={\bf R}^{2m}$, equipped with an inner product $g$ and a
compatible connection $D$ verifying: 
\be d(g(s,t))=g(Ds,t)+g(s,Dt),\,\,\,\,\,\ s,t \in \Gamma(E).
\label{comp}
\ee As our vector bundle is oriented and has  an inner product, we can reduce
the structure group to $G=SO(2m)$. Let $P$ be the principal 
$G$-bundle over $M$ such that $E$ is an associated vector bundle: 
\be E=P\times_G V.
\label{aso}
\ee In particular, $P \times V$ can be considered as a principal $G$-bundle
over 
$E$. Given a principal $G$-bundle $\pi: P \rightarrow M$, a differential form 
 $\phi$ on
$P$ descends to a form ${\bar \phi}$ on $M$ if:

a) $\phi (X_1, \cdots , X_q)=0$ whenever one of the $X_i$ is vertical. In
this case, 
$\phi$ is called horizontal.

b) $R^{*}_g\phi =\phi$, {\it i.e.}, $\phi$ is invariant under the $G$ action. 

These forms are called {\it basic}. The projected form $\bar \phi$ verifies:
\bea {\bar \phi}(V_1, \cdots, V_{q})&=&\phi(X_1, \cdots, X_q), \nonumber\\
(d{\bar
\phi})(V_0, \cdots, V_{q})&=&(d\phi)(X_0, \cdots, X_q),
\label{pro}
\eea where the $X_i$ are such that $\pi_{*}X_i=V_i$. We can apply the same
procedure to  the principal bundle $P\times V \rightarrow P\times_G V$, and
we have in fact the  isomorphism:
\be
\Omega^{*}( P\times_G V) \simeq \Omega^{*}( P\times V)_{\rm basic}.
\label{ulala}
\ee  Suppose now that $P$ is endowed with a connection $\theta \in
\Omega^1(P, {\bf g})$ and  associated curvature $K$,  and consider the Weil
algebra of $G$, ${\cal W}(\bf g)$. As ${\bf g}={\bf so}_{2m}$,  the
generators are antisymmetric matrices
$\theta_{ij}$ (of degree $1$) and 
$K_{ij}$ (of degree $2$) (notice that we use the same notation for the
connection and  curvature of $P$ and the generators of ${\cal W}(\bf g)$). 
The fact that
${\cal W}(\bf g)$ provides  a universal realization of the relations defining
the curvature and connection on 
$P$ gives the {\it Chern-Weil homomorphism}:
\be w: {\cal W}({\bf g}) \too \Omega^{*}(P),
\label{chernweil}
\ee defined in the natural way according to the expansion of $\theta$ and $K$
in  (\ref{lilith}) (for $G=SO(2m)$, the map $w$ is just the correspondence 
between the generators of ${\cal W}(\bf g)$ and the  entries of the
antisymmetric matrices  for the curvature and connection in $P$). The
Chern-Weil  homomorphism maps the universal connection and  curvature in the
Weil algebra to the actual connection and curvature in $P$.  Combined with
the lifting of forms from $V$ to $P\times V$, we obtain another  homomorphism:
\be w \otimes \pi_2^{*}: {\cal W}({\bf g}) \otimes \Omega^{*}(V) \too 
\Omega^{*}(P \times V),
\label{maismapa}
\ee where $\pi_2: P\times V \rightarrow V$ is the projection on the second
factor. It is  easy  to see that the basic subalgebra $\Omega_{G}(V)$ maps to
the basic differential forms  on $P\times V$, and therefore to the
differential forms of the associated vector bundle.  This is the geometric
context of the Mathai-Quillen construction.
 
The universal Thom form $U$ of Mathai and Quillen is an element in ${\cal
W}({\bf g}) \otimes \Omega^{*}(V)$ given by:   
\be U= (2 \pi)^{-m}{\rm Pf}(K){\rm exp}\{ -x_ix_i-(dx_i +\theta
_{il}x_l)(K^{-1})_{ij} (dx_j + \theta_{jm}x_m)\}. \label{Thomform}
\ee In this expression  
$x_i$ are orthonormal coordinate functions on $V$, and $dx_i$ are  their
corresponding differentials. $K$ and $\theta$ are the antisymmetric matrices 
of generators in ${\cal W}(\bf g)$. Notice that this expression  includes the
inverse of $K$, and in fact it should be properly understood, once the
exponential is expanded, as:
\be {\pi}^{-m}{\rm e}^{ -x_ix_i}\sum_{I} \epsilon(I,I'){\rm Pf}({1 \over 2}
K_{I}) (dx +\theta x)^{I'},
\label{Uromano}
\ee where $I$ denotes a subset with an even number of indices, $I'$ its
complement and $\epsilon(I,I')$ the signature of the corresponding
permutation. The equivalence of the two representations is easily seen using
Berezin integration. To this end, introduce a Grassmann orthonormal
coordinate for the fibre 
$V$, $\rho_i$. Then the universal Thom form can be written as:
\be
 U={\rm e}^{-|x|^2} \int D \rho \,\ {\rm exp}\Big( {1 \over 4} 
\rho_i K_{ij} \rho_j  + i\rho_i(dx_i+\theta_{ij} x_j)   \Big), 
\label{mathai}
\ee  and the expansion of this expression leads precisely to
(\ref{Uromano}).  Of course, the expression (\ref{Thomform}) is easier to
deal with, and in fact we can check its properties taking $K^{-1}$ as a
formal  inverse of
$K$. This is because we can consider (\ref{Thomform}) as an element of the 
ring of fractions with ${\rm det}(K)$ in the denominator. As ${\rm det}(K)$
is closed, we can extend the exterior derivative as an algebraic operator to
this localization
\cite{mq}. We will use this  principle in section 3 to check the equivariantly
closed character of the equivariant extension of the Thom form with  respect
to a vector field action. 

One can check that $U$ in (\ref{Thomform}) is a basic form, so it belongs to 
$\Omega_G(V)$, and also that it is equivariantly closed. The image of $U$
under the  map (\ref{maismapa}) is a closed differential form in
$\Omega^{2m}(E)$. Notice that the  resulting form  has no compact  support in
the fibre, but a Gaussian decay along it. We can define a cohomology  of
rapidly decreasing forms $H^{*}_{rd}(E)$ on the fibre $V$ analogous to the
 cohomology of forms with  compact support on $V$. The usual results about
the Thom class also hold in this  slightly generalized setting \cite{mq}. In
particular, the Thom class can be uniquely  characterized as a cohomology
class in  $H^{2m}_{rd}(E)$ such that its integration
 along  the fibre is $1$ \cite{botttu}. All this can be easily checked for the
universal  representative given in (\ref{Thomform}), and therefore the image
of $U$ in 
$H^{2m}_{rd}(E)$ is in fact a differential form representing the Thom class
of $E$.  
 
One can also obtain a  universal Thom form in the Cartan model of the
$G$-equivariant cohomology, using the  algebraic isomorphism between the
respective complexes. This  amounts to put the generator $\theta$ equal to
zero. But if we apply the Chern-Weil  homomorphism to the resulting
equivariant differential form, we don't obtain a  basic differential form in
$\Omega(P\times V)$. This is because the horizontal  character is lost. One
must then enforce a horizontal projection, and this gives  an alternative
representative which  is useful in Topological Gauge Theories \cite{aj, cmr}.
In the  next section we will present in detail this construction,
generalizing it to  the equivariant extension associated to a vector field
action.

\section{Equivariant extension of the Thom form}

The motivation for the equivariant extension of the Thom form in the
framework of the  Mathai-Quillen formalism is twofold. On one hand, many
interesting moduli problems  involve a vector field action or a circle action
which can be very useful to obtain  geometrical information and to relate
different moduli spaces. This is the case  of the Hitchin equations on a
Riemann  surface \cite{jichi}. The same ideas were used by Thaddeus \cite{th}
to relate  the moduli space of flat connections on a Riemann surface to the
moduli space of abelian  vortices, and recently it has been argued \cite{pt,
don}  that they can shed light on the  relation between the Seiberg-Witten
and the Donaldson invariants. This last point will  be developed in Chapter
5.  It should then be very interesting to have an equivariant extension of
the Thom form  with respect to a vector field action. This extension would
naturally arise in these
 kinds  of moduli problems and, combined with fixed point theorems for
equivariant cohomology,  could give new tools to compute invariants
associated to these moduli spaces.

The second motivation to perform this extension comes from the construction 
and interpretation of Topological Quantum Field Theories. As we have
mentioned,  many twisted $N=2$ supersymmetric theories can be interpreted  in
terms of the Mathai-Quillen  formalism. We will present an example in Chapter
5.  However, there are some twisted $N=2$  theories which do not  have a
clear formulation in the Mathai-Quillen framework, and therefore their 
geometrical structure is not very well understood. One should then look  for
generalizations of this formalism to take into account the rich  topological
structures hidden in the supersymmetry algebra. In particular, the $N=2$ 
algebra with central charges leads to twisted models which has been
sometimes  misleading, because the additional terms in the Lagrangian and in
the  BRST transformations seem to spoil the topological invariance of the
theory.  The equivariant extension that we will construct in this section is
the appropriate  framework to understand the geometry involved in these
twisted models. 
 
From a mathematical point of view, this construction can be regarded as a 
generalization of the equivariant  extensions of the curvature considered in
\cite{ab, bottone, botttwo, berlin}.
 It is clear that, as the Mathai-Quillen representative involves the
curvature of the vector bundle, we need an explicit expression for the
equivariant extension of the curvature form. In the first  subsection we will
review this construction for general vector bundles from the point of view of
equivariant cohomology \cite{ab, bottone, botttwo}, and in subsection 2  we
will proceed in  the same way to obtain the equivariant extension of the 
curvature for principal bundles
\cite{berlin}. In subsections 3, 4, 5 we construct the eauivariant extension
of the  Thom form with respect to a vector field action in three different
situations. Finally,  in subsection 6 we give an illustrative,
two-dimensional application: topological  sigma models with a Killing, almost
complex action on an almost Hermitian target space.  We recover in this way
the topological sigma model with potentials of \cite{pot},  obtained from the
twisting of the $N=2$ supersymmetric sigma models with potentials  formulated
in \cite{agone}.   
     
\subsection{Equivariant curvature for vector bundles} 

Let $\pi: E \rightarrow M$ be a real vector bundle.  We suppose that there is
a vector field $X$ acting on $M$, and also an ``action" of this field on  $E$
compatible with the action on $M$. With this we mean \cite{ab, botttwo} that
there is a differential operator  $\Lambda$ acting on the space of sections of
 $E$, $\Gamma (E)$: 
\be
\Lambda: \Gamma (E) \rightarrow \Gamma (E),
\label{opebot}
\ee that satisfies the derivation property
\be
\Lambda (fs)=(Xf)s+f \Lambda s,\,\,\,\,\ f \in C^{\infty}(M), \,\ s \in
\Gamma (E).   
\label{deriv} 
\ee We will be particularly interested in the case in which there is a vector
field
$X_E$  acting on $E$ in a compatible way with the action of $X$ on
$M$. With this we mean the following:  let $\hat \phi_t$, $\phi_t$ be the
one-parameter flows corresponding to $X_E$, $X$, respectively. Then the 
following conditions are verified:

i) $\pi \hat \phi_t = \phi_t \pi$, {\it i.e}, the one-parameter flows
intertwine with the  projection map of the bundle;

ii) the map $E_m \rightarrow E_{\phi_t m}$ between fibres is a vector space
homomorphism.

Notice that, if $X_E$, $X$ are associated to circle actions on $E$, $M$, the
above conditions  simply state that $E$ is a $G$-bundle over the
$G$-space $M$, with $G=U(1)$ (for $G$-bundles, see for instance \cite{kth}).
 An obvious consequence of (i) is that $X_E$ and $X$ are $\pi$-related: 
\be
\pi_{*}X_E = X.
\label{pirel}
\ee When there is a vector field $X_E$ acting on $E$ in the above way the
operator
$\Lambda$ is naturally defined as: 
\be (\Lambda s)(m) = \lim _{t \rightarrow 0} {1 \over t}[s(m) -\hat \phi_t
s(\phi_{-t}(m)) ].  
\label{opetwo}
\ee  It's easy to see that, because of condition (i) above, $\hat \phi_t
s(\phi_{-t}(m))$ is in fact  a section of $E$, and using (ii) one can check
that the derivation property (\ref{deriv}) holds.  We say that the section $s
\in
\Gamma (E)$ is {\it invariant} if $\hat \phi_t s(\phi_{-t}(m))= s(m)$, for
all $t\in I$, $m \in M$. This is equivalent to $\Lambda s =0$. If $s$ is an
invariant  section, $X_E$ and $X$ are also $s$-related: 
\be s_{*}X = X_E.
\label{srel}
\ee

Consider now a connection $D$ on the real vector bundle $E$ of rank $q$.  We
say that $D$ is {\it equivariant} if it commutes with the operator $\Lambda$.
Let's write this condition with  respect to a frame field
$\{ s_i\}_{i=1, \cdots, q}$ on an open set $U \subset M$. We define the
matrix-valued function $\Lambda^{j}_i$ and one-form $\theta^{j}_{i}$ on
$U$ by:
\be
\Lambda s_i =\Lambda^{j}_is_j,\,\,\,\,\ Ds_i=\theta^{j}_{i}s_j.
\label{matrices}
\ee Of course, $\theta^{j}_{i}$ is the usual connection matrix. Under a
change of frame $s'=sg$,  where $g \in Gl(q,{\bf R})$, we can use the
derivation property of
$\Lambda$ (\ref{deriv}) to obtain the matrix  with respect to the new local
frame:\be
\Lambda'=g^{-1} \Lambda g +g^{-1}Xg.
\label{changeone}
\ee Imposing $\Lambda D = D\Lambda$ on the local frame $\{ s_i\}$ one gets 
\be d\Lambda^{j}_i+\theta^{j}_{k}\Lambda^{k}_i={\cal L} (X) \theta^{j}_{i}+
\Lambda^{j}_k\theta^{k}_{i}.
\label{conm}
\ee The next step to construct the equivariant curvature is to define an
operator
$L_{\Lambda}: \Gamma(E) \rightarrow \Gamma(E)$ given by 
\be L_{\Lambda}s=\Lambda s -\iota (X)Ds, \,\,\,\,\,\  s \in \Gamma(E).
\label{equiope}
\ee The matrix associated to this operator with respect to a local frame on
$U$ is 
\be (L_{\Lambda})^{j}_i=\Lambda^{j}_i-\theta^{j}_{i}(X).
\label{opelocal}
\ee Using (\ref{changeone}) and the usual transformation rule for the
connection matrix  it is easy to check that $(L_{\Lambda})^{j}_i$ is a
tensorial matrix of the adjoint type: under a change of local frame one has 
\be L_{\Lambda}'=g^{-1}L_{\Lambda}g.
\label{sinombre}
\ee We will compute now the covariant derivative of the matrix $L_{\Lambda}$.
Using (\ref{conm}) we get:
 \bea DL_{\Lambda} &=& dL_{\Lambda}+[\theta, L_{\Lambda}]\nonumber \\
             &=&d\Lambda + [\theta, \Lambda]-({\cal L}(X)-\iota(X) 
d)\theta-[\theta,\iota(X)\theta]\nonumber \\ &=&\iota(X) (d\theta +\theta
\wedge
\theta) =\iota(X) K,
\label{cov}
\eea where $K$ is the curvature matrix.

We can introduce now the {\it equivariant curvature} $K_X$ for the vector
bundle case, defined as follows: 
\be K_X=K+uL_{\Lambda}.
\label{eqcurv}
\ee This is not an equivariant differential form, not even a global
differential form on $M$. To achieve this we have to introduce a symmetric
invariant polynomial  with $r$ matrix entries, $P(A_1, \cdots, A_r)$.
Consider  then the following quantity \cite{bottone}:
\be P_X = P(K_X, \cdots, K_X) = \sum_{i=0}^{r} u^{i} P_{K}^{(i)},
\label{eqpol}
\ee where
\be P_{K}^{(i)}= {r \choose i} P(\overbrace{L_{\Lambda}, \cdots,
 L_{\Lambda}}^{i};K, \cdots, K).
\label{noname}
\ee Notice that, as $L_{\Lambda}$ is a tensorial matrix of the adjoint type,
$P_X$ is a globally defined differential form in $\Omega^{*}_{X}(M)[u]$. 
Using (\ref{cov}) and the properties of symmetric invariant polynomials it is
easy to prove that 
 \be
\iota(X) P_{K}^{(i)}=dP_{K}^{(i+1)},
\label{basic}
\ee and from this it follows that $P_X$ is  an equivariantly closed
differential form on $M$.

\subsection{Equivariant curvature for principal bundles}

Let $\pi: P \rightarrow M$ be a principal bundle with group $G$. We suppose
that  we have two vector fields $X_P$, $X$ acting on $P$ and $M$,
respectively. We will require that the one-parameter flow associated to $X_P$,
${\hat \phi_t}$, commutes with the right action of $G$ on $P$: 
\be {\hat \phi_t}(pg)=({\hat \phi_t} p)g, \,\,\,\,\ p\in P,\,\ g \in G.
\label{pfbcomp}
\ee In this case, if $\phi_t$ is the one-parameter flow associated to $X$ on
$M$, we have $\pi {\hat \phi_t}= \phi_t \pi$, and $X$ and $X_P$ are
$\pi$-related. The vector field $X_P$  is in addition right invariant: 
\be (X_P)_{pg} = (R_g)_{*p}(X_P)_p.
\label{rightinv}
\ee Let $\theta$ be a connection one-form on $P$, and consider the function
with values in the  Lie algebra of $G$, $\bf g$, given by $\theta
(X_P)=\iota(X_P)
\theta$. Using  (\ref{rightinv}) and the properties of the connection it is
immediate to see that $\theta(X_P)$ is a tensorial zero-form of the adjoint
type, {\it i.e.}
\be
\theta (X_P)_{pg} = \theta_{pg} ((R_g)_{*p}(X_P)_p)= ({\rm ad}_g^{-1}) \theta
(X_P)_p. \label{pfbadj}\ee Suppose now that the connection one-form verifies:
\be {\cal L}(X_P) \theta = 0.
\label{pfbeq} 
\ee This is the analog of having an equivariant connection for a vector
bundle. When (\ref{pfbeq}) holds we can construct an equivariant curvature
for the principal bundle in a natural way. First, notice that the covariant
derivative of
$\theta (X_P)$ is given by: \be D \theta (X_P) = -\iota (X_P)K,
\label{pfbcov}
\ee where $K$ is the curvature associated to $\theta$.  The equivariant
curvature is defined as:
\be K_{X_P}=K-u\theta (X_P).
\label{pfbeqcurv}
\ee With the help of an invariant symmetric  polynomial we can construct the
form in 
$\Omega^{*}(P)[u]$:
\be P_{X_P}=P(K_{X_P}, \cdots, K_{X_P}).
\label{pfbinv}
\ee Taking into account (\ref{pfbcov}) we can proceed as in the vector bundle
case and show  that $ P_{X_P}$ is an equivariantly closed differential form
on $P$ with respect to the action of $X_P$.  On the  other hand, because of
(\ref{pfbadj}) and the usual arguments in Chern-Weil theory, $P_{X_P}$ 
descends to a form in
$\Omega^{*}(M)[u]$, ${\overline P_{X_P}}$. When a form $\phi$ on
$P$ descends to a form ${\bar \phi}$ on $M$, and the vector fields $X_P$ on
$P$ and $X$ on $M$ are $\pi$-related, we can obtain from (\ref{pro})  the
following identities:
\bea &(\iota(X){\bar \phi})(V_1, \cdots, V_{q-1})=(\iota(X_P)\phi )  (X_1,
\cdots, X_{q-1}),\nonumber \\ &({\cal L}(X){\bar \phi})(V_1, 
\cdots, V_{q})=({\cal L}(X_P)\phi ) (X_1, \cdots, X_{q}),
\label{ident}
\eea where the $X_i$ are such that $\pi_{*}X_i=V_i$. It follows from
(\ref{ident}) and  (\ref{pro}) that, if $P_{X_P}$  is equivariantly closed on
$P$, then
${\overline P_{X_P}}$ is equivariantly closed on $M$. We have therefore
obtained an appropriate equivariant extension of the curvature of a principal
bundle.

\subsection{Equivariant extension of the Thom form: general case}

Now we want to find an equivariant extension of the Thom form for  a vector
field action, in the framework of the Mathai-Quillen formalism. If we look at
the expressions for  the equivariant extension of the curvature,
(\ref{eqcurv}) and (\ref{pfbeqcurv}), we see that they  involve  the
contraction of the connection form with a vector field. It is clear that for 
the algebraic elements in the Weil algebra this operation is not defined, and
therefore we won't work with the universal Thom form, but with the explicit
Thom form as an element of
$\Omega^{2m}_{rd}(E)$. This has also the advantage of showing explicitly the
geometry involved in the Mathai-Quillen formalism, which is sometimes hidden
behind the use of $G$-equivariant cohomology.

Recall that we defined an ``action" of a vector field on a vector bundle $E$
as an operator acting on the space of sections of this bundle, and therefore
not necessarily induced by an action of $X_E$ on $E$. In this case we cannot
consider the complex $\Omega^{*}_{X_E}(E)[u]$. However, given an invariant
section $s$ of this bundle,  we can construct an equivariant extension of the
pullback $s^{*}\Phi (E)$ on $M$. 

In the framework of the Mathai-Quillen formalism we need an inner product on
$E$, $g$, and a compatible  connection $D$.  Once we take into account the
action of a vector field $X$ on $M$, and the compatible  operator on sections
$\Lambda$, we need additional assumptions to  construct our equivariant
extensions.  First of all, we assume, as in the previous subsection, that the
connection $D$ is equivariant.  We also assume that the inner product is
invariant with respect to the compatible actions: 
\be {\cal L}(X)(g(s,t))=g(\Lambda s, t)+ g(s, 
\Lambda t),\,\,\,\,\,\ s,t \in \Gamma(E)
\label{cometrica}
\ee From (\ref{comp}) and (\ref{cometrica}) one gets the following  identity
for the operator $L_{\Lambda}$ defined in (\ref{equiope}):
\be g(L_{\Lambda}s,t)+g(s,L_{\Lambda}t)=0,\,\,\,\,\,\ s,t \in \Gamma(E)
\label{antiope}
\ee We suppose that our bundle $E$ is orientable,  and therefore we can
reduce the structural group to $SO(2m)$ and consider orthonormal frames 
$\{s_i\}_{i=1, \cdots, 2m}$ such that $g(s_i, s_j)=
\delta_{ij}$. With respect to an orthornormal frame,  the connection and
curvature matrices are antisymmetric, and because of (\ref{antiope}) 
$L_{\Lambda}$ and $K_X$ are antisymmetric too. 

Consider now a trivializing open covering of $M$, $\{U_{\alpha} \}$,  and the
corresponding orthonormal frames $\{ s_i^{\alpha}\}$. Let $s \in \Gamma(E)$
be an invariant section.  Then, the following form  is an equivariantly closed
differential form on $M$ and is an equivariant extension of the pullback of
the Thom class by $s$: 
\be
 s^{*} \Phi(E)^{\alpha}_{X}= (2 \pi)^{-m}{\rm Pf}(K_X){\rm exp}
\{ -\xi^{\alpha}_i\xi^{\alpha}_i -(d\xi^{\alpha}_i +\theta ^{\alpha}_{il}
\xi^{\alpha}_l)(K_{X}^{\alpha})^{-1}_{ij} (d\xi^{\alpha}_j +
\theta^{\alpha}_{jm}\xi^{\alpha}_m)\},
\label{eqoneThom}
\ee where $s=\xi_i s_i^{\alpha}$ is the local expression of $s$ on
$U_{\alpha}$, and
$\theta^{\alpha}$ and $K_{X}^{\alpha}$ are respectively the connection and
the  equivariant curvature matrices  (the equivariant curvature is the one
given in (\ref{eqcurv})). Both are defined with respect to the orthonormal
frame $\{ s_i^{\alpha}\}$.  

To prove our statement, we will show first of all that the  $ s^{*}
\Phi(E)^{\alpha}_{X}$ define a global differential form on $M$, {\it i.e.},
we will consider a change of  trivialization on the intersections $U_{\alpha}
\cap U_{\beta}$.  The transformations of the different functions  appearing
here are: 
\bea s^{\beta}&=&s^{\alpha}g_{\alpha \beta}, 
\,\,\,\,\,\ \xi^{\beta}=g^{-1}_{\alpha \beta}\xi^{\alpha},\nonumber \\
\theta^{\beta}&=&g_{\alpha \beta}^{-1} \theta^{\alpha} g_{\alpha \beta} 
+g_{\alpha
\beta}^{-1}dg_{\alpha \beta},\nonumber \\ K_{X}^{\beta} &=& g_{\alpha
\beta}^{-1}K_{X}^{\alpha}g_{\alpha \beta},
\label{trans}
\eea where $g_{\alpha \beta}$ are the transition functions and take values in
$SO(2m)$.  To check the invariance of (\ref{eqoneThom}) under this
transformation,  notice that ${\rm Pf}(K_X)$ is an invariant symmetric
polynomial for  antisymmetric matrices and therefore the results of section 2
hold. Also notice that 
$d\xi^{\alpha}_i +\theta ^{\alpha}_{il}\xi^{\alpha}_l$ transforms as a
tensorial matrix of the adjoint type  (because it is the local expression of
the covariant derivative $Ds$). It is easily checked that  $ s^{*}
\Phi(E)^{\alpha}_{X}$ equals $ s^{*} \Phi(E)^{\beta}_{X}$ on the intersections
$U_{\alpha} \cap U_{\beta}$,  and therefore the expression (\ref{eqoneThom})
defines a global differential form on $M$.

To prove that this differential form is in the kernel of $d_{X}$ it  is 
enough to do it for the local expression in (\ref{eqoneThom}), as $d_X$ is a
local operator. Again, by the results of section 2, ${\rm Pf}(K_X)$ is
already equivariantly closed, and we only need to check this property for the
exponent in (\ref{eqoneThom}). The computation is lengthy but
straightforward. Recall that $s$ is an invariant section, and locally this
can be written as: 
\be
\Lambda (\xi_i s_i)=X(\xi_i)s_i+\xi_i\Lambda_{ji}s_j=0.
\label{lolita}
\ee It follows then that
\be d_{X}d\xi_i=-uX(\xi_i)=u\Lambda_{ij} \xi_j,
\label{deman}
\ee and we get the following expression:
\bea & &d_{X} \{  \xi_i \xi_i+(d \xi_i +\theta _{il}\xi_l)(K_{X}^{-1})_{ij}
(d\xi_j + \theta_{jm}\xi_m) \} \nonumber \\ &=& 2 d\xi_i \xi_i + [u (
\Lambda_{il}\xi_l -\theta_{il}(X) \xi_l)+ d\theta_{il}\xi_l-\theta_{il}d
\xi_l] (K_{X}^{-1})_{ij}(d\xi_j + \theta_{jm}\xi_m)\nonumber \\ &-&(d\xi_i
+\theta _{il}\xi_l)(d_{X}K_{X}^{-1})_{ij}  (d\xi_j +
\theta_{jm}\xi_m)\nonumber \\ &-&(d\xi_i +\theta _{il}\xi_l)(K_{X}^{-1})_{ij}
[u (\Lambda_{jm}\xi_m -\theta_{jm}(X) \xi_m)+
 d\theta_{jm}\xi_m-\theta_{jm}d \xi_m].
\label{calculo}
\eea If we add to this $(\theta_{il}+\theta_{li})(d\xi_l  +\theta
_{lp}\xi_p)(K_{X}^{-1})_{ij} (d\xi_j +\theta _{jm}\xi_m)=0$, and we take into
account that
\be D(K_{X}^{-1})_{ij}=d(K_{X}^{-1})_{ij}+
\theta_{il}(K_{X}^{-1})_{lj}-(K_{X}^{-1})_{il}\theta_{lj},
\label{uapiti}
\ee then (\ref{calculo}) reads:
\bea  & &2 d\xi_i \xi_i+(K_{X})_{il}\xi_l  (K_{X}^{-1})_{ij}(d\xi_j +
\theta_{jm}\xi_m)\nonumber \\ &-&(d\xi_i +\theta
_{ip}\xi_p)[D(K_{X}^{-1})_{ij}-u
\iota(X) (K_{X}^{-1})_{ij}] (d\xi_j + \theta_{jm}\xi_m)\nonumber \\ &-&(d\xi_i
+\theta _{ip}\xi_p)(K_{X}^{-1})_{il}(K_{X})_{lm}\xi_m.
\label{marisa}
\eea We can compute $DK_{X}^{-1}-u\iota(X) K_{X}^{-1}$ by considering
$K_{X}^{-1}$ a formal inverse of $K_X$. Notice first that, because  of the
Bianchi identity and (\ref{cov}),   we have:
\be DK_X=u\iota(X) K, \,\,\,\,\ d_{X}K_X=-[\theta, K_X].
\label{covx}
\ee As $d_X$ extends to the ring of fractions with ${\rm det}( K_X)$  in the
denominator (because
${\rm det}( K_X)$ is $d_X$-closed), we have:
\be d_{X}K_{X}^{-1}=K_{X}^{-1}[\theta, K_X]K_{X}^{-1}=-[\theta, K_{X}^{-1}],
\label{judith}
\ee and finally we get:
\be DK_{X}^{-1}-u\iota(X) K_{X}^{-1}=d_{X}K_{X}^{-1}+[\theta, K_{X}^{-1}]=0
\label{clarita}
\ee Using (\ref{clarita}) and the antisymmetry of the matrices 
$(K_{X})_{ij}$, $\theta_{ij}$, we see that (\ref{marisa}) equals zero.
Therefore, (\ref{eqoneThom}) is  in the kernel of $d_X$, and according to
(\ref{nilp})  it is an equivariantly closed differential form.  It is clear
that it is an equivariant extension of the pullback $s^{*} \Phi(E)$, because
if we put $u=0$  we recover the pullback of the Mathai-Quillen form. 

\subsection{Equivariant extension of the Thom form: vector bundle case}

Now we will consider the case in which we have a vector field 
$X_E$ acting on the vector bundle $E$, and the action $\Lambda$ is the one
induced from it.  In this case it makes sense to construct an  equivariant
extension of the Thom form with respect to  the $X_E$ action. Again we will
proceed  locally and we will construct the extension on  trivializing open
sets $U_{\alpha} \times V$.

Let $\pi: E \rightarrow M$ be an orientable real vector bundle of rank $2m$
with  an action of a vector field $X_E$ compatible with an action of $X$ on
$M$ in the sense of  subsection 2.2. On the fibre $V={\bf R}^{2m}$ we choose
an orthonormal basis $\{e_i\}$ with respect to the  standard inner product
$(\,\ , \,\ )$ on it,  and we denote by $x_i$ the coordinate functions with
respect to this basis. Let $\{U_{\alpha}\}$ be a trivializing open covering
of $M$, with attached diffeomorphisms 
\be
\phi_{\alpha}:U_{\alpha} \times V \rightarrow \pi^{-1} (U_{\alpha}).
\label{trivial}
\ee If $g$ is the metric on $E$, we can reduce  the structural group in such
a way that 
\be g(\phi_{\alpha}(m,v),
\phi_{\alpha}(m,w))=(v,w)
\label{metricaigual}
\ee. This also gives an  orthonormal frame for each $U_{\alpha}$ in the
standard way:
\be s_i^{\alpha}(m)=\phi_{\alpha}(m,e_i).
\label{frame}
\ee We want to define a vector field action
${\hat X}_{\alpha}$ on each $U_{\alpha} \times V$ such that
\be (\phi_{\alpha}^{-1})_{*}(X_E)={\hat X}_{\alpha}.
\label{abre}
\ee To do this we will define a one-parameter flow ${\hat \phi_t}$ inducing
${\hat X}_{\alpha}$.  The natural way is to use the conditions of 
compatibility of the vector field actions. On the first factor, $U_{\alpha}$,
we use the restriction of the one-parameter flow associated to $X$,  and we
take the appropriate $t$-interval for this map to be well defined. On the
second factor we use the homomorphism between fibres given by the
one-parameter flow associated to $X_E$, $\phi_t^{E}$. Written in a local
trivialization, this homomorphism  means that, if $p \in E_m$,
$\phi_t^{E}p \in E_{\phi_t m}$, then
\be (\pi_2 \phi_{\alpha}^{-1})(\phi_t^{E}p)=\lambda (t, m) 
\pi_2\phi_{\alpha}^{-1}(p),
\label{homo}
\ee where $\pi_2$ denotes the projection of $\phi_{\alpha}^{-1}$  on the
second factor and $\lambda$ is an endomorphism of $V$ which depends on $t$,
the basepoint 
$m$ and the trivialization. Now  we can define:
\be {\hat \phi_t} (m,v) =(\phi_t (m),\lambda (t, m) v), 
\,\,\,\,\ (m,v) \in U_{\alpha} \times V.
\label{flujo}
\ee Notice that the endomorphism $\lambda$ verifies:
\be
\lambda (s,\phi_t(m)) \lambda (t,m)=\lambda (s+t, m).
\label{compositio}
\ee From the definition of ${\hat \phi_t}$ it follows that 
\be
\phi_{\alpha}^{-1}\phi_t^{E}= {\hat \phi_t} \phi_{\alpha}^{-1},
\label{phiconm}
\ee and this in turn implies (\ref{abre}). 

The procedure is now similar to the one presented in the preceding  section.
We define the  following form on $\Omega^{*}(U_{\alpha} \times V)[u]$:
\be
 \Phi(E) ^{\alpha}_{X}= (2 \pi)^{-m}{\rm Pf}(K_X){\rm exp}
\{ -x_i x_i -(dx_i +\theta_{il}x_l)(K_{X}^{\alpha})^{-1}_{ij} (dx_j +
\theta_{jm}x_m)\},
\label{eqdosThom}
\ee where $\theta_{ij}$, $(K_{X})_{ij}$ denote respectively the connection and
equivariant  curvature matrices associated to the orthonormal frame defined in
(\ref{frame}). The index $\alpha$ labeling the trivialization is understood. 
We want to check that (\ref{eqdosThom}) defines a global differential form on
$E$. First we will consider the behavior of $\omega^{\alpha}=
 \Phi(E) ^{\alpha}_{X}$ under a change of trivialization. The transition
functions  for the vector bundle are defined as  $g_{\beta
\alpha}=\phi_{\beta}^{-1} \phi_{\alpha}$, restricted as usual to  $\{ x
\}\times V$. The behavior of the connection and curvature matrices under the
change of trivialization is given in (\ref{trans}), and the gluing 
conditions for the elements in the trivializing open sets are 
\be (m, v)^{\beta}=(m, g_{\alpha \beta}^{-1} (v))^{\alpha}.
\label{gluon}
\ee The coordinate functions then transform as 
$x_i \rightarrow (g_{\alpha \beta}^{-1})_{ij}x_j$. Following the same steps
as in the preceding  section we see that the forms $\omega^{\alpha}$ do not
change when we go from the $\alpha$ description to  the $\beta$ description:
\be g_{\alpha \beta}^{*} \omega^{\alpha} =\omega^{\beta}.
\label{cambia}
\ee The forms $\omega^{\alpha}$ define the corresponding forms on 
$\pi^{-1}(U_{\alpha})$ by taking
$(\phi_{\alpha}^{-1})^{*}\omega^{\alpha}$ on these open sets.  On the
intersections we have, because of (\ref{cambia}),
\be
(\phi_{\alpha}^{-1})^{*}\omega^{\alpha}=(\phi_{\beta}^{-1})^{*}\omega^{\beta},
\label{nocambia}
\ee and therefore they define a global differential form on $E$. Now it is
clear that, if the $\omega^{\alpha}$ are in the kernel of $d_{\hat X}$, the
$(\phi_{\alpha}^{-1})^{*} \omega^{\alpha}$ are in the kernel of $d_{X_E}$. 
This is a consequence of the following simple result: if $f: M \rightarrow N$
is a differentiable map, $\omega \in \Omega^{*}(N)$, and $X_M$, $X_N$ are two
vector fields which are $f$-related, then
\be
\iota(X_M) f^{*}\omega = f^{*} \iota(X_N) \omega.
\label{simple}
\ee Using (\ref{simple}) and (\ref{abre}) we see that
\be d_{X_E}(\phi_{\alpha}^{-1})^{*}
\omega^{\alpha}=(\phi_{\alpha}^{-1})^{*}
\big(d-\iota ((\phi_{\alpha}^{-1})_{*}(X_E))\big)\omega^{\alpha}=
(\phi_{\alpha}^{-1})^{*}(d_{\hat X} \omega^{\alpha}).
\label{tudobem}
\ee To prove that the $\omega^{\alpha}$ are in the kernel of $d_{\hat X}$,
notice that the computation  is very similar to the one presented in the
preceding section. The only new thing we must compute is $d_{\hat X}(dx_i) =
-u {\cal L}({\hat X})x_i$. Using the definition of Lie  derivative and the
action of the one-parameter group associated to ${\hat X}$ and given in 
(\ref{flujo}), we get:
\be ({\cal L}({\hat X})x_i)(m,v)= -{d \over dt}\lambda_{ij}(-t,
m)\Big|_{t=0}x_j(v).
\label{lie}
\ee The matrix appearing in this expression is not new. To see it, notice
that the matrix representation of the operator $\Lambda$ with respect to the
orthonormal frame (\ref{frame}) is given by 
\be (\Lambda s_i^{\alpha})(m)= 
\lim_{t \rightarrow 0} { s_i^{\alpha}(m)-\phi_t^{E}  s_i^{\alpha}(\phi_{-t}
m) \over t}.
\label{opeframe}
\ee Using (\ref{frame}) and (\ref{phiconm}) we obtain:
\be
\phi_t^{E}  s_i^{\alpha}(\phi_{-t} m)=\phi_{\alpha}  {\hat \phi_t}
(\phi_{-t}m, e_i)=s_j^{\alpha}(m)
\lambda_{ji}(t,\phi_{-t} m),
\label{calculillo}
\ee and this gives
\be (\Lambda s_i^{\alpha})(m)=-s_j^{\alpha}(m) {d \over dt}
\lambda_{ji}(t,\phi_{-t} m) \Big|_{t=0}
\label{culillo}
\ee Finally, using (\ref{compositio}) and comparing (\ref{lie}) and
(\ref{culillo}) we get
\be ({\cal L}({\hat X})x_i)(m,v)=-\Lambda_{ij}(m) x_j(v).
\label{final}
\ee If we compare this expression to (\ref{deman}) we see  that the
computation of the equivariant  exterior derivative of $\omega^{\alpha}$
with  respect to ${\hat X}$ simply mimicks the one we did in the preceding
section. Therefore, the forms defined in (\ref{eqdosThom}) are in the kernel 
of $d_{\hat X}$ and the global differential form $\Phi(E)_{X}$ they induce on
$E$ is an equivariantly closed differential form because of (\ref{tudobem}).
It clearly equivariantly extends  the Mathai-Quillen expression for the Thom
form of the bundle.

Consider now an  invariant section $s \in \Gamma (E)$. Because of
(\ref{srel}) and (\ref{simple}) it is easy to see that  $s^{*}\Phi(E)_X$ is
an equivariantly closed differential form on the base manifold $M$. Of 
course the local expression of this form coincides with (\ref{eqoneThom}):
the map  $\phi^{\alpha}s: U_{\alpha}
\rightarrow U_{\alpha} \times V$ is given by  
\be (\phi^{\alpha}s)(m) = (m, \xi_i^{\alpha}(m)e_i),
\label{localmap}
\ee where we wrote $s=\xi_i^{\alpha}s^{\alpha}_i$.  As the local expression of
$s^{*}\Phi(E)_X$  is
\be (s^{*}\Phi(E)_X)^{\alpha}=(\phi^{\alpha}s)^{*} \Phi(E) ^{\alpha}_{X},
\label{maslocal}
\ee and from (\ref{localmap}) this amounts to substitute $x_i$  by
$\xi_i^{\alpha}$ in (\ref{eqdosThom}), we recover precisely (\ref{eqoneThom}).

Finally, we will give a Field Theory expression for $\Phi(E)_X$ using Berezin
integration. Introduce Grassmann variables $\rho_i$ for the local coordinates
of the fibre. The standard rules of Berezin integration \cite{mq, cmr} give
the following representative for the local expression (\ref{eqdosThom}): 
\be
\Phi(E)^{\alpha}_X = \pi^{-m} {\rm e}^{-x_i x_i} \int {\cal D} \rho \,\ {\rm
exp}
\Big( {1 \over 4} \rho_i K_{ij} \rho_j + {u \over
4}\rho_i(L_{\Lambda})_{ij}\rho_j +i\rho_i(dx_i +\theta_{ij}x_j) \Big), 
\label{campos}
\ee which is the equivariant extension of (\ref{mathai}).  With this
expression at hand, one can also introduce  the standard objects in
Cohomological Field Theory, namely a gauge fermion and a BRST complex. The
gauge fermion approach was first used in 
\cite{labper} (for a review on subsequent developments, see \cite{blth}). 
Here we follow the reformulation given in \cite{cmr}, in the context of the 
Mathai-Quillen formalism. We introduce an auxiliary field
$\pi_i$ with the meaning of a basis of differential  forms $dx_i$ for the
fibre. The BRST operator is given by the $d_{\hat X}$ cohomology, and
therefore we have:
\be Q\rho_i =\pi_i, \,\,\,\,\,\,\ Q\pi_i=u\Lambda_{ij}\pi_j.
\label{brstfibra}
\ee On the original fields $x_i$ and the matrix-valued functions on 
$U_{\alpha}$, $\theta_{ij}$, 
$K_{ij}$, $(L_{\Lambda})_{ij}$, $Q$ acts  again as $d_{\hat X}$. The gauge
fermion is the same than the gauge fermion in the Weil model for the
Mathai-Quillen  formalism \cite{cmr}:
\be
\Psi = -\rho_i (ix_i +{1 \over 4}\theta_{ij} \rho_j +{1 \over 4}\pi_i), 
\label{gauge}
\ee and it is easily checked that $Q \Psi$ gives,  after integrating out the
auxiliary field $\pi_i$, the exponent in (\ref{campos}). This representative
will be useful to construct the equivariant  extension for topological sigma
models. Notice that in the expression (\ref{campos}) we can  work with a
non-orthonormal metric on
$V$ by introducing the corresponding jacobian in the  integration measure.

\subsection{Equivariant extension of the Thom form: principal bundle case}

 We will consider, finally, the case in which the vector bundle $E$ is
explicitly given as an associated vector bundle to a principal bundle $\pi: P
\rightarrow M$, {\it i.e.}, we consider the action of the  structural group
$G$ on
$P\times V$ given by $(p,v)g=(pg, g^{-1}v)$, and we form the quotient
$E=(P\times V) / G$. Notice that $P\times V$ can be considered as a principal 
bundle over $E$. We assume that we have a vector field action on $P\times V$
whose one-parameter flow  $\mu_t$ has the following structure: 
\be
\mu_t (p,v)=(\phi_t^{P}p, \lambda (t,p)v) \,\,\,\,\,\ p \in P, v \in V,
\label{masflujo}
\ee where $\lambda (t,p)$ is an endomorphism of $V$.  We also assume that
this flow conmutes  with the $G$-action on $P\times V$:
\be (\phi_t^{P}p)g=\phi_t^{P}(pg), \,\,\,\,\,\ \lambda  (t,pg)=g^{-1}\lambda
(t,p) g.
\label{conmutamas}
\ee Because of the above condition, a vector field action on $E$ is induced
in the natural way, and the one-parameter flow $\phi_t^{P}$ gives in turn a
vector field action on $M=P/G$ in the way considered in subsection 2, with
one-parameter flow
$\phi_t$. In addition, with  these assumptions, the vector field action on
$E$ is compatible with the vector field action on $M$ according to our
definition  in subsection 1. Condition (i) is immediate, and to see that
condition (ii) holds consider  a trivializing open covering for $M$,
$\{U_{\alpha}\}$, and the corresponding map $\nu_{\alpha}:
\pi^{-1}(U_{\alpha}) \rightarrow G$.  If $m \in U_{\alpha}$, $\phi_t(m) \in
U_{\beta}$,  the map between the fibres $E_m$, $E_{\phi_t m}$ is given by the
homomorphism 
\be
\nu_{\beta}(\phi_t^{P}p) \lambda (t,p) \nu_{\alpha}(p)^{-1}, \label{endob}
\ee where
$p \in \pi^{-1}(m)$. Using (\ref{conmutamas}) it is easy to see that
(\ref{endob})  only depends on the basepoint $m$ and $t$. The vector fields
on $P$, $E$ and $M$ will  be denoted, respectively,  by $X_{P}$, $X_E$ and
$X$.  Our last assumption is that there is an inner product $( \,\  , \,\ )$
on $V$ preserved  by both the action  of $G$ and the endomorphisms
$\lambda(t,p)$. As usual, this means that the matrix  
\be
\Lambda_{ij}(p)=\lim_{t \rightarrow 0}{1 \over t}  [\delta_{ij}-\lambda (t,
p)_{ij}]
\label{openuevo}
\ee is antisymmetric, where the components are taken with
 respect to an orthonormal basis 
${e_i}$ of $V$. If we regard $P\times V$ as a principal bundle, the second 
condition in (\ref{conmutamas}) imply that $\Lambda$ is a tensorial matrix of
the adjoint type. 

We will be particularly interested in the case in which $\lambda (t,p)$
doesn't depend on $p$. In this case $\Lambda$ is a constant matrix  commuting
with all the
$g \in G$ (and then with all the elements in the Lie algebra $\bf g$).  This
happens, for instance, if $G=U(m) \subset SO(2m)$ and $\Lambda$ has the
structure:
\be
\Lambda=\left(\begin{array}{cccc} 0&1 &{\ldots}&{} \\
                      {-1}&0&{\ldots}& \\
                      {\vdots}&{\vdots}&{\ddots}& {} \\
                        {} &{\ldots}&0&{1} \\
                        {} &{\ldots}&{-1}&0 \end{array}
\right).                
\label{matriz}
\ee This is in fact the situation we will find  in the application of our
formalism to  non-abelian monopoles on four-manifolds, in Chapter 5.

Let $\theta$ and $K$ be respectively the connection and curvature of $P$.
Assume now, as 
 in subsection 2, that ${\cal L}(X_P)\theta=0$, 
 and that $\Lambda$ is a constant matrix commuting with
 all the $A \in {\bf g}$. Then $D\Lambda =0$. 
 We want to construct an equivariantly closed differential
 form on $P \times V$ with respect to the vector field action ${\hat X}=(X_P,
X_V)$, where $X_V$ 
 is associated to the flow $\lambda(t)$. First of all we define an equivariant
curvature on 
 $P \times V$:
 \be
 K_X=K+u(\Lambda-\theta(X_{P})).
 \label{claudia}
 \ee Notice that $\Lambda -\theta(X_{P})$ is a tensorial matrix of  the
adjoint type, and if $P(A_1, \cdots, A_r)$ is an invariant symmetric 
polynomial for the adjoint action of ${\bf
 g}$, then we can go through the arguments of subsection 2.3 to show that 
 $P( K_X, \cdots, K_X)$ defines an equivariantly 
 closed differential form on $P \times V$. 
 The construction of the equivariant extension 
 of the Thom class is very similar to the ones
 we have done before, but now we define a form on 
 $P\times V$ and we will show that it descends
 to $E$. Consider then the following element in $\Omega^{*}(P \times V)[u]$:
 \be
 \Phi (P\times V)=(2 \pi)^{-m}{\rm Pf}(K_X){\rm exp}
\{ -x_i x_i -(dx_i +\theta_{il}x_l)(K_{X}^{\alpha})^{-1}_{ij} (dx_j +
\theta_{jm}x_m)\},
\label{eqtresThom}
\ee where $x_i$ are, as before, orthonormal coordinates  on the fibre $V$.
First we will check that the above form descends to $E$. For this we must 
check that it is right invariant and that it vanishes on vertical fields.  
The first property is easily verified using the expressions:
\be (R_g^{*} x_i)(v)=x_i(g^{-1}v)=g_{ij}^{-1} x_j(v), 
\,\,\,\,\,\,\ R_g^{*} dx_i=g_{ij}^{-1} dx_j.
\label{traslada}
\ee To check the horizontal character,  notice that  $K_X$ is horizontal (for
$K$ is horizontal and
$\Lambda- \theta (X_P)$ is a zero-form), and then we only have to check it for
$dx_i +\theta_{il}x_l$, as in \cite{mq}. Notice that we are considering $P
\times V$ as a principal bundle over $E$, and therefore a  fundamental vector
field
$A^{*}$ (corresponding to $A \in {\bf g}$) is induced by the  $G$-action on
both factors. Using the
 properties of the connection one-form and the action of $G$ on $V$, one
immediately gets:
 \be
 \iota(A^{*})\theta_{ij}=A_{ij}, \,\ \iota(A^{*})dx_i=  {\cal
L}(A^{*})x_i=-A_{ij}x_j.
 \label{hop}
 \ee
 We see then that $\Phi (P\times V)$ descends to $E$. If we set $u=0$,  this
computation can  be also understood in the framework of $G$-equivariant
cohomology in the  Weil model, substituting $\theta$ and $K$ by their
universal representatives  in the Weil algebra ${\cal W}({\bf g})$. However,
when we consider a vector field  action it is better to work with the
geometric realizations in order to  make sense of the inner products and Lie
derivatives with respect to the  vector field $X$. 

The fact that $\Phi (P\times V)$ descends to $E$ simplifies the computation of
$d_{\hat X}\Phi (P\times V)$. First, 
 we define a connection on $P\times V$ by pulling-back 
 the connection on $P$. The horizontal subspace at $(p,v)$ is given by $H_{p}
\oplus V$, 
 where $H_p$ is the horizontal subspace of $T_pP$. If we denote by $\Phi h$
the horizontal 
 projection of a form $\Phi$ on $P \times V$ that descends to $E$, we have:
 \be
 d\Phi=d\Phi h =D\Phi,
 \,\,\,\,\,\ \iota({\hat X})\Phi =\big(\iota({\hat X})\Phi \big)h.
 \label{desciende}
 \ee
 As $\theta$ vanishes on horizontal vectors, we can put it 
 to zero after computing the exterior
 derivative of (\ref{eqtresThom}). Also notice that the covariant derivative 
defined by the 
 pullback connection on $P \times V$ acts
 as the covariant derivative of $P$ on the differential
 forms in $\Omega^{*}(P)$, and as the usual exterior
 derivative on the forms in $\Omega^{*}(V)$. 
 
 Now we can compute $d_{\hat X}\Phi (P\times V)$ in a simple way.  Again we
only need to 
 compute
 the equivariant exterior derivative of the exponent:
 \bea 
 & &d_{\hat X} \{ x_i x_i + (dx_i +\theta_{il}x_l)(K_{X})^{-1}_{ij} (dx_j +
\theta_{jm}x_m)\}\nonumber \\ &=&2dx_ix_i +[K_{il}x_l-u({\cal
L}(X_V)x_i+\theta_{il}(X_P) x_l)] (K_{X})^{-1}_{ij}dx_j\nonumber \\
&-&dx_i[(DK_{X}^{-1})_{ij}-u\iota(X_P)(K_{X})^{-1}_{ij}]dx_j\cr
&-&dx_i(K_{X})^{-1}_{ij}[K_{jp}x_p-u({\cal L}(X_V)x_j+
\theta_{jp}(X_P) x_p)].
\label{eles}
\eea The computation of ${\cal L}(X_V)x_i$ is straightforward from the
definition (\ref{masflujo}) and one  obtains $-\Lambda_{ij}x_j$ as in
(\ref{final}).  Assuming (\ref{pfbeq}) we get $DK_{X}=u\iota(X_P)K$ and 
therefore, using the same arguments leading to (\ref{clarita}), we see that
(\ref{eles}) is zero. If we denote   by
${\tilde \pi}$ the projection of $P \times V$ on $E$, it follows from our
assumptions that 
${\tilde \pi}_{*}{\hat X} =X_E$, and therefore, using (\ref{ident}) we see
that the form induced by  (\ref{eqtresThom}) on $E$ is equivariantly closed
with respect to
$X_E$.

The above computation also shows the possibility of introducing a Cartan-like
formulation of the equivariant extension we have obtained. Consider the form
on
$\Omega^{*}(P\times V)[u]$  given by 
\be
 \Phi (P\times V)_{C}=(2 \pi)^{-m}{\rm Pf}(K_X){\rm exp}
\{ -x_i x_i -dx_i(K_{X}^{\alpha})^{-1}_{ij} dx_j \}.
\label{cartanlike}
\ee Clearly it is still invariant under the action of $G$, but the horizontal 
character fails. However we  can consider the  horizontal projection of this
form,
$\Phi(P\times V)_{C}h$, where the horizontal subspace is defined as before by
the pullback connection. This form coincides in fact with (\ref{eqtresThom}),
because the horizontal projection only applies to $dx_i$ and gives \be
(dx_i)h=dx_i+\theta_{ij} x_j.
\label{horizontes}
\ee When $u=0$, (\ref{cartanlike}) gives precisely the representative of the 
Thom form that one obtains from the Cartan model of $G$-equivariant
cohomology.  The interesting thing about this expression    is that when one
enforces the horizontal projection as in \cite{aj}, one obtains the adequate
formalism to Topological Gauge Theories. We will then follow this procedure
to obtain a representative which will be useful later. 

We will suppose, as in Chapter 1, section 4, that the principal bundle $P$ is
endowed  with a $G$-invariant Riemannian metric $g$, and the connection
$\theta$ is  given by (\ref{conecta}). With the assumptions we have made
concerning $P$,  the condition ${\cal L}(X_P)\theta=0$ is 
 equivalent to the metric being invariant under the vector field action. Now
we will write
 (\ref{cartanlike}) as a fermionic integral over Grassmann variables:
 \be
 \Phi (P\times V)_{C}=(\pi)^{-m}{\rm e}^{-x_ix_i}
\int {\cal D}\rho \,\ {\rm exp}
\Big({1 \over 4} \rho_i (K_X)_{ij} \rho_j +i\rho_idx_i \Big).
\label{fermi}
\ee As we want to make a horizontal projection of this form,  we can write
$K=d\theta=R^{-1}d\nu$,
 and for the equivariant curvature defined in (\ref{claudia}) we have:
 \be
 K_X=R^{-1}(d\nu-u\nu(X_P))+u\Lambda .
 \label{liv}
 \ee
 If we introduce Lie algebra variables $\lambda$, $\phi$  and use the Fourier
inversion formula of
 \cite{aj}, we get the expression:
 \bea
 \Phi (P\times V)_{C}&=&(2\pi)^{-d}(\pi)^{-m}{\rm e}^{-x_ix_i}\int  {\rm exp}
\Big({1 \over 4} \rho_i (\phi_{ij}+u\Lambda_{ij}) \rho_j
+idx_i\rho_i\nonumber \\ &+& i\langle d\nu-u\nu(X_P), \lambda \rangle_g
-i\langle \phi, R 
\lambda \rangle_g \Big){\rm det}R
\,\ {\cal D} \rho{\cal D}\phi {\cal D}\lambda,
\label{fourier}
\eea where $\langle \,\ , \,\ \rangle_g$ denotes the Killing form of ${\bf
g}$,  and
$d={\rm dim}(G)$.  Notice that in this expression the integration  over
$\lambda$ gives a $\delta$-function constraining $\phi$  to be
$K-u\theta(X_P)$, which  is precisely (\ref{pfbeqcurv}),  the equivariant
curvature of the principal bundle $P$. To enforce the horizontal projection,
we multiply by the normalized invariant  volume form ${\cal D}g$ along  the
$G$-orbits, and we can write \cite{aj, cmr}:
\be ({\rm det}R) {\cal D}g=\int {\cal D}\eta \,\ {\rm exp}i(\langle \eta, \nu
\rangle_g) ,
\label{medida}
\ee where $\eta$ is a fermionic Lie algebra variable.  Putting everything
together we  obtain a representative for the horizontal projection:
\bea
\Phi (P\times V)_{C}h&=&(2\pi)^{-d}(\pi)^{-m}{\rm e}^{-x_ix_i}\int  {\rm exp}
\Big({1 \over 4} \rho_i (\phi_{ij}+u\Lambda_{ij}) \rho_j\nonumber\\
 &+&i\rho_idx_i +i\langle d\nu-u\nu(X_P), \lambda \rangle_g \nonumber\\ &-&
i\langle
\phi, R \lambda \rangle_g+ i \langle \eta, \nu \rangle_g \Big) {\cal D}\eta
{\cal D}\rho {\cal D} 
\phi {\cal D} \lambda, 
\label{lagrangiano}
\eea where integration over the fibre is understood.  

We will introduce now a BRST complex in a geometrical way.
 As in the preceding section, we  introduce auxiliary fields $\pi_i$ with the
meaning of a basis of differential forms for  the fibre. The natural BRST
operator is precisely the $d_{\hat X}$ operator, but we must  take into
account that in
$\Phi(P \times V)_{C}h$ we have the horizontal projection  of $dx_i$, given in
(\ref{horizontes}). Acting with the equivariant exterior derivative and 
projecting horizontally, as we did in (\ref{eles}), we get: 
\be d_{\hat X}(dx_ih)=u\Lambda_{ij}x_j +(K_{ij}-u\theta_{ij}(X_P))x_j.
\label{misterios}
\ee Remembering that $\phi$ is equivalent to the equivariant curvature of
$P$,  the BRST operator for the fibre is naturally given by:
\be Q\rho_i =\pi_i, \,\,\,\,\,\,\ Q\pi_i=(u\Lambda_{ij}+\phi_{ij})\rho_j.
\label{alsacia}
\ee Following \cite{cmr} we introduce a ``localizing" and a ``projecting" 
gauge fermion:
\be
\Psi_{\rm loc}=-\rho_i(ix_i+{1 \over 4}\pi_i),\,\,\,\,\,\,\ 
\Psi_{\rm proj}=i\langle \lambda  ,\nu \rangle.
\label{fermiones}
\ee On the Lie algebra elements the BRST operator acts as:
\be Q\lambda =\eta, \,\,\,\,\,\,\ Q\eta=-[\phi, \lambda].
\label{mascus}\ee In order to obtain (\ref{lagrangiano}) from
(\ref{fermiones}) using the BRST  complex, we must also  take into account
the horizontal projection of forms on $P$, like in (\ref{misterios}),  and
the equivariant  exterior derivative is then given as
\be d-\iota(C\phi)-u\iota(X_P).
\label{masmisterio}
\ee Notice that $\phi$ is an element of the Lie algebra ${\bf g}$, and
therefore
$C\phi$ is  a fundamental vector field on $P$. Using (\ref{mascus}) and
(\ref{masmisterio}) as BRST operators acting on the gauge  fermions
(\ref{fermiones}),
  the topological Lagrangian (\ref{lagrangiano}) corresponding  to an
equivariant extension of the Thom form is recovered. 

The BRST complex we have introduced looks like a $G\times X_P$ equivariant 
cohomology, but one shouldn't take this analogy too seriously. If one
formulates
 this equivariant cohomology in the Weil model, the relation
$\iota(X_P)\theta =0$ should be introduced, as in (\ref{inner}).
 Clearly, this is not true  geometrically unless $X_P$ is horizontal. In
fact, this term appears in the equivariant curvature of the principal bundle,
and therefore  in the expression for $\phi$ once the $\delta$-function 
constraint has been taken into account. 

The last point we would like to consider is the pullback of the equivariant
extension we have  obtained for this case. As (\ref{eqtresThom}) descends to a
equivariantly closed differential form on $E$, we can pull it back through an
invariant section ${\hat s}: M \rightarrow E$ as we did in subsection 4. But 
every section of $E$ is associated to a $G$-equivariant map 
\be s:P \rightarrow V, \,\,\,\,\,\ s(pg)=g^{-1}s(p).
\label{eqsect}
\ee If ${\hat s}$ is invariant, then  the corresponding $s$ in (\ref{eqsect})
verifies:
\be s\phi_t^{P} = \lambda(t) s.
\label{dobleq}
\ee Consider now the map ${\tilde s}:P \rightarrow P\times V$ given by
${\tilde s}(p)=(p, s(p))$.  From the above it follows that  ${\tilde
s}^{*}\Phi(P\times V)$ is a closed equivariant differential form on $P$ with
respect to $X_P$, and in fact it descends to $M$, producing the same form we
would get had we used the section
${\hat s}$. We have then the commutative diagram: 
\be
 \matrix{ \Omega^{*}(P \times V)_{{\rm basic}, E} & {\longrightarrow}   &
\Omega^{*}(E)\cr                  
                {\tilde s}^{*} \Big \downarrow & &\Big \downarrow{\hat
s}^{*}\cr
                {\Omega^{*}(P )_{{\rm basic}, M}}&{\longrightarrow} 
&\Omega^{*}(M)\cr}
\label{diagramaconm}
\ee This diagram should be kept in mind in Topological Gauge Theories, where
the topological Lagrangian is usually a basic form on $P$ descending to $M$.
When considering the equivariant  extension of the Mathai-Quillen form we
will have the same situation, with an equivariantly  closed differential form
on $P$ descending to
$M$.

\subsection{An application: topological sigma models with potentials} 

Applying the previous formalism to the topological sigma model \cite{tsm} we
will obtain  the model of \cite{pot}, which was constructed by twisting an
$N=2$ supersymmetric sigma model  with potentials \cite{agone}. The
Mathai-Quillen formalism for usual sigma models can be found in \cite{am,
wu,cmr}. Topological sigma models involving group actions  on the target
manifold were also considered in
\cite{tsm}.  

Let $M$ be an almost Hermitian manifold on which a vector field $X$ acts
preserving the almost  complex structure $J$ and the Hermitian metric $G$:
\be {\cal L}(X)J={\cal L}(X)G=0.
\label{variedad}
\ee We have then a one-parameter flow $\phi_t$  associated to $X$ which is
almost complex with respect  to $J$:
\be
\phi_{t*}J=J\phi_{t*}.
\label{casicomplejo}
\ee

Let $\Sigma$ be a Riemann surface with a complex structure $\epsilon$ and
metric $h$ inducing  $\epsilon$. In the topological sigma model, formulated
in the framework of the Mathai-Quillen  formalism, one takes as the base
manifold
${\cal M}$ the space of maps  \be {\cal M}={\rm Map}(\Sigma, M)=\{ f: \Sigma
\rightarrow M, f \in C^{\infty} (\Sigma, M) \}.
\label{base}
\ee Given a $f \in {\cal M}$ we can consider the bundle over $\Sigma$ given
by 
$T^{*}\Sigma \otimes  f^{*}TM$, and define a bundle over ${\cal M}$ by 
giving the fibre on $f \in {\cal M}$:
\be {\cal V}_f=\Gamma(T^{*}\Sigma \otimes f^{*}TM)^{+},
\label{fibra}
\ee where $+$ denotes the self-duality constraint for the elements
 $\rho \in {\cal V}_f$:
\be J\rho\epsilon =\rho.
\label{selfdual}
\ee  There is a natural way to define a vector field action on ${\cal M}$ 
induced by the action of 
$X$ on $M$:
\be (\phi_t f)(\sigma)=\phi_t(f(\sigma)),
\label{arriba}
\ee and similarly we can define an action on the fibre ${\cal V}_f$:
\be ({\tilde \phi}_t \rho)(\sigma)=\phi_{t*}(\rho(\sigma)).
\label{alafibra}
\ee This action is well defined, {\it i.e.}, ${\tilde \phi}_t \rho$ verifies
the self-duality  constraint (\ref{selfdual}) when $\rho$ does, due to 
(\ref{casicomplejo}). It is also clear that the  compatibility conditions of
subsection 1 hold: first, $({\tilde \phi}_t \rho)(\sigma)$ takes values in 
$T^{*}_{\sigma}\Sigma \otimes T_{\phi_tf(\sigma)}M$,  therefore ${\tilde
\phi}_t
\rho \in  {\cal V}_{\phi_tf}$; second, the map (\ref{alafibra}) is clearly a
linear map  between fibres, as  it is given by the action of $ \phi_{t*}$.

Now we will define metrics on ${\cal M}$ and ${\cal V}$. Let $Y$, $Z$ vector
fields on ${\cal M}$.  We can formally define a local basis on $T{\cal M}$
from a local basis on $M$, given by functional  derivatives with respect to
the coordinates:
$\delta /\delta f^{\mu}(\sigma)$ \cite{wu}.  A vector field  on
${\cal M}$ will be written locally as: 
\be Y= \int_{\Sigma} d^2\sigma Y^{\mu}(f(\sigma)) {\delta \over \delta
f^{\mu}(\sigma) }.
\label{campo}
\ee    With respect to this local coordinate description we define the
metric  on
${\cal M}$ as:
\be (Y,Z)=\int_{\Sigma} d^2 \sigma  {\sqrt h}G_{\mu \nu}
Y^{\mu}(f(\sigma))Z^{\nu} (f(\sigma)).
\label{metricaloca}
\ee In a similar way, if $\rho$, $\tau \in {\cal V}_f$ have local coordinates 
$\rho^{\mu}_{\alpha}$, $\tau^{\nu}_{\beta}$, the metric on ${\cal V}_f$  is
given by:
\be (\rho, \tau)=\int_{\Sigma} d^2 \sigma {\sqrt h}G_{\mu \nu} h^{\alpha
\beta}
\rho^{\mu}_{\alpha} \tau^{\nu}_{\beta}.
\label{metricafibra}
\ee As $X$ is a Killing vector for the Hermitian metric $G$, both
(\ref{metricaloca})  and (\ref{metricafibra}) verify (\ref{cometrica}). Now
we will define a connection on
${\cal V}$ compatible  with (\ref{metricafibra}). In analogy with the local
basis for
$T{\cal M}$, we can construct a local basis of differential forms on
$\Omega^{*}({\cal M})$, ${\tilde d}f^{\mu}(\sigma)$, which is dual to $\delta
/\delta f^{\mu}(\sigma)$ in a functional sense: \be ({\tilde
d}f^{\mu}(\sigma))\big({\delta \over \delta f^{\nu}(\sigma ')}
\big) =\delta_{\nu}^{\mu}
\delta (\sigma-\sigma ').
\label{dualidad}
\ee Let $s$ be a section of ${\cal V}$, with local coordinates
$s_{\alpha}^{\mu}$. We will define the  connection by the local expression:
\be Ds_{\alpha}^{\mu} ={\tilde d} s_{\alpha}^{\mu} +
\Big( \Gamma_{\nu \lambda}^{\mu}+{1 \over 2} 
D_{\nu}J_{\kappa}^{\mu}J^{\kappa}_{\lambda}
\Big) s_{\alpha}^{\lambda}{\tilde d}f^{\nu},
\label{cabraloca}
\ee where ${\tilde d}$ is the exterior derivative on 
${\cal M}$, with local expression:
\be {\tilde d} s_{\alpha}^{\mu}=\int_{\Sigma} d^2 \sigma {\sqrt h} {\delta
s_{\alpha}^{\mu} 
\over \delta f^{\nu}(\sigma )}{\tilde d}f^{\nu}( \sigma).
\label{estereo}
\ee The connection defined in this way is induced by the connection on $M$
given  by:
\be D=D_{G}+ {1 \over 2}D_{G}J \cdot J,
\label{covam}
\ee where $D_G$ is the Riemannian connection canonically associated to the
Hermitian metric $G$ on $M$. The corresponding connection matrix on $M$ is
\be
\theta =\theta_G+ {1 \over 2}D_{G}J\cdot J,
\label{conexion}
\ee where $\theta_G$ is the Levi-Civita connection associated to the metric
$G$. Notice  that, if $M$ is K\"ahler, then $D_{G}J=0$ and the covariant
derivative reduces to the usual  form. It is easy to see that
(\ref{cabraloca}) is compatible both with the self-duality constraint and
with the metric (\ref{metricafibra}). The curvature of the connection
(\ref{conexion}) can be  easily computed using $J^2=-1$,
$D_G J \cdot J =-J \cdot D_G J$:
\be R={1 \over 2}(R_G-JR_GJ)-{1\over 4}DJ \wedge DJ,
\label{ccc}
\ee where $R_G$ is the curvature of the Levi-Civita connection.  

To define the usual topological sigma model we also need a section of ${\cal
V}$.  This section is essentially  the Gromov equation for pseudoholomorphic
maps $\Sigma
\rightarrow M$ \cite{gr}, and can be written as:
 \be s(f)={1 \over 2}(f_{*}+Jf_{*}\epsilon).
\label{seccionsigma}
\ee Using (\ref{selfdual}) it is easy to show that $s$ is invariant under the
vector field action on ${\cal M}$. The last ingredient we need to construct
the equivariant extension of the  Thom form is to check the equivariance of
the connection (\ref{cabraloca}). As the action of the  vector field $X$ on
${\cal M}$ is induced by the corresponding action on $M$, it is  sufficient
to prove the equivariance of the connection (\ref{covam}) (equivalently, if
we check the  equivariance in local coordinates for ${\cal M}$, ${\cal V}$,
we are reduced to a computation  involving the local coordinate expressions
of $X$ and $D$ on
$M$). If $X$ is a Killing vector  field for the metric $G$ one has ${\cal
L}(X)D_G=D_{G}{\cal L}(X)$. Using (\ref{variedad}) it is  clear that ${\cal
L}(X)$ commutes with $D$, hence $D$ is equivariant and also the connection on 
${\cal V}$ defined in (\ref{cabraloca}).

Therefore, we are in the conditions of subsection 4, and   we can construct
the equivariant extension of the Thom form introduced there. To do this  we
must first of all compute the operator $L_{\Lambda}=\Lambda-\theta(X)$ in
local coordinates.  As before, the computation reduces to a local coordinate
computation on the  target manifold $M$. Fist we will obtain $\Lambda$
through the equation (\ref{final}). Take as local  coordinates on the fibre
$\rho_{\alpha}^{\mu}(\sigma) $. We have: 
\be ({\cal L}(X)\rho_{\alpha}^{\mu})(\sigma)=\lim_{t \rightarrow 0}
{(\phi_{t*}\rho)_{\alpha}^{\mu}(\sigma)-\rho_{\alpha}^{\mu}(\sigma) \over t} =
\lim_{t \rightarrow 0}{1 \over t}\Big(  {\partial (u^{\mu}\phi_t) \over
\partial u^{\nu}}-
\delta_{\nu}^{\mu} \Big) \rho_{\alpha}^{\nu}(\sigma),
\label{maslie}
\ee where  $u^{\mu}$ are local coordinates on $M$ and we explicitly wrote the
jacobian matrix  associated to $\phi_{t*}$. The limit above is easily
computed once we take into account  that the one-parameter flow in local
coordinates 
$(u^{\mu}\phi)(t, u)= g^{\mu}(t,u)$ verifies the differential system: 
\be {\partial g^{\mu}(t,u) \over \partial t} =X^{\mu}(g(t,u)), 
\,\,\,\,\,\ g^{\mu}(0,u)=u^{\mu},
\label{sistema}
\ee where $X^{\mu}(g(t,u))$ is the local coordinate  expression of the vector
field
$X$  associated to the flow. Using (\ref{sistema}) we get:
\be ({\cal L}(X)\rho_{\alpha}^{\mu})(\sigma)= (\partial_{\nu}
X^{\mu})(f(\sigma))
\rho_{\alpha}^{\nu}(\sigma).
\label{otrolio}
\ee Taking into account that the indices for local  coordinates on ${\cal
V}_f$ are
$\mu$, $\alpha$,  we finally obtain:
\be
\Lambda^{\mu \alpha}_{\nu \beta}(f(\sigma))=-(\partial_{\nu} 
X^{\mu})(f(\sigma)) 
\delta_{\beta}^{\alpha}.
\label{mimatriz}
\ee Next we compute $\theta(X)$. Again, by (\ref{cabraloca}), we can compute
it for the  connection  matrix on $M$ (\ref{conexion}). Using
(\ref{variedad}) we get:
\be
\theta (X) ={1 \over 2} \Big( \theta_G (X)-J\theta_G(X)J \Big), 
\label{contraete}
\ee We can now write the BRST complex and the topological Lagrangian for this
theory,  following the prescriptions in (\ref{campos}) and (\ref{brstfibra}).
First of all,  we introduce the field content. We will be interested in the
pullback of the  Thom form by the section (\ref{seccionsigma}), and then we
have as fields the  map $f$, denoted now in coordinates by $x^{\mu}$, the
basis of differential forms 
${\tilde d}f^{\mu}=\chi^{\mu}$, the Grassmann coordinate on the fibre 
$\rho_{\alpha}^{\mu}$, and the auxiliary field on the fibre
$\Pi_{\alpha}^{\mu}$. The  BRST cohomology follows from the expression
(\ref{mimatriz}) and the  Lie derivative on the coordinate $x^{\mu}$ on $M$:
\bea {[}Q, x^{\mu}] &=& \chi^{\mu},\nonumber\\
\{Q, \chi^{\mu} \}&=&-u X^{\mu}, \nonumber\\
\{Q, \rho^{\mu}_{\alpha} \}&=&\Pi_{\alpha}^{\mu}, \nonumber\\ {[}Q,
\Pi_{\alpha}^{\mu}]&=&-u\rho^{\nu}\partial_{\nu}  X^{\mu}.
\label{tsmbrs}
\eea  To compute the terms in the Lagrangian, we must act on coordinate
fields for the fibre which are self-dual and verify (\ref{selfdual}). Using
this constraint it is easy to see that (\ref{contraete}) is  equivalent to
$\theta_G(X)$, and that the first  term in (\ref{ccc}) is equivalent to
$R_G$. Notice also that to compute the  covariant derivative of the section
we must apply (\ref{estereo}). In this case we  obtain the derivative with
respect to the Riemann surface coordinate  of the delta function
distribution, and then we obtain the  derivative of the field $\chi^{\mu}$.
The final expression is then:
\bea S&=& \int_{\Sigma}d^2\sigma {\sqrt h} \Big[{1\over 2}G_{\mu \nu}h^{\alpha
\beta} 
\partial_{\alpha}x^{\mu} \partial_{\beta}x^{\nu}+  {1\over 2}J_{\mu
\nu}\epsilon^{\alpha \beta}
\partial_{\alpha}x^{\mu} \partial_{\beta}x^{\nu} \nonumber\\ & &
\,\,\,\,\,\,\,\,\,\,\,\,\,\,\,\,\,\ -iG_{\mu \nu}h^{\alpha
\beta}\rho_{\alpha}^{\mu}
\Big(D_{\beta} \chi^{\nu} + {1\over
2}D_{\lambda}J^{\mu}_{\kappa}\partial_{\gamma}
x^{\kappa}\epsilon_{\beta}^{\gamma}\chi^{\lambda} \Big) \nonumber\\  & &
\,\,\,\,\,\,\,\,\,\,\,\,\,\,\,\,\,\ -{1\over 8}G_{\mu \nu}h^{\alpha
\beta}\rho_{\alpha}^{\mu}
\Big(R^{\nu}{}_{\lambda \kappa \tau}-{1\over
2}D_{\kappa}J^{\nu}_{\phi}D_{\tau}  J^{\phi}_{\lambda} \Big)
\chi^{\kappa}\chi^{\tau} \rho_{\beta}^{\lambda} \nonumber\\ & &
\,\,\,\,\,\,\,\,\,\,\,\,\,\,\,\,\,\ -{u \over 4} h^{\alpha \beta}
\rho^{\nu}_{\beta} D_{\nu}X_{\mu}\rho^{\nu}_{\alpha} \Big].
\label{tsmtodo}
\eea where $D_{\nu}$ is the Levi-Civita covariant derivative on $M$, and
$D_{\alpha}$ is its  pullback to $\Sigma$. The last term is $-u\rho_i
(L_{\Lambda})_{ij} \rho_j/4$ and comes  from the equivariant extension of the
curvature. This is precisely one of the extra terms obtained  in \cite{pot}
after the twisting of the $N=2$ supersymmetric sigma model with potentials.
In the  topological action of \cite{pot} there are also two additional terms
that in the topological  model come from a $Q$-exact fermion and have a
counterpart in the non-twisted action. Remarkably, these two terms can be
interpreted as the
$d_{X}$-exact equivariant differential form  that is added to prove
localization in equivariant integration \cite{berlin, blau}. We will present
the general setting and  then apply it to the equivariant extension of the
topological sigma model. As we will see in Chapter 5, the same construction
holds for non-abelian monopoles on four-manifolds. Notice  that, this
additional term being $d_{X}$-exact, we can multiply it by an  arbitrary
parameter $t$ without changing the equivariant cohomology class.  This can be
exploited to give a saddle-point-like proof of  localization of equivariant
integrals on the critical
 points of the vector field (or, equivalently, on  the fixed points of the
associated one-parameter action).  Suppose then that on the base manifold $M$
there is a metric $G$ and that the vector field $X$ acts as a Killing vector
field with respect to $G$. Consider the differential form given by 
\be
\omega_{X}(Y)=G(X,Y),
\label{laforma}
\ee $Y$ a vector field on $M$. As $X$ is Killing,  we have ${\cal L}(X)
\omega_X= 0$, and acting with $d_{X}$ gives the equivariantly exact
differential form 
\be d_{X}\omega_{X}=d\omega_{X}-uG(X,X).
\label{exacta}
\ee The appearance of the norm of the vector field $X$ in (\ref{exacta}) is
what gives  localization on the critical points of the vector field. In the
topological
 sigma model there is a metric on ${\cal M}$  given in (\ref{metricaloca})
which  is Killing with respect to the action of $X$ on ${\cal M}$, and
therefore we  can add the exact form (\ref{exacta}) to our  equivariantly
extended topological action. In fact (\ref{laforma}) is explicitly  given on
${\cal M}$ by:
\be
\omega_{X}=\int_{\Sigma} d^2 \sigma  {\sqrt h}G_{\mu \nu} X^{\mu}(f(\sigma))
\chi^{\nu}.
\label{explicito}
\ee We can then obtain (\ref{exacta}) in this case as
\be d_{X} \omega_{X} =\int_{\Sigma} d^2 \sigma {\sqrt h}\Big(\chi^{\mu} 
\chi^{\nu} D_{\mu}X_{\nu}- uG_{\mu \nu} X^{\mu}X^{\nu} \Big).
\label{masterminos}
\ee With (\ref{masterminos}), we recover all the terms of  the sigma model of
\cite{pot} beside  the usual ones, as we will see in a moment. As a last
remark, notice that the  observables of this Topological Quantum Field Theory
are naturally associated to  the equivariant cohomology classes on $M$ with
respect to the action of $X$ (see also \cite{tsm}). The equivariant 
extension of the topological sigma  model is thus the natural framework to
study quantum equivariant cohomology. 

Now we want to study this model as a twisted version of the non-linear 
topological sigma model with potentials. Twisted $N=2$ supersymmetric sigma 
models with potential terms associated to holomorphic Killing vectors have
been considered in
\cite{pot}. As in the ordinary case, the K\"ahler condition on $M$ can be
relaxed and the topological model exist for any almost Hermitian manifold
admitting at least one holomorphic Killing vector. Although most of what
comes out in our analysis is already in \cite{pot}, we will describe the
construction in some detail, in the  simple case of target manifolds that
are  K\"ahler.

Let $M$ be a $2d$-dimensional K\"ahler manifold endowed with a Hermitian
metric
$G$ and a complex structure $J$. This complex structure verifies
$D_\rho J^\mu{}_\nu=0$, where $D_\rho$ is the covariant derivative with the
Riemann connection canonically associated to the hermitian metric
$G$ on $M$. The action which results after performing the twist of the $N=2$
supersymmetric action given in \cite{agone} (with the functions $h$ and
$G^i$ set to zero) is \cite{pot}:
\bea S_1 &=& \int_\Sigma d^2 \sigma \sqrt{h}
\Big[{1\over 2} G_{\mu\nu}h^{\alpha\beta}\partial_\alpha x^\mu \partial_\beta
x^\nu - ih^{\alpha\beta}G_{\mu\nu}\rho^\mu_\alpha D_\beta\chi^\nu  -{1\over
8}h^{\alpha\beta}
R_{\mu\nu\sigma\tau}\rho^\mu_\alpha\rho^\nu_\beta\chi^\sigma\chi^\tau
\nonumber\\ &  &
\,\,\,\,\,\,\,\,\,\,\,\,\,\,\,\,\,\,\,\,\,\,\,\, + G_{\mu\nu}X^\mu X^\nu +
\chi^\mu\chi^\nu D_\mu X_\nu + {1\over 4}h^{\alpha\beta}
\rho_\alpha^\mu\rho^\nu_\beta D_\mu X_\nu\Big], 
\label{dos}
\eea We follow the notation in (\ref{tsmtodo}): $h$ is the metric on the 
Riemann surface $\Sigma$, $x^{\mu}$, ${\mu}=1,\dots,2d$, are bosonic fields
which  describe locally a map $f: \Sigma \rightarrow M$, as before, and 
$\rho_\alpha^\mu$,
$\mu=1,\dots,2d$, are anticommuting fields which are sections of ${\cal V}_f$
in (\ref{fibra}). The fields $\chi^\mu$ constitute a basis of differential
forms 
$\tilde d f^\mu$ and $X^\mu$ is a holomorphic Killing vector field on $M$
which besides preserving the Hermitian metric $G$ on $M$ it also preserves
the complex structure $J$. These two features are contained in the conditions
(\ref{variedad}) which in local coordinates are:
\be D_\mu X_\nu + D_\nu X_\mu =0,
\,\,\,\,\,\,\,\,\,\,\,\, J^\mu{}_\nu J^\nu{}_\rho  D_\mu X_\nu = D_\mu X_\nu.
\label{manzana}
\ee Notice that we are considering the model presented in \cite{agone} with
only one holomorphic Killing vector. This is the situation which leads to the
Topological Quantum Field Theory constructed in the previous section.

An important remark in the twisting of the topological sigma model leading
to  the action (\ref{dos}) is the following. $N=2$ supersymmetric sigma
models exists for flat two-dimensional manifolds. Their formulation on curved
manifolds implies the introduction of $N=2$ supergravity. The twisting is
indeed done on a flat two-dimensional manifold. Once the flat action is
obtained one keeps only one half of the two initial supersymmetries and
studies if the model exist for curved manifolds. It turns out that it exists
endowed with that part of the supersymmetry which is odd and scalar and often
called topological symmetry, $Q$, and that the resulting action is
(\ref{dos}). This procedure is standard in any twisting process. One might
find, however, that in order to have invariance under the topological
symmetry, $Q$, it is necessary to add extra terms involving the curvature to
the covariantized twisted action. As we will discuss in Chapter 5,
 this will be the case when considering non-abelian monopoles.

The $Q$-transformations of the fields are easily derived from the $N=2$
supersymmetric transformations in \cite{agone}. They turn out to be:
\bea [Q,x^\mu] &=& \chi^\mu, \nonumber\\
\{Q,\chi^\mu\} &=& X^\mu, \nonumber\\
\{Q,\rho_\alpha^\mu\} &=&-i(\partial_\alpha x^\mu+\epsilon_\alpha{}^\beta
J^\mu{}_\nu\partial_\beta x^\nu) -\Gamma_{\nu\sigma}^\mu
\chi^\nu\rho_\alpha^\sigma,\label{cuatro}
\eea where $\epsilon$ is the complex structure induced by $h$ on $\Sigma$. As
it is the case for the $N=2$ supersymmetric transformations in \cite{agone},
this symmetry is realized on-shell. After using the field equations one finds:
\bea [Q^2, x^\mu] &=& X^\mu, \nonumber\\ {[}Q^2, \chi^\mu]& =& \partial_\nu
X^\mu
\chi^\nu, \nonumber\\ {[}Q^2, \rho_\alpha^\mu] &=& \partial_\nu X^\mu
\rho_\alpha^\nu. 
\label{cinco}
\eea From these relations one can read off the action of the central-charge
generator in (\ref{orangedos}): $Z$ acts as a Lie derivative with respect to
the vector field $X^\mu$. This is exactly the action one finds for $Q^2$ in
(\ref{tsmbrs})  with $u=-1$.  In addition, the first two transformations in 
(\ref{cuatro}) are the same as the ones generated by $d_{\hat X}$, for this
value  of $u$. In order to compare the transformation for $\rho_\alpha^\mu$ in
(\ref{cuatro}) to the one in (\ref{tsmbrs}), we need first to introduce
auxiliary fields to reformulate the twisted theory off-shell. As shown in
\cite{tsm,tmtwo}, this is easily achieved by twisting the off-shell version
of the $N=2$ supersymmetric theory. In the twisted theory these auxiliary
fields, which will be denoted as $H_\alpha^\mu$, can be understood as a basis
on the fibre
${\cal V}_f$. Coming from an off-shell untwisted theory,  they enter in the
twisted action quadratically. As expected, after adding the topological
invariant term, 
\be S_2= {1\over 2} \int_\Sigma d^2 \sigma \sqrt{h}
 \epsilon^{\alpha\beta} J_{\mu\nu} \partial_\alpha x^\mu
\partial_\beta x^\nu,
\label{seis}
\ee the action of the off-shell twisted theory can be written in a $Q$-exact
form:
\bea & &\Big\{Q, \int_\Sigma \sqrt{h}\Big[ {1\over 2}h^{\alpha\beta}G_{\mu\nu}
\rho_\alpha^\mu(i\partial_\beta x^\nu +{1\over 2}H^\nu_\beta) + 
G_{\mu\nu}X^\mu\chi^\nu\Big]\Big\}\nonumber\\ & & \,\,\,\,\,\,\,\,\,\,\,\ =
S_1 + S_2 + {1\over 4} \int_\Sigma
\sqrt{h}h^{\alpha\beta}G_{\mu\nu}H^\mu_\alpha H^\nu_\beta,
\label{siete}
\eea where one has to take into account the
$Q$-transformation of the auxiliary field $H_\alpha^\mu$ and the corresponding
modifications of the  third $Q$-transformation in (\ref{cuatro}):
\bea
\{Q,\rho_\alpha^\mu\} &= & H^\mu_\alpha - i(\partial_\alpha
x^\mu+\epsilon_\alpha{}^\beta   J^\mu{}_\nu\partial_\beta x^\nu) -
\Gamma_{\nu\sigma}^\mu\chi^\nu\rho_\alpha^\sigma, \nonumber\\
{[}Q,H^\mu_\alpha] &= &i D_\alpha\chi^\mu +i\epsilon_\alpha{}^\beta  
J^\mu{}_\nu D_\beta\chi^\nu +\Gamma_{\nu\sigma}^\mu\chi^\nu H_\alpha^\tau
\nonumber\\ &+& {1\over 2}
R_{\sigma\nu}{}^\mu{}_\tau\chi^\sigma\chi^\nu\rho_\alpha^\tau +D_\tau X^\mu
\rho_\alpha^\tau. 
\label{ocho}
\eea

The auxiliary field $H^\mu_\alpha$ entering (\ref{siete}) and (\ref{ocho}) is
not the same as the one in (\ref{tsmbrs}) and (\ref{gauge}). Notice that in
the action resulting after computing $Q\Psi$ in (\ref{gauge}) the auxiliary
field does not enter only quadratically in the action. A linear term is also
present. In  (\ref{siete}), however, only a term quadratic in $H^\mu_\alpha$
appears. Also the  transformations (\ref{ocho}) and (\ref{tsmbrs}), as well
as the gauge fermion in (\ref{siete}) and (\ref{gauge}), are different.
Redefining the auxiliary field 
$H^\mu_\alpha$ as:
\be
\Pi^\mu_\alpha = H^\mu_\alpha + i(\partial_\alpha
x^\mu+\epsilon_\alpha{}^\beta   J^\mu{}_\nu\partial_\beta x^\nu) -
\Gamma_{\nu\sigma}^\mu\chi^\nu\rho_\alpha^\sigma,
\label{laurita}
\ee one finds that:
\bea
\{Q,\rho_\alpha^\mu\} &= & \Pi^\mu_\alpha, \nonumber\\ {[}Q,\Pi^\mu_\alpha]&
= & {\partial}_\tau X^\mu \rho_\alpha^\tau, 
\label{laspis}
\eea and  the resulting action has the form:
\be
\Big\{Q, \int_\Sigma \sqrt{h}\Big[ h^{\alpha\beta}G_{\mu\nu}
\rho_\alpha^\mu(i\partial_\beta x^\nu +{1\over
4}\Gamma_{\sigma\tau}^\nu\chi^\sigma\rho_\beta^\tau+{1\over 4}\Pi^\nu_\beta)
+  G_{\mu\nu}X^\mu\chi^\nu\Big]\Big\}.
\label{martia}
\ee Notice that the first term in (\ref{martia}) has the structure of the
gauge fermion  (\ref{gauge}). Now we have all the ingredients to compare the
theory constructed  geometrically  as an equivariant extension of the Thom
form, and the twisted $N=2$ supersymmetric  sigma model. The final form of
the BRST algebra of the twisted theory, after  the redefinition of the
auxiliary field, is identical to (\ref{tsmbrs}), with  the parameter $u=-1$.
This action differs from (\ref{tsmtodo}) in the terms which  originated from 
$ G_{\mu\nu}X^\mu\chi^\nu$ in (\ref{martia}).  These are precisely the terms
obtained in (\ref{masterminos}) in the previous section. Thus the twisted
theory corresponds to the one obtained from the equivariant extension of the
Mathai-Quillen formalism once the localization term (\ref{masterminos}) is
added. Notice that from the point of view of the equivariant extension of the
Mathai-Quillen formalism this additional term can be introduced with an
arbitrary multiplicative constant $t$.  Since the dependence on the parameter
$u$ of section 2 can be reabsorbed in the vector field $X$, one has a
one-parameter family of actions for a fixed Killing vector $X$. Since this
parameter enters only in a $Q$-exact term one expects that no dependence on
it appears in vacuum expectation values, at least if some  requirements on
compactness are fulfilled. This opens new ways to compute topological
invariants by considering different limits of this parameter, and the
resulting approach corresponds mathematically  to localization of  integrals
of equivariant forms. The simplest case, the homotopically  trivial maps from
the Riemann surface $\Sigma$ to the target space $M$, was  explicitly
considered in
\cite{pot}, and some classical localization results like  the Poincar\'e-Hopf
theorem were rederived in this framework.

This topological theory, as the non-extended one, can be generalized to the
case of an almost Hermitian manifold. We will not describe this
generalization here. The existence of this generalization was first discussed
in \cite{pot} and, as shown before, it can also be formulated from an
equivariant extension of the Mathai-Quillen formalism.

\chapter{Non-abelian monopole equations}

In this capter we will present a moduli problem which can be understood as a 
generalization of both Donaldson theory and of the Seiberg-Witten monopole
equations:  the non-abelian monopole equations. They involve the coupling of a
non-abelian  Yang-Mills field to a spinor with values in the Yang-Mills
bundle, and give  rise to a moduli space which includes in a natural way the
moduli space of ASD  connections. Indeed, the conditions for having a well
defined moduli space of  non-abelian monopoles are essentially the same as
those in the case of Donaldson  theory. In this chapter we give a preliminary
analysis of this moduli space and  we also study the equations on K\"ahler
manifolds, where they reduce essentially  to the well-known non-abelian
vortex equations.  The chapter is organized as follows: In section 1  we
briefly review the Seiberg-Witten equations. In section 2, the non-abelian
monopole  equations are introduced and the associated moduli space is
studied. In section 3,  we consider the equations on K\"ahler manifolds and
we give a description of the  space of solutions in terms of holomorphic
objects.

\section{The Seiberg-Witten monopole equations}

The Seiberg-Witten equations, introduced in \cite{mfm}, constitute perhaps
the  major breakthrough in the study of  the geometry of  four-manifolds
since Donaldson work. The theory is equivalent, in a very precise  way, to
Donaldson theory. Actually one can recover the Donaldson invariants from the 
Seiberg-Witten ones, but one can also take another point of view and study
the  Seiberg-Witten equations and their implications for four-manifold
topology without  taking into account  this connection. By now there is no a
precise mathematical proof of the equivalence,  and the link still is
provided by the Quantum Field Theory point of view. In this  section we will
introduce these equations from a mathematical point of view,  and in Chapter
7 of this work we will explain its origin in the realm of the 
Seiberg-Witten  exact solution of $N=2$ supersymmetric Yang-Mills theory. For
a more
 precise treatment of  the Seiberg-Witten equations, and some of the results
that have been obtained, one can  see \cite{don, morgan, marcolli, kmthree,
taubes, taubesdos,taubestres}. The point of  view of  Topological Quantum
Field Theory can be found in \cite{lmab, tribu}. 

Let $X$ be an oriented, closed, Riemannian four-manifold. If $X$ is Spin, we 
introduce a complex line bundle $L$ and a section of $S^+ \otimes L$, denoted
by
$M$. Recall  that $S^+$ is endowed with a Hermitian metric. Then, we can 
consider the trace-free endomorphism of $S^+$ 
\be i(M {\overline M})_0=i M \otimes {\overline M}-{i \over 2}|M|^2 {\rm Id}
\label{endo}
\ee It is easy to see that this endomorphism is also skew-adjoint. With
respect to a local  basis for $S^+$, the matrix associated to $i(M {\overline
M})_0$ is given by 
\be i(M_\alpha {\overline M}^{\beta})_0=
\left(\begin{array}{cc}{i\over 2}(|M_1|^2-|M_2|^2)& iM_1M^{*}_2\\
              iM_2M^{*}_1 & -{i\over 2}(|M_1|^2-|M_2|^2)\end{array} \right).
\label{sepuede}
\ee The Seiberg-Witten equations are a system of partial differential
equations for a pair  consisting of a connection on $L$, $A$, and a section
of $S^+ \otimes L$,
$M$. They are:
\bea
\rho(F^{+}_A)+2i(M {\overline M})_0&=&0,\nonumber\\ D_{L} M&=&0,
\label{swmon}
\eea where the map $\rho$ is given in the Appendix, (\ref{casi}), and 
$D_L$ is the twisted Dirac operator for the Dirac bundle $S^+\otimes L$.  In
this chapter, the Lie algebra of the unitary group consists of the 
self-adjoint matrices, as it is usual in Physics. In particular, the Lie
algebra of 
$U(1)$  is ${\bf R}$. The connection and curvature of $L$ are then real 
differential forms. It is  illuminating to write in components the first
equation in (\ref{swmon}). Using the  explicit representative of the $\rho$
map given in (\ref{mias}), we get the equations:
\bea F_{03}+F_{12}&=&|M_1|^2-|M_2|^2,\nonumber \\
F_{01}+F_{23}&=&M_1M^{*}_2+M^{*}_1M_2, \nonumber\\
F_{02}+F_{31}&=&i(M_1M^{*}_2-M^{*}_1M_2).
\label{swcompo}
\eea Another useful way of writing the Seiberg-Witten equations uses the
spinor notation  more familiar in Physics (Appendix, section 1). We represent
the self-dual part of  the field strength by the  symmetric matrix $F_{\alpha
\beta}$ given in (\ref{sime}) and (\ref{expli}). The  equations in this
notation read:
\bea F^{+}_{\alpha \beta}+i M_{(\alpha} {\overline M}_{\beta )}&=&0,
\nonumber \\  D^{{\dot \alpha}\alpha} M_{\alpha}&=&0,
\label{ab}
\eea    where ${\overline M}_{(\alpha}  M_{\beta )}={\overline
M}_{\alpha}M_{\beta} +{\overline M}_{\beta}  M_{\alpha}$ (notice that we do
not include the usual 
$1/2$ factor in this symmetrization). 

In case $X$ is not Spin, we choose on $X$ a $\sp^c$-structure (see Appendix, 
section 3) with associated  positive-chirality spinor bundle $S^+_{L^2}=S^+
\otimes L$, where the bundles $S^+$ and 
$L$ only exist locally. The determinant line bundle of this $\sp^c$-structure
is  then $L^2$, and is endowed with a $U(1)$ connection $A_{L^2}$ and
curvature 
$F_{L^2}$. In this case,  the Seiberg-Witten equations can be written as: 
\bea {1\over 2}\rho(F^{+}_{L^2})+2i(M {\overline M})_0&=&0,\nonumber\\
D_{L^2} M&=&0.
\label{swspinc}
\eea  One can wonder why we choose these normalizations for the
Seiberg-Witten  equations, following \cite{mfm}. From a mathematical point of
view, the  normalizations are irrelevant to the analysis of the moduli space.
But these equations have their origin  in a twisted $N=2$ supersymmetric
gauge theory, as we will see in next Chapter, and  the above choice of
factors is the standard one in Field Theory.  

The analysis of the above equations from the point of view of the definition
of  topological invariants follows the same lines of Donaldson theory. An
important  difference, and one of the main advantages of the Seiberg-Witten
equations, is that  the associated moduli space is compact. We won't review
here the construction of the  Seiberg-Witten moduli space and the definition
of the Seiberg-Witten invariants.  One can find them in \cite{mfm, morgan,
marcolli}.  Many of the relevant aspects will appear in the analysis of the
non-abelian  generalization.

\section{Non-abelian monopole equations}

In this section, we will study the non-abelian monopole equations and some 
aspects of their moduli space, following \cite{lmna}. We will consider the
equations  defined on an oriented, Riemannian, closed four manifold $X$. We
will also  suppose that $X$ is Spin, and we will denote the positive and
negative  chirality spinor bundles, as in the Appendix, by $S^+$ and $S^-$,
respectively.  The generalization to arbitrary four-manifolds can be done 
using a $\sp^c$-structure \cite{tqcd}. The obvious way to  construct the
non-abelian  moduli problem consists of considering, instead of a complex
line bundle, as in the Seiberg-Witten equations,
 a principal fibre bundle $P$ with some compact, connected, simple Lie group
$G$. The Lie algebra of $G$ will be denoted by ${\bf g}$. For the monopole
part, we need a vector  bundle $E$ associated to the principal bundle $P$ by
means of a  representation $R$ of the Lie group $G$, of rank $d_R$ (the
dimension of the representation), and endowed  with a $R(G)$-invariant metric
$g$.  The representation of the Lie algebra of $G$ associated to $R$  will be
denoted by ${\bf g}_E$. As in the previous section, the Lie algebra is
defined  with a $i$ factor in the exponential map. In this way, the Lie
algebra of the  unitary group is taken as the set of Hermitian matrices.  
The data for  the non-abelian monopole equations are a connection $A$ on $E$
and a  section of the twisted bundle $M 
\in \Gamma(S^+ \otimes E)$. They can be written in a compact form as:
\bea F_{\alpha\beta}^{+a}+{i}{\overline M}_{(\alpha} (T^{a}) M_{\beta
)}&=&0,\nonumber\\ D_{E}M&=&0,
\label{compacto}
\eea where the $T^a$ are the generators of the Lie algebra taken in the 
representation $R$, $F_{\alpha\beta}=F_{\alpha \beta}^{+a} (T^a)^{ij}$, and
 ${\overline M}_{(\alpha} (T^{a}) M_{\beta )}$ is a shortened form for
${\overline M}^i_{(\alpha} (T^{a})^{ij} M^j_{\beta )}$ (this convention will
be used throughout this work). $D_E$ is the Dirac
 operator for the twisted bundle $S^+\otimes E$, {\it i.e}, the Dirac operator
coupled to the gauge connection in the corresponding representation.  Notice
that we have used the metric on $E$ to identify ${\overline E} \simeq E^{*}$
(if  the representation is complex) or $E \simeq E^{*}$ (if the
representation is real).      Within this general framework it is easy to
write out explicitly the non-abelian monopole equations for the case in which
Donaldson theory has been proved more useful: $G=SU(N)$ and $E$ associated to
the fundamental representation,  with ${\rm rk}(E)= N$. In this case, $E$ is
endowed with a Hermitian metric 
$h$ and ${\bf g}_E$ are the trace-free, self-adjoint  matrices. The first
monopole equation can be written as: 
\be  F_{\alpha\beta}^{+ij}+{i}({\overline M}^j_{(\alpha} M^i_{\beta
)}-{\delta^{ij}
\over N}{\overline M}^k_{(\alpha} M^k_{\beta )})=0,
\label{nonab}
\ee In this equation (and similar ones in this chapter), a sum in the
repeated index
$k$ is understood. It is easy to check that the term  involving the monopole
fields is a Hermitian matrix.  The expression in (\ref{nonab}) is obtained
from (\ref{compacto}) after contracting it with $T^a$ and using the fact that
the normalization of the generators can be chosen such that for the
fundamental representation of $SU(N)$ one has 
$(T^a)^{ij} (T^a)^{kl} = \delta^{il}\delta^{jk}-{1\over N}\delta^{ij}
\delta^{kl}$. Notice that  in the $SU(N)$ case, there is a topological
restriction on the  Hermitian bundle $E$: it has a trivial determinant line
bundle. The equation  (\ref{nonab}) can be written in the same notation that
(\ref{swmon}): 
\be
\rho(F_A^+)+2i\Big[(M \otimes_h {\overline M})_0- {1\over N}{\rm Tr}_h(M
\otimes{\overline M})_0\Big]=0,
\label{noesabel}
\ee where the Hermitian metric $h$ on $E$ is explicitly introduced, and the
trace is taken  with respect to $h$ ({\it i.e.}, in the $E$ indices). In the
rest  of this chapter we will mainly focus on the $SU(N)$ case, with the
monopoles  in the fundamental representation, although many expressions  will
be valid for a general gauge group and representation.

The field space is in this case ${\cal M}={\cal A} \times \Gamma (M, S^{+}
\otimes E)$, where ${\cal A}$ is the moduli space of $G$-connections on $E$.
The group
${\cal G}$ of gauge transformations of the bundle $E$ acts on this
configuration  space according to (\ref{ramon}):    
\bea  g^{*}(A_\mu)&=&gA_\mu g^{-1}-ig\partial_\mu g^{-1},\\ \nonumber
g^{*}(M_{\alpha})&=& g M_{\alpha}   \label{mongauge}
\eea      where $M \in \Gamma (X, S^{+} \otimes E)$ and $g$ takes values in
the group
$G$ in the representation $R$. Notice that in terms of the covariant
derivative
$d_A=d+i[A,\;\;]$ the infinitesimal form of the transformations
(\ref{mongauge}) becomes, after considering
$g=\exp(i\phi)$: 
\bea
\delta A &=& -d_A \phi, \nonumber \\
\delta M_\alpha &=& i\phi M.
\label{mongaugedos}
\eea The moduli space of solutions to the non-abelian  monopole equations,
modulo gauge transformations, is denoted by ${\cal M}_{\rm NA}$.  An
important point is that the non-abelian monopole equations always have the
solution $M=0$, $F_A^{+}=0$, and therefore ${\cal M}_{\rm ASD} \subset {\cal
M}_{\rm NA}$.

The tangent space to the configuration space at the point $(A,M)$ is just
$T_{(A,M)}{\cal M}=T_{A}{\cal A} \oplus T_{M}\Gamma (X, S^{+} \otimes
E)=\Omega^{1}(X,{\bf g}_E)
\oplus \Gamma (X, S^{+} \otimes E)$, for $\Gamma (X, S^{+}
\otimes E)$ is a vector space. We can define a gauge-invariant Riemannian 
metric on
${\cal M}$ from the metric on the forms $\Omega^*({\bf g}_E)$ and the 
Riemannian metric on $S^+\otimes E$:   
\be  
\langle (\psi, \mu), (\theta, \nu) \rangle={1 \over 2}\int_{M} \tr(\psi
\wedge *\theta) +{1 \over 2} \int_{X} e ({\bar \mu}^{\alpha i}
\nu_{\alpha}^i+ \mu_{\alpha}^i {\bar \nu}^{\alpha i}),  
\label{pina}
\ee    where $e=\sqrt g$. Notice that we have introduced a $1/2$ factor in
the first piece  of the metric. This is again to match Field Theory
normalizations. 

A first step to understand the structure of the moduli space of solutions to
the non-abelian monopole equations modulo gauge transformations is to
construct a slice of the gauge action, as we did in Donaldson theory. For
this we need an explicit construction of the gauge orbits, given by the map
${\cal G}\times  {\cal M}
\rightarrow {\cal M}$. The tangent space to these orbits is the  image of a
map from the Lie algebra of the group
${\cal G}$ to the tangent space to ${\cal M}$,  
\be C:{\rm Lie}({\cal G})  \too T{\cal M}.
\label{monope}  
\ee It can be obtained from (\ref{mongaugedos}) and reads, in local
coordinates:
\be   C(\phi)=(-d_{A}\phi ,i\phi^{ij} M^j) \in \Omega^{1}({\bf g}_E) \oplus
\Gamma (X, S^{+}
\otimes E),\,\,\,\,\ \phi \in \Omega^{0}({\bf g}_E). 
\label{flor} 
\ee  Using the natural metrics, we can compute the adjoint operator
$C^{\dagger}$ (which will be needed later to obtain the topological
Lagrangian of the theory). Let us consider $(\psi,\mu) \in T_{(A,M)}{\cal M}=
\Omega^{1}(X,{\bf g}_E) \oplus \Gamma (X, S^{+} \otimes E)$. One finds, 
\be    C^{\dagger}(\psi,\mu)^{ij}=-(d_A^{*}\psi)^{ij} + {i\over 2} \big({\bar
\mu}^{\alpha j}M_{\alpha}^i-{\overline M}^{\alpha
j}\mu_{\alpha}^i-{\delta^{ij}\over N} ({\bar \mu}^{\alpha
k}M_{\alpha}^k-{\overline M}^{\alpha k}\mu_{\alpha}^k)\big),
\label{marga}  
\ee       which is an element in ${\rm Lie}({\cal G})=\Omega^0({\bf g}_E)$.
The linear  version of the slice is just the kernel of this mapping, and the 
quotient of  this kernel by the corresponding isotropy group describes a
neighbourhood of 
$[(A,M)]$ in ${\cal M}/{\cal G}$. 

We must then analyze the type of isotropy groups or stabilizers that appear
in the  action of ${\cal G}$ on ${\cal M}$. First of all, it is clear that if
we have a  non-minimal isotropy group $\Gamma_{(A,M)} \not=C(G)$, then the
connection $A$ must be  reducible. If we restrict ourselves to the case of
$G=SU(2)$, and monopoles  in the fundamental representation, we know that
reductions correspond to splittings of 
$E=L\oplus L^{-1}$, and stabilizers correspond to the subgroup $U(1) \subset
SU(2)$  given by
\be
\left(\begin{array}{cc}{\rm e}^{i \theta}&0\\  0&{\rm e}^{-i
\theta}\end{array}
\right).
\label{estabilizate}
\ee Clearly, these gauge transformations don't leave $M$ invariant unless
$M=0$.    Then, the only elements in ${\cal M}$ having a non-minimal isotropy
group,  for
$G=SU(2)$ and $E$ in the  fundamental representation, are of the form
$(A,0)$, where 
$A$ is a reducible $SU(2)$ connection. Notice that the center of 
$SU(2)$ acts in a non-trivial way on the monopole fields (living in the
fundamental  representation of $SU(2)$), therefore pairs with $M \not=0$ have
a trivial isotropy  group. A reducible solution to the  non-abelian monopole
equations is then a reducible ASD connection, {\it i.e.}, an abelian
instanton, like in $SU(2)$ Donaldson theory.   

The local model of the moduli space is given by the zero locus in 
$\ker \,\ C^{\dagger}$  of the following map $s:{\cal M}\rightarrow
\Omega^{2,+}({\bf g}_E)$:
\be s(A,M)=\Big({1 \over \sqrt 2} \big(F^{+ ij}_{\alpha \beta}+i ({\overline
M}_{(\alpha}^j  M_{\beta )}^i-{\delta^{ij}\over N} {\overline M}_{(\alpha}^k 
M_{\beta )}^k)\big), (D^{{\dot \alpha}\alpha }M_{\alpha})^i\Big).  
\label{section}
\ee  The normalization factor ${1/ \sqrt 2}$ is introduced in order to obtain
a  topological action with the standard normalizations in Field Theory, as we
will  see in the next Chapter. Notice that the map (\ref{section}) is gauge
equivariant,  and if we consider the vector space 
${\cal F}=\Omega^{2,+} ({\bf g}_E) \oplus \Gamma (X, S^{-}
\otimes E)$, we can form, as in (\ref{mivector}), the associated vector
bundle:
\be {\cal E}={\cal M} \times_{\cal G} {\cal F}. 
\label{otrovector}
\ee The map (\ref{section}) descends to a section ${\hat s}:{\cal M}/{\cal G} 
\rightarrow {\cal E}$, and the moduli space of non-abelian monopoles ${\cal
M}_{\rm NA}$.   is the zero locus of this section. As in Donaldson theory, we
study the linearization  of this map, $ds:T_{(A,M)}{\cal M} \too {\cal F}$.
The explicit expression is:       
\bea ds(\psi,\mu) &= & \bigg( {1 \over \sqrt
2}\Big(\big(p^+(d_A\psi)\big)_{\alpha
\beta}^{ij}+ i\big({\overline M}_{(\alpha}^j \mu_{\beta )}^i+{\bar
\mu}_{(\alpha}^j M_{\beta )}^i -{\delta^{ij}\over N}({\overline
M}_{(\alpha}^k \mu_{\beta )}^k+{\bar
\mu}_{(\alpha}^k M_{\beta )}^k)\big)\Big),\nonumber \\ &
&\,\,\,\,\,\,\,\,\,\,\, (D^{{\dot \alpha}\alpha  }
\mu_{\alpha})^i+i\psi^{{\dot
\alpha}\alpha }_{ij}M_{\alpha}^ j  \bigg), \label{disco}
\eea    where $p^+$ is the projector on SD forms. Instead of studying the
restriction of  this map to $\ker \,\ C^{\dagger}$, we can consider the
instanton deformation complex:
\be    0 \too \Omega^{0}({\bf g}_E)
\buildrel C \over \too \Omega^{1}({\bf g}_E) \oplus \Gamma (X, S^{+} \otimes
E)
\buildrel ds \over \too
 \Omega^{2,+} ({\bf g}_E) \oplus \Gamma (X, S^{-} \otimes E) \too 0. 
\label{file}   
\ee To check that this is in fact a complex, we compute the composition
$ds\cdot C$:
\be (ds\cdot C)(\phi)=\Big({1 \over \sqrt 2}(-p^+d_Ad_A \phi+ [M_{(\alpha}
{\overline M}_{\beta )}, \phi]),  iD^{\dot\alpha \alpha} (\phi M_{\alpha})-
i(d_A\phi)^{{\dot
\alpha}\alpha } M_{\alpha} \Big),
\label{compruebo}
\ee where $\phi \in \Omega^0({\bf g}_E)$, and the $E$ indices are understood.
An  explicit computation shows that
\be (ds\cdot C)(\phi)=\Big(-{i \over \sqrt 2} [F^{+}_{\alpha \beta}+i
(M_{(\alpha} {\overline M}_{\beta )}-{1\over N} {\overline M}_{(\alpha}^k 
M_{\beta )}^k), \phi], i\phi D^{{\dot\alpha} \alpha} M_{\alpha}\Big)
\label{calcuplaya}
\ee which equals zero on a solution to the non-abelian monopole equations. To
see  that this complex is elliptic, we can deform it dropping terms of order
zero in the operators $C$ and $ds$ (as their leading symbol is not changed).
In this way, the complex (\ref{file}) splits into the complex associated to
the anti-self-dual (ASD) connections of Donaldson theory (\ref{asd}) and the
complex of the  twisted Dirac operator. Both are elliptic, proving that
(\ref{file}) is elliptic  too. This shows that the map (\ref{section}),
restricted to $\ker \,\ C^{\dagger}$,  is Fredholm. The index will be simply
the sum of the index of (\ref{asd}) (which equals the virtual dimension of
the moduli  space of ASD instantons, given in (\ref{bunito})), and of twice
the index of the  twisted  Dirac complex (for we are considering $S^{+}
\otimes E$,
$S^{-} \otimes E$ as real vector bundles, in order to obtain the real
dimension of the moduli space). This is easily computed and gives: 
\be {\rm index}\,\ D_E=\int_X {\rm ch}(E) {\hat A}(X)=-{N \over 8}\sigma
-c_2(E). 
\label{indice}
\ee The index of (\ref{file}) is: 
\bea {\rm dim} \,\ {\cal M}_{\rm NA}&=&{\rm dim} \,\ {\cal M}_{\rm ASD}+2 \,\
{\rm index}\,\ D_E\nonumber\\ &=&(4N-2)c_2(E)-{N^2-1 \over 2}(\chi
+\sigma)-{N \over 4}\sigma,
\label{dim}
\eea This is the virtual dimension of the moduli space ${\cal M}_{\rm NA}$,
in the
$SU(N)$ case, and for the monopoles in the fundamental representation.
 The generalization of this expression to other gauge groups is
straightforward.    

Orientability of the moduli space is easily checked. As we have seen, the
operator 
$T=C^{\dagger}
\oplus ds$ can be deformed to a direct sum of the Dirac operator and the
elliptic operator for the complex of ASD connections (\ref{asd}). The
determinant line bundle of the Dirac operator, when  regarded as a real
operator, has a natural trivialization coming from its underlying complex
structure. The determinant line bundle of the operator  appearing in the
Atiyah-Hitchin-Singer complex, as we saw in Chapter 1, is orientable 
according to the  results of Donaldson \cite{donor, dk}. Therefore, ${\cal
M}_{\rm NA}$ is orientable  and the orientation is induced by a
trivialization of 
${\rm det}\,\ {\rm ind} \,\ \delta_A$, as in Donaldson theory. 

These analysis show that the non-abelian monopole theory appears as a rather
natural generalization of Donaldson theory: the moduli space of solutions
contains ${\cal M}_{\rm ASD}$ as a subset, and the conditions for having a
well defined moduli problem are essentially the same, at least  in the
$SU(2)$ case. There are two points concerning the construction of the moduli 
space that we have not discussed: transversality and compactness. ${\cal M}_
{\rm NA}$ is not compact and must be compactified in an appropriate way.
Both  problems have been adressed in a related context in \cite{pt,
okotwo}.    

A first constraint on the solutions to the non-abelian monopole equations is
obtained  by using classical vanishing theorems in spin geometry \cite{sg},
and is a  generalization  of the vanishing theorem derived for the
Seiberg-Witten equations in
\cite{mfm}.  There are two ways to obtain this result. In the first one, one
considers
 the Weitzenb\"ock formula (\ref{miwei}) for $D_E$: 
\be D_E D_E M=\nabla^*_E \nabla_E M+{1 \over 4}R M + i\rho(F_A^+)M.
\label{weiotravez}
\ee Using now the first non-abelian monopole  equation for $\rho(F_A^+)$, one
obtains for the last term,  in components,
\be
\Bigg((1 + {1 \over N})|M|^2\delta^{ij} -{2\over N}(M^i_{\beta}{\overline
M}^{j
\beta}) 
\Bigg) M^j_{\alpha}.
\label{cosalinda}
\ee Taking now the Hermitian inner product with $M$ in the equation
(\ref{weiotravez}),  we have:
\bea (M,D_E D_E M)&=&(M, \nabla^*_E \nabla_E M)+{1 \over 4}R |M|^2 
\nonumber\\ &+&  (1 + {1 \over N})|M|^4-{ 2 \over N} \sum_{ij}|{\overline
M}^{i1}M_1^j+  {\overline M}^{i2}M_2^j|^2,
\label{seguimos}
\eea As the last term is $\le |M|^4/2N$, the terms in the second line are $\ge
(1-1/N)|M|^4$,  which is greater than or equal to zero. As $D_E M=0$, if the
scalar curvature of $X$ is  positive, then $M=0$. In this case, ${\cal
M}_{\rm NA}={\cal M}_{\rm ASD}$. 

Another way to reach the same result is to compute the norm squared of the
section 
$s(A,M)$,  with the natural Riemannian metric on the fibre ${\cal F}$, given
by the integration  over $X$ of the expression  given in (\ref{hula}). This
norm will become the  bosonic sector of the topological action of the theory,
as we will see in the next  Chapter. We then have: 
\bea |s(A,M)|^{2}&=& \int _{X} e (D_E M,D_E M) \nonumber\\  &-&{1
\over 4}\int _{X}e \big(F^{+\alpha \beta ji} +i{\overline M}^{i(\alpha} 
M^{\beta )j} \big)  \big(F^{+ ij}_{\alpha \beta} +i({\overline
M}^{j}_{(\alpha} M^{i}_{\beta )}-{\delta^{ij}\over N}{\overline
M}^{k}_{(\alpha} M^{k}_{\beta )})\big) \nonumber
\\ &=&\int_{X} e \big[ g^{\mu\nu}D_\mu\overline M^\alpha D_\nu M_\alpha +
{1\over 4} R |M|^2 - {1\over 4} \tr (F^{+\alpha\beta} F_{\alpha\beta}^+)
\nonumber\\ & &
\,\,\,\,\,\,\,\  +{1\over 4} (\overline M^{i(\alpha} M^{\beta)j} \overline
M_{(\alpha}^j M_{\beta)}^i -{1\over N}
\overline M^{i(\alpha} M^{\beta)i} \overline M_{(\alpha}^j
M_{\beta)}^j\big)].    
\label{melon}   
\eea   After using the fact that for
$SU(N)$ the normalization of the generators $T^a$ can be chosen in such a way
that for the fundamental representation $(T^a)^{ij} (T^a)^{kl} =
\delta^{il}\delta^{jk}-{1\over N}\delta^{ij} \delta^{kl}$, the last term in
(\ref{melon}) can be written as: 
\be {1\over 4}(\overline M^{(\alpha} T^a M^{\beta)}) (\overline M_{(\alpha}
T^a M_{\beta)}),
\label{current}
\ee where a sum over $a$ must be understood. This term can be computed
explicitly in local coordinates with the result,
\be {1\over 2}\Big( (1-{1 \over N})|M|^4 +(1+{2
\over N})\sum_{ij}|M_1^{[i}M_2^{j]}|^2 \Big),
\label{melondos}         
\ee and therefore it is positive definite. The factor $i\overline M^\alpha
F^{+\beta}_{\alpha} M_\beta$ has cancelled in the sum, and then each term in
the second expression for $|s(A,M)|^2$ in (\ref{melon}) is positive definite
except the one involving the scalar curvature. For a solution of the
non-abelian monopole  equations, $|s(A,M)|^2$ is zero, and therefore if $R>0$
we obtain again $M=0$. We can  also get from this expression an upper bound
for the squared norm of the self-dual part of the curvature on solutions of
the monopole equations: 
\be   I^{+}=\int _{X}e F_a^{+\alpha\beta} F_{a,\alpha\beta}^+ \le {1 \over
16(1-1/N)}\int_{X}eR^{2}.
\label{melontres}
\ee

\section{Non-abelian monopoles on K\"ahler manifolds.} In this section we will
analyze in more detail the non-abelian monopole equations on a compact
K\"ahler, Spin manifold $X$ for the case in which the gauge group is $SU(N)$
and the monopoles are in the fundamental  representation. We will obtain in
this case a description in terms  of holomorphic objects as well as a
relation with the vortex equations. This  analysis confirms that the moduli
space of non-abelian monopoles  has a rich structure and can give an
interesting information in four-manifold  topology.

On a K\"ahler manifold the spinor bundle $S^{+}$ splits into $K^{1/2}\oplus
K^{-1/2}=(\Omega^0(X) \oplus \Omega^{2,0}(X) )\otimes K^{1/2}$, where
$K^{1/2}$  is a square root of  the canonical bundle $K$ (see Appendix,
section 3). Let
$E$ be the vector bundle associated to the fundamental representation of
$SU(N)$, and denote by $\alpha^i$ and $i{\overline
\beta}^i$ the components of $M^i_\alpha$ in $K^{1/2}\otimes E$ and 
$K^{-1/2}\otimes E$, respectively. We also have the  decomposition of SD
forms given in (\ref{desco}). If we express $F_A^+$ as an endomorphism of 
the positive spinor bundle, following (\ref{endokal}), we can write the first
$SU(N)$ monopole equation as: 
\bea F_{2,0}^{ij}&=&\alpha^i \beta^j-{1 \over N}\delta^{ij}\alpha^k
\beta^k,\nonumber\\ F_{\omega}^{ij}&=&{\omega \over 2}\Big( \alpha^i
{\overline
\alpha}^j-{\overline \beta}^i \beta^j-{\delta^{ij} \over
N}(|\alpha^k|^2-|\beta^k|^2) \Big),\nonumber\\ F_{0,2}^{ij}&=&{\overline
\alpha}^j {\overline \beta}^i-{1 \over N}\delta^{ij}{\overline \alpha}^k
{\overline \beta}^k. 
\label{kal}
\eea The Dirac operator $D_E$ can be written in terms of Dolbeault operators
for the  bundles $K^{\pm1/2}\otimes E$, according to (\ref{diropcan}), and
the second  monopole equation reads:
\be {\sqrt 2}\Big({\overline \partial}_A \alpha + i{\overline \partial}^*_A 
{\overline \beta}\Big)=0. 
\label{didol}
\ee Now we can use the expression (\ref{melon}) to obtain vanishing results
for the solutions of (\ref{kal}), as in \cite{mfm,vw}. Suppose
$(A,\alpha,\beta)$ is a solution to (\ref{kal}), and hence (\ref{melon})
vanishes. Then $(A,\alpha, -\beta)$ makes  (\ref{melon}) vanish too, and we
obtain another solution to (\ref{kal}). Therefore, any solution of these
equations verifies: 
\be F_{2,0}^{ij}=F_{0,2}^{ij}=0.
\label{hol}
\ee This tells us that the connection $A$ endows $E$ with the structure of a
holomorphic bundle, as it happens in Donaldson theory and in the abelian
theory
\cite{mfm}.  

The most general solution to the equation 
\be
\alpha^i \beta^j-{1 \over N}\delta^{ij}\alpha^k \beta^k=0,
\label{lisa}
\ee is $\alpha \not= 0$, $\beta = 0$ or $\alpha=0$, $\beta \not= 0$ (with
$\alpha$, $\beta$ understood as vectors). Of course we also have the solution
$\alpha =\beta = 0$, which corresponds to an ASD instanton. Let's  consider
the first kind of solutions. Suppose $\alpha \not= 0$, $\beta = 0$. The Dirac
equation (\ref{didol}) for this kind of solution is simply ${\overline
\partial}_{A}\alpha=0$, with ${\overline \partial}_{A}$ the twisted Dolbeault
operator on $K^{1/2} \otimes E$. As $A$ defines a holomorphic structure on
$E$,  according to (\ref{hol}), and $K^{1/2}$ is clearly holomorphic, the
Dirac equation simply tells us  that $\alpha$ is a holomorphic section of the
bundle ${\cal E}=K^{1/2}\otimes E$, {\it i.e.}, 
$\alpha \in H^0(X,{\cal E})$. For $\alpha=0$, $\beta \not= 0$ we have the
symmetric situation, with $\beta$ a holomorphic section of the bundle
${\cal E}'=K^{1/2}\otimes {\overline E}$. Let's  change slightly our
conventions and  take the curvature $F_A$ as a skew-Hermitian  matrix,
following the usual ones in geometry. Then, when $\beta=0$, the first 
non-abelian monopole equation reads:
\be i\Lambda F_E +\alpha \otimes_h {\overline \alpha}-{1 \over
N}|\alpha|^2_h=0,
\label{casiaborto}
\ee where $\alpha$ is a holomorphic section of the holomorphic bundle 
$K^{1/2}\otimes E$. To put this equation in  the standard framework of vortex
equations, the curvature must be that  of the holomorphic section of which
$\alpha$ is a section. As $K$ is endowed with the  canonical Riemannian
connection, and the equality $i\Lambda F_{K}=R$ holds, where 
$R$ denotes, as before, the scalar curvature of $X$, one can write:
\be i\Lambda F_E = i\Lambda F_{K^{1/2}\otimes E}-{1\over 2} R.
\label{cambiacurva}
\ee The non-abelian monopole equation (\ref{casiaborto}) then reads:
\be i\Lambda F_{K^{1/2}\otimes E}+\alpha \otimes_h {\overline \alpha}=
{1\over 2} R+{1 \over N}|\alpha|^2_h.
\label{projvortex}
\ee This equation is closely related to the vortex equations on K\"ahler
manifolds studied  in \cite{brad, gpuno, gpdos}, and can be analyzed using
similar methods. This version  of the vortex equations involves a {\it
holomophic pair} $({\cal E},
\alpha)$  consisting of a  holomorphic bundle ${\cal E}=K^{1/2} \otimes E$
together  with a holomorphic section, 
$\alpha$. In addition, ${\cal E}$ has a {\it fixed} determinant bundle 
${\rm det}({\cal E})=K$ with a fixed connection (the canonical one). 

Now we briefly consider the case $\alpha=0$. Using the K\"ahler identity
(\ref{kalidem}),  we can write (\ref{didol}) as 
\be
\Lambda \partial_A {\overline \beta}=0. 
\label{alterne}
\ee If we denote by $L$ the operator given by $\cdot \wedge \omega$, it is
easy to check  that $[L, \Lambda]=p+q-2$ acting on $(p,q)$-forms \cite{gh},
and this in turn implies  that $\Lambda$ is an isomorphism on $(1,2)$-forms.
We can then write 
$\partial_A {\overline \beta}=\overline {{\overline \partial_A} \beta}=0$, 
and
$\beta$ is a holomorphic section of ${\cal E}'=K^{1/2}\otimes {\overline E}$. 
Conjugating the second equation in (\ref{kal}), we then get:
\be i\Lambda F_{K^{1/2}\otimes {\overline E}}+\beta \otimes_h {\overline
\beta}= {1\over 2} R+{1 \over N}|\beta|^2_h,
\label{projvortexdos}
\ee      and therefore we have the same vortex equation but for the pair
$({\cal E}'=  K^{1/2} \otimes {\overline E}, \beta)$. 

The relevant stability notion here involves {\it stable pairs}, and has been
analyzed  in \cite{okotwo} in the $SU(2)$ case. The equation
(\ref{projvortex}) has a  solution if and only if the  pair $({\cal E},
\alpha)$ is stable, in the sense, essentially, of $\alpha$-stability.  A
related stability notion was introduced in the context of self-dual equations
on  Riemann surfaces in \cite{jichi}, and was studied for general vortex 
equations on K\"ahler manifolds in \cite{brad, gpuno, gpdos}. It can be
roughly  described as follows:  let ${\cal F}$ be a proper subbundle of
${\cal E}$, such that the  image of $\alpha$ lies on 
${\cal F}$. Then, the pair $({\cal E}, \alpha)$ is said to be $\alpha$-stable
if for  any such a subbundle ${\cal F}$, $\mu({\cal F})<\mu({\cal E})$. This
stability notion  applies to the two cases, (\ref{projvortex}) and
(\ref{projvortexdos}). More  information on stability conditions in relation
to non-abelian monopoles and vortices  can be found in
\cite{bradgp,gptres}.   

To complete the picture of the moduli space of non-abelian monopoles in
terms  of holomorphic objects, we will use the techniques of symplectic
geometry that  we introduced in Chapter 1, and that have also been applied in
the abelian  case
\cite{mfm}. Suppose again we are in the case $\alpha \not= 0$,
$\beta = 0$. We define a symplectic structure on ${\cal M}_{\beta =0}={\cal
A}^{1,1}\times \Gamma (X,K^{1/2}\otimes E)$ according to: 
\be
\Omega ((\psi, \mu), (\theta, \nu) )=\int_{X}  {\rm Tr} (\psi \wedge
\theta)\wedge \omega -{i
\over 2} \int_{X} \omega \wedge \omega ({\bar \mu}^i \nu^{i}- \mu ^{i} {\bar
\nu}^i),
\label{sym}\ee where $\psi$, $\theta$ are in $\Omega^{1}(X,{\bf g}_E)$ and
$\mu$,
$\nu$ in
$\Gamma (X,K^{1/2}\otimes E)$. This symplectic form is obviously preserved by
the action of the group of gauge transformations. We consider
$\Omega^{4}(X,{\bf g}_E)$ as the dual of
${\rm Lie}({\cal G})=\Omega^{0}(X,{\bf g}_E)$, and the pairing is given by the
integration over $X$ of the trace of the wedge product, as we did in the
Donaldson  case. $C$ is the map given in (\ref{flor}). The brackets denote
the dual pairing.  The explicit expression of this map is given by:
\be m(A,
\alpha)=F^{ij} \wedge \omega - {\omega \wedge \omega \over 2}(\alpha^i
{\overline
\alpha}^j-{\delta^{ij} \over N}|\alpha^k|^2).
\label{map}
\ee        The first piece of this map is just the corresponding map for
Donaldson theory (\ref{mediolimon}), and the second piece contains the
dependence on the  monopole part. The property (\ref{mome}) is easily
verified from the expression for the differential of (\ref{map}): 
\be (dm)_{(A, \alpha)}(\psi, \mu)=(d_A \psi)^{ij} \wedge \omega-{\omega \wedge
\omega
\over 2}\Big(\mu^i {\overline \alpha}^j+\alpha^i {\overline
\mu}^j-{\delta^{ij} \over N}(\mu^k {\overline \alpha}^k+\alpha^k {\overline
\mu}^k) \Big).
\label{luisa}
\ee The solutions of (\ref{casiaborto}) are precisely the zeroes of the
moment map (\ref{map}), as it happens in Donaldson theory. This indicates
that the moduli space of solutions of the $SU(N)$ monopole equations with
$\beta=0$ can be identified with the symplectic quotient 
$m^{-1}(0)/{\cal G}$. Following the general remarks given in Chapter I, we can
identify  the moduli space of non-abelian monopoles with $\beta=0$, for a
fixed topological
 $SU(N)$ bundle $E$, with the quotient of  the stable orbits of ${\cal
M}_{\beta=0}$ under the complex gauge group 
${\cal G}_{\bf C}$. The stable orbits in this case correspond to  stable pairs
$({\cal E}, 
\alpha) \in {\cal M}_{\beta=0}$ (where we have identified holomorphic
structures on 
$E$ with integrable connections, and $\alpha$ is a holomorphic section).  A
similar picture  holds for $\alpha=0$. 

If we restrict ourselves to the $SU(2)$ case (as reducibility and stability
are  under control), we then have the following result: on a compact,
K\"ahler, Spin  four-manifold,  the moduli space of solutions to the $SU(2)$
monopole equations has three branches: the first one corresponds to the
irreducible ASD connections with
$M=0$, and can be identified with the equivalence classes of stable
holomorphic
$Sl(2,{\bf C})$ bundles $E$. The second branch corresponds to pairs 
consisting of an equivalence class of holomorphic $Sl(2,{\bf C})$ bundles $E$
together with a holomorphic section of $K^{1/2}\otimes E$ modulo $Sl(2,{\bf
C})$ gauge transformations (the case $\alpha \not= 0$, $\beta = 0$ discussed
before), and such that the pair $({\cal E}, \alpha)$ is stable. The third
branch is  similar to the second branch, but now
$\alpha=0$, $\beta \not= 0$, and consequently we must consider instead
holomorphic sections of $K^{1/2}\otimes {\overline E}$. Notice that the first
branch is the  intersection of the second and third ones, and that the second
and third branch  can be identified, because $E$ corresponds to the
fundamental representation of 
$SU(2)$ and $E\simeq {\overline E}$. The non-abelian  monopole theory has
then a richer content than Donaldson theory.

\chapter{Non-abelian monopoles and $N=2$ SUSY}

One of the fundamental aspects of the abelian and non-abelian monopole
equations  that we analyzed in the previous Chapter is that they can be
obtained from the twist  of supersymmetric gauge theories in four dimensions.
More precisely, the Thom  form in the  Mathai-Quillen formalism associated to
these moduli problems can be understood as  the topological action of a
twisted supersymmetric $N=2$ theory.  In this Chapter we will  analyze this
connection and formulate the Field Theory framework to understand these 
equations. In section 1 we construct the topological action of the
non-abelian  monopole theory in purely geometrical terms, using the
Mathai-Quillen formalism.  We also  construct the observables of the theory
and we define the polynomial invariants for 
$SU(2)$ monopoles as correlation functions of this Topological Quantum Field
Theory. In  section 2 we show how this action appears after twisting $N=2$
supersymmetric QCD with  one flavour, $N_f=1$. In section 3, we consider the 
equivariant extension of the Thom form  associated to a $U(1)$ action on the
moduli space of non-abelian monopoles and we  show that it coincides with the
twisted $N=2$ theory with one massive hypermultiplet.  Finally, in section 4
we show how to include
$\sp^c$-structures by twisting the 
$U(1)_B$ global symmetry of the supersymmetric theory with $N_f=1$. 

\section{The topological action and the observables} 

In the first subsection we will build the topological action  corresponding
to the non-abelian  monopole equations of the previous chapter, using the
Mathai-Quillen formalism. In  other words, we will construct the Thom form of
the associated vector bundle  (\ref{otrovector}). One of the advantages of
this formalism is that it provides a  procedure to construct the action of a
Topological Quantum Field Theory starting from a moduli problem formulated in
purely geometrical terms. We follow again \cite{lmna}. The topological action
for the  abelian case, corresponding to the Seiberg-Witten equations, was
constructed along  the same lines in \cite{lmab}. In the second subsection we
construct the observables of  the theory, stressing the similarities with
Donaldson-Witten theory.

\subsection{The topological action}  

As the vector bundle (\ref{otrovector}) is constructed as  an associated
vector bundle to a principal bundle ${\cal M} \rightarrow  {\cal M}/{\cal
G}$, the adequate formulation of the Thom form is the one given  in \cite{aj}
and reformulated in
\cite{cmr}. All the necessary ingredients have been  already presented in
Chapter 3, section 5. First of all, we need a connection in the  principal
bundle. As ${\cal M}$ is endowed with the Riemannian metric  given in
(\ref{pina}), we can obtain this connection according to the general 
construction given in Chapter 1, section 4. Recall that, if the tangent
space  to the gauge orbits is given by the image of the map $C$ in
(\ref{flor}), then the  connection is 
\be
\theta =  R^{-1} C^{\dagger}.
\label{lima}
\ee The operator $R=C^{\dagger}C$ can be explicitly computed from 
(\ref{monope}) and (\ref{marga}),  and reads
\be  R(\phi)^{ij}=(d_A^{*}d_A \phi)^{ij} +  {1\over 2}({\overline M}^{\alpha
k}\phi^{kj}M_{\alpha}^i +{\overline M}^{\alpha j}\phi^{ik}M_{\alpha}^k) -
{\delta^{ij}\over N}{\overline M}^{\alpha k}\phi^{kl} M_\alpha^l,      
\label{tren}
\ee  where $\phi \in \Omega^{0}(X,{\bf g})$. We will write the pullback of
the  Thom form of ${\cal E}$ to ${\cal M}/{\cal G}$ using  the section ${\hat
s}$ induced by (\ref{section}). Taking into account the commutative  diagram
(\ref{diagramaconm}), the most useful representative of this form is  a
horizontal, basic form on 
${\cal M}$. This is obtained through the pullback of the form given  in
(\ref{lagrangiano}) by  the equivariant section $s$ (\ref{section}). As we
are not yet considering  any vector field action, we put $u=0$ and pullback
the form. One obtains in this way:   
\bea  
\int  D\eta D\rho D\phi D \lambda \,\   {\rm exp}\Big(& &-|s|^2 + {1 \over 4}
\langle \rho , \phi \rho \rangle + i\langle \rho  , ds \rangle \nonumber\\ & &
+i{\langle dC^{\dagger}, \lambda \rangle }_{g}- i{\langle \phi,R \lambda
\rangle }_{g} +i\langle \eta, C^{\dagger} 
\rangle_{g} \Big).
\label{mqaj}   
\eea  We recall that $\phi$, $\lambda$ are conmuting Lie algebra variables and
$\eta$ is a Grassmann one. We have introduced coordinates  in the fibre not
necessarily orthonormal, and we denote the metric on the  fibre by $\langle
\,\ ,
\,\ \rangle$. The bracket with the subscript $g$ is the Cartan-Killing form
of ${\rm Lie}({\cal G})$, given by the trace together  with the product of
forms, defined as in (\ref{pina}) with a $1/2$ factor.  We know explicit
expressions  for all  the operators appearing in this expression, for $ds$
has been already computed in  (\ref{disco}). 

Before writing the action, we would like to indicate the field content  and
the topological symmetry, {\it i.e.}, the BRST complex. These are determined
by the  geometrical structure that we have been developing. We also introduce
ghost numbers for the different fields involved in the  model. From the
mathematical point of view, this can be interpreted as a
 grading for the different differential forms appearing here, although it has
a  physical origin in the $U(1)_{\cal R}$ symmetry of the untwisted theory.
For the  configuration  space we have commuting fields $P=(A,M) \in {\cal M}=
{\cal A} \times
\Gamma (X, S^{+} \otimes E)$, with ghost number $0$ and their superpartners,
representing a basis of differential forms on ${\cal M}$, $dP=(\psi, \mu)$,
with ghost number $1$. Now, we must introduce fields for the fibre  which we
denote by
$(\chi_{\mu\nu}, v_{\dot \alpha}) \in \Omega^{2,+} (X, {\bf g}_E)
\oplus \Gamma (X, S^{-} \otimes E)$, with ghost number $-1$. We also know
that  in the construction of the action from gauge fermions it is useful to
introduce auxiliary commuting fields for the fibre, with the same geometrical
content,
$(H_{\mu\nu}, h_{\dot \alpha})$, and ghost number $0$. The field 
$\phi\in \Omega^0(X,{\bf g}_E)$,  with ghost number
$2$, corresponds to the curvature of the principal bundle ${\cal M}$ and  can
be understood as the parameter $u$ in the ${\cal G}$-equivariant cohomology.
The  fields $\lambda$ (commuting) and  $\eta$ (anticommuting), also in 
$\Omega^0(X,{\bf g}_E)$ with ghost number $-2$ and $-1$, respectively, come
from the projection form, as we explained in Chapter 3 following \cite{cmr}.
The BRST cohomology of the model  is obtained simply from the cohomology of
${\cal M}$ and
${\cal F}$. As we emphasized  in Chapter 3, one must consider the horizontal
projection we are performing  in order to descend to the quotient ${\cal
M}/{\cal G}$, or, alternatively, we can  directly work in the Cartan model
for the ${\cal G}$-equivariant cohomology. We must take into account, with
respect  to our previous conventions, that in this case the group ${\cal G}$
acts on the  left on ${\cal M}$ and ${\cal F}$, as one can see in
(\ref{mongauge}). Therefore there  are some slight differences with respect
to the expressions given in Chapter 3.  Essentially we must act on the fibre
with the transpose of the connection and curvature  matrices, as one can
easily check. On the configuration space ${\cal M}$, the 
$Q$ operator is simply given by (\ref{masmisterio}) with $u=0$. On $\phi$,
$Q$ gives  obviously zero, and this can be understood as an implementation of
the Bianchi identity.  On the fields associated to the fibre, $Q$ acts
according to the expressions in  (\ref{alsacia}), but with $-\phi$ instead of
$\phi$. Finally, for  the remaining Lie algebra variables, we have the
equations (\ref{mascus}). We then  obtain the BRST complex for this model:
\be   
\begin{array}{cclcccl}  [Q,A] &=& \psi,&
\,\,\,\,\,\,\,\,\,\,\,\,\,\,\,\,\,\,\,\,\,\,\,\,\,\,\,\ &  [Q,M_{\alpha}^i]
&=&\mu_{\alpha}^i,  \nonumber\\
\{ Q,\psi \} &=& d_A \phi,&
\,\,\,\,\,\,\,\,\,\,\,\,\,\,\,\,\,\,\,\,\,\,\,\,\,\,\,\ &
\{Q, \mu_{\alpha}^i \} &=& -i\phi^{ij} M_{\alpha}^j, \nonumber \\ 
{[}{Q},{\phi}{]} &=&0, &
\,\,\,\,\,\,\,\,\,\,\,\,\,\,\,\,\,\,\,\,\,\,\,\,\,\,\,\ &
 \{Q, v_{\dot \alpha}^i \} &=&h_{\dot \alpha}^i, \nonumber\\
\{ Q,\chi_{ \mu\nu} \} &=&H_{\mu\nu},&
\,\,\,\,\,\,\,\,\,\,\,\,\,\,\,\,\,\,\,\,\,\,\,\,\,\,\,\ & {[} {Q},{h_{\dot
\alpha}^i} {]}&=&-i \phi^{ij} v_{\dot \alpha}^j, \nonumber\\  {[}
{Q},{H_{\mu\nu}}{]} &=&i[\chi_{\mu\nu},\phi ],& 
\,\,\,\,\,\,\,\,\,\,\,\,\,\,\,\,\,\,\,\,\,\,\,\,\,\,\,\ &
 \{Q, \eta \} &=&i[\lambda,\phi], \nonumber\\  {[}Q, \lambda] &=& \eta,  &
\,\,\,\,\,\,\,\,\,\,\,\,\,\,\,\,\,\,\,\,\,\,\,\,\,\,\,\ & 
\end{array}
\label{pera}
\ee     This BRST gauge algebra closes up to a gauge transformation generated
by 
$-\phi$, according to the fact that the squared of the differential operator
in  the Cartan model of equivariant cohomology is given by (\ref{nonil}). It
is perhaps  interesting to analyze how the question of reducibility arises in
this framework 
\cite{icft}. Using the underlying fermionic symmetry, one can prove  Witten's
localization  theorem \cite{wzw}: the path integral associated  to any
Cohomological Field Theory is localized on the fixed point locus (in the 
configuration space) of the BRST operator $Q$. If  a connection $A$ is
reducible, then $d_A\phi=0$ has non-trivial solutions. This  means that  the
space of $Q$-fixed points is larger than the space to which we want to
localize the path integral (the solutions to  the non-abelian monopole
equations, which appear as a fixed point of the $Q$ symmetry  on-shell, as we
will see in a moment).

We are now in the position to write down the action of the theory. Let us 
consider first the last five terms in the exponential of the Thom form
(\ref{mqaj}), 
\bea   -i{\langle \phi, R\lambda \rangle }_{g} &=&-{i\over 2} \int _{X} \tr(
\lambda  \wedge *d_A^{*}d_A\phi) -{i\over 2}\int_{X}e {\overline
M}^{\alpha}\{\lambda,\phi\} M_{\alpha},   
\nonumber\\   i\langle  (\chi , v), ds \rangle &=& -{i \over 2 {\sqrt2}}
\int _{X} \tr \Big(\chi^{\alpha \beta} (d_A \psi)_{\alpha \beta} \Big)  +{1
\over {\sqrt2}}
\int _{X}e ({\overline M}_{\alpha}{\chi}^{\alpha \beta} \mu_{\beta}- {\bar
\mu}_{\alpha}{\chi}^{\alpha \beta} M_{\beta })  
\nonumber\\  
  &+&{i \over 2}\int _{X} e (\bar v_{\dot \alpha} D^{{\dot\alpha} \alpha }
{\mu}_{\alpha} + {\bar
\mu}^{\alpha } D_{\alpha {\dot\alpha}}v^{\dot \alpha}) -{1 \over 2}\int _{X} e
({\overline M}^{\alpha }{\psi}_{\alpha {\dot \alpha}}v^{\dot\alpha }+{\bar
v}_{\dot
\alpha }{\psi}^{{\dot \alpha}\alpha }M_{\alpha }),
\nonumber\\       i \langle C^\dagger (\psi,\mu), \eta \rangle_g   &=& - {i
\over 2}
\int_{X} \tr (\eta \wedge *d_A^*\psi ) +{1\over2} \int _{X} e  ({\bar
\mu}^{\alpha }
\eta M_{\alpha}+{\overline M}^{\alpha } \eta \mu_{\alpha}), 
\nonumber\\   {1 \over 4}{\langle (\chi , v), \phi (\chi , v) \rangle} &=&
{i\over 8}\int_X e \tr (\chi^{\alpha \beta}[\phi,\chi_{\alpha \beta}])
 - {i \over 4} \int_{X} e  {\bar v}^{\dot\alpha } \phi v_{\dot\alpha},
\nonumber\\   i{\langle \lambda,  dC^{\dagger}\rangle}_{g}&=&{1 \over
2}\int_X \tr (\lambda\wedge * [\psi,*\psi]) +
\int _{X} e  {\bar \mu}^{\alpha } \lambda \mu_{\alpha}.  
\label{fresa} 
\eea To obtain these expression, one must use the spinor conventions compiled
in the  Appendix, specially (\ref{bajo}) and (\ref{unodiracs}). The section
term in  (\ref{mqaj}) has been computed in (\ref{melon}), and the action
resulting after adding to it all the terms in  (\ref{fresa}) constitutes the
field theoretical representation of  the Thom form of the bundle ${\cal E}$.
This action is invariant under the transformations (\ref{pera}) once the
auxiliary field $H_{\alpha\beta}$ and $h_{\dot\alpha}$ are introduced. It can
be obtained in its off-shell form using the nilpotent transformations
(\ref{pera}) (up to a gauge transformation)  and the gauge invariant gauge
fermions (\ref{fermiones}). In our case,  the localizing gauge fermion is:
\be   
\Psi _{\rm loc}= -i\langle (\chi, v),s(A,M) \rangle-{1 \over 4}\langle  (\chi,
v),(H,h) \rangle,  
\label{tigre}
\ee   while the projection gauge fermion, which implements the horizontal
projection, is,  
\be 
\Psi _{\rm proj}=i{\langle \lambda , C^{\dagger} (\psi ,
\mu ) \rangle}_{g}.  
\label{oso} 
\ee
  
Making use of the $Q$-transformations (\ref{pera}) one easily computes  the 
localization  and the projection Lagrangians:  
\bea
 \{Q,\Psi _{\rm loc}\} &=&
\Big\{ Q, \int _{X} e \Big[{i\over 2} \chi^{\alpha \beta ji}\Big({1 \over
{\sqrt 2}}
\big(F^{+ij}_{\alpha \beta} +i ({\overline M}_{(\alpha}^j M_{\beta)}^i
-{\delta^{ij}\over N}{\overline M}_{(\alpha}^k M_{\beta)}^k)\big) - {i \over
4} H_{\alpha \beta}^{ji} \Big)
\nonumber\\  & & \,\,\,\,\,\,\,\,\,\,\,\,\,\,\,\,\,\,\,\,\,\,\,\,\,
 -{i\over 2} ({\bar v}_{\dot \alpha}D^{\dot \alpha \alpha }
M_{\alpha}-{\overline M}^{\alpha}D_{\alpha \dot \alpha}v^{\dot \alpha})-{1
\over 8} ({\bar v}_{\dot \alpha} h^{\dot \alpha}+{\bar h}_{\dot
\alpha}v^{\dot \alpha} ) \Big] \Big\}
\nonumber\\  &=&\int _{X}e \Big[{i \over 2{\sqrt 2}}
 H^{\alpha \beta ji} \big(F^{+ij}_{\alpha \beta} +i ({\overline M}_{(\alpha}^j
M_{\beta)}^i -{\delta^{ij}\over N}{\overline M}_{(\alpha}^k M_{\beta)}^k)\big)
\nonumber\\
  & &\,\,\,\,\,\,\,\,\,\,\,\,\,\, -{i \over 2{\sqrt 2}}\tr\big(\chi ^{\alpha
\beta}(d_A\psi)_{\alpha\beta}\big)  -{1\over \sqrt{2}}({\bar
\mu}_{\alpha}\chi^{\alpha\beta} M _{\beta}- {\overline
M}_{\alpha}\chi^{\alpha\beta}\mu _{\beta})
 \nonumber\\  & &\,\,\,\,\,\,\,\,\,\,\,\,\, +{1 \over 8} \tr(H^{\alpha \beta}
H_{\alpha \beta}) -{i\over 8}
\tr(\chi^{\alpha\beta}[\chi_{\alpha\beta},\phi])  -{i\over 2} ({\bar h}_{\dot
\alpha}D^{\dot \alpha \alpha } M_{\alpha}-{\overline M}^{\alpha}D_{\alpha
\dot \alpha}h^{\dot \alpha})
\nonumber\\  & &\,\,\,\,\,\,\,\,\,\,\,\,\,\, +{i \over 2}({\bar v}_{\dot
\alpha}D^{\dot \alpha\alpha } \mu_{\alpha}+{\bar 
\mu}^{\alpha}D_{\alpha \dot \alpha}v^{\dot \alpha}) -{1 \over 2}
 ({\overline M}^{\alpha} \psi_{\alpha \dot
\alpha}v^{\dot \alpha}+ {\bar v}^{\dot \alpha} \psi_{\dot
\alpha \alpha }M_{\alpha})\nonumber\\  & &\,\,\,\,\,\,\,\,\,\,\,\,\,\, -{1
\over 4} ({\bar h}_{\dot\alpha} h^{\dot\alpha} +i{\bar v}_{\dot\alpha} \phi
v^{\dot\alpha})
\Big], 
\label{manza} 
\eea
\bea  
\{Q,\Psi _{\rm proj}\}&= &\{ Q, -{ 1\over 2}\int _{X} \big[i
 \tr(\lambda \wedge *d_A^{*} \psi) +e  ({\bar
\mu}^{\alpha}\lambda M_{\alpha}- {\overline M}^{\alpha} \lambda
\mu_{\alpha}) \big] \}\nonumber\\
 & =& \int _{X} \big[\tr\big(- { i\over 2}\eta \wedge *d_A^{*} \psi -{i \over
2}
\lambda \wedge *d_A^{*}d_A \phi -{1 \over 2} 
\lambda\wedge *[*\psi,\psi] \big) 
\nonumber\\ & & \,\,\,\,\,\,\, + {1 \over 2} e  ({\bar \mu}^{\alpha}\eta
M_{\alpha} + {\overline M }^{\alpha} \eta \mu_{\alpha}) \nonumber\\ & &
\,\,\,\,\,\,\,+ e  ({\bar
\mu}^{\alpha}\lambda\mu_{\alpha} -{i\over 2} {\overline M}^{\alpha}
\{\phi,\lambda\} M_{\alpha} \big)\big]. \label{limon} 
\eea

The sum of (\ref{manza}) and (\ref{limon}) is just the same as the  sum of
the terms in (\ref{fresa}) plus $-|s(A,M)|^2$ as given in (\ref{melon}) once
the auxiliary fields 
$H_{\alpha\beta}$ and $h_{\dot\alpha}$ have been integrated out. This is
indeed the exponent appearing in the Thom form (\ref{mqaj}) which must be
identified as minus the action, $-S$, of the Topological Quantum Field
Theory. After carrying out the integration of the auxiliary fields the
resulting action turns out to be:  
\bea S&=&
\int_{X} e \big[ g^{\mu\nu}D_\mu\overline M^\alpha D_\nu M_\alpha + {1\over
4} R
\overline M^\alpha M_\alpha \nonumber\\ & & \,\,\,\,\,\,\,\,\ -{1\over 4}
\tr (F^{+\alpha\beta} F_{\alpha\beta}^+)  +{1\over 4}(\overline M^{(\alpha}
T^a M^{\beta)}) (\overline M_{(\alpha} T^a M_{\beta)})]\nonumber\\
 &+& \int_X \tr\big({ i\over 2} \eta \wedge *d_A^{*} \psi +{i \over 2{\sqrt
2}}\chi ^{\alpha
\beta}(p^+(d_A\psi))_{\alpha\beta} +{i\over
8}\chi^{\alpha\beta}[\chi_{\alpha\beta},\phi]\nonumber\\ & &
\,\,\,\,\,\,\,\,\ + {i
\over 2} \lambda \wedge *d_A^{*}d_A \phi - { 1\over 2}
\lambda\wedge *[*\psi,\psi] \big) 
\nonumber\\  &+&\int_{X}e\Big({i\over 2}{\overline M}^{\alpha}\{\phi,
\lambda\} M_{\alpha}  -{1
\over {\sqrt2}} ({\overline M}_{\alpha} {\chi}^{\alpha \beta} \mu _{\beta
}-{\bar
\mu}_{\alpha}{\chi}^{\alpha \beta} M_{\beta}) \nonumber\\ & &
\,\,\,\,\,\,\,\,\ -{i\over 2} ({\overline v}_{\dot
\alpha} D^{{\dot\alpha} \alpha } {\mu}^{\alpha}+{\bar \mu}^{\alpha} D_{\alpha
{\dot\alpha}} v^{\dot \alpha}) +{1 \over 2}({\overline
M}^{\alpha}{\psi}_{\alpha {\dot \alpha}}v^{\dot \alpha}+{\bar v}_{\dot
\alpha}{\psi}^{{\dot \alpha}\alpha }M_{\alpha}) \nonumber\\ & &
\,\,\,\,\,\,\,\,\ -{1\over2}  ({\bar
\mu}^{\alpha}  \eta M_{\alpha}+{\overline M}^{\alpha}
 \eta\mu_{\alpha}) +{i \over 4}{\bar v}^{\dot\alpha} \phi v_{\dot\alpha}
-{\bar
\mu}^{\alpha} \lambda \mu _{\alpha}\Big), 
\label{action}
\eea   where we have used (\ref{current}) to write the term quartic in the
fields
$M_\alpha$. Although this action has been computed considering the fundamental
 representation of the gauge group $SU(N)$, its form is also valid for any
other gauge  group and representation. This action is invariant under the
modified BRST transformations which are obtained from (\ref{pera}) after
taking into account the modifications which appear once the auxiliary fields
have been integrated out. As we will see in the next  section, it contains
the standard gauge fields of a twisted
$N=2$ vector multiplet, or Donaldson-Witten fields, coupled to the matter
fields of the twisted $N=2$ hypermultiplet.

\subsection{The observables}

We want to compute now the observables of the theory, {\it i.e.}, the
cohomology
 of the BRST operator $Q$, in the $SU(2)$ case. As we will see, they take the
same  form that in Donaldson-Witten  theory \cite{tqft}. From the
transformations (\ref{pera}) it follows that any polynomial  in the field
$\phi$ gives an observable. But we saw in Chapter 3 that the  natural
interpretation  of $\phi$ is as the curvature of the principal bundle ${\cal
M} \rightarrow {\cal M}/ {\cal G}$. If we recall the construction of
cohomology classes in Donaldson theory  presented in Chapter 1, section 4,
and the equation (\ref{ponti}), we see that the  natural  starting point for
the cohomology, in the $SU(2)$ case, is the operator:
\be {\cal O}= {1\over 8 \pi^2} \tr (\phi^2),
\label{mioperador}
\ee just like in Donaldson-Witten theory. It it important to notice that,
although  the structure of the observable is the same, it is different from
the one  appearing in Donaldson theory. This is because the curvature of the
corresponding  principal bundle is different and includes a contribution form
the monopole fields,  as it is clear from (\ref{curvatriz}), (\ref{marga})
and (\ref{tren}). 
 
Starting from the operator (\ref{mioperador}) one can find the  other
generators of the cohomology ring using the {\it descent procedure},
introduced in 
\cite{tqft}. Starting from ${\cal O}^{(0)}={\cal O}$, one recursively finds
$k$-forms  on the manifold 
$X$, ${\cal O}^{(k)}$, such that the following descent equations are verified:
\be d{\cal O}^{(k)}= \{ Q, {\cal O}^{(k+1)}\}.
\label{descent}
\ee  One then finds:
\bea {\cal O}^{(1)} &=& {1\over 4 \pi^2} \tr (\phi\psi), \nonumber\\ {\cal
O}^{(2)} &=& {1\over 4\pi^2} \tr (\phi F + {1\over
2}\psi\wedge\psi),\nonumber\\ {\cal O}^{(3)} &=& {1\over 4\pi^2} \tr
(\psi\wedge F),\nonumber\\ {\cal O}^{(4)} &=& {1\over 8\pi^2} \tr (F\wedge F).
\label{obser}
\eea From the descent equations (\ref{descent}) follow that if $\Sigma$ is a
$k$-dimensional homology cycle, then
\be I(\Sigma) = \int_\Sigma {\cal O}^{(k)},   
\label{moreobser}
\ee is in the cohomology of $Q$. For simply connected four-manifolds, which
is the case we will be interested in, $k$-dimensional homology cycles only
exist for
$k=0,2,4$. For $k=4$ the cycle $\Sigma$ is the four-manifold $X$ and $I(X)$
is the instanton number. The only operators which are relevant are then
(\ref{mioperador})  and the $I(\Sigma)$ associated to two-cycles. Notice that
this operator has the same  structure than (\ref{dosclase}), but involves a
different curvature. As a final remark,  the degrees of the above observables
regarded as differential forms on the configuration  space are precisely
their ghost numbers, according to the assignment above. 

Now we can define polynomial invariants for this theory, mimicking the
Donaldson  construction. Of course, we have  not provided all the ingredients
from the mathematical  point of view, but we can just define them as the 
correlation functions of the Topological Quantum  Field Theory we have
constructed, which is equivalent, as we will see in next section,  to the
twisted $N=2$ QCD with one massless hypermultiplet. At least in the $SU(2)$
case,  the moduli space of non-abelian monopoles can be defined  in a proper
way, as we have seen  in last Chapter. One may hope that the topological
correlation functions of the  twisted theory will be equivalent to the
properly defined invariants associated to  the
$SU(2)$ monopole equations. 

As usual in Quantum Field Theory, we will group all the correlation functions
in a generating function, as in (\ref{corre}):
\be 
\langle {\rm exp} (\sum_{a}\alpha_{a}I(\Sigma_{a})+\mu {\cal O}) \rangle, 
\label{generatriz}
\ee  summed over instanton numbers of the bundle
$E$. In (\ref{generatriz}), as in (\ref{corre}), the $\Sigma_{a}$ denote a 
basis of the two-dimensional homology of $X$, and therefore $a=1, \cdots,
{\rm dim} \,\ H_2(X,{\bf Z})$.

In (\ref{generatriz}), the terms appearing in the expansion have the form:
\be {\cal O}(x_1)\cdots{\cal O}(x_r) I(\Sigma_1)\cdots I(\Sigma_s).
\label{general}
\ee Correlation functions involving operators of the form (\ref{general})
vanish unless the following selection rule holds:
\be 4r+2s={\rm dim} \, {\cal M}_{\rm NA}.
\label{selrule}
\ee where ${\rm dim} \, {\cal M}_{\rm NA}$ is given in (\ref{dim}) with 
$N=2$. As we know  from the discussion on Donaldson theory, and the
expressions (\ref{corre}) and  (\ref{otrodon}), correlation functions of the
topological  field theory are interpreted mathematically as intersection
forms in the moduli space. The operator
$\cal O$ represents a cohomology class of degree four, and $ I(\Sigma)$
represents a cohomology class of degree two. If we assume regularity, or if
the obstruction cohomology  can be considered as a vector bundle over the
moduli space, the condition  (\ref{selrule}) simply says that the integral of
these differential forms vanishes unless the total degree equals the
dimension of the moduli space. This condition has a natural interpretation in
field-theoretical terms \cite{tqft}. The dimension of the moduli space
corresponds to the index of the operator $T=ds \oplus C^{\dagger}$, which
gives the two-term  form of the deformation complex (\ref{file}). But this is
also the operator  associated to the grassmannian fields in (\ref{action}),
and its index gives the anomaly in the ghost number. The selection rule
(\ref{selrule}) is therefore the 't Hooft rule which says that fermionic zero
modes in the path integral measure should be soaked up in the correlation
functions. 

A consequence of (\ref{selrule}) is that, in order to have non-zero
correlation  functions, the dimension of the moduli space must be even.
Looking at (\ref{dim}),  it follows that the quantity
\be
\Delta={ \chi + \sigma \over 4}
\label{ladelta}
\ee must be an integer. Notice that on a Spin manifold (the case we are
considering)  the signature verifies 
\be
\sigma  \equiv 0 \,\ {\rm mod}\,\ 8,
\label{rollin}
\ee as the index of the Dirac operator is $-\sigma/8$. In particular, on a
four-dimensional K\"ahler manifold, one has
\be
\chi+\sigma=2-2b_1+b_2^{+}=4(1-h^{1,0}+h^{2,0}).
\label{laura}
\ee where $h^{1,0}$, $h^{2,0}$  denote Hodge numbers.  Therefore,
(\ref{ladelta}) is always an integer. These conditions  appear also in
Donaldson theory, and they will have an interesting interpretation in the 
light of the equivalence of both theories to the Seiberg-Witten monopole
theory, as  we will see in Chapter 7.

\section{Twist of the $N=2$ theory}

The action (\ref{action}) can be obtained from the twist of $N=2$
supersymmetric Yang-Mills theory with gauge group $SU(N)$ coupled to an $N=2$
hypermultiplet in the fundamental  representation. The twist of $N=2$ matter
superfields coupled to $N=2$ supersymmetric  Yang-Mills theory has been
studied from different perspectives in 
\cite{kr, alone, altwo,  afone, aftwo, af, tqcd}. Here we follow the approach
in
\cite{lmpol}. For an  exhaustive study of the twisting procedure for $N=2$
and $N=4$ theories,  see \cite{high}.  The basic idea involved in the
twisting has been introduced in Chapter 2.  Recall that in the proccess of
twisting the isospin index
$i$ becomes a spinor index, and the supercharges become 
$Q_{i \alpha} \rightarrow Q_{\beta \alpha}$ and ${\overline Q}^{i \dot\alpha }
\rightarrow {\overline Q}^{\beta \dot\alpha}$.  The trace of 
$Q_{\beta \alpha}$, $Q=Q_\alpha{}^\alpha=Q_{{\bar 1}2}-Q_{{\bar 2}1}$,
becomes a
$(0,0)$ rotation invariant operator (we put bars on the isospin indices).
 If there is no 
$N=2$ central extension, from the supersymmetry algebra follows that $Q$ obeys
$Q^2=0$. This operator can be regarded as a BRST operator and the
$U(1)_{\cal R}$ charges as ghost numbers.

The twist of $N=2$ supersymmetric Yang-Mills theory has been already
described in Chapter 2. We will consider  the case of its coupling to an
$N=2$ hypermultiplet.
$N=2$ matter is usually represented by $N=2$  hypermultiplets. The
hypermultiplet contains a complex scalar isodoublet
$q=(q_1,q_2)$, fermions $\psi_{q \alpha}$, $\psi_{\tilde q \alpha}$,
$\overline \psi_{q \dot\alpha}$, $\overline \psi_{\tilde q 
\dot\alpha}$, and a complex scalar isodoublet auxiliary field
$F_i$. The fields
$q_i$, $\psi_{q \alpha}$, $\overline \psi_{\tilde q 
\dot\alpha}$, and $F_i$ are in the fundamental representation of the gauge
group, while the fields 
$q^{i\dagger}$, $\psi_{\tilde q \alpha}$,
$\overline \psi_{q \dot\alpha}$, and $F^{i\dagger}$ are in the conjugate
representation. From the point of view of $N=1$  superspace, this multiplet
contains two $N=1$ chiral multiplets and therefore it can be described by two
$N=1$ chiral superfields
$Q$ (this $Q$ should not be confused with the BRST operator) and $\widetilde
Q$, {\it i.e.}, these superfields satisfy the constraints $\overline
D_{\dot\alpha} Q=0$ and
$\overline D_{\dot\alpha} \widetilde Q=0$. They have 
$U(1)_{\cal R}$ charge $0$. While the superfield $Q$ is in the fundamental
representation of the gauge group, the superfield $\widetilde Q$ is in the
corresponding conjugate representation. The component fields of these
$N=1$ superfields are:
\bea Q, \;\; Q^\dagger
 \;\; & \longrightarrow \;\;  q_1, \;\;   \psi_{ q\alpha}  \;\;  F_2, \;\;
q^{1\dagger} \;\; 
\overline\psi_{ q \dot\alpha}, \;\;  F^{2\dagger}, \nonumber\\
\widetilde Q , \;\; \widetilde Q^\dagger 
\;\;& \longrightarrow \;\;  q^{2\dagger}, \;\; \psi_{\tilde q \alpha}, 
\;\; F^{1\dagger}, \;\; q_2, \;\;
\overline \psi_{\tilde q\dot\alpha},   \;\;  F_1.
\label{compas}
\eea The $U(1)_{\cal R}$ transformations of the
$N=1$ superfields are
\be Q \rightarrow Q({\rm e}^{i\phi}\theta),\;\;\; {\rm and} \;\;\;
\widetilde Q \rightarrow \widetilde Q({\rm e}^{i\phi}\theta),
\label{pertiga}
\ee

In $N=1$ superspace the action for the $N=2$ hypermultiplet coupled to $N=2$
supersymmetric Yang-Mills takes the form:
\bea & & \int d^4x\,d^2\theta\,d^2\overline\theta\,  (Q^\dagger {\rm e}^V Q +
\widetilde Q^\dagger {\rm e}^{-V} \widetilde Q)
\nonumber\\ & &-i\sqrt{2} \int  d^4x\,d^2\theta\, \widetilde Q \Phi Q +
i\sqrt{2}\int d^4x\,d^2\overline\theta\, 
\widetilde Q^\dagger \Phi^\dagger Q^\dagger.
\label{silvana}
\eea  Notice that the last two terms are consistent with the fact that while
$\Phi$ is in the adjoint representation of the gauge group, the superfields
$Q$ and $\widetilde Q$ are in the fundamental and in its conjugate,
respectively.
 
The $SU(2)_I$ current includes now a  contribution from the bosonic part of
the hypermultiplet:
\be j^{\mu}_a={\overline
\lambda}{\sigma}^{\mu}{\sigma_a}\lambda-iq^{\dagger}\sigma_aD^{\mu}q 
+iD^{\mu}q^{\dagger}\sigma_a q.
\label{corredos}
\ee The twist of the theory can also be understood as a gauging of the
$SU(2)_I$ current,  as it happens with the pure $N=2$ Yang-Mills theory.
This  is achieved by  adding to the original Lagrangian  a term
$-\omega_{\mu}^a j^{\mu}_a + q^{\dagger}\omega_{\mu}^a\omega^{\mu
b}{\sigma_a}  {\sigma_b}q$. The kinetic term for the bosons in the resulting
theory is then
\be (D_{\mu}+ i\omega_{\mu}^a{\sigma_a})q^{\dagger}(D_{\mu}-i\omega^{\mu
a}{\sigma_a})q,
\label{kinbos}
\ee where $D_{\mu}$ is the covariant derivative acting on scalars in  the
fundamental of $SU(N)$. We then see that, after the twisting, the bosonic
fields 
$q$ become positive-chirality spinors. This is the result we get after
redifining the  rotation group as in (\ref{nuevasim}). Therefore, under the
twisting, the fields in  the $N=2$ hypermultiplet become:
\bea q_i\;\;  (0,0,1/2)^0 \;\; & \longrightarrow& M_\alpha \;\;
(1/2,0)^0,\nonumber\\
\psi_{q \alpha} \;\; (1/2,0,0)^1 \;\; & \longrightarrow&  -\mu_\alpha /{\sqrt
2}
\;\; (1/2,0)^1,\nonumber\\
\overline \psi_{\tilde q}^{ \dot\alpha} \;\; (0,1/2,0)^{-1} \;\; &
\longrightarrow&  v^{\dot\alpha}/{\sqrt 2}  \;\; (0,1/2)^{-1}, \nonumber\\
F_i \;\; (0,0,1/2)^2 \;\; &
\longrightarrow & K_\alpha \;\; (1/2,0)^{2},\nonumber\\ q^{\dagger i} \;\; 
(0,0,1/2)^0 \;\; & \longrightarrow& 
\overline M^\alpha \;\; (1/2,0)^0,  \nonumber\\
\overline \psi_{q  \dot\alpha} \;\; (0,1/2,0)^{-1} \;\; & \longrightarrow & 
\overline v_{\dot\alpha}/{\sqrt 2} \;\; (0,1/2)^{-1}, \nonumber\\
\psi_{\tilde q}^ \alpha \;\; (1/2,0,0)^1 \;\; & \longrightarrow & 
 \overline  \mu^\alpha/{\sqrt 2} \;\; (1/2,0)^1,\nonumber\\ F^{\dagger i}\;\;
(0,0,1/2)^{-2} \;\; & \longrightarrow&  
\overline K^\alpha \;\; (1/2,0)^{-2}.
\label{paloma}
\eea Again, the assignment of ghost numbers is consistent with (\ref{salto}).

The $Q$-transformations of the twisted fields can be obtained very simply
from the
$N=2$ supersymmetry transformations. These last transformations are generated
by the operator
$\eta^{i\alpha} Q_{i \alpha} + \overline \eta_{i\dot\alpha }
\overline Q^{i \dot\alpha }$ where $\eta^{i\alpha}$ and 
$\overline \eta_{i\dot\alpha }$ are anticommuting parameters. To get the
$Q$-transformations of the fields one must consider $\overline
\eta^{\dot\alpha i}=0$ and replace
$\eta_{i\alpha} \rightarrow \rho \epsilon_{\beta \alpha}$, being
$\rho$ an arbitrary scalar anticommuting parameter, and where $\epsilon_{21}=
\epsilon^{12}=1$ raises and lowers isospin indices. Let's consider first the 
action and the $Q$-symmetry in the on-shell theory. The $Q$-transformations
of the  fields are: 
\be
\begin{array}{cclcccl} [Q, A_\mu] &=& \psi_\mu,&
\,\,\,\,\,\,\,\,\,\,\,\,\,\,\,\,\  & [Q, M_\alpha] &=& \mu_\alpha, \nonumber\\
\{Q,\psi_\mu \} &=& D_\mu \phi,& 
\,\,\,\,\,\,\,\,\,\,\,\,\,\,\,\,\  & [ Q, \overline M_\alpha] &=& \bar
\mu_\alpha,
\nonumber\\ {[}Q, \lambda] &=& \eta,& \,\,\,\,\,\,\,\,\,\,\,\,\,\,\,\,\  &
\{Q,\mu_\alpha \} &=& - i \phi M_\alpha, \nonumber\\
\{Q,\eta\} &=& i[\lambda,\phi],& 
\,\,\,\,\,\,\,\,\,\,\,\,\,\,\,\,\  &
\{Q, \bar\mu_\alpha \} &=& i \overline M_\alpha \phi,
\nonumber\\  {[} Q, \phi] &=& 0,& \,\,\,\,\,\,\,\,\,\,\,\,\,\,\,\,\  &
\{Q,v^{\dot\alpha}\} &=& -2iD^{\dot\alpha\alpha} M_\alpha,\nonumber\\
\{Q,\chi^a_{\alpha\beta} \} &=& -i\sqrt{2}  (F^a_{\alpha\beta}+{i}\overline
M_{(\alpha} T^a M_{\beta)}), &  
\,\,\,\,\,\,\,\,\,\,\,\,\,\,\,\,\  &
\{Q,\bar v_{\dot\alpha}\} &=& 
 -2iD_{\alpha\dot\alpha}\overline M^\alpha,
\end{array}
\label{dadi}
\ee and for the twisted action in Euclidean  space one gets:
\bea S_1&=&\int_X \sqrt{g}  \Big[\tr\big( {1\over 4} F_{\mu\nu}
F^{\mu\nu}-{i\over  2\sqrt{2}}\chi^{\alpha\beta}  D_{\alpha \dot \alpha}
\psi_{\beta}{}^{\dot \alpha} -{i\over 8} \chi^{\alpha\beta}
[\phi,\chi_{\alpha\beta}] \nonumber\\ & &
\,\,\,\,\,\,\,\ +{i\over 2}D_{\mu}\lambda  D^{\mu} \phi + {i\over 2}\eta 
D^{\mu}
\psi_{\mu}
 - {1\over 2}\psi_{\mu} [\psi^{\mu},\lambda] - {1\over 8}
[\phi,\lambda]^2+{i\over 8}
\eta[\phi,\eta] \big) \nonumber\\ & &\,\,\,\,\,\,\,\ +D_\mu \overline
M^\alpha D^\mu M_\alpha +{1\over 4} R \overline M^\alpha M_\alpha +{1\over 4}
\overline M^{(\alpha} T^a M^{\beta )}
\overline M_{(\alpha} T^a M_{\beta )} \nonumber\\ & &\,\,\,\,\,\,\,\ -{i\over
2} (\bar v_{\dot\alpha}D^{\dot\alpha \alpha}\mu_\alpha +\bar\mu^\alpha
D_{\alpha\dot\alpha}v^{\dot\alpha}) +{i\over 2}\overline M^\alpha
\{\phi,\lambda\} M_\alpha \nonumber\\ & &\,\,\,\,\,\,\,\ -{1\over
\sqrt{2}}(\overline M_\alpha
\chi^{\alpha\beta} \mu_\beta -\bar \mu_\alpha \chi^{\alpha\beta} M_\beta)
+{1\over 2}(\overline M^\alpha \psi_{\alpha\dot\alpha} v^{\dot\alpha} +\bar
v_{\dot\alpha}\psi^{\dot\alpha \alpha} M_\alpha ) \nonumber\\ &
&\,\,\,\,\,\,\,\ -{1\over 2}(\bar \mu^\alpha \eta M_\alpha + 
\overline M^\alpha\eta\mu_\alpha) +{i\over 4} \bar v^{\dot\alpha} \phi
v_{\dot\alpha} - \bar\mu^\alpha \lambda
\mu_\alpha \Big].
\label{taction}
\eea Notice that the matter fields with bars carry a representation
$\overline R$ conjugate to $R$, the one carried by the matter fields without
bars. Notice also the presence of a term involving the curvature of the 
four-manifold
$X$, as in (\ref{melon}). This term must enter the twisted action in order to 
preserve invariance under the topological symmetry $Q$ on curved manifolds.
This  possibility was already envisaged in \cite{tqft}, but it was found
there that, in  the case of $N=2$ Yang-Mills theory, the
 gauging of the $SU(2)_I$ symmetry was enough to gurantee $Q$-invariance on a
curved  four-manifold. In the case at hand, it originates in the following 
way. After gauging the internal $SU(2)_I$ symmetry, the kinetic terms in the 
Lagrangian read
\be g_{\mu \nu}D^{\mu}{\overline M}^{\alpha}D^{\nu}M_{\alpha}-  { i \over
2}\Big({\bar v}_{\dot \alpha}D^{\dot \alpha \alpha}\mu_{\alpha}+ {\bar
\mu}^{\alpha}D_{ \alpha \dot \alpha}v^{\dot \alpha} \Big),
\label{kinbar}
\ee where $D^{\mu}$, $D^{\dot \alpha \alpha}$ are respectively the covariant 
derivative and the Dirac operator associated to the tensor product connection
on 
$S^{+}\otimes E$. If we compute $[Q,{\cal L}]$ with the Lagrangian obtained
by just  gauging the $SU(2)_I$ symmetry, the kinetic terms (\ref{kinbar})
give a non-zero  result. After using the Weitzenb\"ock formula (\ref{miwei}),
we  obtain 
\be [ Q,{\cal L} ] =-{R \over 4}({\overline M}^{\alpha}\mu_{\alpha}+ {\bar
\mu}^{\alpha}M_{\alpha}),
\label{noex}
\ee where $R$ is the scalar curvature of the manifold. It then follows that
the modified  Lagrangian
\be {\cal L}_{\rm top}={\cal L}+{R \over 4}{\overline M}^{\alpha}M_{\alpha}
\label{totex}
\ee is $Q$-closed on a general four-manifold. This then gives the topological
action in  (\ref{taction}), which coincides with (\ref{action}), and shows
that the twisted
$N=2$  QCD theory with $N_f=1$ is equivalent to the non-abelian monopole
theory considered in  the last Chapter. Notice that, as we indicated  in the
previous section, Witten's fixed point theorem implies  that the path
integral of this theory is localized on $Q$-invariant configurations.  The
$Q$-transformations of $\chi$ and
$v^{\alpha}$ in (\ref{dadi}) indicate that these  invariant configurations
verify the non-abelian monopole equations. 

The $Q$-transformations close on-shell up to a gauge transformation whose
gauge parameter is the scalar field $\phi$:
\be
\begin{array}{cclcccl} [Q^2, A_\mu] &=& D_\mu \phi,&
\,\,\,\,\,\,\,\,\,\,\,\,\,\,\,\,\,\,\,\,\,\,\,\,\,\,\,\,\,\,\  & [Q^2,
M_\alpha] &=& - i \phi M_\alpha, \nonumber\\ 
\{Q^2,\psi_\mu \} &=& i[\psi_\mu,\phi],&
\,\,\,\,\,\,\,\,\,\,\,\,\,\,\,\,\,\,\,\,\,\,\,\,\,\,\,\,\,\,\ & [Q^2,
\overline M_\alpha] &=&  i \overline M_\alpha
\phi, \nonumber\\   {[}Q^2, \lambda] &=& i[\lambda,\phi], &
\,\,\,\,\,\,\,\,\,\,\,\,\,\,\,\,\,\,\,\,\,\,\,\,\,\,\,\,\,\,\ &
\{Q^2,\mu_\alpha \} &=&- i \phi \mu_\alpha, \nonumber\\  
\{Q^2,\eta\} &=& i[\eta,\phi],& 
\,\,\,\,\,\,\,\,\,\,\,\,\,\,\,\,\,\,\,\,\,\,\,\,\,\,\,\,\,\,\ &
\{Q^2, \bar\mu_\alpha \} &=&  i \bar\mu_\alpha \phi, \nonumber\\ {[}Q^2,
\phi] &=& 0,& 
\,\,\,\,\,\,\,\,\,\,\,\,\,\,\,\,\,\,\,\,\,\,\,\,\,\,\,\,\,\,\  &
\{Q^2,v_{\dot\alpha}\} &=& - i \phi v_{\dot\alpha},\nonumber\\ 
\{Q^2,\chi_{\alpha\beta} \} &=& i[\chi_{\alpha\beta},\phi],& 
\,\,\,\,\,\,\,\,\,\,\,\,\,\,\,\,\,\,\,\,\,\,\,\,\,\,\,\,\,\,\ & 
\{Q^2,\bar v_{\dot\alpha}\} &=&i \bar v_{\dot\alpha} \phi.
\end{array}
\label{dadidos}
\ee Notice that for the last transformation in the first set and for the last
two in the second set we have made use of the field equations. To obtain the 
$Q$-transformations in (\ref{pera}), we must first of all realize that the
auxiliary  fields of the twisted theory in (\ref{paloma}) are different than
the auxiliary fields in (\ref{pera}). This is a first hint on the existence
of some differences between the theory presented in the previous section and
the twisted theory. Auxiliary fields are useful in supersymmetry because they
permit to close the supersymmey off-shell. In this section we have considered
a version of the $N=2$ hypermultiplet which contains a minimal set of
auxiliary fields.  If one considers this off-shell version of $N=2$
supersymmetry \cite{kr,alone,altwo},  it is possible to show that it does not
lead to a formulation whose action is $Q$-exact.  The theory presents a
non-trivial central charge $Z$, and this  is an inconvenient for the twisting
because then one finds $Q^2=Z$ instead of $Q^2=0$. On the other hand, if one
disregards this problem and  goes along considering the twisted theory, it
turns out that after integrating the auxiliary fields the action of the
twisted theory (\ref{taction}) is just the action (\ref{action}), plus some 
additional terms that will be analyzed in a moment. This  equivalence proves
that in the twisted theory the auxiliary content of the theory and the
$Q$-transformations involving these fields can  be changed in such a way that
an off-shell action can be written as a $Q$-exact quantity and, furthermore,
$Q^2=0$. In other words, one can indeed affirm that the twisted theory is
topological. This was observed for the first time in \cite{kr}.
 
Our next goal is, as in the case of topological sigma models in Chapter 3, to
 construct an off-shell version of the twisted model. In this approach the
steps to  be followed are the same ones as in the case of the topological
sigma models: introduce auxiliary fields $K_{\alpha\beta}$,  $k_{\dot\alpha}$
and $\bar k_{\dot\alpha}$  in the transformations of
$\chi_{\alpha\beta}$, $v_{\dot\alpha}$ and $\bar v_{\dot\alpha}$
respectively, and define the transformations of these fields in such a way
that $Q^2$ on 
$\chi_{\mu\nu}$, $v_{\dot\alpha}$ and $\bar v_{\dot\alpha}$ closes without
making use of the field equations. Following this approach one finds:
\bea
\{ Q , \chi_{\alpha\beta}^a \} &=& K^a_{\alpha\beta} -i\sqrt{2}(
F^a_{\alpha\beta}+{i}\overline M_{(\alpha}  T^a M_{\beta)}), \nonumber\\
\{ Q , v^{\dot\alpha} \} &=& k^{\dot\alpha} -2i D^{\dot\alpha\alpha} M_\alpha,
\nonumber\\
\{ Q , \bar v_{\dot\alpha} \}& = & \bar k_{\dot\alpha} -2i
D_{\alpha\dot\alpha}
\overline M^\alpha, \nonumber\\ {[} Q , K_{\alpha\beta}^a ] &=&
i[\chi_{\alpha\beta}, \phi]^a -{i\sqrt{2}}(p^{+}( D
\psi))_{\alpha\beta}^a + \sqrt{2}(\bar\mu_{(\alpha} T^a M_{\beta)} + \overline
M_{(\alpha} T^a \mu_{\beta)}),\nonumber\\ {[} Q , k^{\dot\alpha} ] &=&- i\phi
v^{\dot\alpha} - 2
\psi^{\dot\alpha \alpha}M_\alpha + 2i D^{\dot\alpha\alpha} \mu_\alpha,
\nonumber\\ {[} Q , \bar k_{\dot\alpha} ] &=& 
 i\phi \bar v_{\dot\alpha} - 2
\psi_{\alpha\dot\alpha}\overline M^\alpha + 2i D_{\alpha\dot\alpha}\bar
\mu^\alpha.  
\label{pendulo}
\eea The non-trivial check now is to verify that $Q^2$ on the auxiliary fields
closes properly. One easily finds that this is indeed the case:
\bea [ Q^2 , K_{\alpha\beta} ] &=& i[K_{\alpha\beta},\phi],\nonumber\\ {[}
Q^2 , k_{\dot\alpha} ] &=& - i\phi k_{\dot\alpha},\nonumber\\ {[} Q^2 , \bar
k_{\dot\alpha} ] &=& i\bar k_{\dot\alpha}\phi.
\label{pendulodos}
\eea It is important to remark that these relations imply that $Q$ closes
off-shell. Our next task is to show that $S_1$ is equivalent to a $Q$-exact
action.

After adding the topological invariant term involving the Chern class,
\be S_2 = {1\over 4} \int_X F\wedge F,
\label{cherndos}
\ee one finds that the off-shell twisted action of the model can be written
as a
$Q$-exact term:
\be
\{Q, \Lambda_0 \} =   S_1+ S_2 + {1\over 4}\int_X \sqrt{g} \big(\bar
k^{\dot\alpha} k_{\dot\alpha}-K^{\alpha\beta} K_{\alpha\beta} \big),
\label{burbuuno} 
\ee where,
\bea
\Lambda_0 &=& \int_X \sqrt{g} \Big[ -{1\over 8}\chi^{\alpha\beta a}
\big(i\sqrt{2}(F^a_{\alpha\beta} +{i} \overline M_{(\alpha}T^a M_{\beta)})+
K^a_{\alpha\beta} \big)\nonumber\\ & & \,\,\,\,\,\,\ +{1\over 8} \bar
v_{\dot\alpha} ( 2 i D^{\dot\alpha\alpha} M_\alpha +k_{\dot\alpha})  +{1\over
8}( 2 i D_{\alpha\dot\alpha} \overline M^\alpha +\bar k_{\dot\alpha})  
v^{\dot\alpha}
\nonumber\\ & & \,\,\,\,\,\,\ +\tr\big({i\over 2} \lambda
D^{\mu}\psi_{\mu}-{i\over 8}
\eta [\phi,\lambda]\big) -{1\over 2}(\bar\mu^\alpha\lambda M_\alpha -\overline
M^\alpha\lambda \mu_\alpha) \Big].
\label{burbu}
\eea The auxiliary field entering (\ref{burbu}) is not the same  as the one
entering (\ref{alsacia}). Again, the auxiliary fields $K_{\alpha\beta}$,
$k_{\dot\alpha}$ and
$\bar k_{\dot\alpha}$ in (\ref{burbu}) appear only quadratically in the
action, contrary to the way they appear in the Mathai-Quillen formalism. The
relation between these two sets of fields can be easily read comparing
(\ref{fermiones}) and (\ref{burbu}), or  (\ref{alsacia}) and (\ref{pendulo}).
Redefining the auxiliary fields as
\bea H_{\alpha\beta}^a &=& K^a_{\alpha\beta} -i\sqrt{2}(
F^a_{\alpha\beta}+i\overline M_{(\alpha}  T^a M_{\beta)}), \nonumber\\
h^{\dot\alpha} &=& k^{\dot\alpha} -2i D^{\dot\alpha\alpha} M_\alpha,
\nonumber\\
\bar h_{\dot\alpha} &=& \bar k_{\dot\alpha} -2i D_{\alpha\dot\alpha} \overline
M^\alpha, 
\label{jas}
\eea one finds that,
\bea [ Q , H_{\alpha\beta} ] &=& i[H_{\alpha\beta},\phi],\nonumber\\ {[} Q ,
h^{\dot\alpha} ] &=&- i\phi v^{\dot\alpha},\nonumber\\ {[} Q , \bar
h_{\dot\alpha} ] &=& i \bar v_{\dot\alpha}\phi ,
\label{muelle}
\eea and the resulting action takes the form:
\be
\{Q, \Lambda \}
\label{osci}
\ee where,
\bea
\Lambda &=& \int_X \sqrt{g} \Big[- {1\over 4}\chi^{\alpha\beta
a}\big(i{\sqrt{2}} (F^a_{\alpha\beta} +{i} \overline M_{(\alpha}T^a
M_{\beta)}) + H^a_{\alpha\beta}\big) \nonumber\\ & & \,\,\,\,\,\,\ +{1\over
8}\bar v_{\dot\alpha} (4iD^{\dot\alpha\alpha}  M_\alpha
+h_{\dot\alpha})+{1\over 8}(4i D_{\alpha\dot\alpha} 
\overline M^\alpha +
\bar h_{\dot\alpha}) v^{\dot\alpha}
\nonumber\\ & &\,\,\,\,\,\,\ +\tr \big({i\over 2}\lambda D^{\mu}
\psi_{\mu}-{i\over 8}
\eta [\phi,\lambda]\big) -{1\over 2}(\bar\mu^\alpha \lambda M_\alpha
-\overline M^\alpha \lambda \mu_\alpha) \Big].
\label{burbutres}
\eea

The action (\ref{osci}) differs from the one that follows after acting with
$Q$ on the gauge fermions (\ref{tigre}) and (\ref{oso}) in the terms which
are originated from
$-\tr({i\over 8} \eta [\phi,\lambda])$. These terms comes form the potential 
energy of the $B$, $B^{\dagger}$ in the untwisted theory, and their absence 
in the Mathai-Quillen formalism is a well known fact in  Donaldson-Witten
theory
\cite{tqft}. As they are $Q$-exact, their presence does not play any important
r\^ole towards the computation of topological invariants. This  proves the 
equivalence between the topological action, obtained through the
Mathai-Quillen  formalism, and the action obtained from the twist of $N=2$
supersymmetric Yang-Mills  coupled to one matter hypermultiplet ($N_f=1$). 

\section{The massive theory}

The purpose of this section is twofold. First of all, we will exploit  the
fact that the non-abelian monopole equations, for $SU(N)$ and the  monopole
fields in the fundamental representation, have a $U(1)$ symmetry 
\cite{pt, tqcd, parktwo}. We will then obtain an equivariant extension of 
the Thom form in this case, providing in this way the second example of the
construction in  Chapter 3. After this construction, we will show that the
equivariant extension  constructed in this way is the twisted $N=2$
Yang-Mills theory coupled to a  {\it massive} hypermultiplet. The connection
between the $U(1)$ equivariant cohomology and the massive theory was  pointed
out in \cite{parktwo}. 

First of all, we define vector field actions on ${\cal M}$ and ${\cal F}$
associated to a $U(1)$ action  as follows:
\bea
\phi_t^{\cal P} (A, M_{\alpha} ) =& (A, {\rm e}^{it}M_{\alpha}), \nonumber\\
\phi_t^{\cal F} (\chi, M^{\dot \alpha}) =& (\chi, {\rm
e}^{it}M^{\dot\alpha}), 
\label{circulos}
\eea where, as usual, $M_{\alpha} \in \Gamma (X, S^{+} \otimes E)$, $M^{\dot
\alpha}
\in \Gamma (X, S^{-} \otimes E)$ and $\chi \in \Omega^{2,+} (X, {\bf g}_E)$. 
It is clear that these actions commute with the action of the group of gauge 
transformations on both ${\cal M}$ and ${\cal F}$. Furthermore, the metrics
 on these spaces are preserved by the $U(1)$ action.  The section $s:{\cal M}
\rightarrow {\cal F}$ defined in (\ref{section})  is clearly equivariant with
respect  to the $U(1)$ actions given in (\ref{circulos}). Namely,
\be s\big(\phi_t^{\cal P} (A, M_{\alpha} )\big)=\phi_t^{\cal F}s(A,
M_{\alpha}).
\label{equimono}
\ee We are in the conditions of Chapter 3, section  3.5, and therefore we can
construct the equivariant  extension of the Thom form of  the associated 
vector bundle ${\cal E}=({\cal M} \times {\cal F})/ {\cal G}$.  First we
compute the 
${\Lambda}$ matrix on the fibre according to (\ref{openuevo}).  In local
coordinates we get:
\be
\Lambda \chi =0, \,\,\,\,\,\,\ \Lambda M_{\dot \alpha}^{j}= -iM_{\dot
\alpha}^{j}.
\label{lieataca}
\ee  Notice that, if we split $M_{\dot \alpha}^{j}$ in its real  and imaginary
parts, 
$\Lambda$ is given by the matrix (\ref{matriz}). From (\ref{circulos}) and 
(\ref{vector}) we can also obtain the local  expression of the associated
vector field $X_{{\cal M}}$ in $(A,M_{\alpha})$:
\be X_{\cal M}=(0, iM_{\alpha}) \in \Omega^{1}(X,{\bf g}_E) \oplus 
\Gamma (X, S^{+} \otimes E).
\label{campillo}
\ee  The additional terms we get in the topological Lagrangian
(\ref{lagrangiano}) after the equivariant  extension are associated to
$\Lambda$, which has already been computed, and to $\nu(X_{\cal M})$. The
explicit expression of $\nu=C^{\dagger}$  is given in  (\ref{marga}), and
together with (\ref{campillo}) we get:
\be
\nu(0, iM^{i}_{\alpha})={\overline M}^{\alpha
j}M_{\alpha}^i-{\delta^{ij}\over N} {\overline M}^{\alpha k}M_{\alpha}^k.
\label{masexplicito}
\ee The additional terms in the topological lagrangian due to the equivariant 
extension are  then given by:
\be u \int_{X} e \big( -{i \over 4}{\bar v}^{\dot\alpha} v_{\dot\alpha}-
i{\overline M}^{\alpha}\lambda M_{\alpha} \big)
\label{masillas}
\ee where we have deleted the $SU(N)$ indices, and $v_{\dot\alpha}$ is the
auxiliary field  associated to the monopole coordinate on the fibre. The BRST
cohomology of the resulting model follows from the expressions in Chapter 3, 
(\ref{alsacia}) and (\ref{masmisterio}), and the explicit expressions for the
vector  field action (\ref{campillo}) and (\ref{lieataca}). The only changes
are in the  transformations for the monopole fields, as $U(1)$ does not act
on ${\cal A}$, 
$\Omega^{2,+}({\bf g}_E)$. We get:
\bea [Q, M_\alpha] &=& \mu_\alpha, \nonumber\\   
\{Q,\mu_\alpha \} &=& -iu M_\alpha - i \phi M_\alpha, \nonumber\\ 
\{Q,v^{\dot\alpha}\} &=& h^{\dot\alpha},\nonumber\\  {[} Q,h^{\dot\alpha} ]
&=& -iu v^{\dot\alpha}- i \phi v^{\dot\alpha}.
\label{brsmasa}
\eea   As in our discussion on the topological sigma model in Chapter 3, we
can add a 
$d_{X_{\cal M}}$-exact piece to the action starting  with a differential form
like the one in (\ref{laforma}). Now we must take into account that  we can
only add to the topological lagrangian basic forms on ${\cal M}$ which
descend to  ${\cal M}/{\cal G}$. If we define a differential form on ${\cal
M}$ starting from (\ref{pina}) as
\be
\omega_{X_{\cal M}}(Y)=g_{\cal M}(X_{\cal M}, Y),
\label{formilla}
\ee we can use invariance of the metric on ${\cal M}$, $g_{\cal M}$ and
$X_{\cal M}$  under the action of the gauge group  to see that the above form
is in fact invariant. But the  horizontal character of (\ref{formilla}) is
only guaranteed   if
$X_{\cal M}$ is horizontal. This is in fact not true in our case, as it
follows from  (\ref{masexplicito}). Therefore we must enforce a horizontal
projection of
$\omega_{X_{\cal M}}$ using  the connection on ${\cal M}$, and consider the
form
$\omega^{h}_{X_{\cal M}}= \omega_{X_{\cal M}}h$. Actually we are interested
in  
\be d_{X_{\cal M}}\omega^{h}_{X_{\cal M}}=d\omega^{h}_{X_{\cal M}}-
u\iota(X_{\cal M})\omega^{h}_{X_{\cal M}},
\label{derivando}
\ee which also descends to ${\cal M}/{\cal G}$. In computing the above
equivariant exterior  derivative we must be careful, as in (\ref{misterios}).
This can be easily done using the BRST  complex that we motivated
geometrically in (\ref{alsacia}) and (\ref{masmisterio}). Of course, from
(\ref{pina})  and (\ref{campillo})
 we can give an explicit expression of (\ref{formilla}). Introducing the
basis of differential forms for
$\Gamma(X,S^{+} \otimes E)$, $\mu_{\alpha}$, we get: 
\be
\omega_{X_{\cal M}}={i \over 2} \int_{X} e\big({\bar \mu}^{\alpha
}M_{\alpha}-{\overline M}^{\alpha }\mu_{\alpha}\big).
\label{masmasilla}
\ee Acting with $d_{X_{\cal M}}$ or, equivalently, with the BRST operator, we
get:
\be Q\omega_{X_{\cal M}}=-i \int_{X} e{\bar \mu}^{\alpha
}\mu_{\alpha}-\int_{X} e{\overline M}^{\alpha }\phi M_{\alpha}-  u\int_{X} e
{\overline M}^{\alpha } M_{\alpha}.
\label{masotta}
\ee As we will see, with (\ref{masillas}) and (\ref{masotta}) we  reconstruct
all the terms appearing in the  twisted theory with a massive hypermultiplet.

The observables of the non-abelian monopole  theory are differential forms on
the corresponding moduli spaces, and they  are constructed from the 
horizontal projections of differential forms on the principal bundle
associated to the  problem, as we saw in Chapter 1 in the  case of Donaldson
theory. They involve the curvature form of this bundle. In the equivariant
extension of the monopole theory these observables have the  same form, but
one must use instead the equivariant curvature of the bundle, given in
(\ref{pfbeqcurv}). From the point of view of the BRST complex they have the
form given in (\ref{obser}), and for simply-connected  manifolds we are
interested only in  
\be {\cal O}={1 \over 8\pi^2}\tr 
\phi^2,\,\,\,\,\ I(\Sigma)={1 \over 4\pi^2}\int_{\Sigma}
\tr \big(\phi F+{1 \over 2} \psi \wedge \psi \big).
\label{observa}
\ee As we have pointed out, the $\delta$-function involved in
(\ref{lagrangiano}) constrains $\phi$ to be the equivariant curvature  of the
bundle ${\cal M}$,
$K_{X_{\cal M}}$. To check that the forms in (\ref{observa}) are closed one
must be careful with  the horizontal projection involved in the computation.
Although the vector field $X_{\cal M}$ doesn't act  on ${\cal A}$, the
contraction $\iota (X_{\cal M})\psi$ is not zero, as
$\psi$ must be horizontally  projected and the field $X_{\cal M}$ must be
substituted by  $X_{\cal M}h=X_{\cal M}-R_{p*}\theta (X_{\cal M})$.  Of
course, using the BRST complex this verification is automatic, but one should
not forget the geometry hidden inside it.

We will show now that the above equivariant extension correspond to the
inclusion  of mass terms for the matter hypermultiplet. These terms are
obtained with the  following superpotential:
\be -{m\over 4}\int  d^4x\,d^2\theta\, \widetilde Q  Q  -{\rm h.c.},
\label{supermasa}
\ee where $m$ is a mass parameter.  The twisted action in euclidean space,
including (\ref{supermasa}), has the  additional terms:
\be  S_{m}=\int_X {\sqrt g} \Big[ {1\over 4} m^2 \overline M^\alpha M_\alpha
+  {1\over 4} m \bar \mu^\alpha
\mu_\alpha - {1\over 4} m \bar v^{\dot \alpha}v_{\dot\alpha} - m \overline
M^\alpha
\lambda M_\alpha - {i\over 4} m \overline M^\alpha\phi M_\alpha \Big].
\label{maccion}
\ee The $Q$-transformations for the matter fields, once the mass terms are
added, have  a non-trivial central charge:
\bea [Q, M_\alpha] &=& \mu_\alpha, \nonumber\\  {[}Q, \overline M_\alpha] &=&
\bar
\mu_\alpha, \nonumber\\  
\{Q,\mu_\alpha \} &=& m M_\alpha - i \phi M_\alpha, \nonumber\\ 
\{Q, \bar\mu_\alpha \}&=&- m \overline M_\alpha + i \overline M_\alpha \phi,
\nonumber\\ 
\{Q,v^{\dot\alpha}\} &=& -2iD^{\dot\alpha\alpha} M_\alpha,\nonumber\\ 
\{Q,\bar v_{\dot\alpha}\} &=  & -2iD_{\alpha\dot\alpha}\overline M^\alpha.
\label{dadimasa}
\eea These transformations close on-shell up to a gauge transformation,  as in
(\ref{dadidos}),  but now they include a central charge transformation of the
type presented in  (\ref{orange}), and whose parameter is proportional to the
mass of the field  involved:
\bea [Q^2, M_\alpha] = & m M_\alpha - i \phi M_\alpha, \nonumber\\  {[}Q^2,
\overline M_\alpha] = & -m \overline M_\alpha + i \overline M_\alpha
\phi, \nonumber\\ 
\{Q^2,\mu_\alpha \} = & m \mu_\alpha - i \phi \mu_\alpha, \nonumber\\
\{Q^2, \bar\mu_\alpha \} = &-m \bar\mu_\alpha + i \bar\mu_\alpha \phi,
\nonumber\\
\{Q^2,v_{\dot\alpha}\} = & m v_{\dot\alpha} - i \phi
v_{\dot\alpha},\nonumber\\
\{Q^2,\bar v_{\dot\alpha}\} = &- m \bar v_{\dot\alpha}  + i \bar
v_{\dot\alpha} \phi
\label{dadidosmasa}.
\eea For the last two transformations we have made use of the field
equations.  The central charge acts trivially on the pure Yang-Mills fields
or Donaldson-Witten fields but non-trivially on the matter fields. As it will
become clear in the forthcoming discussion this symmetry is precisely the
$U(1)$ symmetry entering the equivariant extension carried out at the 
beginning of this section. Notice also that the transformations in
(\ref{dadi})  agree with (\ref{brsmasa}), with $im$ playing the role of the
parameter $u$, except for the  fields $v_{\dot\alpha}$ and $\bar
v_{\dot\alpha}$. In fact, the mass terms in (\ref{maccion}) are precisely 
(\ref{masillas}) and (\ref{masotta}). The terms coming from the 
$d_{X_{\cal M}}$-exact term  can have an arbitrary multiplicative parameter
$t$, {\it i.e.}, they enter in  the exponential of (\ref{lagrangiano}) as $t Q
\omega_{X_{\cal M}}$. This parameter  must be $t=-im/4$ in order to recover
the twisted theory (notice that the exponential of (\ref{lagrangiano}) has to
be compared to minus the action of the twisted theory).

To construct an off-shell version of the twisted model, we follow the
procedure  we applied in the massless case. The auxiliary fields are again
$K_{\alpha\beta}$, 
$k_{\dot\alpha}$ and $\bar k_{\dot\alpha}$, as  in (\ref{pendulo}), but  the
transformations for the last two are modified to: 
\bea [ Q , k^{\dot\alpha} ] &= & m v^{\dot\alpha} - i\phi v^{\dot\alpha} - 2
\psi^{\dot\alpha\alpha}M_\alpha + 2iD^{\dot\alpha\alpha}
\mu_\alpha,\nonumber\\ {[}Q , \bar k_{\dot\alpha} ]& = &  - m \bar
v_{\dot\alpha} + i\phi \bar v_{\dot\alpha} - 2
\psi_{\alpha\dot\alpha}\overline M^\alpha - 2i D_{\alpha\dot\alpha}\bar
\mu^\alpha, 
\label{pendulomasa}
\eea and for $Q^2$ one has:
\bea [ Q^2 , k_{\dot\alpha} ] &= & m k_{\dot\alpha} - i\phi
k_{\dot\alpha},\nonumber\\ {[} Q^2 , \bar k_{\dot\alpha} ] &= & - m \bar
k_{\dot\alpha} + i\bar k_{\dot\alpha}\phi.
\label{pendulodosmasa}
\eea Introducing now the gauge fermion corresponding to (\ref{masmasilla}),
\be
\Lambda_m=-{1\over 8} m\int_X \sqrt{g} (\bar\mu^\alpha M_\alpha -\overline
M^\alpha
\mu_\alpha ),
\label{burbumasa} 
\ee one finds that the off-shell twisted action of the model can be written
again as a
$Q$-exact term:
\be
\{Q, \Lambda_0 + \Lambda_m \} =   S_1+ S_2 + S_{m}+{1\over 4}\int_X \sqrt{g}
\big(\bar k^{\dot\alpha} k_{\dot\alpha}-K^{\alpha\beta} K_{\alpha\beta} \big),
\label{burbuunomasa} 
\ee where $S_1$ is (\ref{taction}), $S_2$ is the topological term
(\ref{cherndos}). To  recover  now the action in the Mathai-Quillen formalism
and the last two transformations in  (\ref{brsmasa}), we redefine the
auxiliary fields as in (\ref{jas}) to obtain:
\bea [ Q , h_{\dot\alpha} ] &= & m v_{\dot\alpha} - i\phi v_{\dot\alpha}, \\
{[} Q ,
\bar h_{\dot\alpha} ] &=  & - m \bar v_{\dot\alpha} + i \bar
v_{\dot\alpha}\phi ,
\label{muellemasa}
\eea and the resulting action takes the form:
\be
\{Q, \Lambda + \Lambda_m \}
\label{oscimasa}
\ee with $\Lambda$ given in (\ref{burbutres}). As we mentioned before, the
term 
$\Lambda_m$ is precisely the localization term discussed in  (\ref{masotta})
and from a geometrical point of view it has the same origin as
(\ref{masterminos}). Again, this term can be introduced with an arbitrary
constant providing a model in which an additional parameter can be
introduced. As in the case of topological sigma models one would expect that
the vacuum expectation values of the observables of the theory are
independent of this parameter, and therefore that one can localize this
computation  to the fixed points of the $U(1)$ symmetry, as it has been
argued in
\cite{pt} from a  different point of view.

As a general conclusion to this section, we can make some observations about 
twisted $N=2$ theories with central charge. We have analyzed two cases, the 
topological sigma model in Chapter 3, and the massive non-abelian monopoles.
In general,  these Topological Quantum Field Theories possess a non-trivial
parameter. It is  likely that the vacuum expectation values of the
observables, {\it i.e.}, the topological invariants, are functions of this
parameter. This is a very surprising feature, specially if one thinks that
the origin of that parameter is a mass in the four-dimensional case, but, at
the same time,  very appealing. Recall that in ordinary Donaldson-Witten
theory, as well as in its extensions involving twisted massless matter
fields, the action of the theory turns out to be
$Q$-exact and therefore no dependence on the gauge coupling constant appears
in the vacuum expectation values. This is also the case in the presence of a
non-trivial central charge, for the action can again be written in a
$Q$-exact form and therefore there is no dependence on the gauge coupling
constant. However, one can not argue so simply independence of the 
parameter  originated from the mass or central charge of  the physical
theory. In this case the parameter not only enters in the 
$Q$-exact action but also in the $Q$-transformations. Notice that vacuum
expectation  values in these topological theories should be  interpreted as
integrals of  equivariant extensions of differential forms. From the 
equivariant  cohomology point of view, the parameter  of the central charge
is the generator of the  cohomology ring, which we have  denoted by $u$, and
the integration of an equivariant extension of a differential form can give
additional  contributions because of the new terms needed in the extension.
These  contributions have the form of a polynomial in $u$. Therefore, we
should expect  a dependence of the vacuum expectation values of the twisted
theory with  respect to this parameter. A different  situation arises when
one considers the addition of  equivariantly exact  forms like (\ref{exacta})
or (\ref{derivando}) multiplied by another  parameter $t$. If some 
requirements of compactness  are fulfilled, the topological invariants don't
depend on this $Q$-exact piece,  and we can compute them for different values
of $t$.  This is precisely the usual  way to prove localization of
equivariant integrals. It is likely that a  rigorous application of this
method to the models considered in this work can provide new ways to compute
the corresponding topological invariants.
 
\section{Twisting the $U(1)_B$ current in $N=2$ QCD}

We have seen in section 2 that, after the twisting, the bosonic fields 
$q$ in the hypermultiplet become positive-chirality spinors: the gauging of 
the
$SU(2)_I$ current makes possible to define $N=2$ Yang-Mills theory on a 
curved manifold, but the  obstruction associated to $w_2(X)$ reappears when
matter  hypermultiplets are introduced. In previous sections we have then
restricted ourselves  to the $\sp$ case, but from the point of view of
four-dimensional geometry it would be  desirable to construct twisted $N=2$
QCD on a general four-manifold. In the case of the  Seiberg-Witten monopole
equations, the issue is precisely to consider 
${\rm Spin}^c$-structures, and one would like to extend this possibility to
the  non-abelian generalization of these equations. This problem has been
addressed  in
\cite{tqcd}, where the non-abelian monopole equations and the topological 
action were obtained in the $\sp^c$ case.  Here we will propose a different
way to couple matter hypermultiplets  to ${\rm Spin}^c$-structures in
non-abelian monopole theories \cite{toap}. The  idea is to consider  an
extended twisting procedure by gauging an additional global symmetry of the 
physical theory. As the pure Yang-Mills sector is already well-defined with 
the usual gauging of the $SU(2)_I$ isospin group, this symmetry can only act 
on the matter sector. Morevover, ${\rm Spin}^c$-structures involve a $U(1)$
gauge group  associated to  a line bundle $L$ over the four-manifold $X$. In
fact, in four dimensions we have that 
\be {\rm Spin}^c_4=\{(A,B) \in U(2)\times U(2): {\rm det}(A)={\rm det}(B) \},
\label{spincfour}
\ee (see Appendix), and the structure groups of the complex spinor bundles 
$S^{\pm}\otimes L^{1/2}$ are $SU(2)_{L,R}\times U(1)$, with the same $U(1)$
action in both  sectors. We should then gauge a global, non-anomalous $U(1)$ 
symmetry in the original $N=2$ theory, acting solely on the matter
hypermultiplets. The  required symmetry is precisely the baryon number. Let's
see this in some detail. 

The global anomaly-free symmetry of $N=2$ QCD with $N_f$ hypermultiplets is 
\be SU(N_f)\times U(1)_{B}\times SU(2)_I.
\label{fsime}
\ee For $N_f=1$, the baryon number $U(1)_B$ acts on the hypermultiplet as 
\bea Q& \rightarrow& {\rm e}^{i\phi}Q, \,\,\,\,\,\,\,\,\ {\widetilde Q} 
\rightarrow {\rm e}^{-i\phi}{\widetilde Q},\nonumber\\ Q^{\dagger}
&\rightarrow &{\rm e}^{-i\phi}Q, \,\,\,\,\,\,\,\,\ {\widetilde Q}^{\dagger} 
\rightarrow {\rm e}^{i\phi}{\widetilde Q}^{\dagger}.
\label{barion}
\eea As it is a vector symmetry, it is non-anomalous. In components it reads:
\bea   q& \rightarrow& {\rm e}^{i\phi}q, \,\,\,\,\,\,\,\,\ q^{\dagger} 
\rightarrow {\rm e}^{-i\phi}q^{\dagger}, \nonumber\\
\psi_{q \alpha}&\rightarrow &{\rm e}^{i\phi}\psi_{q \alpha},
\,\,\,\,\,\,\,\,\  {\overline \psi}_{{\tilde q} \dot  \alpha} 
\rightarrow {\rm e}^{i\phi}{\overline \psi}_{{\tilde q} \dot 
\alpha},\nonumber\\ {\overline \psi}_{ q \alpha}&\rightarrow &{\rm
e}^{-i\phi}{\overline \psi}_{ q
\alpha}, 
\,\,\,\,\,\,\,\,\ 
\psi_{ {\tilde q} \alpha}\rightarrow  {\rm e}^{-i\phi}\psi_{{\tilde q}
\alpha}.
\label{barioncom}
\eea The $U(1)_B$ current associated to this symmetry is 
\be j^{\mu}_{B}=-iD_{\mu}q^{\dagger}q+iq^{\dagger}D^{\mu}q+ {\overline 
\psi}_{{ q}
\dot  \alpha} (\sigma^{\mu})^{\dot \alpha \alpha}{\psi}_{{ q} \dot  \alpha}-
{\overline  \psi}_{{\tilde q} \dot  \alpha}(\sigma^{\mu})^{\dot \alpha \alpha}
{\psi}_{{\tilde q} \dot  \alpha}.
\label{barcor}
\ee To gauge this $U(1)$ symmetry, consider the determinant line bundle $L$
associated to a
 ${\rm Spin}^c$-structure on $X$, endowed with a connection $b_{\mu}$, and
add to the  Lagrangian the term 
\be {1 \over 2}j^{\mu}_B b_{\mu} -{1 \over 4}q^{\dagger}b_{\mu}b^{\mu}q.
\label{copla}
\ee If we gauge both the $SU(2)_I$ and the $U(1)_B$ symmetries, the covariant
derivatives  acting on the components  of the matter hypermultiplet in the
resulting Lagrangian are the appropriate ones  for complex spinors taking
values  in
$S^{\pm}\otimes L^{1/2}\otimes E $. To further analyze the consistency of the 
procedure, it is  useful to consider the correspondence between the fields in
the original $N=2$ theory  and the fields appearing in the Mathai-Quillen
formulation of the moduli problem.  This correspondence is given in
(\ref{paloma}). Taking into acount the baryon number  assignment in
(\ref{barioncom}), we see that the  fields
$M_{\alpha}$, $\mu_{\alpha}$ are sections of $S^{+}\otimes L^{1/2} \otimes
E$, 
$v^{\dot \alpha}$ is a section of $S^{-}\otimes L^{1/2}\otimes E$, ${\overline
M}^{\alpha}$, 
${\bar \mu}^{\alpha}$ are sections of $S^{-}\otimes L^{-1/2}\otimes
{\overline E} $, and 
${\bar v}_{\dot \alpha}$ is a section of $S^{-} \otimes L^{-1/2}\otimes
{\overline E}$.  This is then  consistent with the expected structure of the
complex spinor bundles. The  kinetic terms in the twisted theory are then
\be g_{\mu \nu}D^{\mu}_{L}{\overline M}^{\alpha}D^{\nu}_{L}M_{\alpha}-  { i
\over 2}\Big({\bar v}_{\dot \alpha}D_{L}^{\dot \alpha \alpha}\mu_{\alpha}+
{\bar
\mu}^{\alpha}D_{L \alpha \dot \alpha}v^{\dot \alpha} \Big),
\label{kinbardos}
\ee where $D^{\mu}_{L}$, $D_L^{\dot \alpha \alpha}$ are respectively the
covariant  derivative and the Dirac operator associated to the tensor product
connection on 
$S^{+} \otimes L^{1/2}\otimes E $. 

As in the case of the usual twisting, the Lagrangian obtained after the
gauging of the 
$SU(2)_I$ and the $U(1)_B$ symmetries is not $Q$-closed. If we compute
$[Q,{\cal L}]$,  using now the Weitzenb\"ock formula  for the ${\rm
Spin}^c$-case, (\ref{weic}), we obtain 
\be [ Q,{\cal L} ]=-{R \over 4}({\overline M}^{\alpha}\mu_{\alpha}+{\bar
\mu}^{\alpha}  M_{\alpha})- {i \over 2}\big( {\overline
M}^{\alpha}\Omega_{\alpha}{}^{\beta}\mu_{\beta}+ {\bar
\mu}^{\alpha}\Omega_{\alpha}{}^{\beta} M_{\alpha}\big),
\label{noexdos}
\ee where $R$ is the scalar curvature of the manifold and
$\Omega_{\alpha}{}^{\beta}$ is the  self-dual part of the curvature of the
line bundle $L$. It then follows that  the modified  Lagrangian
\be {\cal L}_{\rm top}={\cal L}+{R \over 4}{\overline
M}^{\alpha}M_{\alpha}+{i \over 2} {\overline
M}^{\alpha}\Omega_{\alpha}{}^{\beta}M_{\alpha}
\label{totexdos}
\ee is $Q$-closed on a general four-manifold. This Lagrangian was obtained in 
\cite{tqcd}, and an analysis following the lines developed in this and the
previous  Chapter shows that the resulting  Topological Quantum Field Theory
corresponds  to the moduli problem encoded in the equations
\bea F_{\alpha\beta}^{+a}+{i}{\overline M}_{(\alpha} (T^{a}) M_{\beta
)}&=&0,\nonumber\\ D_{L}M&=&0.
\label{compactoc}
\eea These are equations for a pair $(A,M)$ consisting of a connection 
$A$ on $E$ and a section $M$ of the twisted complex spinor bundle $S^+\otimes
L^{1/2} 
\otimes E$, and the connection on the determinant line bundle is fixed. The
operator 
$D_{L}$ is just the Dirac operator for this twisted bundle. Similar
equations  have been considered in the mathematical literature, see 
\cite{okone,okotwo,bradgp,gptres,pt}. 

Usually, the fact that the theory is topological means that correlation
functions  do not depend on the Riemannian metric of the four-manifold. In
the same way one can  easily check that the theory is topological with
respect to the ${\rm Spin}^c$-connection:  the correlation functions do not
depend on the choice of the connection $b_{\mu}$  on the line  bundle $L$,
but only on the topological class of the ${\rm Spin}^c$-structure. To see
this,  notice that the Mathai-Quillen formulation of the model coupled to a
$\sp^c$-structure is  almost identical to the one presented in section 1 of
this Chapter, the only difference  being that in the localizing gauge fermion
(\ref{tigre}) we must consider instead the  expression for the Dirac operator
$D_L$ (including the connection $b_{\mu}$ on the  determinant line bundle
$L$). The $Q$-transformations of  the fields in this off-shell formulation
are the ones in (\ref{pera}) and do not depend on  the connection on
$L$. As the full Lagrangian is 
$Q$-exact, we have
\be {\delta S \over \delta b_{\mu}}=\{Q, {1 \over 2}\int_X {\sqrt g} ({\bar
v}_{\dot
\alpha}  (\sigma^{\mu})^{\dot \alpha \alpha}M_{\alpha}-{\overline M}^{\alpha}
(\sigma^{\mu})_{\dot \alpha \alpha}v^{\dot \alpha}) \}
\label{topscon}
\ee This in turn guarantees that the twisted theory is independent of the
choice of
$b_{\mu}$.  Notice that the metric (or Spin connection) and the ${\rm
Spin}^c$-connection enter  the construction on the same footing.

We have then seen that $N=2$ QCD has the possibility of coupling the matter
fields to 
$\sp^c$-structures once the adequate symmetry has been identified. The
gauging of the 
$U(1)_B$ current can be generalized to theories with more than one
hypermultiplet. In this  case there are $N_f$ $U(1)$ symmetries that can be
gauged, and this makes possible to  consider $N_f$ different
$\sp^c$-structures, as it has been already noticed in \cite{tqcd}.  The
analysis of the moduli problem associated to the equations (\ref{compactoc})
is very  similar to the one presented in the previous Chapter. The virtual
dimension of the  moduli space depends now on the first Chern class of $L$,
and other features (like the  orientability, the analysis of reducibles and
the structure of the observables) are almost  identical. The topological
correlation functions are now asociated to the four-manifold 
$X$ together with the topological class of the $\sp^c$-structure chosen to
gauge the 
$U(1)_B$ symmetry.

\chapter{Exact results in $N=1$ and TQFT}

The main interest of having an equivalence between a moduli problem and a
Topological  Quantum Field Theory lies in the possible application of physical
methods to obtain  new mathematical results. As topological invariants are
simply correlation functions  of the twisted theory, one can try to compute
these functions using techniques in  Quantum Field Theory and then, under
suitable modifications, extract the topological  information. This line of
work was successfully applied in two-dimensional gravity  and Chern-Simons
theory, but supersymmetric gauge theories remained rather elusive.  This is
because the computation of the correlation functions involve a precise 
knowledge of the non-perturbative behaviour of the theory. For gauge theories
in four  dimensions, this behaviour is far from being completely understood.
However, {\it  supersymmetric} gauge theories offer a different situation.
The extra  (super)symmetry is such a powerful constraint that many exact
results can be obtained.  Some of these results have been known for some time
\cite{amati, ads}, but the  situation radically  improved with the work of
Seiberg on $N=1$ theories and Seiberg and Witten on $N=2$  theories. By now
we know the exact low-energy superpotential of many $N=1$ theories  as well
as their vacuum structure. We also know the exact low-energy effective 
action of a family of $N=2$ QCD theories. These results are sufficient to
make many  predictions about the twisted theories that we have been
discussing in previous  Chapters, starting from the seminal work by Witten
and Vafa and Witten in 
\cite{sym,vw,  mfm, abelia}.  As the Topological Field Theories we have been
studying are twisted versions of $N=2$  supersymmetric theories, one can
think that
$N=1$ theories are not relevant for this  problem. However, as Witten showed
in
\cite{sym} for Donaldson-Witten theory,  on K\"ahler manifolds one can  break
supersymmetry from $N=2$ down to $N=1$ in a topologically trivial way, and 
use non-perturbative results for $N=1$ theories to obtain information about
the  topological correlation functions of the twisted theory. Although some
of the  ingredients  in this $N=1$ approach are not so clear in comparison to
the $N=2$ results (and  this is perhaps the reason Witten calls it the
``abstract" approach), in our opinion  it is extremely illuminating to
complement the ``concrete" $N=2$ approach with the 
$N=1$ information. In this Chapter we develop the techniques to analyze
twisted  theories with non-perturbative $N=1$ results. In section 1 we
present a brief overview  of non-perturbative results in $N=1$ theories. In
section 2 we apply these methods  to an example which will be relevant in the
analysis of the $SU(2)$ monopole theory.  In section 3, we introduce the
special twisting one can perform in K\"ahler  manifolds and analyze from this
point of view the breaking of $N=2$ to
$N=1$ in the  case of the non-abelian monopole theory. Finally, in section 4
we use all this  information to compute topological correlation functions for
the $SU(2)$ monopole  theory.
 
\section{Exact results in $N=1$ SUSY gauge theories}

We will begin considering a simple case, namely $N=1$ supersymmetric
Yang-Mills theory,  where many of the relevant patterns already appear, and
then  we will present some general techniques that will be useful to analyze
the $SU(2)$  monopole theory. 

$N=1$ super Yang-Mills with gauge group $SU(N)$ contains a gauge field
$A_{\mu}$ in  the adjoint representation  of the gauge group $SU(N)$ and a
gluino field
$\lambda_{\alpha}$, a Majorana spinor also  in the adjoint representation. In
$N=1$ superspace the action for this  theory is given by:
\be S_{N=1}=\int d^4x \, d^2\theta \, \tr\Big( W^\alpha W_\alpha \Big) +
\int d^4x \, d^2\overline\theta \, \tr\Big(\overline  W^{\dot\alpha} 
\overline W_{\dot\alpha}\Big), 
\label{nunoym}
\ee and in components it reads:
\be S_{N=1}=\int d^4x \tr \bigg( -{1 \over 4}F_{\mu \nu}F^{\mu
\nu}-i{\overline
\lambda}^ {\dot \alpha} D_{\alpha \dot\alpha}\lambda^{\alpha} \bigg).
\label{componentes}
\ee The $U(1)_{\cal R}$ symmetry of this theory is given by:
\be
\lambda^{\alpha} \rightarrow {\rm e}^{i\phi}\lambda^{\alpha},
\,\,\,\,\,\,\,\,\,\  {\overline \lambda}^ {\dot \alpha} \rightarrow {\rm
e}^{-i\phi}{\overline \lambda}^ {\dot \alpha}.
\label{uuno}
\ee This symmetry is anomalous in an instanton background. As the fermion is
in the adjoint  of $SU(N)$, the anomaly is $2N$ and the 't Hooft rule implies
that the
$U(1)_{\cal R}$  is broken down to ${\bf Z}_{2N}$. Much of the knowledge of
the  vacuum structure of this theory comes from  the computation of the
Witten index $\tr (-1)^F$ \cite{csb}. In this case, $\tr (-1)^F =  N$ and
this indicates that the theory has $N$ vacua with a spontaneous  breaking of
the chiral symmetry (\ref{uuno}) down to a ${\bf Z}_2$ subgroup:
\be
\lambda^{\alpha} \rightarrow -\lambda^{\alpha}, \,\,\,\,\,\,\,\,\,\ 
{\overline
\lambda}^ {\dot \alpha} \rightarrow -{\overline \lambda}^ {\dot \alpha}.
\label{serompe}
\ee  This symmetry breaking has two important features: first, the residual
subgroup allows  fermion masses. Second, it is dynamically generated by
gluino condensation.  In other words, the glueball superfield $S=-\tr
(W_{\alpha}^2)$ gets a VEV. This  can be checked by instanton computations
(see \cite{amati} and references therein). This  theory also has a mass gap
and it is believed to be confining (in fact, evidence for  these results have
been obtained from the exact results for $N=2$ supersymmetric  Yang-Mills 
theory in \cite{swone}). 

In general, non-perturbative results in supersymmetric theories have been
obtained by  direct dynamical computations or trying to find effective
(Wilsonian) actions to  describe  the relevant, light degrees of freedom,
once the heavy fields  have been integrated out. A powerful method to obtain
results on the vacuum  structure for a wide class of theories  has been
developed by Seiberg and collaborators \cite{natu, vacua, super, inin,
insei}  (see also the reviews \cite{holom, nunolect}), extending previous
work along  the lines of the effective action approach \cite{ads}. Seiberg's
idea is to exploit  the symmetries of supersymmetric theories to constrain
the effective superpotential,  and in many cases one is able to find it
exactly with some dynamical information.  Let us briefly review the main
points of this powerful method. We mainly follow
\cite{super}, where the general strategy is clearly formulated. 

Suppose we are given an $N=1$ supersymmetric gauge theory with gauge group
$G$  and matter  superfields $\phi_i$ transforming in representations $R_i$
of $G$. We suppose that  the theory is asymptotically free and is
characterized by a dynamical scale 
$\Lambda$. At the classical  level, and with no tree-level superpotential, the
ground states are  determined by the vanishing of the $D$ terms in the
Lagrangian. This determines vacuum  expectation  values (VEVs) for the fields
$\phi_i$ which in general spontaneously break  the gauge symmetry  down to
some subgroup of $G$. The configuration space for these VEVs is called  the
{\it classical moduli space} (CMS) of the theory. It turns out that the light
fields
 are given classically by the gauge invariant polynomials $X^r$ in the matter
 superfields,  obeying some classical constraint. These polynomials can be
used  as coordinates for the CMS.  A good example of this construction is
given by 
$N=1$ QCD, analyzed in \cite{vacua}. 

In the quantum theory, the Wilsonian description by an effective Lagrangian
involves  three ingredients: the light superfields $X^r$, the dynamical scale
$\Lambda$ and the  couplings in the tree level superpotential:
\be W_{\rm tree}= \sum_r g_r X^r(\phi_i).
\label{arbol}
\ee We want to compute the {\it exact} quantum effective superpotential of
this theory, 
$W_{\rm eff}$. The principles that one may use to obtain 
$W_{\rm eff}$ are \cite{holom}:

1) Holomorphy: the effective potential is a holomorphic function of the gauge
invariant  superfields $X^r$, the scale $\Lambda$ and the coupling constants
$g_r$. 

2) Symmetries: apart form the explicit symmetries of the theory, we can use 
{\it selection  rules}. These are obtained as follows: when $g_r=0$, the
theory has in  principle a larger  global symmetry group $H$ broken by the
terms in (\ref{arbol}). We can give quantum  numbers to the coupling
constants $g_r$ with respect to $H$ such that the total  Lagrangian
(including (\ref{arbol})) is invariant under $H$. Then, the quantum 
effective superpotential $W_{\rm eff}$ is invariant under the combined $H$ 
transformations of $X^r$ and $g_r$. 

3) Asymptotic behaviours: the analysis of $W_{\rm eff}$ at various limiting
behaviours  (weak coupling, big masses, ...) give additional constraints that
often completely  determine $W_{\rm eff}$.

In addition to these general principles, there are also specific techniques
that  can be extremely useful to obtain the quantum superpotential. Among
them the  most useful to us will be the {\it integrating in} procedure
\cite{super, inin}. The  situation in which this technique applies is the
following. Suppose that we are given an 
$N=1$ supersymmetric theory with the above inputs, which will be called the 
``downstairs" theory, and that we know the  effective superpotential for
$g_r=0$,
$W_d(X^r, \Lambda_d)$. Consider now another  theory, the ``upstairs" theory,
which differs from the downstairs theory only in  that it contains an
additional matter field $\phi$ in a representation $R$ of $G$. The  upstairs
theory has, apart from the gauge invariant polynomials $X^r$, additional 
polynomials $X^{\hat r}$ including the field $\phi$. The  integrating in
procedure allows one to derive the effective superpotential for the  upstairs
theory $W_u(X^r,X^{\hat r}, 
\Lambda)$ starting from the superpotential $W_{d}(X^r, \Lambda_d)$. The steps
are as  follows: first of all, we consider the upstairs theory after turning
on a tree level  superpotential including all the ``extra" polynomials
$X^{\hat r}$ with couplings 
$g_{\hat r}$. In particular, we turn on a mass ${\hat m}$ for $\phi$. Then we
assume two  principles (which have been proved in many particular cases):

1) Principle of linearity: it simply states that the full superpotential of
the  upstairs theory is given by
\be W_f(X^r, X^{\hat r}, \Lambda, g_{\hat r})= W_u(X^r,X^{\hat r}, 
\Lambda) +\sum_{\hat r}g_{\hat r}X^{\hat r}.
\label{linealoye}
\ee

2) Principle of simple thresholds: if we integrate out the fields $X^{\hat
r}$ from  (\ref{linealoye}), we will  obtain a theory closely related to the
original downstairs theory, in the sense that  the  corresponding
superpotential is given by
\be W_l(X^r, \Lambda, g_{\hat r})=W_d(X^r, \Lambda_d) + W_I(X^r, \Lambda,
g_{\hat r}),
\label{inout}
\ee and the additional piece $W_I$ is irrelevant from the RG point of view. In
particular, 
\be 
 W_I \rightarrow 0 \,\,\,\,\,\,\,\,\  {\rm for} \,\,\,\,\,\,\,\,\ {\hat m}
\rightarrow \infty.
\label{masagorda}
\ee The principle states that the  relation between the scales is given by
\be
\Lambda_d^{n_d}=\Lambda^n {\hat m}^{\mu/2},
\label{escalas}
\ee where $n_d$, $n$ are the coefficients of the one-loop $\beta$-function of
the  downstairs and  upstairs theory, respectively, and $\mu$ is the index of
the representation $R$ of $G$.  Of course, the relation between the scales is
obtained by matching the  corresponding running coupling  constants at the
scale ${\hat m}$. What the principle states is that this relation is  {\it
exact} and independent of all other couplings. 

Recall now that our unknown is $W_u$. The key point is to realize that $W_l$
is a  Legendre transform of $W_u$, and then $W_u$ can be obtained by an
inverse  Legendre transform. This is done as follows. Consider the auxiliary
superpotential:
\be W_n(X^r, X^{\hat r}, \Lambda, g_{\hat r})=W_l(X^r, \Lambda, g_{\hat
r})-\sum_{\hat r}  g_{\hat r}X^{\hat r}. 
\label{auxilio}
\ee By integrating out the $g_{\hat r}$, we obtain the superpotential $W_u$.
As it is clear  from the above analysis, the only non-trivial step in this
procedure is to find the  additional piece $W_I$. In theories where the gauge
invariants $X^{\hat r}$ are  quadratic in the field $\phi$, the tree-level
additional terms in (\ref{linealoye})  are just mass terms, and the condition
(\ref{masagorda}) implies that $W_I=0$. In more  general theories, one must
carefully analyze the theory along the lines explained before.  In general,
we can decompose $W_I$ in two pieces:
\be W_I=W_{{\rm tree},d}+W_{\Delta}.
\label{cosasraras}
\ee The first term in (\ref{cosasraras}) appears when integrating 
$\phi$ from $W_{\rm tree}$ in (\ref{linealoye}). The other piece is unknown,
but  in many cases it can be shown to be zero because of the symmetries. 

This ends our brief introduction to some non-perturbative methods in $N=1$
theories.  We have deliberately restricted ourselves to the techniques and
principles that will  be useful to us. In the next section we will apply
these principles to determine the  vacuum structure and pattern of chiral
symmetry breaking of the $N=1$ theory which is  relevant  to $SU(2)$
monopoles. 

\section{Vacuum structure of broken $N=2$ QCD with $N_f=1$}

The theory we will analyze in this section is an $N=1$ theory  obtained from
the  soft breaking of an $N=2$ theory. The original $N=2$ theory is  the one
we studied in the  previous Chapter to formulate the non-abelian monopole 
theory in terms of a  twisted theory, namely $N=2$ QCD with $N_f=1$. The
gauge group is $SU(2)$ and the  hypermultiplet is in the fundamental
representation. The  soft breaking down to
$N=1$  is achieved by adding a mass term for the chiral multiplet $\Phi$ in
(\ref{trapecio}).  In the next section we will explain why this perturbation
is  topologically trivial when the  twisted theory is formulated on a
K\"ahler manifold, and therefore the vacuum structure  we will obtain is
relevant to the analysis of
$SU(2)$  monopoles on K\"ahler manifolds.  This analysis has been presented in
\cite{lmpol}, starting  from results in \cite{swtwo,  insei}.  

First of all, as in the $N=1$ pure Yang-Mills case, we will describe the 
$U(1)_{\cal R}$ symmetry  of the theory. $N=2$ supersymmetric QCD with  gauge
group
$SU(N_{c})$ and $N_{f}$ hypermultiplets in the fundamental representation of
the  gauge group has the  $U(1)_{\cal R}$ symmetry  described in
(\ref{salto}) and (\ref{pertiga}):
\bea W_{\alpha} &\longrightarrow  {\rm e}^{-i\phi}W_{\alpha}({\rm e}^{i\phi}
\theta ),\,\,\,\,\,\,\,\,\,\,\,\,\,\,\,\
 Q^{i}  &\longrightarrow Q^{i}({\rm e}^{i\phi}\theta),
\nonumber \\ 
\Phi &\longrightarrow  {\rm e}^{-2i\phi} \Phi ({\rm
e}^{i\phi}\theta),\,\,\,\,\,\,\,\,\,\,\,\,\,\,\,\
 {\widetilde Q}_{\tilde i} &\longrightarrow {\widetilde Q}_{\tilde i} ({\rm
e}^{i\phi}\theta).
\label{lola}
\eea In component fields the corresponding transformations can be read from
(\ref{twist}) and (\ref{paloma}):
\bea
\lambda_{1},\; \lambda_{2}  &\longrightarrow& {\rm e}^{-i\phi} \lambda_{1},\;
{\rm e}^{-i\phi}  \lambda_{2}, \nonumber\\ B &\longrightarrow &{\rm
e}^{-2i\phi} B,\nonumber\\
\psi_{qi},\; \psi_{\tilde q \tilde i} &\longrightarrow&   {\rm e}^{i\phi}
\psi_{qi},\; {\rm e}^{i\phi} \psi_{\tilde q \tilde i}.
\label{maslola}
\eea
 This symmetry is anomalous because of instanton effects. The anomaly is 
$4N_{c}-2N_{f}$ ($2N_{c}$ from $\lambda_1$  and $\lambda_2$, which live in the
adjoint representation of the gauge group,  and $2$ from each couple of
fermions
$\psi_{q}$,
$\psi_{\tilde q}$ in the hypermultiplet). In the case we are dealing with,
namely
$N_{c}=2$ and $N_{f}=1$  (which gives the $SU(2)$ monopole equations) the
anomaly is
$6$ and we should expect the 
${\bf Z}_{6}$ anomaly-free discrete subgroup:
\bea 
\lambda_{1},\; \lambda_{2} &\longrightarrow& {\rm e}^{-{i\pi \over 3}} 
\lambda_{1}, \; {\rm e}^{-{i\pi \over 3}} \lambda_{2}, \nonumber\\  B
&\longrightarrow& {\rm e}^{-{2i\pi \over 3}}  B,\nonumber\\ 
\psi_{q},\; \psi_{\tilde q}  &\longrightarrow & {\rm e}^{i\pi\over 3}
\psi_{q},
\;{\rm e}^{i\pi\over 3} \psi_{\tilde q}. 
\label{enedos}
\eea However, since we are considering one hypermultiplet in the  fundamental
representation of $SU(2)$, we must take into account that the quark $Q$ and
the antiquark ${\widetilde Q}$  live in isomorphic representations of the
gauge group.  Let's explain this in some detail. If we denote by $a$ the
color index, we can  define the fields:
\bea { Q}^1_a&=&Q_a,\nonumber\\ { Q}^2_a&=&(\sigma_2)_{ab}\widetilde Q_b, 
\label{primtrans}
\eea where $\sigma_2$ is the Pauli matrix. If $U \in SU(2)$, as $\sigma_2
U^{*}\sigma_2=U$,  the field ${ Q}^2_a$ transform also in the 
${\bf 2}$. This is an explicit realization of the isomorphism ${\bf 2} 
\simeq {\bf {\bar 2}}$. We must also redefine the chiral superfield $\Phi$,
which  lives in the adjoint  representation, in the following way:
\be
\hat {\Phi} =(\sigma_2)^{\rm T} \Phi.
\label{sombrero}
\ee The new field $\hat {\Phi}$ is a symmetric matrix because ${\rm Tr}
\Phi=0$.  The $N=2$ coupling 
$\widetilde {Q}^{\rm T} \Phi Q$ appearing in (\ref{silvana}) is written in
terms of  the new variables as:
\be {1 \over 2}({ Q}^2_a
\hat {\Phi}_{ab}{ Q}^1_b +{ Q}^1_a
\hat {\Phi}_{ab}{ Q}^2_b).
\label{acopla}
\ee The $N=2$ mass term for the matter fields involves the gauge invariant
quantity 
$X=\widetilde {Q}_a Q_a$, which in the new variables is written as:
\be X=-i({ Q}^2_2 { Q}^1_1 -{ Q}^1_2 { Q}^2_1).
\label{masacambia}
\ee  As a  consequence of the isomorphism ${\bf 2}\simeq {\bf {\bar 2}}$, 
there is a parity transformation which interchanges the quark and the
antiquark:
\be
\rho : { Q}^1 \leftrightarrow { Q}^2.
\label{parida}
\ee As the term (\ref{acopla}) is invariant under (\ref{parida}), this is a 
symmetry of $N=2$ QCD with massless matter  fields. Notice however  that the
$SU(2)$ singlet $X$ changes its sign under (\ref{parida}),
 as it is obvious from (\ref{masacambia}).  Therefore the $N=2$ mass term for
the quark and the antiquark changes its  sign accordingly. 

Another set of variables which is useful to take into account the 
$\rho$ symmetry is the following:
\bea {\hat Q}^1&=&{1 \over 2i}({ Q}^1-{ Q}^2),\nonumber\\ {\hat Q}^2&=&{1
\over 2}({ Q}^1 +{ Q}^2). 
\label{variadef}
\eea The $N=2$ coupling in these new variables reads as:
\be
 {\hat Q}^1 \hat {\Phi} {\hat Q}^1+ {\hat Q}^2 \hat {\Phi} {\hat Q}^2,
\label{leame}
\ee and the singlet $X$ as:
\be X=2({\hat Q}_1^1 {\hat Q}_2^2-{\hat Q}_2^1 {\hat Q}_1^2).
\label{meame}
\ee The parity symmetry in terms of these variables is
\be {\hat Q}^1 \rightarrow -{\hat Q}^1, \,\,\,\,\,\,\,\,\,\ {\hat Q}^2
\rightarrow {\hat Q}^2.
\label{paridafinal}
\ee Using the variables defined in (\ref{variadef}) it is easy to see that the
flavour  symmetry for $N=2$  QCD with gauge group $SU(2)$ and $N_f$
hypermultiplets is $O(2N_f)$. 

The parity symmetry is anomalous, as can be seen from  the 't Hooft
interaction term:
\be
\psi_{\tilde q}^{\alpha}\psi_{q \alpha}= (\psi^2_2)^{\alpha}
(\psi^1_1)_{\alpha}-(\psi^1_2)^{\alpha} (\psi^2_1)_{\alpha}.
\label{tof}
\ee Here, $\psi^1$ and $\psi^2$ are the fermion components of $Q^1$ and $Q^2$,
respectively.  Nevertheless, one can combine the $\rho$ symmetry with the
square  root of ${\bf Z}_{6}$ in (\ref{enedos}) to obtain an anomaly-free
${\bf Z}_{12}$ subgroup. 

This analysis has been done for the $N=2$ theory, but we must consider this
theory  perturbed by a mass term for $\Phi$ and broken down to $N=1$. This
mass term has the  form, in $N=1$ superspace, 
\be m \int d^4x \, d^2 \theta \tr(\Phi^2) + {\rm h.c.}.
\label{masa}
\ee Notice that the mass term for the $\Phi$ field breaks the second
transformation in (\ref{lola}) due to the presence of the fermionic fields
$\lambda_2$. Thus under the new $U(1)_{\cal R}$ symmetry we must have: 
\be 
\Phi  \longrightarrow {\rm e}^{-i\phi} \Phi ({\rm e}^{i\phi}\theta),
\label{orense}
\ee  and this in turn imposes, because of the superpotential term, the
following  transformation for the matter fields: 
\bea Q &\longrightarrow &{\rm e}^{-i\phi/2}  Q({\rm
e}^{i\phi}\theta),\nonumber\\  {\widetilde Q} &\longrightarrow& {\rm
e}^{-i\phi /2}{\widetilde Q}  ({\rm e}^{i\phi}\theta).\label{uonemat}
\eea  Rescaling the charges to make them integers, we have the following
$U(1)_{\cal R}$ symmetry for  the perturbed theory in terms of components
fields:
\bea
 \lambda_{1},\; B &\longrightarrow& {\rm e}^{-2i\phi}
 \lambda_{1},\; {\rm e}^{-2i\phi}  B,
\nonumber\\  q, \; {\tilde q} &\longrightarrow & {\rm e}^{-i\phi} q, 
\; {\rm e}^{-i\phi} {\tilde q},
 \nonumber\\
\psi_{q}, \; \psi_{\tilde q} &\longrightarrow&  {\rm e}^{i\phi} \psi_{q},
\; {\rm e}^{i\phi} \psi_{\tilde q}.\label{uonefin} 
\eea

The anomaly-free discrete subgroup of the transformations (\ref{uonefin}) is
${\bf Z}_{6}$. However,  one must take into account the $\rho$ symmetry
(\ref{parida}), as the addition of the mass term for $\Phi$ doesn't break it.
Again, we have an enhancement of the discrete symmetry to ${\bf Z}_{12}$. The
resulting transformations are: 
\bea
 \lambda_{1} &\longrightarrow& {\rm e}^{-\pi i/3} \lambda_1, \,\,\,\,\ B
\longrightarrow {\rm e}^{-\pi i/3}  B, \nonumber\\
 q^1 &\longrightarrow& {\rm e}^{-\pi i/6}{ q}^2 ,\,\,\,\,\ { q}^2
\longrightarrow {\rm e}^{-\pi i/6}q^1,\nonumber\\
 \psi^{1} &\longrightarrow&  {\rm e}^{\pi i/6} \psi^{2},\,\,\,\,\
 \psi^{2} \longrightarrow  {\rm e}^{\pi i/6} \psi^{1}.
\label{uonedef}
\eea   These transformations leave invariant the 't Hooft  term $(\lambda_1)^4
\psi_{\tilde q} \psi_{q}$.

The question now is: is there additional breaking of the $U(1)_{\cal R}$
symmetry in  this theory?  To answer this question, we will obtain its exact
superpotential following the  techniques presented in section 1. More
precisely, we will use the integrating in  procedure. In our case we can take
as the downstairs theory the
$SU(2)$, $N=1$ theory  with a quark and an antiquark, whose exact
superpotential is known \cite{ads, vacua},  and as the additional field for
the upstairs theory the chiral superfield in the  adjoint representation,
$\Phi$. To ``integrate in" the field $\Phi$ we must consider  the
gauge-invariant  polynomials including it (the fields $X^{\hat r}$ in the
previous section), which  in our case are simply:
\be U={\rm Tr} \Phi ^2,\,\,\,\,\,\ Z=\sqrt 2 {\widetilde Q } \Phi Q,
\label{gaugepol}
\ee and we must turn on a tree-level superpotential:
\be W_{\rm tree}=mU+\lambda Z.
\label{miarbol}
\ee The scales $\Lambda_{d}$ of the downstairs theory and $\Lambda$ of the
upstairs  theory with the  mass term in (\ref{arbol}) are related according
to the principle of simple  thresholds:
\be
\Lambda_{d}^5={1\over 4}m^2\Lambda^3.
\label{escala}
\ee (Recall that, in an $N=1$ gauge theory with gauge group $N_c$  and $N_f$
hypermultiplets in the fundamental representation, $n=3N_c-N_f$, and a
matter  field in the adjoint representation has index $\mu=2N_c$ and
contributes $-\mu/2$  to
$n$). The $1/4$ in (\ref{escala}) is introduced to match the normalizations 
in
\cite{swtwo}. The full superpotential of the upstairs theory  with the
additional tree-level  term (\ref{miarbol}) is given by the principle of
linearity (\ref{linealoye}):
\be W_f(X, U, Z, \Lambda^3, m, \lambda)=W_u(X, U, Z, \Lambda^3)+mU+\lambda Z,
\label{full}
\ee where $X$ is the gauge-invariant polynomial of the downstairs theory, 
$X={\widetilde Q} Q$,  and $W_u$ is the exact superpotential of the upstairs
theory we are looking for.  If we integrate  out the field $\Phi$ and,
correspondingly, the fields $U$, $Z$, we obtain  the superpotential
(\ref{inout}) in this case:
\be W_l(X, \Lambda^3, m, \lambda)=W_d(X, \Lambda_{d}^5)+W_{I}(X,\Lambda^3, m,
\lambda).
\label{abajo}
\ee In this equation, $W_d$ is the dynamically generated superpotential of
the  downstairs theory and is  given by \cite{ads, vacua}:
\be W_d(X, \Lambda_{d}^5)= { \Lambda_{d}^5 \over X}.
\label{supabajo}
\ee
$W_{I}$ must be determined using the symmetries of the problem and can be
split  as in (\ref{cosasraras}). The first contribution to this piece comes
from integrating out $\Phi$ from $W_{\rm tree}$. In this case  the result is
\cite{insei}
\be W_{\rm tree, d}=-{\lambda^2 \over 4m} X^2.
\label{fasesin}
\ee The upstairs theory has two non-anomalous symmetries which can be used to 
constrain the form of $W_{I}$,  following the methods of \cite{holom}. The
first one is a $U(1)$ symmetry under
 which $Q$, $\widetilde Q$, $\Phi$, $m$ and $\lambda$ have charges $2$, $2$,
$-1$,
$2$ and $-3$, respectively.  The other one is a $U(1)_{\cal R}$ symmetry with
charges $1$, $1$, $-1$, 
$0$ and $-3$. Invariance of the superpotential  under these symmetries as
well as holomorphy determine the form of $W_{I}$ to be:
\be W_{I}={  X^2 \lambda^2 \over m} f\Big( {\Lambda^3 m^3 \over X^3 \lambda^2}
\Big),f
\label{seriehol}
\ee where $f(u)=\sum_{n=0}^\infty a_n u^n$ is an analytic function. Notice
that  the first term of this expansion corresponds to 
$W_{\rm tree, d}$. Now, in the $m \rightarrow \infty$ limit, only $W_d$
survives,  and this implies that the coefficients $a_n$ in the expansion of
$f(u)$ must be zero for $n>0$. Therefore $W_{I}=W_{\rm tree, d}$  and the
superpotential  (\ref{abajo}) is given by:
\be W_l(X, \Lambda^3, m, \lambda)={ m^2 \Lambda^3  \over 4X}-{\lambda^2 \over
4m} X^2.
\label{respuesta}
\ee To perform the inverse Legendre transform, we introduce the auxiliary
superpotential  (\ref{auxilio}), which in this case reads:
\be W_n=W_l(X, \Lambda^3, m, \lambda)-mU-\lambda Z. 
\label{wcasi}
\ee If we integrate out $m$ and $\lambda$, we obtain the following expectation
 values:
\be m={ 2 X U \over  \Lambda^3}\Big( 1-{Z^2 \over X^2 U} \Big),\,\,\,\,\,\,\
\lambda= -{4 Z U \over X \Lambda^3}\Big( 1-{Z^2 \over X^2 U} \Big),
\label{param}
\ee and substituting these values in (\ref{wcasi}) one gets the
superpotential of  the upstairs theory \cite{insei}:
\be W_u=-{ X U^2 \over  \Lambda^3}\Big( 1-{Z^2 \over X^2 U} \Big)^2.
\label{suparriba}
\ee Now we want to obtain the vacua of the $N=2$ theory perturbed by the
$N=1$ mass term  for $\Phi$. Because  of the principle of linearity, the
superpotential of this theory is given by  (\ref{suparriba}) plus
(\ref{arbol}) with $\lambda=1$ because of the 
$N=2$ supersymmetry (we follow now the conventions in \cite{swtwo}):
\be W=-{ X U^2 \over  \Lambda^3}\Big( 1-{Z^2 \over X^2 U} \Big)^2+mU+ Z.
\label{supdef}
\ee The equation $\partial W /\partial X =0$ gives
\be {Z^2 \over X^2 U}=-{1 \over 3},
\label{mitercio}
\ee which together with $\partial W /\partial Z =0$ gives
\be U^3=-{27 \over 256}\Lambda^6.
\label{uy}
\ee This theory has therefore three vacua, corresponding to the three roots
of  this equation. As we will see in the next Chapter, this is  in agreement
with the results obtained from the $N=2$ point of view. Finally,  we have
VEVs for the field $X$ in these vacua given by the roots of:
\be X^3=-{1 \over 2}m^3 \Lambda^3.
\label{quarks}
\ee Clearly, these non-zero VEVs spontaneously break the chiral symmetry in 
(\ref{uonedef}), as it happens  in \cite{ads}. The subgroup of ${\bf Z}_{12}$
which preserves the VEV of these
 gauge-invariant order parameters is easily obtained:
\bea
 \lambda_{1} &\longrightarrow& -\lambda_1, \,\,\,\,\ B \longrightarrow -B,
\nonumber\\
 q^1 &\longrightarrow& -i{ q}^2 ,\,\,\,\,\ { q}^2 \longrightarrow
-iq^1,\nonumber\\
 \psi^{1} &\longrightarrow&  i \psi^{2},\,\,\,\,\
 \psi^{2} \longrightarrow  i\psi^{1}.
\label{martin}
\eea The residual chiral symmetry is in this case ${\bf Z}_4$. This is
precisely the maximal  subgroup of (\ref{uonedef}) which allows fermion
masses for
$\lambda_1$ and for $\psi^{1}$, $\psi^{2}$ (recall that, according to
(\ref{masacambia}),  the mass term for the matter fields changes its sign
under the  parity symmetry). There are two comments about this result that
are worth  mentioning. First, general arguments \cite{insei} suggest that the
above vacua,  where the gauge-invariant field $U$ is locked at some fixed
values, are in a  confinig phase. We will have additional support for this in
the next Chapter. The important  point is that these vacua present a mass
gap. Second, the spontaneous chiral symmetry  breaking  in (\ref{martin}) can
also be associated to gluino condensation, as in the pure $N=1$  Yang-Mills
case. The gluino condensate can be computed in many ways. We can use  the
general  expression given in \cite{super} to obtain:
\be S=\Lambda^3 {\partial W \over \partial \Lambda^3}={2 \over 3} mU.
\label{gluino}
\ee   As a matter of fact, one can check with the above VEVs the Konishi 
anomaly in this model:
\be 4S=2mU+Z.
\label{konishi}
\ee Therefore, the gluino condensate is an order parameter for  confinement,
as it happens in $N=1$ supersymmetric Yang-Mills.

The $N=2$ theory with one hypermultiplet and perturbed by the mass term for
$\Phi$  has an additional 
${\bf Z}'_4$ symmetry given by:
\bea
 \lambda_{2} &\longrightarrow& -\lambda_2, \,\,\,\,\ B \longrightarrow -B,
\nonumber\\
 q^1 &\longrightarrow& i{ q}^2 ,\,\,\,\,\ { q}^2 \longrightarrow i
q^1,\nonumber\\
 \psi^{1} &\longrightarrow&  i \psi^{2},\,\,\,\,\
 \psi^{2} \longrightarrow  i \psi^{1},
\label{extraz}
\eea   which is non anomalous. In terms of $N=1$ superfields, this symmetry 
corresponds to 
\be
\Phi \rightarrow -\Phi, \,\,\,\,\,\ Q^1 \rightarrow iQ^2, \,\,\,\,\,\  Q^2
\rightarrow iQ^1. 
\label{extrasuper}
\ee Notice that the diagonal in the product of the discrete  symmetries
(\ref{martin}), (\ref{extraz}), ${\bf Z}_4 \times{\bf Z}'_4$, is precisely 
the fermion number $(-1)^{F}$.

We should also point out that the above superpotential is only valid for the
confining  phase, once the mass term for the chiral multiplet $\Phi$ has been
added. When $m=0$  the theory is in the Coulomb phase and the results
obtained for the VEVs are no  longer valid,  as it has been emphasized in
\cite{inin, insei}. In the Coulomb phase there are  additional  masless
degrees of freedom that are not taken into account in the confining 
superpotential  description. This will be important to understand the cosmic
string theory on  K\"ahler manifolds.  
 
\section{Twist on K\"ahler manifolds}

In \cite{sym}, Witten showed that on a four-dimensional K\"ahler manifold the
twisting  procedure has an important property: the topological charge $Q$ can
be decomposed in two  pieces, and both of them can be regarded as topological
symmetries. If we recall  that 
$Q$ corresponds to a differential operator on the configuration space,  this
decomposition  is related to the fact that, when the manifold is K\"ahler, the
configuration space  is K\"ahler too (as we mentioned in Chapter 1 in
relation to Donaldson theory). Then  one can decompose the de Rham operator
in two Dolbeault operators. From the physical  point of view, the
decomposition corresponds to two different $N=1$ subalgebras in the 
$N=2$ algebra. This indicates that one can provide $N=1$ supersymmetry with a
topological  interpretation, and this will imply, as we will see, that terms
softly breaking $N=2$  down to $N=1$ supersymmetry are topologically trivial
with respect to one of the new  topological symmetries. Work on
Donaldson-Witten theory on K\"ahler manifolds can  be found in \cite{parkdos,
parktres}, and a detailed analysis of Witten's  construction for  pure $N=2$
Yang-Mills theory has been done in \cite{high}. 

Let us explain the basic points of the construction in \cite{sym}, and extend
it  to the $N_f=1$ theory. When the metric on the  four-manifold $X$ is
K\"ahler the global holonomy group $SU(2)_L\times SU(2)_R$ becomes
$U(1)_L\times SU(2)_R$, $U(1)_L$ being  a subgroup of $SU(2)_L$. The two
dimensional representation of $SU(2)_L$ decomposes under $U(1)_L$ as a sum of
one dimensional representations.  This means that the components $M_1$ and
$M_2$ of a spinor transform in definite representations of $U(1)_L$ with
opposite  charges. In other words, $S^+\otimes E$ has a decomposition  into
$(K^{1/ 2}\otimes E) \oplus (K^{-{1/2}}\otimes E)$,  where
$K$ is the canonical bundle. This is the same fact we took into account to 
analyze the non-abelian monopole equations on K\"ahler manifolds (of course,
we are  assuming  that $X$ is Spin, and then $K^{1/2}$ exists).   The complex
structure on
$X$ allows to have well defined complex forms of type $(p,q)$. We define this
complex structure stating the following assignment:
\bea (\sigma_m)_{1 \dot\alpha}\, d x^m,  
\;\;\;\;\; & {\rm type} \;\; (1,0), \nonumber\\ (\sigma_m)_{2 \dot\alpha}\, d
x^m,  
\;\;\;\;\; & {\rm type} \;\; (0,1).
\label{complex}
\eea This implies that $(\sigma_{mn})_{\alpha\beta}\,dx^m\wedge dx^n$ can be
regarded as a $(2,0)$ form when $\alpha=\beta=1$, as $(0,2)$ form when
$\alpha=\beta=2$, and as a $(1,1)$ form when $\alpha=1,\;\beta=2$.

Let us recall that in the process of twisting the BRST operator
$Q$ was obtained from the supersymmetric charge $Q_{i\alpha}$ after
identifying
$Q_{i\alpha} \longrightarrow Q_{\beta\alpha}$ and then performing the sum
$Q=Q_1{}^1+Q_2{}^2$. In the K\"ahler case, each of the components, $Q_1{}^1$
and
$Q_2{}^2$, transforms under definite $U(1)_L$ representations and therefore
one can define two BRST charges $Q_1=Q_1{}^1$ and $Q_2=Q_2{}^2$. Of course,
from  the supersymmetry algebra follows that $Q_1^2=0$ and $Q_2^2=0$.
Furthermore, from their construction: $Q=Q_1+Q_2$. The action of each of
these two operators on the fields is easily obtained from the supersymmetry
transformations. One just have to set
${\overline \eta}^{i\dot\alpha}=0$ and, for $Q_1$
$\eta_2^\alpha=\rho_1 \delta_2^\alpha$  and $\eta_1^\alpha=0$, while, for
$Q_2$,
$\eta_2^\alpha=0$ and $\eta_1^\alpha=\rho_2\delta_2^\alpha$.  From the point
of view of $N=2$ superspace the operators $Q_1$ and $Q_2$ can be regarded as
a specific derivative respect to some of the $\theta$'s. In the formulation
of the theory on
$N=1$ superspace the operators
$Q_{1,2}$ can be identified as the derivative respect to $\theta_{1,2}$. This
observation will be very helpful in proving the invariance under $Q_{1,2}$ of
the twisted theories.

On a K\"ahler manifold each of the fields on the right hand side of 
(\ref{twist}) splits into fields which can be thought as components of forms
of type $(p,q)$. For the matter fields on the right hand side  of
(\ref{paloma}) one just has the standard decomposition of $S^+\otimes E$ into
$(K^{1/ 2}\otimes E) \oplus (K^{-{1/ 2}}\otimes E)$. For  example for the
field
$M_\alpha$ one has:
\bea M_\alpha & \rightarrow M_1 \in \Gamma(K^{1/ 2}\otimes E),\;
\;\; M_2 \in \Gamma(K^{-{1/ 2}}\otimes E), \nonumber\\
\overline M^\alpha & \rightarrow \overline M^1 \in
\Gamma(K^{-{1/ 2}}\otimes \overline E), \;\;\;
\overline M^2 \in
\Gamma(K^{{1/ 2}}\otimes \overline E).
\label{rosi}
\eea A similar decomposition holds  for the rest of the fields in
$\Gamma(S^+\otimes E)$ on the right hand side of (\ref{paloma}). Notice that
the product of an element of
$\Gamma(K^{1/ 2}\otimes E)$ times an element of 
$\Gamma(K^{{1/ 2}}\otimes \overline E)$ is a gauge  invariant form of type
$(2,0)$. From the identifications in (\ref{paloma}) and (\ref{compas})
follows that the first component of $\widetilde Q Q$, {\it i.e.},
$\widetilde Q  Q| = q^{2\dagger}  q_1 = 
\overline M^2   M_1$ is a $(2,0)$ form. Therefore, superpotentials of the form
$\widetilde Q Q$, or $\widetilde Q \Phi Q$ as the one in (\ref{silvana}) can
be regarded as $(2,0)$-forms. This is consistent with the observation made in
\cite{sym, cv} that superpotential terms of a twisted theory on a K\"ahler
manifold must transform as $(2,0)$-forms. This requirement guarantees that
the terms  in the Lagrangian are scalars after the twisting.

Since the twisted theory obtained from (\ref{superespacio}) and
(\ref{silvana}) and the topological theory (\ref{taction}) are equivalent
on-shell, we will work out  the on-shell 
$Q_1$-transformations for this case. For the twisted fields in the 
$N=2$ vector multiplet they turn out to be:
\be
\begin{array}{cclcccl}  [Q_1,A_{1\dot\alpha}] &=& \psi_{1\dot\alpha},& 
\,\,\,\,\,\,\,\,\,\,\,\,\,\,\ & [Q_1, \lambda] &=& {1\over
2}\eta-i{\chi_{12}\over {\sqrt 2}}, \nonumber\\ {[}Q_1,A_{2\dot\alpha}] &=&
0,&
\,\,\,\,\,\,\,\,\,\,\,\,\,\,\ & 
\{Q_1, {i\over 2}\eta -{\chi_{12}\over {\sqrt 2}} \} &=&  {1\over
2}[\lambda,\phi]^a , \nonumber\\
 & & & & &+&i(F^{a+}_{12}+{i}(\overline M_1 T^a M_2 +\overline M_2 T^a M_1)
)\\
\{Q_1,\psi_{1\dot\alpha}\}&=&0, & \,\,\,\,\,\,\,\,\,\,\,\,\,\,\ &
\{Q_1, {i\over 2}\eta^a+{\chi_{12}^a\over {\sqrt 2}}\}&=& 0,
 \nonumber\\ 
\{Q_1,\psi_{2\dot\alpha}\}&=&D_{2\dot\alpha}\phi,&
\,\,\,\,\,\,\,\,\,\,\,\,\,\,\ &
\{Q_1,\chi_{11}^a \}&=& 2{\sqrt 2} \overline M_1 T^a M_1 , \nonumber\\ 
{[}Q_1,
\phi]&=&0, & \,\,\,\,\,\,\,\,\,\,\,\,\,\,\ &
\{Q_1,\chi_{22} \}&=& -i{\sqrt 2}F^{+}_{22},
\end{array}
\label{sandia}
\ee where we have used that the generators of the gauge group are normalized
in such a way that $\tr(T^aT^b)=\delta^{ab}$. For the matter fields one finds:
\be
\begin{array}{cclcccl}
 {[}Q_1, M_1]   &=& 0, & \,\,\,\,\,\,\,\,\,\,\,\,\,\,\,\,\,\,\,\,\,\ & [Q_1,
\overline M_1]   &=&  0 , \nonumber\\
 {[}Q_1, M_2]   &=& \mu_2, & \,\,\,\,\,\,\,\,\,\,\,\,\,\,\,\,\,\,\,\,\,\ &
{[}Q_1,
\overline M_2]   &=& \overline \mu_2, \nonumber\\
 \{ Q_1, \mu_1 \}   &=& -i\phi M_1 ,& 
\,\,\,\,\,\,\,\,\,\,\,\,\,\,\,\,\,\,\,\,\,\     & \{ Q_1, \overline \mu_1
\}   &=& i
\overline M_1 \phi, \nonumber\\
 \{ Q_1, \mu_2 \}   &=&0 ,& \,\,\,\,\,\,\,\,\,\,\,\,\,\,\,\,\,\,\,\,\,\ & \{
Q_1,
\overline \mu_2 \}   &=& 0 ,      \nonumber\\
 \{ Q_1, v^{\dot\alpha}\}   &=& -2i D^{\dot\alpha 1}M_1,&
\,\,\,\,\,\,\,\,\,\,\,\,\,\,\,\,\,\,\,\,\,\ & \{ Q_1, \overline
v_{\dot\alpha}\}   &=& -2i D_{1\dot\alpha}
\overline M^1.
\end{array}
\label{elipsekal}                     
\ee The $Q_2$-transformations are easily computed from (\ref{sandia}),
(\ref{elipsekal}) and (\ref{dadi}) after using $Q=Q_1+Q_2$. They read,  for
the
$N=2$ vector multiplet:
\be
\begin{array}{cclcccl}  [Q_2,A_{1\dot\alpha}]&=&0, &
\,\,\,\,\,\,\,\,\,\,\,\,\,\,\  &[Q_2, \lambda] &=&{1\over
2}\eta+i{\chi_{12}\over {\sqrt 2}},\nonumber\\ {[}Q_2,A_{2\dot\alpha}]&=&
\psi_{2\dot\alpha},& 
\,\,\,\,\,\,\,\,\,\,\,\,\,\,\ &\{Q_2, {i\over 2}\eta -{\chi_{12}\over {\sqrt
2}}
\}&=& 0,\nonumber\\
\{Q_2,\psi_{1\dot\alpha}\}&=&D_{1\dot\alpha}\phi, & 
\,\,\,\,\,\,\,\,\,\,\,\,\,\,\ &\{Q_2, {i\over 2}\eta^a+{\chi_{12}^a\over
{\sqrt 2}}\}&=& {1\over 2}[\lambda,\phi]^a\\ & & & &
&-&i(F^{a+}_{12}+{i}(\overline M_1 T^a M_2 +\overline M_2 T^a M_1)
),\nonumber\\ 
\{Q_2,\psi_{2\dot\alpha}\}&=&0,& 
\,\,\,\,\,\,\,\,\,\,\,\,\,\,\ &\{Q_2,\chi_{11} \}&=&-i{\sqrt 2}F^{+}_{11} ,
\nonumber\\  {[}Q_2, \phi]&=&0,& \,\,\,\,\,\,\,\,\,\,\,\,\,\,\
&\{Q_2,\chi^a_{22}
\}&=& 2{\sqrt 2} \overline M_2 T^a M_2,
\end{array}
\label{sandiados}
\ee and for the matter hypermultiplet,
\be
\begin{array}{cclcccl}
 [Q_2, M_1]   &=& \mu_1, & \,\,\,\,\,\,\,\,\,\,\,\,\,\,\,\,\,\,\,\,\,\ & [Q_2,
\overline M_1]   &=& {\overline \mu}_1 , \nonumber\\
 {[}Q_2, M_2]   &=& 0, & \,\,\,\,\,\,\,\,\,\,\,\,\,\,\,\,\,\,\,\,\,\ & [Q_2,
\overline M_2]   &=& 0, \nonumber\\
 \{ Q_2, \mu^1 \}   &=& 0,  &\,\,\,\,\,\,\,\,\,\,\,\,\,\,\,\,\,\,\,\,\,\    
& \{ Q_2, \overline \mu_1 \}   &=& 0, \nonumber\\
 \{ Q_2, \mu_2 \}   &=& -i\phi M_2,
&\,\,\,\,\,\,\,\,\,\,\,\,\,\,\,\,\,\,\,\,\,\ &
\{ Q_2, \overline \mu_2 \}   &=& i \overline M_2 \phi ,      \nonumber\\
 \{ Q_2, v^{\dot\alpha}\}   &=& -2i D^{\dot\alpha 2 }M_2, & 
\,\,\,\,\,\,\,\,\,\,\,\,\,\,\,\,\,\,\,\,\,\ & \{ Q_2, \overline
v_{\dot\alpha}\}   &=& -2i D_{2\dot\alpha}
\overline M^2.
\end{array}
\label{elipsekaldos}                     
\ee It is straightforward to verify that indeed $Q_1^2=Q_2^2=0$  on-shell
after working out the transformations of the different components of
$F_{\alpha\beta}^+$. For $Q_1$ we have,
\be [Q_1, F_{11}^+] =i D_{1\dot\alpha} \psi_1{}^{\dot\alpha}, \;\;\;  [Q_1,
F_{12}^+]={i\over 2} D_{2\dot\alpha}
\psi_1{}^{\dot\alpha}, \;\;\; [Q_1, F_{22}^+]=0.          
\label{otromelon}
\ee and for $Q_2$,  
\be [Q_2, F_{11}^+] = 0, \;\;\; [Q_2, F_{12}^+]={i\over 2} D_{1\dot\alpha}
\psi_2{}^{\dot\alpha}, \;\;\; [Q_2, F_{22}^+]=i D_{2\dot\alpha}
\psi_2{}^{\dot\alpha}.          
\label{otromelondos}
\ee  The action $S$ in (\ref{taction}) is invariant under both, $Q_1$ and
$Q_2$ symmetries. This can be verified explicitly or just using the following
argument based on $N=1$ superspace. On the one hand, the topological action
(\ref{taction}) can be regarded as  a twisted version of the sum of the $N=1$
superspace actions (\ref{superespacio}) and (\ref{silvana}). On the other
hand, the $Q_{1,2}$ operators are equivalent to a $\theta_{1,2}$-derivative.
Acting with this derivative on (\ref{superespacio}) and (\ref{silvana}) one
gets zero: for the terms involving chiral fields one ends with two many
$\theta$-derivatives, while for the other terms one just gets a total
derivative after using the fact that 
$[D_\alpha,\overline D^2] = i \partial_{\alpha\dot\alpha}
\overline D^{\dot\alpha}$.

It is often convenient to regard the  the observables $I(\Sigma)$ in 
(\ref{moreobser}) in terms of the Poincar\'e dual of the homology cycle
$\Sigma$:
\be I(\Sigma) = \int_\Sigma {\cal O}^{(k)} =
\int_X {\cal O}^{(2)}\wedge [\Sigma],
\label{leon}
\ee where $[\Sigma]$ denotes the Poincar\'e dual. On K\" ahler manifolds,
$I(\Sigma)$ can be decomposed in three different types of operators depending
on which holomorphic part of ${\cal O}^{(2)}$ is taken into account. If only
the
$(p,q)$ part ($p+q=2$) of ${\cal O}^{(2)}$ is considered we will denote the
corresponding operator by
$I^{p,q}(\Sigma)$. For example, for the $(2,0)$ part (which is SD):
\be I^{2,0}(\Sigma) = {1\over 4 \pi^2}\int_X e \tr ( \phi F^+_{11} - {1\over
2}
\psi_{1\dot\alpha}
\psi_1^{\dot\alpha}) [\Sigma]_{22} 
\label{launouno}
\ee where we have denoted by $\Sigma_{22}$ the $(0,2)$ part of $\Sigma$.
Notice the following: the $(2,0)$ part of  the two-form observable only
includes the conjugate of the fermionic field $\lambda_2$  in the $N=1$
multiplet $\Phi$.  In the same way, the $(0,2)$ part only includes the
conjugate of the $N=1$ gluino field 
$\lambda_1$  in the vector  superfield $W_{\alpha}$, and the $(1,1)$ part
includes the wedge product of both fields.                            

One of the main ingredients in the analysis made by Witten in \cite{sym} is
the existence of a perturbation of the twisted $N=2$ Yang-Mills theory on
K\"ahler manifolds  which  while preserving the topological character of the
twisted theory it  allows to regard the theory from an untwisted point of
view as an $N=1$ supersymmetric theory. Witten achieved this by demonstrating
that on a K\"ahler manifold it is possible to add an $N=1$ supersymmetric 
 mass-like term for the chiral superfield $\Phi$ while keeping the topological
character of the theory. We will show that this is also possible for the
Topological Quantum Field Theory which describes non-abelian monopoles.
Notice that, as we need the superpotentials to transform as $(2,0)$-forms, to
generate a mass term for
$\Phi$ we must pick a holomorphic $(2,0)$-form on $X$. This is not always
possible on an arbitrary K\"ahler manifold, but we will assume $b_2^+ >1$. 
On a K\"ahler manifold, $b_2^+=2h^{2,0}+1$, and this condition guarantees
that 
$H^{2,0}(X)
\not= 0$ and hence that such a form exists. Recall that, as we showed in
Chapter 4, for $SU(2)$ monopoles in the fundamental  representation,
$b_2^+>1$ implies that there are no reducible solutions in  the moduli space
${\cal M}_{\rm NA}$. This is then a natural requirement from the  point of
view of the moduli problem and the computation of the corresponding
invariants.   

Let us consider a holomorphic $(2,0)$ form $ \omega$ on $X$. Its only 
non-vanishing component is:
\be
\omega_{11} = (\sigma_{lk})_{11} \omega_{mn} \epsilon^{lkmn}.
\label{forma}
\ee We will denote the unique non-vanishing component of the
$(0,2)$ form $\overline\omega$, conjugate to $\omega$, by
$\overline\omega_{22}$ (
$\overline\omega_{22}= (\omega_{11})^*$)

Following \cite{sym} we begin making a perturbation of the action $S$ in
(\ref{taction}) by adding a term of the form,
\be I(\overline \omega) = \int_X {\cal O}^{(2)} \wedge  {\overline
\omega},\label{primer}
\ee where $I(\overline \omega)$ is the observable defined in
(\ref{moreobser}), but without  the 
$1/(4\pi^2)$ factor. Using the $Q_2$-transformations (\ref{sandiados}) and 
(\ref{elipsekaldos}), this term can be written as:
\be I({\overline \omega})= -{1\over 2}\int_X d^4 x \, e \, {\overline
\omega}_{22}
\tr(\psi_{1\dot\alpha}\psi_1^{\dot\alpha}) +
\big\{ Q_2, {i\over {\sqrt 2}}  \int d^4 x \, e \, {\overline \omega}_{22}
\tr(\phi\chi_{11}) \big\}.
\label{tomate}
\ee                          The first part of this term indicates some
progress towards the construction of an $N=1$ mass-like term. However,
(\ref{tomate}) is not invariant under $Q_2$. Contrary to the case of the
theory without matter fields this term is not even $Q_2$-invariant on-shell.
One can  remedy this problem if instead of introducing $I({\overline
\omega})$ one considers:
\be
\tilde I({\overline \omega}) = I({\overline \omega}) + 2i\int_X d^4 x \, e \, 
{\overline \omega}_{22}
\overline M_1 \phi M_1.
\label{flordos}
\ee Indeed, $\{ Q_2, \tilde I({\overline \omega}) \}$ turns out to be 
proportional to the field equation resulting after making a variation respect
to $\chi^{11}$ in the twisted action
$S$ in (\ref{taction}).

The term $\tilde I({\overline \omega})$ implies further progress towards the
perturbation by an $N=1$ supersymmetric mass term. Notice that from an $N=1$
superspace point of view we intend to obtain a term like (\ref{masa}). This
type of term, added to a  theory which already has the last two terms in
(\ref{silvana}), leads, when written in component  fields, to the mass terms 
for the $N=1$ superfield:
\be m{\overline m}B B^{\dagger}+i{\sqrt 2}m B {\overline M}^1M_2- i{\sqrt 2}m
B^{\dagger} {\overline M}^2M_1 +m \lambda_2{}^{\alpha} \lambda_{2 \alpha} -
{1 \over 2}{\overline m}{\overline \lambda}^2{}_{\dot \alpha} {\overline
\lambda}^{2\dot
\alpha},
\label{masaexplicita}
\ee that includes terms like the one added to $I({\overline \omega})$ in
(\ref{flordos}).

To make further progress in the perturbation towards an $N=1$ supersymmetric
mass term while maintaining the $Q_2$ symmetry we will modify the
$Q_2$-transformation of
$\chi_{22}$ in the following way:
\be
\{Q_2, \chi_{22}^a \} \longrightarrow
\{Q_2', \chi_{22}^a \} = 2{\sqrt 2}\Big(\overline M_2 T^a M_2+ \, i{\overline
\omega}_{22} \phi^a\Big),
\label{tinte}
\ee while for the rest of the fields the action of $Q_2$ remains the same. 
Notice that still one has
$({Q_2'})^2=0$ on-shell.

Under $Q_2'$ the action $S$ is not invariant. However, one can verify that
now the perturbed action
$S+\tilde I({\overline \omega})$ is invariant and does not close on a field
equation as before. On the other hand,  adding a $Q_2'$-exact term will keep
the
$Q_2'$-invariance of the theory. It is rather remarkable that adding just the
term,
\be 2{\sqrt 2}\Big\{Q_2', \int d^4 x\,e \, 
\omega_{11} \tr(\lambda \chi_{22}) \Big\},
\label{lugo}
\ee one finds that the perturbed action is just the action $S$ plus an
$N=1$ supersymmetric mass term for the chiral superfield $\Phi$:
\bea S&+&\tilde I(\overline \omega) + \{ Q_2',...\}\nonumber\\ &=&S+\int_X
d^4 x\, e
\Big(4 \omega_{11}
\tr\big(({1\over 2}\eta+i{\chi_{12}\over {\sqrt 2}}){\chi_{22}\over {\sqrt
2}}\big)- {\overline \omega}_{22} \tr( {1\over 2}\psi_{1\dot\alpha}
\psi^{\dot\alpha}{}_1))
\nonumber\\ & &+  {i \over 2} \int_X d^4 x\, e\,
\omega_{11}\overline\omega_{22} 
\tr(\lambda \phi)\nonumber\\ & &+\int_X d^4 x \, e  (8 \omega_{11} \overline
M_2
\lambda M_2 + 2i\overline\omega_{22} \overline M_1 \phi M_1).
\label{marmol}
\eea This perturbation of the action $S$  contains all the terms present in
the
$N=1$ supersymmetric mass term (\ref{masa}) after setting $\omega_{11}=im/4$
and writing the twisted fields in terms of the untwisted ones, using
(\ref{twist}) and (\ref{debbie}).  For the matter fields one can read their
untwisted counterparts from (\ref{paloma}).

Our analysis implies that if one denotes correlation functions of observables
in the twisted theory by 
$\langle A_1\cdots A_n\rangle$, and in the perturbed  theory by  $\langle
A_1\cdots A_n\rangle_1$, the relation between them is:
\be
\langle A_1\cdots A_n\rangle_1 =\langle A_1\cdots A_n {\rm e}^{\tilde
I({\overline
\omega})} \rangle.
\label{lawi}
\ee As argued in \cite{sym}, given some  homology cycles $\Sigma$,  it can be
assumed that near their intersection  they look like holomorphically embedded
Riemann surfaces. This means that actually the only relevant part of the
two-form operators entering (\ref{lawi}) are of type $(1,1)$. Precisely those
are the two-form operators invariant under $Q_2'$.  As the zero-form
observables in (\ref{lawi}) are also invariant under $Q_2'$ one can regard
the right hand side of (\ref{lawi}) as a topological quantum field theory
whose BRST operator is $Q_2'$ and its action  is $S+\tilde I({\overline
\omega})$.

The effect of  an extra term $I({\overline \omega})$ in the action  of
Donaldson-Witten theory was studied by Witten in \cite{sym}. He showed that
its effect on correlators of  observables can be described as a shift on the
parameters corresponding  to the observables containing two-form operators. 
We will finish this section showing that the relevant contribution from
$\tilde I({\overline
\omega})$ in (\ref{lawi}) and from $I({\overline \omega})$ in the case of
Donaldson-Witten theory is the same. Therefore, in our theory the effect of 
the presence of $\tilde I({\overline \omega})$ in (\ref{lawi}) is also a
shift in those parameters.

The quantity $\tilde I({\overline \omega})$ can be written as,
\bea
\tilde I({\overline \omega}) &=&
\int_X d^4 x\, e\, {\overline \omega}_{22}
\Big(\phi^a(F^{a+}_{11}+2i\overline M_1 T^a M_1)-\tr({1\over
2}\psi_{1\dot\alpha}\psi_1{}^{\dot\alpha})\Big)\nonumber\\ &=&\int_X d^4 x\,
e\, {\overline \omega}_{22} \Big(\tr(
\{Q,i\phi{\chi_{11}\over {\sqrt 2}}\} -{1\over
2}\psi_{1\dot\alpha}\psi_1{}^{\dot\alpha})\Big),
\label{loro}
\eea after using (\ref{dadi}). As $Q$ is still a topological symmetry of the
theory, and  the observables are $Q$-invariant, this means that the vacuum
expectation values on the right hand side of (\ref{lawi}) can be written as
\be
\langle A_1\cdots A_n {\rm e}^{J({\overline \omega})} \rangle.
\label{lawidos}
\ee where,
\be J({\overline \omega}) =  -{1\over 2}
\int_X d^4 x\, e\, {\overline \omega}_{22} 
\tr(\psi_{1\dot\alpha}\psi_1{}^{\dot\alpha}). 
\label{francesca}
\ee This is precisely the same expression that one obtains in the case of
Donaldson-Witten theory. Notice that in that case
$F_{11}^+$ is $Q$-exact and one has the same $Q$-transformations as in our
theory for the field $\psi_{\alpha \dot\alpha}$.  

Another argument to show that the presence of the term involving the massive
fields in $\tilde I ({\overline \omega})$ is irrelevant is just to point out
that the contributions from the functional integral on the right hand side of
(\ref{lawi}) are localized  on  configurations satisfying the non-abelian
monopole equations. As  shown in Chapter 4, the $(2,0)$  part of  those
equations implies $\overline M_1 T M_1=0$. This can be also seen from
Witten's  fixed point theorem for the $Q_{1,2}$ transformations: a BRST
invariant  configuration for  the $Q_1$ topological symmetry has a vanishing
$\{Q_1, \chi_{11} \} \sim 
\overline M_1 T M_1$.  

\section{The polynomial invariants in the K\"ahler case}

In this section we compute the topological invariants corresponding to $SU(2)$
monopoles on K\"ahler, Spin manifolds, following the  abstract approach
introduced by Witten in \cite{sym} and further developed in 
\cite{vw}. The strategy is the following: in the previous section,  we showed
that the correlation functions in the twisted theory can be computed in
terms  of the theory perturbed by a mass term (see (\ref{lawi}) and the
following discussion).  The vacuum structure of this theory was analyzed in
section 2, and we  determined the pattern of chiral symmetry breaking. We
also deduced that  the vacua have a mass gap and gluino condensation. The
main assumption we will  make is that this vacuum structure is the same on
any manifold, in other words,  that it doesn't depend on  global features of
the space-time and coincides with the one we obtained before.  With this
assumption, we can use two properties of the theory under consideration to 
constrain the form of the correlation functions and determine them up to
some  universal constants \cite{sym}. These properties are the mass gap and 
the topological invariance  (more precisely, invariance under rescalings of
the metric). This can be regarded as  a generalization of the properties of
correlation functions in supersymmetric  gauge theories,  leading to
clustering \cite{amati}. There are of course some  subtleties due to the
structure  of the mass perturbation on a K\"ahler manifold, and we will
consider them later.   

The clustering property and the invariance under rescaling imply that the
correlation  functions have the form:
\be
\langle {\rm exp} (\sum_{a}\alpha_{a}I(\Sigma_{a})+\mu {\cal O})
\rangle=\sum_{\rho}C_{\rho}  {\rm exp} (\gamma_{\rho}v^2 +\mu \langle \rho
|{\cal O}|
\rho \rangle). \label{corr}
\ee In this expression the sum is  over the three vacua $|\rho \rangle$,
associated to the three roots of (\ref{uy}),  and labeled by the index
$\rho =1,2,3$; $v=\sum_{a}\alpha_{a}[\Sigma_{a}]$, where $[\Sigma_{a}]$ is the
cohomology class Poincar\'e dual to $\Sigma_{a}$, and 
$v^2=\sum_{a,b}\alpha_{a}\alpha_{b}\sharp(\Sigma_{a}\cap \Sigma_{b})$, where
$\sharp (\Sigma_{a}\cap \Sigma_{b})$  is the intersection number of
$\Sigma_{a}$ and
$\Sigma_{b}$. The constant 
$C_{\rho}$ is the partition function in the $\rho$ vacuum. For the partition 
function, again the mass gap and the invariance under rescalings imply that
it must be a  combination of topological invariants of the four-manifold that
can be written in  terms of the integral of a local density. The only such
invariants are the signature  and the Euler characteristic, and then we have: 
\be  C_{\rho}={\rm exp}(a_{\rho}\chi + b_{\rho}\sigma).
\label{asturias}
\ee The constants which appear in (\ref{corr}) are not independent because the
theory has a ${\bf Z}_{3}$ broken symmetry which relates  the three vacua and
is given by (\ref{uonedef}).  First let us work out  the relation between the
$C_{\rho}$.  As these constants are given by the partition function of the
theory at different vacua,  and the vacua are related by a non-anomalous
symmetry, one should think  that they are equal. But actually, as we are
working now on a curved four-manifold,  the anomalies have gravitational
contributions which were not taken into account in section 1  (where $X={\bf
R}^4$), and the path integral measure does change. We must take into  account
also the new geometrical content of the fields after twisting. The field
$\lambda_{1}$ contains a $(2,0)$-form part,  a $(1,1)$ part and a scalar
part, and couples to ${\overline \lambda^1}$ (a 
$(0,1)$-form) through a twisted Dirac operator $D_{K^{1/2}\oplus K^{-1/2}}$,
which is  the  adjoint of the operator:
\be {\partial}^{\dagger}_ A\oplus {\partial}_ A:
\Omega^{1,0}({\rm End}_0({\cal E})) \too \Omega^{0}({\rm End}_0({\cal E})) 
\oplus \Omega^{2,0} ({\rm End}_0({\cal E})),
\label{conjelipse}
\ee  This is nothing but the complex conjugate of the deformation complex
(\ref{otraelipsis}),  and its index (equal to $-{\rm index} \,\
D_{K^{1/2}\oplus K^{-1/2}}$) is given by  half the  dimension of ${\cal
M}_{\rm ASD}$. This simply follows from the fact that,  as we saw in Chapter
1, section 3, this complex is isomorphic to  the ASD complex as a real
complex. Therefore the anomaly due to the first transformation in
(\ref{uonedef}) is:
\be {\rm e}^{ {\pi i \over 3}\{ 4k-{3 \over 4}(\chi +\sigma)\} }.
\label{fase}
\ee We must take into account also the  transformation of the matter
fermions. After twisting they are spinors, and we have the correspondence
$\psi_{q}\rightarrow 
\mu$, $\psi_{\tilde q} \rightarrow {\bar
\mu}$ (see (\ref{paloma})).  Notice that, due to the $\rho$ symmetry in
(\ref{parida}), there is an additional contribution to the  anomaly coming
from this transformation, as we saw in (\ref{tof}). Now we must also compute
the gravitational part and obtain the total anomaly (a similar problem is
addressed in section 4.4 of
\cite{abelia}). The path  integral measure  for the twisted matter fermions
can be written as:
\be
\prod _{I} d\mu_{I} d{\bar \mu}_{I} \prod _{J} dv_{J} d{\bar v}_{J},
\label{tinta}
\ee where the index $I=1, \cdots, \nu_{+}$  refers to the $\mu^{\alpha}$ zero
modes (of positive chirality) and the index 
$J=1, \cdots, \nu_{-}$ to the $v_{\dot\alpha}$ zero modes (of negative
chirality).  Under the transformation in (\ref{uonedef}), the measure
(\ref{tinta}) transforms as: 
\be (-1)^{\nu_{+}+\nu_{-}}{\rm e}^{ -{\pi i \over 6}2 (k+ {\sigma \over 4})}
=(-1)^{-k-{\sigma \over 4}}{\rm e}^{ -{\pi i \over 3} (k+ {\sigma \over 4})},
\label{discman}
\ee where we have taken into account  that $\nu_{+}-\nu_{-}={\rm index}\,\ D =
-k-\sigma / 4$, according to (\ref{indice}). Putting together both factors we
obtain: 
\be (-1)^{\Delta}{\rm e}^{-{\pi i \over 12}\sigma},
\label{anomaly}
\ee where the integer $\Delta$ was introduced in  (\ref{ladelta}) and we have
used that $\sigma \equiv 0
\,\ {\rm mod}\,\ 8$.  Notice that the $k$ dependence has dropped out, because
the symmetry in (\ref{uonedef}) is not anomalous under Yang-Mills instantons,
and (\ref{anomaly}) contains only the gravitational contribution to the
anomaly. The result, in terms of the constants $C_{\rho}$, is: 
\be C_{2}=(-1)^{\Delta}{\rm e}^{-{\pi i \over 12}\sigma} C_{1},\,\,\,\,\
C_{3}= {\rm e}^{-{\pi i \over 6}\sigma}C_{1}.
\label{zvacios}
\ee

We would also like to relate the constants 
$\gamma_{\rho}$ and the expectation values
$\langle \rho |{\cal O}| \rho \rangle$ in  (\ref{corr}) for the different
vacua. This is easily done by taking into account the  transformations of the
corresponding observables under the symmetry (\ref{uonedef}). As is argued in
\cite{sym}, for the observables
$I(\Sigma)$ on a K\"ahler manifold  one need consider only the $(1,1)$ part.
This part transforms under (\ref{uonedef})  as $B^{\dagger}$, because its
fermionic part includes the wedge product of 
${\overline \lambda}^1$ and ${\overline \lambda}^2$. If we call
$\alpha$ the generator of the discrete  symmetry in (\ref{uonedef}) in the
operator formalism, one finds  the following relations:
\bea
 \alpha|\rho \rangle &=& |\rho +1
\rangle, \,\,\,\,\,\ \rho \equiv 1 \,\ {\rm mod} \,\ 3, \nonumber\\
\alpha {\cal O} \alpha^{-1} &=&  {\rm e}^{2\pi i \over 3} {\cal O},\,\,\,\,\
\alpha I^{1,1}(\Sigma)  \alpha^{-1} = {\rm e}^{\pi i \over 3} I^{1,1}(\Sigma),
 \label{obanomalia}
\eea which lead to:
\bea
\gamma_{2}&=&{\rm e}^{-{2\pi i \over 3}} 
\gamma_1, \,\,\,\,\ \gamma_{3}={\rm e}^{-{4\pi i
\over 3} }\gamma_1, \nonumber\\
\langle 2 |{\cal O}|2 
\rangle &=&{\rm e}^{-{2\pi i \over 3}}\langle 1 |{\cal O}|1 \rangle,
\,\,\,\,\ \langle 3 |{\cal O}|3 
\rangle={\rm e}^{-{4\pi i \over 3}}\langle 1 |{\cal O}|1
\rangle. 
\label{relaciones}
\eea With these relations we have determined completely  the bulk structure
of the vacua, which comes from the underlying $N=1$ theory.

One has to take into account however that
 the mass perturbation which gives this theory was done with a $(2,0)$
holomorphic form $\omega$, and the mass will vanish when this form does. One
must then be careful because the assumption  of a mass gap fails on the
vanishing locus. In general, $\omega$ vanishes  on a divisor $C$ representing
the canonical class of
$X$.  The simplest case is the one in which $C$ is a union of disjoint Riemann
surfaces 
$C_y$ of multiplicity $r_{y}=1$ ({\it i.e.},
$\omega$ has simple zeroes along this components),  and therefore the
canonical divisor of $X$ can be written as,
\be c_{1}(K)=\sum_{y}[C_y].
\label{canonico}
\ee What are the consequences of this fact? When the mass $m$ vanishes,  one
doesn't expect the vacuum structure to be given by the superpotential
(\ref{supdef}),  for the untwisted physical theory is in the Coulomb phase
and one has additional  massless degrees of freedom. As discussed in
\cite{sym, vw}, near the  surfaces
$C_{y}$, these additional degrees of freedom  are described by an effective 
two-dimensional  theory (the cosmic string theory), and the vacuum structure
along these surfaces  has additional symmetry breaking. The cosmic string
theory has a chiral masslees fermion coming  from the field $\lambda_2$ of
the underlying perturbed $N=2$ theory. We then expect,  as in \cite{sym},
dynamical breaking of the
${\bf Z}'_4$ symmetry given in  (\ref{extraz}), in such a way that the
diagonal
$(-1)^F \subset {\bf Z}_4 \times  {\bf Z}'_4$ is left unbroken. In principle,
the new pattern of symmetry breaking  produces four cosmic string vacua
associated to
${\bf Z}'_4$. However, both the cosmic  string theory and the observables of
the microscopic theory depend solely on the fields  coming from the Yang-Mills
multiplet, and the restriction of ${\bf Z}'_4$ to this subset  of fields is a
${\bf Z}_2$ symmetry, giving then the pattern of symmetry breaking  analyzed
in \cite{sym, vw}. We will assume that along the worldsheets of  the strings
$C_{y}$ each bulk vacuum bifurcates according the dynamical breaking of 
$\lambda_{2} \rightarrow -\lambda_{2}$. One may include the four vacua
associated to 
${\bf Z}'_4$, but as it will become clear in our computation, this just gives
a global   factor in the correlation functions. 

As we will see,  this assumption on the vacuum structure is the most natural
one from several points of view. First of all,  the contributions from the
new vacua cooperate with the bulk structure in such a way  that the resulting
expression has the adequate properties. Second, with this assumption, the
final  expression can be naturally understood as a consequence of
electric-magnetic  duality of the underlying $N=2$ theory, as we will see in
the next Chapter.  
      
 Let us briefly explain, following \cite{sym},  what is the effect of the new
vacua in the computation of the correlation functions.  Each bulk vacuum
$|\rho \rangle$ leads to two vacua of the cosmic string  theory $|\rho
+\rangle$, $|\rho - \rangle$, which are related by the broken symmetry
$\alpha^{3}$. The surfaces $C_{y}$ give new contributions  to the correlation
functions through their intersection with the surfaces $\Sigma_{a}$.  The
observables $I(\Sigma_a)$ will be described by
 $\sharp (\Sigma_a \cap C_{y}) V_{y}$, where $V_{y}$ is the insertion of a
cosmic string operator 
$V$ on $C_{y}$ which has the same quantum numbers of $I^{1,1}(\Sigma_a)$. From
(\ref{uonedef}) follows that it transforms under $\alpha^3$ as: 
\be     
\alpha^3 V \alpha^{-3}=-V.
\label{santander}
\ee Now, for a given bulk  vacuum $|\rho \rangle$ we must take into account
its bifurcation along the diferent surfaces 
$C_{y}$, and compute the vacuum expectation values of ${\rm exp}
(\sum_{a}\alpha_{a}I(\Sigma_{a}))$. The result is \cite{sym}:
\be
 \prod_{y}\Big( {\rm exp}(\phi_{y}\langle \rho +|V|\rho + \rangle) + t_{y}{\rm
exp}(-\phi_{y}\langle \rho +|V|\rho + \rangle) \Big).
\label{cosmicomicas}
\ee In this expression, 
$\phi_{y}=\sum_{a} \alpha_{a}\sharp (\Sigma_a \cap C_{y})$ and the factor
$t_{y}$ is similar  to (\ref{anomaly}) and comes from an anomaly in the
two-dimensional effective theory. It is given by:
\be t_{y}=(-1)^{\epsilon_{y}},
\label{late}
\ee where ${\epsilon_{y}}$ is $0$  ($1$) if the Spin bundle of $C_{y}$ is even
(odd). The
$\epsilon_{y}$ verify \cite{sym}:
\be
\Delta + \sum_{y} \epsilon_{y} \equiv 0 \,\,\ {\rm mod}\,\,\ 2.
\label{lastes}
\ee

At this point we have all the information that we  need to compute the
polynomial invariants for
$SU(2)$ monopoles on K\"ahler, Spin four-manifolds. Notice that the result
will involve unknown constants which should be fixed by comparing to
mathematical computations of these invariants. These  constants are
universal, in the sense that they depend only on  the dynamics of the
physical theory (as shown in \cite{sym}) and not on the particular manifold
we are considering. It should be mentioned that  these constants are closely
related to VEVs of operators in the physical theory.  The VEV of the operator
${\cal O}$ in the twisted theory depends on the VEV of  the operator $\tr
(B^{\dagger})^2$ of the untwisted theory, as one can easily  see by comparing
(\ref{mioperador}) to (\ref{twist}). This VEV has been computed  in
(\ref{uy}) and is non-zero. It can also be seen \cite{sym} that $\gamma=
\gamma_1$ is related to the gluino condensate, computed in (\ref{gluino}) in
terms  of $U$. This suggests that the exact solution of the physical theory
should provide  relations  between these universal coefficients. As we will
see in the next Chapter,  the $N=2$ theory allows one to relate the VEV
$\langle V \rangle$ to the other operators  in terms of the exact low-energy
solution. It should be extremely interesting  to make more precise the
relation between the untwisted and the twisted  theory in order to relate the
different topological constants appearing in this  kind of expressions. 

If we denote $C=C_1$, $\langle 1  |{\cal O}|1 \rangle = \langle {\cal O}
\rangle$ and $\langle 1 +|V|1+ \rangle = \langle   V\rangle$, the expression
for the polynomial invariants reads:
\bea & & C \Bigg( {\rm exp} (\gamma v^2 +\mu \langle {\cal O} \rangle)
\prod_{y}\Big( {\rm exp}( 
\langle V \rangle \phi_{y}) + t_{y}{\rm exp} (-\langle V
\rangle \phi_{y}) \Big) \nonumber\\  & &+(-1)^{\Delta}{\rm e}^{-{\pi i \over
12}\sigma} {\rm exp}\Big( {\rm e}^{-{2\pi i 
\over 3}}(\gamma v^2 +\mu \langle{\cal O} \rangle) \Big)
\prod_{y}\Big( {\rm exp}( {\rm e}^{-{\pi i \over 3}} \langle V \rangle
\phi_{y}) + t_{y}{\rm exp}( -{\rm e}^{-{\pi i \over 3}}\langle V
\rangle\phi_{y} \Big) \nonumber\\ & &+{\rm e}^{-{\pi i \over 6}\sigma} {\rm
exp}\Big( {\rm e}^{-{4\pi i 
\over 3}}(\gamma v^2 +\mu \langle{\cal O} \rangle) \Big)\nonumber\\ & &
\times\prod_{y}\Big( {\rm exp}( {\rm e}^{-{2\pi i \over 3}} \langle V \rangle
\phi_{y}) + t_{y}{\rm exp}( -{\rm e}^{-{2\pi i \over 3}}\langle V
\rangle\phi_{y} \Big) \Bigg).  
\label{polyone}
\eea

In order to check some of the properties of (\ref{polyone}) we will express
it in a more convenient way. Notice that because of (\ref{late}) and
(\ref{lastes}) we can  extract the factor $t_{y}$ in the second summand of
(\ref{polyone}) and cancel the factor $(-1)^{\Delta}$. Using some
straightforward algebra and the fact that 
$\sigma \equiv 0 \,\ {\rm mod}\,\ 8$, (\ref{polyone}) can be rewritten as:
\bea & &\langle {\rm exp} (\sum_{a}\alpha_{a}I(\Sigma_{a})+\mu {\cal O})
\rangle\nonumber\\ & &=C \Bigg( {\rm exp}  (\gamma v^2 +\mu \langle {\cal O}
\rangle) \prod_{y}\Big( {\rm exp}( \langle V \rangle \phi_{y}) + t_{y}{\rm
exp} (-\langle V \rangle \phi_{y}) \Big) \nonumber\\ & &+{\rm e}^{-{\pi i
\over 6}\sigma} {\rm exp}\Big( -{\rm e}^{-{\pi i 
\over 3}}(\gamma v^2 +\mu \langle{\cal O} \rangle)
\Big) \prod_{y}\Big( {\rm exp}( {\rm e}^{-{2\pi i \over 3}} \langle V \rangle
\phi_{y}) + t_{y}{\rm exp}( -{\rm e}^{-{2\pi i \over 3}}\langle V
\rangle\phi_{y} \Big) \nonumber\\ & &+{\rm e}^{-{\pi i \over 3}\sigma} {\rm
exp}\Big( -{\rm e}^{{\pi i 
\over 3}}(\gamma v^2 +\mu \langle{\cal O} \rangle)
\Big)\nonumber\\ & & \times \prod_{y}\Big( {\rm exp}( {\rm e}^{-{4\pi i \over
3}}
\langle V \rangle \phi_{y}) + t_{y}{\rm exp}( -{\rm e}^{-{4\pi i \over
3}}\langle V
\rangle\phi_{y} \Big) \Bigg), \label{fer}
\eea    where the second summand of  (\ref{polyone}) is now the last one.
This is our final expression for the polynomial invariants associated to
$SU(2)$ monopoles on K\"ahler, Spin manifolds whose canonical divisor can be
written as in (\ref{canonico}). The result is obviously real,  as the first
summand in (\ref{fer}) is real and the second one is the complex conjugate of
the third one. 

Another check of (\ref{fer}) is the following. As we noticed in Chapter 5, a
product of $r$  observables ${\cal O}$ and $s$ observables
$I(\Sigma)$ has ghost number $4r+2s$,  and this must equal the dimension of
the moduli space for some instanton number $k$. In terms of $\Delta$ we have
the selection rule (\ref{selrule}): 
\be 4r+2s={\rm dim} \,\ { \cal M }_{\rm NA}=6(k-\Delta )-{\sigma \over 2},
\label{dimension}
\ee {\it i.e.}, if we suppose that the 
$\alpha_a$ are of degree $2$ and $\mu$ of degree
$4$, in the expansion of (\ref{fer}) we  can only find terms whose degree is
congruent to
$-\sigma /2$ mod $6$.  This is easily checked. If we consider the terms of
fixed degree
$4r+2s$ we see that they can be grouped  in terms with the same coefficient,
given by:
\be 1+{\rm e}^{-{\pi i \over 6}\sigma}{\rm e}^{-{\pi i \over 3} (4r+2s)}+{\rm
e}^{-{\pi i \over 3}\sigma}{\rm e}^{-2{\pi i \over 3}(4r+2s)}.
\label{bilbao}
\ee This is a geometrical series whose sum  is zero unless ${\rm e}^{-{\pi i
\over 6}\sigma-{\pi i
\over 3}(4r+2s)}=1$, which gives  precisely the condition we were looking
for. Notice that to obtain the well-behaved  expression (\ref{fer}) from
(\ref{polyone}) the key point is that the contributions from the  cosmic
string theory have the form (\ref{cosmicomicas}). This is what allows to drop
out the  factor $(-1)^{\Delta}$ which comes from the bulk structure and
suggests that the pattern  of bifurcation of vacua along the cosmic string is
the right one.  
 
Another point of interest is that,  according to our expression (\ref{fer}),
the generating function for the correlation functions $f=\langle {\rm exp}
(\sum_{a}\alpha_{a}I(\Sigma_{a})+\mu {\cal O}) \rangle $ verifies the
equation:
\be {\partial^3 f \over \partial \mu^3}={\langle {\cal O} \rangle }^3 f,
\label{tiposimple}
\ee which seems to be the adequate  generalization to our moduli problem of
the simple type condition which appears in Donaldson theory \cite{kmone,
kmtwo}. Physically,  the order of  this equation is clearly related to the
number of ground states of the $N=1$  theory. It would be interesting to have
a clear picture of the mathematical meaning of this generalized simple type
condition  as well as to know what is the form it takes in  other moduli
problems.

This ends our computation of the polynomial invariants for Spin, K\"ahler
manifolds.  In the next Chapter we will consider this computation in the
$N=2$ framework, and we  will drop the K\"ahler assumption.  

\chapter{Exact results in $N=2$ and
TQFT}                                                 

In this Chapter we take a somewhat more direct approach to the computation of
the  polynomial invariants associated to $SU(2)$ monopoles.  We will analyze
the $N=2$ theory  directly, without any kind of soft breaking to $N=1$. For
this we will use the  results of Seiberg and Witten \cite{swone, swtwo}. In
these by now classical works,  they were able to compute the exact low-energy
effective action of the asymptotically  free $N=2$ QCD theories with gauge
group $SU(2)$. One of the implications of this  fundamental work was
precisely an alternative formulation of Donaldson theory through  the
Seiberg-Witten monopole equations, which were presented from a mathematical
point  of view in Chapter 4, section 1. This in turn allows one to compute
the Donaldson  invariants from the invariants associated to the abelian
monopole equations. We will  here briefly review the non-perturbative results
on $N=2$ theories and apply the 
$N_f=1$ result to the moduli problem of $SU(2)$ monopoles, on general Spin
manifolds  with $b_2^+>1$. The organization of this chapter is as follows: in
section 1 we review  the basics of the Seiberg-Witten solution. In section 2
we consider the
$N_f=1$ theory  and compute the polynomial invariants. In section 3 we
consider again the K\"ahler case  from the point of view of the $N=2$ theory,
and we rederive the expression given in  (\ref{fer}). 

\section{The Seiberg-Witten solution}

We give a brief overview of the Seiberg-Witten solution in the $N_f=0$ case,
as the main  features of the $N_f=1$ case are very similar. For the analysis
of the topological  version of the theory we don't need all the ingredients
of the solution, and therefore  we will restrict ourselves to a general
picture of the quantum moduli. For more details,  one can see the original
papers \cite{swone, swtwo} or the reviews \cite{goh, bilal,  enedosluis}.

To study the vacuum structure of $N=2$ supersymmetric Yang-Mills theory,  one
first studies, as in $N=1$ theories, the classical vacua. $N=2$
supersymmetry  does not allow a superpotential for the theory and therefore
the scalar potential is purely
$D$-term:
\be V(\phi)={1 \over g^2} {\rm Tr} [B, B^{\dagger}]^2
\label{iii}
\ee There is a classical moduli space of vacua, and the minima of (\ref{iii})
 can be taken to be of the form $B ={1\over 2}a\sigma^3$
 with $a$ complex. A gauge invariant description of this
 moduli space is provided by the variable
$u ={\rm Tr}B^2={1\over 2}a^2$ at the classical level.
 Each point in this moduli space represents a different theory.
 For $a \not= 0$ the charged multiplets acquire a mass
$M=\sqrt {2}|a|$, and $SU(2)$ is spontaneously broken to $U(1)$,
 and at $a=0$ the full $SU(2)$ symmetry is restored. Away from the origin we
can integrate out the massive multiplets and obtain a low-energy effective
theory which depends only on the ``photon" multiplet \cite{seidos}. The
theory is fully described in terms of an $N=2$ prepotential
${\cal F}(A)$. The Lagrangian in $N=1$ superspace is
\be {\cal L}={1 \over 4 \pi}{\rm Im} \Big[ \int d^4 \theta {\partial{\cal  F}
\over
\partial A} {\overline A} + {1\over 2} \int d^2 \theta {\partial ^2 {\cal F}
\over
\partial A^2} W_{\alpha}W^{\alpha} \Big].
\label{iv}
\ee The K\"ahler potential and gauge kinetic functions are given in general
by:
$$ K(a, {\bar a})={1 \over 4 \pi}{\rm Im}\,\  a_{D}{\bar a},
$$
$$
\tau={\partial ^2 {\cal F} \over \partial a^2}, $$
\be a_{D} \equiv {\partial {\cal F} \over \partial a}.
\label{v}
\ee

An important point is that effective action (\ref{iv}) has a duality group
which  corresponds physically to electric-magnetic duality. The original
photon variable is 
$a$ (the lowest component of the chiral superfield $A$), and the magnetic
photon is  described by the variable $a^{(m)}=-a_D$. 

In perturbation theory ${\cal F}$ only receives one-loop contributions.
 The important thing is to determine the non-perturbative corrections. This
was done in \cite{swone,swtwo}. Some of the properties of the exact solution
are:

1) The $SU(2)$ symmetry is never restored. The theory stays in the Coulomb
phase throughout the $u$-plane.

2) The moduli space has, for $N_f=0$, a symmetry $u \rightarrow -u$
 (the non-anomalous subset of the $U(1)_R$ group), and at the points
$u=\Lambda^2$,
$-\Lambda^2$ (where $\Lambda$ denotes the scale  of the $N=2$ Yang-Mills
theory) singularities in ${\cal F}$ develop.  Physically they correspond
respectively to a massless monopole and dyon with charges $(n_m,n_e)=(1,0)$,
$(1,1)$, where 
$n_m$ denotes the magnetic charge and $n_e$ the electric one. Hence near
$u=\Lambda^2$, $-\Lambda^2$ the correct effective action should include
together with the photon vector multiplet monopole or dyon hypermultiplets.
In this way,  around the singularities, the effective theory is essentially
$N=2$ QED, and the  matter fields are represented by $N=2$ hypermultiplets
corresponding  to the massless particles. As these particles are monopoles or
dyons, they must be  coupled to the magnetic or ``dyonic" photon. 

3) The function ${\cal F}(a)$ is holomorphic. It is better to think in terms
of the vector
$^{t}v=(a_D, a)$ which defines a flat $Sl(2,{\bf Z})$ vector bundle over the
moduli space ${\cal M}_u$ (the $u$-plane). Its properties are determined by
the singularities and the monodromies around them. Since
${\partial ^2 {\cal F} / \partial a^2}$ or
${\partial a_D/ \partial a}$ is the coupling constant, these data are
obtained from the $\beta$-function in the three patches: large-$u$ (the Higgs
region), the monopole and the dyon regions. In the Higgs, monopole and dyon
patches, the natural independent variables to use are respectively
$a^{(h)}=a$, $a^{(m)}=-a_D$,
$a^{(d)}=a_D-a$. Thus in each patch we have a different prepotential:
\be {\cal F}^{(h)}(a), \,\,\,\ {\cal F}^{(m)}(a^m), \,\,\,\ {\cal
F}^{(d)}(a^d).
\label{xii}
\ee The descriptions given by these prepotentials correspond to mutually
non-local objects.  Therefore, there is no effective lagrangian describing
these three objects  simultaneously (although there are some interesting
phenomena in these kinds of theories  associated to non-local objects).

4) The explicit form of $a(u)$, $a_D(u)$ is given in terms of the
 periods of a meromorphic differential of the second kind on a genus
 one surface described by the equation:
\be y^2=(x^2-\Lambda^4)(x-u),
\label{xiii}
\ee corresponding to the double covering of the plane branched at
${\pm \Lambda^2}$, $u$, $\infty$. The singularities of this curve occur
precisely  at infinity and at the singularities associated to extra massless
particles.

The situation for the $N_f=1$ theory is very similar. There is no Higgs
phase, and the  moduli space of vacua is again parametrized by the VEV of the
operator $u=\tr B^2$. The  anomaly-free discrete subgroup of the $U(1)_{\cal
R}$ symmetry is the square root of  (\ref{enedos}). Under this ${\bf Z}_{12}$
symmetry the quantity
$u={\rm Tr} B^2$ transforms as $u \rightarrow {\rm e}^{-2\pi i /3}u$  and
gives a global ${\bf Z}_{3}$ symmetry on the $u$ plane. Again, in this case
there are singularities in the  moduli space. The elliptic curve describing
the $u$-plane is given by \cite{swtwo}:
\be y^2=x^2(x-u)-{1\over 64}\Lambda_1^6, 
\label{unocurva}
\ee where $\Lambda_1$ is the scale of the $N_f=1$ theory. The singularities
are the points  in the $u$-plane where the discriminant of this polynomial
vanishes. This gives:
\be u^3=-{27 \over 256}\Lambda_1^6,
\label{singula}
\ee and coincides with the result in (\ref{uy}) for the perturbed theory. We
will comment on  this later. Notice that these three singularities are
interchanged by the ${\bf Z}_3$  symmetry. Each of them corresponds to a
single state becoming massless.  The charges of the three different states
are $(n_{m},n_{e})=(1,0)$,
$(1,1)$, $(1,2)$,  {\it i.e.}, one monopole and two dyons. This is,
essentially, all the information  we need from the Field Theory point of
view, although, as we mentioned in the  previous Chapter, to make more
precise predictions the expressions for the $a$, $a_D$  variables in terms of
$u$ would be needed.

\section{The polynomial invariants from $N=2$}

In the previous Chapter we have  computed the polynomial invariants for
$SU(2)$ monopoles on K\"ahler, Spin manifolds.  The fact that we have a
K\"ahler structure allows one to perform the computation in the
$N=1$ theory. In this section we will show that one can use electric-magnetic
duality and the
$U(1)_{\cal R}$ symmetry of the  original $N=2$ theory to obtain expressions
which are valid on a general Spin manifold $X$, as for in Donaldson theory
\cite{mfm}. One of the most interesting features of the $N=2$ point of view
is that  it gives a more concrete picture of the cosmic string theory, and
ultimately  confirms our assumptions about its vacuum structure. 

When dealing with the twisted $N=2$ theory, topological invariance implies 
that the results are formally invariant under rescalings of the metric. For
large  metrics, {\it i.e.}, very low energies, the effective, Wilsonian
description of the  untwisted theory should provide all the relevant
information from the topological  point of view. The basic assumption is then
that one can describe the twisted,  original theory, in terms of a twisted
version of the low-energy effective theory. 
   
As we have seen, the vacuum degrees of freedom are described by the quantum 
moduli space, parametrized by the $u$-plane. At every point in this moduli
space  there is a low-energy abelian $N=2$ effective  theory, given by the
expression (\ref{iv}), which can also be twisted  to give a topological field
theory. At a generic point the only light degree of freedom is the
$U(1)$ gauge field which survives after gauge symmetry breaking, and the
twisting of this theory would corresponds to the moduli problem of abelian
instantons on $X$. At the singularities new massless states (monopoles or
dyons) appear  which must be included in the low-energy lagrangian. For the
theory with $N_f=1$, the resulting effective theory is $N=2$ QED with a
single hypermultiplet at every singularity, as we saw in the previous section
following \cite{swtwo}.  In these cases, the twisted theory near these points
is just the abelian version  of the topological theory considered in Chapter
4. In principle, when computing a  correlation function of the original,
``microscopic" twisted theory, one should  integrate over the $u$-plane, to
take into acount all the relevant degrees of  freedom. However, we are
assuming, in order to avoid reducible solutions, that 
$b_2^{+}>1$. Recall that this condition means that there are  no abelian
instantons on $X$ for a generic metric, and therefore one expects
contributions only from the singularities, as the moduli space of the twisted
effective, ``macroscopic" theory is empty for the other points in the
$u$-plane. This is consistent with the $N=1$ point of view. Once we know the
contribution from one of the singularities, the other contributions can be
obtained through the anomaly-free  subgroup ${\bf Z}_{12}
\subset U(1)_{\cal R}$ of the underlying microscopic 
$N=2$ theory.

One should notice, however, that although with these arguments one captures 
the structure of the invariants, they are rather heuristic. It would be
desirable to  have a more direct proof of these facts, studying the precise 
relation between the path integral of the macroscopic theory and the
underlying  effective theory, at least in the twisted case. Some steps in
this direction have  been given in
\cite{abelia}. A more precise approach should  also give information about 
the  specific factors appearing in the expression of the invariants. This is
because there is  only a free parameter in the exact solution of the $N=2$
$SU(2)$ theories  with $N_f <4$, namely the scale $\Lambda$. Therefore, and
as we  mentioned in the previous  Chapter, the exact solution should fix the
relations between the different universal  constants that also appear in the
$N=2$ approach. 

Before starting the computation, let us mention some of the basic properties 
of Seiberg-Witten invariants. The abelian monopole equations, in the Spin
case,  were written in Chapter 4, (\ref{swmon}), and they require the
introduction  of a line bundle $L$. The deformation complex for these
equations  is obtained from (\ref{file}), with $E=L$ and ${\bf g}_E={\bf R}$.
The virtual  dimension of the  corresponding moduli space can be easily
computed from the index theorem and is given by
\be
\dim \,\ {\cal M}_{\rm SW}=-{2 \chi + 3 \sigma \over 4}+c_1(L)^2.
\label{dimsw}
\ee In the $\sp^c$ case, the line bundle $L$ is not well defined globally,
but $L^2$ is and  its first Chern class is denoted by $c_1(L^2)=-x$. Notice
that, in the Spin case, 
$x=-2c_1(L)$. From (\ref{dimsw}) it follows that, when the moduli space
associated to the abelian monopole  equation has zero dimension, $x$
satisfies:
\be x^{2}=2\chi + 3\sigma. 
\label{basicas}
\ee In this case, generically, the moduli space ${\cal M}_{\rm SW}$ consists
of  isolated points  and one can define a topological invariant in the
standard way, just like in Donaldson  theory. We already know that  this
invariant is the partition function of the corresponding topological field 
theory (for an explicit construction of this theory, see \cite{lmab}). The 
invariant associated to a line bundle $L$ (or to a $\sp^c$-structure $L^2$) 
is denoted by $n_x$. It is called the {\it Seiberg-Witten invariant}. The $x$
such  that $n_x
\not= 0$ are called {\it basic  classes}, and they must verify the condition
(\ref{basicas}). It should be mentioned  that the cohomology classes verifying
(\ref{basicas}) have a standard mathematical  characterization in
four-dimensional topology. First of all, the class $x=-c_1(L^2)$  defines a
$\sp^c$-structure, and therefore it must be characteristic  (see Appendix,
section 3). A characteristic element verifies (\ref{basicas})  if and only if
it defines an almost complex structure on $X$, and $x$ is then the first 
Chern class of the canonical line bundle $K$ associated  to the almost
complex structure.  Conversely, any almost complex structure defines a
$\sp^c$-structure with line bundle 
$L^2=K$.  

Notice that the index of the Dirac operator for the abelian monopole
equations is  given by $-\sigma/8 + c_1(L)^2/2 $. If (\ref{basicas}) holds,
this index has the value: 
\be
\ind \,\ D_L={\chi + \sigma \over 4}=\Delta,
\label{otradelta}
\ee and therefore $\Delta$, already introduced in (\ref{ladelta}), must be an
integer  for the $n_x$ to be non-zero.

In the Topological Field Theory of abelian monopoles, the descent
construction can be  applied to obtain a family of operators which should
give rise to higher dimensional  Seiberg-Witten invariants. One obtains
\cite{lmab}:
\bea
\Theta_0^n&=&{ n \choose 0} \phi^n,\,\,\,\,\,\,\,\,\,\ 
\Theta_1^n={ n \choose 1} \phi^{n-1}\psi,\nonumber\\
\Theta^n_2&=&{ n \choose 1} \phi^{n-1}F+ {n \choose 2} 
\phi^{n-2} \psi\wedge \psi, \nonumber\\
\Theta^n_3&=&{n \choose 3}\phi^{n-3}\psi \wedge \psi \wedge \psi  +2{n \choose
2}\phi^{n-2} 
\psi\wedge F, \nonumber\\
\Theta^n_4&=&{n \choose 4} \phi^{n-4}\psi \wedge \psi \wedge \psi \wedge \psi
+  3{n
\choose 3}\phi^{n-3} \psi \wedge \psi \wedge F\nonumber\\ &+& {n \choose 2}
\phi^{n-2}F\wedge F,
\label{abdes}
\eea  where the fields involved here are simply the abelian counterparts of
the  fields appearing  in the non-abelian monopole theory of Chapter 4. 

Now, we will evaluate the correlation functions of the $SU(2)$ monopole
theory in terms  of the twisted, low-energy effective theory of abelian
monopoles, for $X$  a Spin manifold with $b_2^+>1$. The contributions  come
from the three singularities, and it is sufficient to consider only one of
them,  as the contributions coming from the other singularities can be
obtained through the 
${\bf Z}_3$ symmetry of the $u$-plane. We consider then the singularity
associated to  the magnetic monopole, with charges $(1,0)$. To couple the
monopole in a local way, we  must perform a duality transformation to
magnetic variables. In these variables, the  chiral superfield we have
denoted by $\Phi$ is the dual field $A_D =
\partial {\cal F}/
\partial A$. The field $\phi$ appearing in (\ref{abdes}) corresponds after
twisting to  the complex conjugate of $a_D$. This is because we choose to
twist with the positive  chirality spinor bundle. The opposite choice is
possible \cite{sym}, leading to the  more natural identification $\phi
\rightarrow a_D$, but then the bosonic  components of the  matter
hypermultiplets become negative-chirality spinors. 

The first thing to do is to expand the operators of the microscopic, original 
$SU(2)$ theory, in terms of operators of the macroscopic, magnetic theory. To
obtain  the structure of this expansion we will use the expansion in the
untwisted, physical  theory together with the descent equations in the
topological, abelian  theory
\cite{wper}. At the monopole singularity, located at $u_0$ (where 
$u_0$ is one of the three roots of the equation (\ref{singula})), the monopole
becomes  massless and $a_D(u_0)=0$. Near this point the $u$ variable has an
expansion:
\be u(a_D)=u_0 + \bigg({du \over d a_D}\bigg)_{u_0}a_D+ {\rm higher \,\
order},
\label{clexp}
\ee where $(du/da_D)_{u_0} \not= 0$ (this can be checked from the explicit 
solution for the $N_f=1$ theory given in \cite{bilfer}), and ``higher order"
means  operators of higher  dimension in the expansion. In terms of the
corresponding  observables of the topological theories, the correspondence is
then:
\be {\cal O}=\langle {\cal O} \rangle -{1 \over \pi}
 \langle V \rangle \phi + {\rm higher \,\ order},
\label{topex}
\ee where $\langle {\cal O} \rangle$, $\langle V \rangle$ are real
$c$-numbers which  should be related to the values $u_0$, $({du /d
a_D})_{u_0}$ of the  untwisted theory. They coincide with the universal
constants introduced  in the previous Chapter, as we will see in the
computation of the  invariants. From the observable
${\cal O}$ one can obtain  the observable ${\cal O}^{(2)}$ by the descent
procedure in (\ref{obser}). Applying  this procedure (\ref{abdes}) in the
topological abelian theory  to the right hand side of  (\ref{topex}), we
obtain:
\be {\cal O}^{(2)}=-{1 \over \pi} \langle V \rangle F + {\rm higher \,\
order}, 
\label{dosex}
\ee where $F$ is the dual electromagnetic field. In particular, taking into
account that 
$x=-2c_1(L)=-[F]/\pi$, where $[F]$ denotes the cohomology class of the
two-form $F$, we  finally get:
\be I(\Sigma)=\langle V \rangle (\Sigma \cdot x) + {\rm higher \,\ order},
\label{sigmalow}
\ee where the dot denotes the pairing between $2$-cohomology and
$2$-homology. From the  point of view of Topological Field Theory, we
wouldn't expect contributions  from higher dimensional operators, because of
the invariance of the theory under metric  rescalings (this is in fact the
same argument to constrain correlation functions in the 
$N=1$ theory). Higher order terms in this expansion should correspond to
higher  monopole invariants, just like in the computation of Donaldson
invariants in
\cite{mfm}.   This problem is intimately related to the simple type condition
on four-manifolds.  The analysis in the K\"ahler case presented in the
previous Chapter suggests that the  simple type condition in this case is
given by (\ref{tiposimple}). As in  Donaldson theory,  it follows that, if
the polynomial invariants defined for $SU(2)$ monopoles verify  {\it a
priori} this condition, then the analysis in terms of the $N=2$ theory implies
 that there will be no contribution from higher dimensional operators. 

We can now evaluate the correlator (\ref{corre}) in terms of a path integral
in the  low-energy twisted theory. This includes, of course, a sum over the
topological  classes of dual line bundles $L$. If we assume no contribution
from higher  dimensional operators, the expansion of the observables of the
original $N=2$ theory  only includes $c$-numbers and, possibly, contact terms
for the operators $I(\Sigma)$  giving the intersection form of the manifold,
as in (\ref{corr}). The appearance  of these  terms is more clear from the
$N=1$ point of view, and, as we mentioned in the previous  Chapter, they are
related to gluino condensation. From the $N=2$ point of view their  presence
is not obvious from the above expansions, but, as we will see in the  next
section,  we get a clear picture of the  cosmic string theory vacua. As the
operators are now $c$-numbers, the computation of the  path integral of the
low-energy, effective theory, just gives the partition  funcion of the theory
for each line bundle $L$. This is just the Seiberg-Witten  invariant 
$n_x$, with $x=-2c_1(L)$ verifying (\ref{basicas}) in order to have a 
zero-dimensional moduli space. We then obtain, for the  first singularity:
\be C {\rm exp} (\gamma v^2 +\mu \langle  {\cal O} \rangle) \sum_{x} n_x {\rm
e}^{\langle V
\rangle v \cdot x},
\label{unvacio}
\ee we have included the universal constants which  also appear in
(\ref{fer}). The sum is over all the basic classes. The constant $C$ appears
in the comparison of the macroscopic and  microscopic path integrals, after
fixing the respective normalizations. Some  steps to determine it have been
given in \cite{abelia}. Now, to get  the contributions  from the other two
vacua we can use the $U(1)_{\cal R}$ symmetry given in (\ref{lola}) and
(\ref{maslola}). The resulting ${\bf Z}_6$  anomaly-free discrete subgroup is
given in (\ref{enedos}). Notice that if we use the
${\bf Z}_6$ symmetry,  the transformation of the order parameter $u$ is $u
\rightarrow {\rm e}^{-{4 \pi i \over 3}}u$. This is still a ${\bf Z}_3$
symmetry of the $u$-plane which goes through all the singularities, and we
won't need to implement the additional symmetry (\ref{parida}). 

At this point the computation becomes very similar to the one we did for the
$N=1$ theory. First we must take into account the gravitational contribution
of the anomaly and the geometrical character of the fields after twisting.
The fields
$\lambda_{1}$, $\lambda_{2}$,  ${\overline \lambda^{1}}$ and ${\overline
\lambda^{2}}$ give now the whole ASD complex and the anomaly is the square of
(\ref{fase}). For the matter fermions the anomaly is given by
${2 \pi i \over 3} {\rm index}\,\ D $. The total contribution is:
\be  {\rm e}^{ {\pi i \over 3}\{ 8k-{3 \over 2}(\chi +\sigma)\}- {\pi i \over
3}(2k+{\sigma \over 2})}={\rm e}^{-{\pi i \sigma
\over 6}}.
\label{vitoria}
\ee This is the anomaly we obtained for the third $N=1$ vacuum, for it is the
square of (\ref{anomaly}).  We see that, as happens in the pure $N=2$
Yang-Mills theory
\cite{sym},  the $U(1)_{\cal R}$  symmetries of the $N=1$ and the $N=2$
theory, which are certainly different,  work in such a way that after
twisting one obtains the same contribution for the gravitational  part of the
anomaly. 

To implement the symmetry under consideration in the observables we need the 
action of the generator of (\ref{enedos}), call it $\alpha$ again, on them.
One obtains:
\be
\alpha {\cal O} \alpha^{-1} =  -{\rm e}^{\pi i \over 3} {\cal O},\,\,\,\,\
\alpha I(\Sigma) \alpha^{-1} = {\rm e}^{2\pi i \over 3} I(\Sigma) .
\label{obenedos}
\ee  Now we can apply these transformations to  (\ref{unvacio}), as we did in
the K\"ahler case, to obtain:
\bea & &\langle {\rm exp} (\sum_{a}\alpha_{a}I(\Sigma_{a})+\mu {\cal O})
\rangle =C \Bigg( {\rm exp} (\gamma v^2  +\mu \langle {\cal O} \rangle)
\sum_{x} n_x {\rm exp}(\langle V \rangle v \cdot x)  \nonumber\\ & & +{\rm
e}^{-{\pi i \over 6}\sigma} {\rm exp}\Big( -{\rm e}^{-{\pi i \over 3}}
(\gamma v^2 +\mu \langle{\cal O} \rangle)
\Big) \sum_{x} n_x {\rm exp}({\rm e}^{-{2\pi i \over 3}} \langle V
\rangle v \cdot x) \nonumber\\ & &+{\rm e}^{-{\pi i \over 3}\sigma} {\rm
exp}\Big( -{\rm e}^{{\pi i \over 3}} (\gamma v^2 +\mu \langle{\cal O} \rangle)
\Big) \sum_{x} n_x {\rm exp}({\rm e}^{-{4\pi i \over 3}} \langle V
\rangle v \cdot x) \Bigg). \label{fernanda}
\eea This is our final expression for  the polynomial invariants associated to
$SU(2)$ monopoles on a Spin manifold $X$ with $b_{2}^{+}>1$. This implies
that the  polynomial invariants associated to $SU(2)$ monopoles can be
expressed in terms of  Seiberg-Witten invariants. Notice that the condition
to have a non-trivial polynomial  invariant for $SU(2)$ monopoles is that
$\Delta$, defined in (\ref{ladelta}),  should be an integer, and this is the
same condition to have non-trivial Seiberg-Witten  and Donaldson invariants. 

\section{The K\"ahler case revisited}

Although the $N=2$ and the $N=1$ points of view are rather complementary,  the
former is, in some ways, more fundamental, and in this section we want to 
reobtain the results for K\"ahler manifolds starting from the $N=2$
low-energy  effective theory. 

The perturbation of the $N=2$ theory by a mass term for the 
$N=1$ chiral superfield $\Phi$ is obtained by adding a superpotential 
$W=m \tr \Phi^2$. As shown in \cite{swone, swtwo}, the operator $\tr \Phi^2$ 
can be represented in the low energy  theory by a chiral superfield $U$,
whose first component is the gauge-invariant  parameter for the quantum
moduli space $u$. Near  the point where the monopole becomes  massless, the
total superpotential of the effective theory  is the sum of the $N=2$
superpotential in (\ref{silvana}) and the effective  superpotential coming
from the $N=1$ perturbation, $W_{\rm eff}=mU$: 
\be W_{\rm M}={\sqrt 2}A_D M {\widetilde M}+ mU .
\label{superab}
\ee  The superfields $M$ and ${\widetilde M}$ are the $N=1$ chiral multiplets
which  represent the monopole $N=2$ hypermultiplet in the abelian case, and
they have opposite  charges. We are interested in the vacuum structure of 
this $N=1$ theory. The vacua of the
$N=1$ theory are given by the  critical points of the superpotential, up to
complexified $U(1)$ gauge  transformations (this is equivalent to set the $D$
terms to zero and divide by $U(1)$). The only points in the quantum moduli
space of vacua of the $N=2$ theory which give rise  to $N=1$ vacua are
precisely the singularities where extra particles become massless. Consider
for example the  monopole point, described by (\ref{superab}). The equations
for the critical points are:
\bea {\partial W \over \partial A_D}&=& {\sqrt 2}M {\widetilde M}+ m {d u
\over da_D}=0,\nonumber\\ {\partial W \over \partial M}&=&a_D {\widetilde M}
=0,\nonumber\\ {\partial W \over \partial {\widetilde M}}&=&a_D M =0.
\label{minima}
\eea The last two equations give $a_D=0$, and fix the vacuum at the monopole
singularity,  while the first one implies that the monopoles get a VEV. The
same analysis can be done  near each singularity, and therefore there is one
$N=1$ vacuum at each singularity.   Hence, for the $N_f=1$ theory, the
perturbed theory has three vacua related by the
${\bf Z}_{3}$ symmetry of the $u$-plane. The pattern of chiral symmetry
breaking can be  obtained taking into account that the vacua are labeled by
the gauge-invariant  parameter $u=\tr B^2$, and the unbroken symmetry is then
$B \rightarrow -B$. The  pattern (\ref{martin}) is thus recovered. In these
vacua, the magnetic Higgs mechanism  gives a mass gap and confinement of
electric charge. Hence we rederive all the  results in  Chapter 5, section 2,
and check the confining character of the theory.    

In the twisted topological theory on a K\"ahler manifold with $b_2^+>1$, the
mass  perturbation is done with a $(2,0)$ form $\eta$ (we change notation to 
distinguish it from the K\"ahler form $\omega$). We will now see that  the
mass term gives,  in the low-energy effective theory, the perturbed
Seiberg-Witten monopole equations  on K\"ahler manifolds of \cite{mfm}. To do
this, we compute the bosonic part of the  twisted effective theory once the
superpotential $W_{\rm eff}$ has been added.  If we consider the expansion
(\ref{clexp}) in terms of superfields,  we get the additional  terms in the
Lagrangian:
\be {du \over d a_D} \eta A_D {\Big |}_{{\theta}^2} + {\rm h.c.}= 
\eta {du \over d a_D} D_{22}+ {\overline \eta}{\overline {du \over d a_D}}
D_{11}, 
\label{expandoto}
\ee where $D_{11}$, $D_{22}$ are the auxiliary real fields in the chiral
multiplet
$A_D$.   Now we take into account the decomposition of the monopole fields  in
(\ref{rosi}), which  is just the decomposition we used to study the
non-abelian monopole equations in  Chapter 4, section 3. The bosonic part
involving the auxiliary $D_{ij}$ in the original 
$N=2$ abelian theory is:
\bea & &{1 \over 2}D_{12}^2+D_{11}D_{22}+\Big(|M_1|^2 -|M_2|^2 \Big) D_{12}
\nonumber\\ & & +i{\sqrt 2}{\overline M}^1 D_{11} M_2-i{\sqrt 2}{\overline
M}^2 D_{22} M_1.
\label{calculoto}
\eea Denoting the monopole fields as in (\ref{kal}), $M_1=\alpha$,
$M_2=i{\overline
\beta}$,  and integrating out the auxiliary field $D_{ij}$, the last term in
the bosonic  part of the topological action (\ref{melon}) is no longer
$|M|^4/2$, as we computed  in (\ref{melondos}), but 
\be {1 \over 2}\Big(|\alpha|^2 -|\beta|^2\Big)^2 + 2{ \Big| \alpha \beta -{1
\over {\sqrt 2}} {du\over d a_D}\eta \Big|} ^2.
\label{kalaccion}
\ee As ${du/d a_D}\not= 0$ at the monopole singularity  (this is again the key
point), we can reabsorb the factor ${1 \over {\sqrt 2}} {du\over d a_D}$ in
$\eta$. Now we assume that the first Chern class of 
$L$ is of type $(1,1)$. This just means that  the original monopole equations
have a solution (recall that we are perturbing the  original theory).
Therefore, we have that
\be
\int_{X}F^{2,0}\wedge {\overline \eta}=\int_{X}F^{0,2}\wedge \eta=0.
\label{exige}
\ee Taking into account (\ref{kalaccion}) and (\ref{exige}), we see that the
bosonic terms  in the action involving the field $F$ can be written as:
\be 2|F^{2,0}-\alpha \beta + \eta |^2+2|f-{1\over 2}(|\alpha|^2-|\beta|^2)|^2.
\label{incom}
\ee We used the decomposition (\ref{des}) in the Appendix as well as
(\ref{endometrica}).  We then obtain the perturbed Seiberg-Witten equations
for  the K\"ahler case \cite{mfm}:
\bea F^{2,0}&=&\alpha \beta -\eta,\nonumber\\ F_{\omega}&=& {\omega \over
2}\Big(|\alpha|^2-|\beta|^2 \Big),\nonumber\\ F^{0,2}&=&{\overline \alpha}
{\overline\beta }-{\overline\eta }.
\label{kalpert}
\eea This is the most adequate perturbation to compute the Seiberg-Witten
invariants in  the K\"ahler case. Recall from our general discussion in
Chapter 1 that K\"ahler  metrics are not generic, and therefore one must
either compute the Euler class  of the cokernel of the obstruction
cohomology, or to perturb the  equations in an appropriate way. The $N=1$
mass term clearly suggests (\ref{kalpert}).  Using now these equations it is
easy to show that, when $X$ is K\"ahler and its canonical divisor is of the
form (\ref{canonico}) with disjoint $C_y$, the Seiberg-Witten invariants
give  the pattern of vacuum bifurcation of the cosmic string, and one 
recovers (\ref{fer}) from (\ref{fernanda}). Indeed, in such a situation the
basic classes are given by \cite{mfm}:
\be x_{(\rho_1, \cdots, \rho_n)}= \sum_{y} \rho_{y}[C_y],
\label{burgos}
\ee and each $\rho_{y}=\pm 1$. The corresponding $n_x$ are:
\be n_x=\prod_{y} t_{y}^{s_y} ,
\label{kahlerclases}
\ee where $s_y=(1-\rho_{y})/2$ and  the $t_y$ are given in (\ref{late}). A
simple computation leads to,
\be
\sum_{x} n_x {\rm exp}(\langle V \rangle v \cdot x)=\sum_{\rho_y}\prod_{y}
t_{y}^{s_y}{\rm exp}(\sum_{y}\langle V \rangle \rho_y
\phi_{y})=\prod_{y}\Big( {\rm exp}( \langle V \rangle \phi_{y}) + t_{y}{\rm
exp} (-\langle V \rangle \phi_{y})
\Big),
\label{alava}
\ee which shows that in this case  (\ref{fernanda}) becomes (\ref{fer}).
Using also (\ref{kalpert}) one can show  that invariants associated with
moduli  spaces of higher dimension vanish, and therefore, even if higher
order terms appear,  they don't give contributions in the  K\"ahler case.

\chapter{Conclusions and outlook} 

Some of the results presented in this work are the following:

1) We have obtained equivariant extensions of the Thom form  with respect to 
a vector field action, in the framework of the Mathai-Quillen formalism, and 
we  have shown that this equivariant extension corresponds to the 
topological action of twisted $N=2$ supersymmetric theories with a  central
charge. The formalism we have introduced gives  a unified framework to 
understand the topological structure of this kind  of models in terms of 
equivariant cohomology with  respect to a vector field action. We also have
analyzed  in detail two explicit  realizations of this formalism: topological
sigma models  with a Killing, almost  complex action on an almost Hermitian
target space,  and topological Yang-Mills  theory coupled to twisted massive
hypermultiplets.

2) We have presented the non-abelian generalization of the Seiberg-Witten
 monopole equations, as well as the Topological  Quantum Field Theory
associated to this moduli problem. These equations lead to the study of a new
moduli problem which is a generalization of Donaldson theory. We have
performed a first analysis of the space of solutions and it has been argued
that they constitute an enlarged moduli space. 

3) We have computed the polynomial invariants associated to the moduli space
of
$SU(2)$ monopoles on four-dimensional Spin manifolds, with the monopole
fields in the fundamental  representation of the gauge group. Our computation
is based on the exact results about the quantum moduli space of vacua of the
corresponding $N=2$  and $N=1$ supersymmetric theories. The resulting
expressions (\ref{fernanda})  and (\ref{fer}) can be written in terms of
Seiberg-Witten invariants, and therefore the first conclusion of our analysis
is that  these invariants underlie not only Donaldson theory, but also the
generalization of
 this theory presented here. The picture which emerges from our computation 
is that non-perturbative methods in supersymmetric gauge theories are not
only an  extremely powerfool tool to obtain topological invariants, but also
to relate very different moduli problems in four-dimensional geometry:  it
seems that different four-dimensional moduli problems can be in the same
``universality class"  when considered from the point of view of the
underlying supersymmetric theories.  Therefore, using non-perturbative
results in the physical theories, one should be able to identify truly basic
topological invariants characterizing a whole family of moduli problems.
According to our results, the $SU(2)$ Donaldson invariants and the
$SU(2)$ monopole invariants are both in the same class, which is associated
to the Seiberg-Witten invariants (as the topological information that both
give is encoded in the basic  classes of the manifold). 

The results presented here can be extended in many different directions.
Perhaps the
 most interesting extension of the construction in Chapter 3 is to implement 
the localization theorems of equivariant cohomology in this framework. It
has  been shown in \cite{ab, berlin} that the integral of a closed
equivariant  differential form can be always restricted to the fixed points
of the  corresponding $U(1)$ or vector field action. This can be used  to
relate, for  instance, characteristic numbers to quantities associated to
this fixed-point  locus. The topological invariants associated to topological
sigma models and  non-abelian monopoles on four-manifolds can be understood
as integrals of  differential forms on the corresponding moduli spaces. In
the first case we  get the Gromov invariants, and in the second case a
generalization of the  Donaldson  invariants for four-manifolds. If we
consider  the equivariant extension of  these models, we could compute the
topological  invariants in terms of adequate  restrictions of the equivariant
integration  to the fixed-point locus of the corresponding abelian symmetry.
In fact, it has been  argued in \cite{pt} that localization techniques can 
provide an explicit link  between the Donaldson and the Seiberg-Witten
invariants,  because their  moduli spaces are precisely the fixed points of
the abelian $U(1)$ symmetry  considered in (\ref{circulos}), acting on the
moduli space of $SU(2)$  monopoles. Perhaps the techniques of equivariant
integration,  applied to the equivariant differential forms considered in
this work, can give an explicit proof of this link. However, a key point when
one tries  to apply localization techniques is the compactness of the moduli
spaces. The  vector field action can have fixed points on the
compactification divisors  which give crucial contributions to the
equivariant integration. This situation  arises in both the topological sigma
model and the non-abelian monopoles on  four manifolds. It can be easily seen
that, without taking into account the  compactification of the moduli space,
one doesn't obtain sensible results for  the quantum cohomology rings or the
polynomial invariants of four-dimensional  manifolds. However, we have seen
that the equivariant extension  of the non-abelian monopole theory
corresponds to the twisted $N=2$ Yang-Mills  theory coupled to massive
hypermultiplets. It would be very interesting to  use the exact solution of
the physical theory given in \cite{swtwo} to obtain  the topological
correlation functions of the twisted theory.   It seems that the duality
structure of $N=2$ and 
$N=4$ gauge  theories ``knows" about the compactification of the moduli space
of their  twisted  counterparts, and therefore the physical approach would
shed new light on the  localization problem. 

 Another interesting question is the possible relation with string theory.
Harvey and  Strominger have shown \cite{hs} that Donaldson theory appears
naturally in a certain  compactification of the heterotic string. Perhaps the
improved knowledge of duality  in string theory can give a stringy version of
the relation between Donaldson and  Seiberg-Witten invariants. This string
theory framework can be also useful to  understand from another point of view
the relation between the Seiberg-Witten and  the Gromov invariants discovered
by Taubes \cite{taubesdos}. 

\newpage
\thispagestyle{empty}

\begin{center}
\large{ACKNOWLEDGEMENTS}
\end{center} I would like to thank J.M.F. Labastida for his continuous advice
and an enjoyable  collaboration. I have benefitted from discussions with many
people. I would like to  thank: L. \'Alvarez-Gaum\'e,  O. Garc\'\i a-Prada,
G. Moore, J.A. Oubi\~na, V. Pidstrigach, A.V. Ramallo and  S. Rangoolam.  I
would also like to thank O. Garc\'\i a-Prada and G. Thompson for a careful
reading  of previous drafts, and the Theory Division at CERN, where part of
this work  was done, for its hospitality. This work was supported in part by
DGCIYT under grant  PB93-0344 and by CICYT under grant AEN94-0928. 
\thispagestyle{empty}

\appendix

\chapter{Spinors in four dimensions}

We review in this appendix some facts about Spin and $\sp^c$-structures in 
four dimensions,  and we fix our notations. In section 1 we present some
algebraic results from the  theory of Clifford algebras. In section 2 we
recall the basics of
$\sp$ structures,  Dirac operators and Weitzenb\"ock formulae. In section 3 
we briefly consider ${\rm Spin}^c$-structures. Adequate references are
\cite{sg, morgan,  dk}. The spinor notation is based on \cite{lmna}.

\section{Spinor algebra in four dimensions}

We will consider only Euclidean spaces. The real Clifford algebra in four
dimensions 
${\rm Cl}_4$ is isomorphic to the $2 \times 2$ quaternionic matrices,
denoted  by
${\bf H}(2)$. The complexified Clifford algebra ${\bf  C}{\rm l}_4$  is then
isomorphic to ${\bf C}(4)$. The only irreducible representations of these 
algebras are ${\bf H}^2$ and ${\bf C}^4$, respectively. Recall that Clifford
algebras  are
${\bf Z}_2$ graded, and they decompose as $\cl _n^0\oplus {\rm Cl}_n^1$ (also
in  the complex case).  From the isomorphism $\cl _n^0 \simeq \cl_{n-1}$, and
$\cl _3
\simeq {\bf H} \oplus  {\bf H}$, ${\bf  C}{\rm l}_3 \simeq {\bf C}^2 \oplus
{\bf C}^2$, we see that the real and  complex representation of $\sp _4$
split into two irreducible representations. It is easy to  see from the above
isomorphisms that 
\be
\sp_4 \simeq SU(2)_{L}\times SU(2)_{R},
\label{estr}
\ee {\it i.e.} the complex representation of $\sp _4$ is given by two copies 
of
$SU(2)$ acting  on ${\bf C}^2$. These are the {\it positive} and {\it
negative}  chirality representations,  which can be distinguished by the
action of the complex volume element  in ${\bf  C}{\rm l}_4$:
\be 
\omega_{\bf C}=-e_0e_1e_2e_3.
\label{vol}
\ee This corresponds, under the complex representation of the Clifford
algebra, to the  usual $\gamma_5$ matrix in Field Theory. We will denote the 
two-dimensional complex vector spaces associated to these representations as
$S^{+}$, 
$S^{-}$. The total representation space for ${\bf  C}{\rm l}_4$ is then 
$S=S^{+} \oplus  S^{-}$. From the isomorphisms above we deduce that:
\be {\bf  C}{\rm l}_4 \simeq {\rm End}_{\bf C}(S),
\label{isobelo}
\ee and this isomorphism holds for all even-dimensional complexified Clifford
algebras,  as well as the above facts on their representations. 

An important fact to relate spinors and differential forms is the algebraic
isomorphism 
$\cl_n \simeq \Omega^{*}({\bf R}^n)$, as vector spaces. An explicit
correspondence is  constructed from the map $f: \otimes^{r}{\bf R}^n
\rightarrow \cl_n$ given by 
\cite{sg}:
\be f(v_1, \cdots, v_r)={1 \over r!}\sum_{\sigma} {\rm
sign}({\sigma})v_{\sigma(1)}
\cdots  v_{\sigma(r)}.
\label{mapilla}
\ee which obviously descends to a map ${\hat f}: \Omega^{r}{\bf R}^n
\rightarrow
\cl_n$.  As ${\bf R}^4 =\Lambda^{1}{\bf R}^4$, the representation of the
Clifford algebra 
$\cl _4$ in ${\rm Hom}_{\bf C}(S)$ gives a map \cite{dk}:
\be
\gamma : {\bf R}^4 \too {\rm Hom}_{\bf C}(S),
\label{clave}
\ee and the image of this map is contained in ${\rm Hom}(S^{+}, S^{-})
\oplus  {\rm Hom}(S^{-}, S^{+})$. This is because, as a vector subspace of
$\cl_4$, 
${\bf R}^4 \subset \cl_4^{1}$ and $S$ is a  graded 
${\bf Z}_2$ module. We then have a map $\sigma: {\bf R}^4 \rightarrow  {\rm
Hom}(S^{+}, S^{-})$, and we can give an explicit representation in terms of 
the matrices
\be
\sigma(e_0)={\bf 1}_{2 \times 2}, \,\,\,\,\ \sigma(e_a)=i \tau_{a}, \,\,\
a=1,2,3,
\label{mismatrices}
\ee where $\tau_a$ are the Pauli matrices. The matrices $\sigma^{*}(e_i)$ are
the  adjoint matrices  to (\ref{mismatrices}), $\sigma^{*}(e_0)={\bf 1}_{2
\times 2}$,
$\sigma^{*}_{e_a}= -i \tau_{a}$, $a=1,2,3$. For (\ref{clave}) we have:
\be
\gamma (e_i) = \left(\begin{array}{cc}0&-\sigma^{*}(e_i)\\
               \sigma(e_i)&0 \end{array} \right), \,\,\,\,\ i=0,\cdots, 3.
\label{mimapa}
\ee It is easy to see that the matrices $\gamma(e_i)$ verify  in fact the
Clifford algebra
\be
\gamma(e_i)\gamma(e_j)+ \gamma(e_j)\gamma(e_i)=-2\delta_{ij},
\label{clifford}
\ee which in terms of the $\sigma$ matrices reads:
\be
\sigma^{*}(e_i)\sigma(e_j)+\sigma^{*}(e_j)\sigma(e_i)=2\delta_{ij}.
\label{cliffordos}
\ee In this representation, the volume element (\ref{vol}) is given by:
\be
\omega_{\bf C}=\left(\begin{array}{cc}1&0\\
               0&-1 \end{array} \right), 
\label{chiral}
\ee which is the usual (euclidean) $\gamma_5$ matrix in a Weyl or chiral
representation.  From this representation of the Clifford algebra we can
construct the representation  of $\Omega^{*}({\bf R}^4)$ in ${\rm End}_{\bf
C}(S)$, using (\ref{mapilla}).
 For the two-forms, the map is given  by:
\be e_i \wedge e_j \too \left(\begin{array}{cc}-\sigma^{*}(e_i)\sigma(e_j)&0\\
               0& -\sigma(e_i)\sigma^{*}(e_j)\end{array} \right).
\label{nadiuska}
\ee If we decompose $\Omega^{2}({\bf R}^4)$ into SD and ASD forms, we can
check that  (\ref{nadiuska}) gives an isomorphism
\be
\rho: \Lambda^{+}({\bf R}^n) \too {\bf su}(S^+),
\label{casi}
\ee where ${\bf su}(S^+)$ denotes an algebra isomorphic to the Lie algebra of
$SU(2)$,  and consisting of skew-adjoint,  trace-free  endomorphisms of $S^+$
(with the structure given in (\ref{sudos})). Explicitly, we have:
$$
\rho (e_0 \wedge e_1 + e_2 \wedge e_3)=\left(\begin{array}{cc}0&-2i\\
               -2i&0 \end{array} \right),\,\,\,\,\,\,\
\rho (e_0 \wedge e_3 + e_1 \wedge e_2)=\left(\begin{array}{cc}-2i&0\\
               0&2i \end{array} \right),
$$
\be
\rho (e_0 \wedge e_2 + e_3 \wedge e_1)=\left(\begin{array}{cc}0&-2\\
               2&0 \end{array} \right).
\label{mias}
\ee If we consider the representation of the complexified Clifford algebra 
${\bf  C}{\rm l}_4$ (and  therefore of the complexified exterior algebra), we
get the isomorphisms
\be
\Omega^{1}_{\bf C} \simeq {\rm Hom}_{\bf C}(S^{+}, S^{-}),\,\,\,\,\,\ 
\Omega^{2,+}_{\bf C} \simeq {\rm End}_{0}(S^{+}),
\label{isos}
\ee where ${\rm End}_{0}(S^{+})$ denotes the complex, trace-free
endomorphisms of
$S^{+}$.  These isomorphisms are specializations of the isomorphism
(\ref{isobelo}) and can be  established without using an specific matrix
representation.

Now we will introduce a notation more familiar in Physics. This will just be a
component  notation for some of the basic structures introduced above. An
element
$\Psi \in S= S^{+} \oplus S^{-}$ will be denoted by a column vector of the
form
\be  
\Psi = \left(\begin{array}{c}M_{\alpha}\\
                N^{\dot\alpha}\end{array} \right),
\label{diracspinor}
\ee     where $\alpha$, ${\dot \alpha}=1,2$. Undotted indices correspond to
elements in the  positive chirality bundle $S^{+}$, while dotted indices
correspons to elements in 
$S^{-}$. An important point is that $S^{\pm}$ have a complex symplectic
structure, which  can be implemented by the $\tau_2$ matrix. Following the
conventions in 
\cite{roc} we define the $C$ matrices:
\be C_{\alpha\beta} =  (\tau_2)_{\alpha\beta},
\,\,\,\,\,\,\,\,\,\,\,\,\,\,\, C_{\dot\alpha\dot\beta} = 
(\tau_2)_{\dot\alpha\dot\beta},
\label{lasces}
\ee and their inverses:
\be C^{\alpha\beta} = - (\tau_2)_{\alpha\beta},
\,\,\,\,\,\,\,\,\,\,\,\,\,\,\, C^{\dot\alpha\dot\beta} = -
(\tau_2)_{\dot\alpha\dot\beta},
\label{lascesmas}
\ee so that,
\be C^{\alpha\beta}C_{\gamma\beta} = \delta_\gamma^\alpha,
\,\,\,\,\,\,\,\,\,\,\,\,\,\,\, C^{\dot\alpha\dot\beta}C_{\dot\gamma\dot\beta}
= 
\delta_{\dot\gamma}^{\dot\alpha}.
\label{lascesmasmas}
\ee The symplectic structure provides isomorphisms:
\bea
 C^{+}: S^{+} & \too &(S^{+})^{*} \nonumber \\ M_{\alpha} & \mapsto&
M^{\alpha}=C^{\alpha\beta} M_{\beta},
\label{cuantoiso}
\eea and 
\bea
 C^{-}: S^{-} & \too &(S^{-})^{*} \nonumber \\ N^{\dot \alpha} & \mapsto&
N_{\dot
\alpha}=N^{\dot \beta} C_{\dot \alpha \dot\beta}.
\label{cuantoisodos}
\eea
 With the above conventions, an element $A \in {\rm End}_{\bf C} (S^{+})$
has  the index structure
$A_{\alpha}{}^{\beta}$. With the induced isomorphism ${\rm End}_{\bf C}(
S^{+})
\simeq  {\rm Hom}_{\bf C}((S^{+})^{*}, S^{+})$ we obtain a map with indices
$A_{\alpha \beta}= A_{\alpha}{}^{\delta} C_{\beta \delta}$. It is easy to
check that the trace-free  endomorphisms become {\it symmetric} tensors. 

We can give now explicit expressions for the spinor representation of the
self-dual  forms, and define the projector $p^{+}$. This simply follows from
(\ref{nadiuska}) and  (\ref{casi}). We then have: 
\be F_{\alpha}{}^{\beta} = (p^+(F))_{\alpha}{}^{\beta}= -{1 \over
2}(\sigma^{*}(e_i))_{\alpha\dot\alpha} (\sigma (e_j))^{\dot \alpha \beta}
F_{ij}.
\label{proj}
\ee  It is most usual to consider, instead of (\ref{proj}), the corresponding 
symmetric tensor  induced by the symplectic structure. If we lower the
indices of
$\sigma(e_i)$ we get:
\be (\sigma(e_i))_{\beta \dot \beta}= C_{\beta
\alpha}(\sigma(e_i)^{\top})^{\alpha
\dot \alpha}  C_{\dot \alpha \dot \beta}=(\sigma^{*}(e_i))_{\beta \dot \beta},
\label{bajo}
\ee   where we used the fact that: 
\be
\tau_2\tau_a^\top\tau_2=-\tau_a,\,\,\,\,\,\,\, a=1,2,3.
\label{latreinta}
\ee Finally we obtain:
\be F_{\alpha \beta} = -{1 \over 2}C^{\dot \alpha \dot
\beta}(\sigma^{*}(e_i))_{\alpha\dot\alpha} (\sigma^{*}(e_j))_{\beta\dot\beta}
F_{ij}.
\label{sime}
\ee The matrices $F_{\alpha}{}^{\beta}$ and $F_{\alpha \beta}$ read:
$$ F_{\alpha}{}^{\beta}=\left(\begin{array}{cc}-i(F_{03}+F_{12})&
-F_{02}-F_{31}-i(F_{01}+F_{23})\\
               F_{02}+F_{31}-i(F_{01}+F_{23})& i(F_{03}+F_{12})\end{array}
\right),
$$
\be F_{\alpha \beta}=\left(\begin{array}{cc}
i(F_{02}+F_{31})-F_{01}-F_{23}&F_{03}+F_{12}\\
             F_{03}+F_{12}&F_{01}+F_{23}+i( F_{02}+F_{31})\end{array} \right)
\label{expli}
\ee

We introduce now metrics on the vector spaces $S^{\pm}$ and on the space of
trace-free,  skew-adjoint endomorphisms of $S^{+}$. Given an element of
$S^+$, 
$M_\alpha = (a,b)$, we define $\overline M^{\alpha } = (a^{*},b^{*})$. In a
similar way, given an element  of $S^-$, $N^{\dot\alpha} = (a,b)$, we define
$\overline N_{\dot \alpha } = (a^{*},b^{*})$. A Riemannian metric  on $S^+
\simeq  {\bf R}^4$  is given by:
\be
\langle M, N \rangle_{+} ={1 \over 2}\big(\overline M^{\alpha } N_{\alpha} +
\overline N^{\alpha } M_{\alpha}\big),
\label{positiva}
\ee and a similar expression gives a Riemannian metric on $S^-$. Together they
define  a Riemannian metric on $S$. We also define:
\be {\overline M}_{\alpha}=\overline M^{\beta}C_{\alpha \beta},
\,\,\,\,\,\,\,\,\,\,\ {\overline N}^{\dot \alpha}=\overline N_{\dot
\beta}C^{\dot \beta \dot\alpha },
\label{subebaja}
\ee in such a way that ${\overline M}_{\alpha}M^{\alpha}={\overline
M}^{\alpha}M_{\alpha}$, 
${\overline N}^{\dot \alpha}N_{\dot \alpha}={\overline N}_{\dot \alpha}N^{\dot
\alpha}$. Notice that, with respect to this metric on $S$,  the following 
property holds:
\be
\langle \Psi, \gamma(e_i) \Phi\rangle+\langle \gamma(e_i) \Psi,  \Phi\rangle
=0, 
\,\,\,\,\,\,\,\,\,\ \Psi, \Phi \in S
\label{nuevo}
\ee This is a requirement to have a Dirac bundle \cite{sg}, and guarantees
that the Dirac  operator is formally self-adjoint. 

We can also define a Riemannian metric for the  matrices in ${\bf su}(2)$,
using (\ref{mimetra}):
\bea
\langle A, B \rangle &=& -{1 \over 2}\tr (AB)\nonumber \\ &=& -{1 \over
2}A_{\alpha}{}^{\beta}B_{\beta}{}^{\alpha}= -{1 \over 2}A^{\alpha
\beta}B_{\alpha
\beta}, \,\,\,\,\,\,\,\,\,\,\ A,B \in {\bf su}(2).
\label{hula}
\eea

\section{$\sp$-structures on four-manifolds}     

The $\sp_n$ group is a double covering of the orthonormal group $SO(n)$, and
we have  an exact short sequence
\be 0 \too {\bf Z}_2 \buildrel 
 \over \too \sp_n 
\buildrel  \xi_0 \over \too SO(n)  \too 1.
\label{spinshort}
\ee If $E$ is an  oriented, Riemannian vector bundle of rank $n$ on a
manifold $M$, a natural question  is wether one can lift the principal
$SO(n)$-bundle associated to $E$, 
$P_{\rm SO}(E)$, to a $\sp_n$-bundle $P_{\rm Sp}(E)$, providing in this way a 
bundle realization of the covering in (\ref{spinshort}). This lifting of
bundles is  a double covering 
\be
\xi : P_{\rm Sp}(E) \too P_{\rm SO}(E)
\label{spst}
\ee such that $\xi(pg)=\xi(p) \xi_0(g)$, where $p \in P_{\rm Sp}(E)$, $g \in
\sp_n$.  If this can  be done,  we say that $E$ is endowed with a {\it Spin
structure}. Locally, Spin structures can  always be found, but globally there
are topological obstructions  encoded in the second Stiefel-Whitney  class of
$E$, $w_2(E) \in H^2(M;{\bf Z}_2)$. If $M$ is Riemannian and  oriented, we
say that $M$ admits a Spin structure if $TM$ does, and $M$ is called a  {\it
Spin manifold}. The necessary and sufficent condition for $M$ to be Spin is
then 
$w_2(M)=0$. 

If ${\bf R}^n$ is endowed with the euclidean quadratic form, the universal
property of  Clifford algebras allows us to define a representation of
$SO(n)$ in $\cl_n$,  denoted by ${\rm cl}_n$. If $E$ is  a real vector bundle
of rank $n$ as before, there is a bundle associated to the  representation
${\rm cl}_n$ and called the Clifford bundle of $E$:
\be
\cl (E) =P_{\rm SO}(E) \times_{{\rm cl}_n} \cl_n.
\label{cli}
\ee This bundle can be always defined. If $M$ is Riemannian and oriented, the
Clifford  bundle  of $M$ is $\cl (M) =\cl (TM)$. To define a {\it spinor
bundle} of
$E$ we need a Spin  structure on $E$. Let then $S$ be a Clifford module for
$\cl_n$, and consider the  representation $\rho_{{\rm Sp}_n}$ given by
restricting Clifford multiplication to 
$\sp_n$.  A real spinor bundle of $E$ is the associated vector bundle
\be S(E) = P_{\rm Sp}(E) \times_{\rho_{{\rm Sp}_n}} S.
\label{bunny}
\ee Similarly, one can define complex spinor bundles considering a complex
left module  for $\ccl_n$. Notice that $S(E)$ is a left bundle module for
$\cl (E)$. 

Given a connection on $E$, or equivalently, a connection on $P_{\rm SO}(E)$,
we can use  the covering map (\ref{spst}) to lift it to $P_{\rm Sp}(E)$.
Using now the representation 
$\rho_{{\rm Sp}_n}$ we get a connection on the spinor bundle associated to
it.  In particular,  if $M$ is a Riemannian manifold and $S$ a spinor bundle
over $M$, it inherits a  canonical Riemannian connection from the one on
$M$.  The local expression for the covariant derivative associated to a
connection on $S(E)$ can be written  considering the explicit  realization of
the isomorphism ${\bf so}_n \simeq {\bf {\rm spin}}_n$. Namely, let  
$\{\sigma_{\alpha} \}_{\alpha=1, \cdots, N}$ be a local section of $S(E)$,
induced by  a local section of $P_{\rm SO}(E)$ $\{ e_i \}_{i=1, \cdots, n}$
with associated  connection matrix 
$\omega_{ij}$. $N$ denotes the rank of 
$S(E)$. A local expression for the covariant derivative is given by:
\be
\nabla \sigma_\alpha = {1 \over 2} \sum_{i<j} \omega_{ji}e_i e_j\cdot
\sigma_{\alpha},
\label{cos}
\ee where the dot denotes Clifford multiplication. 

We will define now the Dirac operator. Let $M$ be a Riemannian manifold with
Clifford  bundle $\cl(M)$, and let $S$ be a bundle of left modules over
$\cl(M)$, endowed with a Riemannian metric and connection. The Dirac operator
is defined on a section $\sigma$ of 
$S$ as follows:
\be D\sigma =\sum_{i=1}^n e_i \cdot \nabla_{e_j} \sigma,
\label{dirope}
\ee where $\{e_i\}_{i=1, \cdots n}$ is an orthonormal local section of
$TM$.   An important case of this construction occurs when $S$ has in
addition two properties:

1) Clifford multiplication by unit vectors is orthonormal with respect to the 
Riemannian metric on $S$:
\be
\langle e\sigma_1, e\sigma_2 \rangle =\langle \sigma_1, \sigma_2 \rangle,
\label{dirbuno}
\ee where $e$ is a unit vector in $T_{x}M$, $\sigma_1$, $\sigma_2 \in S_{x}$. 

2) The covariant derivative must be a module derivation with respect to the
Riemannian  connection in $\cl(M)$:
\be
\nabla(\phi \cdot \sigma)=(\nabla \phi)\cdot\sigma +\phi \cdot (\nabla
\sigma), 
\label{dirbdos}
\ee
$\phi \in \Gamma(\cl(M))$, $\sigma \in \Gamma(S)$. 

If $S$ has these properties, it is called a {\it Dirac bundle}. It is not 
difficult to check that the Dirac operator on any Dirac bundle is formally
self-adjoint.  Spinor bundles on Spin manifolds are the most important
examples of Dirac bundles, with  the Riemannian metric they inherit from the
inner product on the vector space
$S$.
 
In the case of a Spin four-manifold, we can give explicit, local expressions
for the  covariant derivative and the Dirac operator. Let $M_{\alpha}$ be
local coordinates  for a section of the positive-chirality spinor bundle
$S^{+}$. Using the map in  (\ref{nadiuska}), we have for the covariant
derivative:
\be {D}_\mu M_\alpha = \partial_\mu M_\alpha^i - {1\over 2}\sum_{i<j}
\omega_{\mu}^{ij} (\sigma^{*}(e_{i})\sigma(e_j))_\alpha{}^\beta M_\beta.
\label{cuatroco}
\ee The structure group of the complex bundle $S^+$ is $SU(2)_L$, and we can
write the  Spin connection as an $SU(2)$ gauge field using the expressions
(\ref{mias}). Then  (\ref{cuatroco}) reads
\be D_{\mu}M_{\alpha}=\partial_{\mu}M_{\alpha}-
i\sum_{a=1}^3\omega_{\mu}^a(\sigma_{a})_{\alpha}{}^{\beta} M_{\beta},
\label{cova}
\ee where $\omega_{\mu}^a$, $a=1,2,3$ are the components the $SU(2)_{L}$ Spin
connection on $X$. 

As $S=S^{+} \oplus S^{-}$ is a ${\bf Z}_2$-graded module for the action of
$\cl_4$,  the Dirac operator can be split in two pieces. One of them is a map
$\Gamma(S^+) 
\rightarrow \Gamma(S^-)$, and is given by:
\be {D}^{\dot\alpha\alpha} = (\sigma^{\mu})^{\dot\alpha \alpha} {D}_{\mu},
\label{dirfour}
\ee where $\sigma^{\mu}= e^{\mu}_i \sigma(e_i)$ and $e^{\mu}_i$ is the
vierbein, {\it i.e.}  the components of the orthonormal basis $e_i$ with
respect to the usual coordinate  basis. The other piece of the Dirac operator
is a map $\Gamma(S^-) 
\rightarrow \Gamma(S^+)$, and is the formal adjoint of (\ref{dirfour}) with
respect to  the metric in (\ref{positiva}):
\be {D^{\dagger}}_{\alpha\dot\alpha} = -({\bar
\sigma}^{\mu})_{\alpha\dot\alpha}{D}_{\mu}  
\label{dirfourm}
\ee where the covariant derivative acting on the negative-chirality spinors
is given by a  expression similar to (\ref{cuatroco}) but with the matrices 
$-\sigma(e_{i})\sigma^{*}(e_j)$, as it is clear from (\ref{nadiuska}). The 
matrices
${\bar \sigma}^{\mu}$ are given by ${\bar \sigma}^{\mu}= e^{\mu}_i
\sigma^{*}(e_i)$. The total Dirac  operator is formally self-adjoint. Notice
that, according to (\ref{bajo}), one has:
\be {D^{\dagger}}_{\alpha\dot\alpha}=-{D}_{\alpha\dot\alpha}.
\label{unodiracs}
\ee   

It is clear that we can ``twist" a Dirac bundle with any riemannian bundle
$E$ with  connection and get again another Dirac bundle $S\otimes E$. This
bundle is naturally  endowed with the tensor product connection, and this
gives a twisted Dirac operator 
$D_{E}$. If 
$S$ is a spinor bundle on a Riemannian manifold $M$, endowed  with its
canonical Riemannian connection, and $E$ is a bundle  with a connection, the
square of the Dirac operator on $S\otimes E$ verifies an  important identity
kwnown as the Weitzenb\"ock formula. First define an endomorphism of 
$S\otimes E$ by the local expression 
\be R^{E}={1\over 2}\sum_{ij}F^{E}_{ij}e_ie_j, 
\label{endocrino}
\ee where $F^{E}$ is the curvature of the connection on $E$ and $e_ie_j$ acts
by  Clifford multiplication on $S$. The Weitzenb\"ock formula states that:
\be D_{E}^2=\nabla^{*} \nabla + {1\over 4}R +R^E. 
\label{wei}
\ee In this expression,$\nabla^{*} \nabla$ is the usual connection laplacian
of 
$S\otimes E$, and $R$ is the scalar curvature of $M$. On a four-dimensional
Spin  manifold, the square of $D_E$ on $S^{+}\otimes E$ involves in an
interesting way the  self-dual part of the curvature. This is because the
image of the map in  (\ref{nadiuska}), when restricted to positive-chirality
spinors, keeps only the  self-dual part of the differential form one starts
from. In this way we get:      
\be D^{\dagger}_{\alpha\dot\alpha} D^{\dot\alpha \beta} = (\nabla^{*} \nabla +
{1\over 4} R)\delta_{\alpha}{}^{\beta} +i F^{E}_{\alpha}{}^{\beta}
\label{miwei}.
\ee

\section{$\sp^c$-structures} Spin structures are difficult to find because of
the topological obstruction associated  to the second Stiefel-Whitney class,
and for this reason one can not define spinor  bundles on  arbitrary
four-manifolds. However, one would like to find a more general framework  in
order to be able to define generalized spinors in an appropriate way. This is
the  main achievement of
$\sp^c$-structures, which in fact can be defined on an arbitrary 
four-manifold.
$\sp^c$-structures also arise in a natural way under  duality
transformations  in the twisted version of the Seiberg-Witten effective
action \cite{abelia}, and have a key r\^ole  to define generalized moduli
spaces. We will present here some basic  facts concerning them. 

The first ingredient is purely algebraic. Consider a complex spinor
representation of  dimension $N$, given by a map
\be
\Delta_{\bf C}: \sp_n \too U(N),
\label{del}
\ee   and consider also the inclusion map of the center $U(1)$ of $U(N)$, 
\be z:U(1) \too U(N).
\label{ucd}
\ee We can form the map:
\be
\Delta_{\bf C} \times z: \sp_n \times U(1) \too U(N),
\label{mixto}
\ee whose kernel is ${\bf Z}_2=\{(1,1), (-1,-1)\}$. The quotient of the
domain by this  kernel  is by definition the group $\sp^c_n$:
\be
\sp^c_n = {\sp_n \times U(1) \over {\bf Z}_2}.
\label{spinc}
\ee There is a natural isomorphism $\sp^c_n \simeq \sp_n \otimes U(1)$ and
this gives a  natural inclusion $\sp^c_n \subset {\bf C}{\rm l}_n$. From the
definitions  above we obtain the  short exact sequence:
\be  1 \too {\bf Z}_2 \buildrel i
 \over \too \sp^c_n 
\buildrel  \xi \over \too SO(n) \times U(1)   \too 1,
\label{short}
\ee where ${\bf Z}_2=\{[(1,1)], [(-1,1)]\}$, $i$ is the inclusion and $\xi$
is given by:
\be
\xi[(v, {\rm e}^{i\theta})]=(\xi_0(v),{\rm e}^{2i\theta}).
\label{explico}
\ee
$\xi_0$ is of course the covering map in (\ref{spinshort}). Obviously, complex
spinor  representations extend to $\sp^c_n$ representations, and for even
dimensions these  representations also split in two. It is easy to check with
the above definitions,  and using the basic isomorphism $\sp_4 \simeq SU(2)
\times SU(2)$, that $\sp^c_4$ is  isomorphic to the subgroup of $U(2) \times
U(2)$ consisiting of pairs of matrices with  the  same determinant.

We would like now to construct $\sp^c$-bundles on manifolds, {\it i.e.}, to
find a  bundle  version of the covering in (\ref{short}). Notice that in this
case we need an auxiliary  geometric structure, due to the $U(1)$ factor in
the last term of (\ref{short}).  Namely, let $P_{\rm SO}$ be a principal
$SO(n)$ bundle over a manifold $M$.  A $Spin^c-${\it structure} on $P_{\rm
SO}$ consists of a principal
$U(1)$-bundle 
$P_{U(1)}$, a $\sp^c_n$-bundle $P_{{\rm Sp}^c}$ and a map 
\be
\Xi: P_{{\rm Sp}^c} \too P_{\rm SO} \times P_{U(1)} 
\label{mapac}
\ee such that $\Xi(pg)=\Xi(p) \xi(g)$, where $p \in P_{{\rm Sp}^c}$, $g \in
\sp^c_n$  and $\xi$ is the map in (\ref{explico}). The first Chern class of
$P_{U(1)}$, 
$c \in H^2(X, {\bf Z})$, is called the canonical class of the
$\sp^c$-structure, and the  complex line bundle associated to $P_{U(1)}$ is
called its determinant line bundle.  A standard obstruction analysis shows
that $P_{\rm SO}$ admits a
$\sp^c$-structure if  its second Stiefel-Whitney class $w_2(P)$ is the mod $2$
reduction of any integral  class in $H^2(M, {\bf Z})$, which is then
associated to the determinant line bundle  of the $\sp_c$ structure. An
oriented, Riemannian manifold with a $\sp^c$-structure on  the bundle of
orthonormal frames is called a
$\sp^c$-manifold.

An important example of bundles admitting a $\sp^c$-structure are complex
vector bundles.  To see this, notice first that, from the theory of
characteristic classes,
$w_2(E)  =c_1(E) \,\ {\rm  mod} \,\ 2$. The $\sp^c$-structure in this case is
canonical: if we endow 
$E$ with a Hermitian metric, there is a $U(n)$ principal bundle of unitary
frames,  denoted by $P_{U(n)}(E)$. There is also a natural morphism: 
\bea U(n) &\too& SO(2n) \times U(1)\nonumber\\ g & \mapsto & (i(g), {\rm
det}\,\ g), 
\label{uen}
\eea where $i:U(n) \rightarrow SO(2n)$ is the natural inclusion. This map can
be lifted to a  map $j:U(n) \rightarrow \sp^c_{2n}$, and this allows one to
construct an associated bundle 
\be P_{{\rm Sp}^c}=P_{U(n)}(E) \times_j \sp^c_{2n},
\label{canon}
\ee and the determinant line bundle of this $\sp^c$-structure is given by
$\Lambda^n(E)$  (the transition functions are the determinant of the
transition functions of $E$,  according to (\ref{uen})). In particular, any
almost complex manifold has a canonical 
$\sp^c$-structure. 

Just like in the $\sp$ case, we can construct associated vector bundles
starting from  complex representations of the Clifford group. Let $M$ be a
$\sp^c$-bundle of  dimension $n$. Let $V$ be a complex $\cl_n$-module.  As
$\sp^c \subset \cl_n \otimes {\bf C}$, we have a representation $\Delta$ of
$\sp^c$  in the  group of complex automorphisms of $V$. We can construct:
\be S=P_{{\rm Sp}^c} \times_{\Delta} V.
\label{sfunda}
\ee This vector bundle is called a complex spinor bundle. In the case of $n$
even, we have,  as in the $\sp$ case, a fundamental spinor bundle associated
to the unique irreducible  complex representation of the complexified
Clifford group. It also splits into a direct  sum, $S=S^+ \oplus S^-$, and we
can also define positive and negative chirality spinors,  as sections of
$S^+$ and $S^-$, respectively.     

A very useful way to approach $\sp^c$-structures is in terms of the
construction  of complex spinor bundles. Suppose that the  second
Stiefel-Whitney cohomology class of a manifold $M$, 
$w_2(M)\in H^2(M; {\bf Z}_2)$, is the ${\rm mod} \,\ 2$  reduction of an
integral class $[w]$. Let $L$ be the line bundle corresponding to 
$[w]$, {\it i.e}, $[c_1(L)]=[w]$. In general, the square root of the line
bundle 
$L$ does not exist, and the topological obstruction to define it globally is
precisely 
$w_2(M)$. In the same way, the spinor bundle constructed in terms of a 
$\sp$-structure and an irreducible complex representation of $\cl_n$ is not 
globally defined if the manifold is not Spin. However, a standard obstruction 
analysis in \v Cech cohomology shows that,  although $L^{1/2}$ and the spinor
bundle
$S(M)$ are not globally defined separately,  the product bundle 
\be S_L(M)=S(M) \otimes L^{1/2}
\label{spincfun}
\ee is well defined, and is the complex  spinor bundle associated to the
$\sp^c$-structure defined by $L$ and to the complex  representation of
$\cl_n$. This expression is useful to understand the action of the  second
cohomology group
$H^2(M;{\bf Z})$ on the set of $\sp^c$-structures. Suppose  that $L$ defines a
$\sp^c$-structure as in (\ref{spincfun}), and let $L_{\alpha}$ be  the line
bundle associated to an element $\alpha \in H^2(M;{\bf Z})$. If we twist 
$S_{L}(M)$ by $L_{\alpha}$, we get a new complex spinor bundle 
\be S_{L\otimes L_{\alpha}^2}(M)=S(M) \otimes \Big(L^{1/2}\otimes
L_{\alpha}\Big).
\label{maspinc}
\ee Notice that, if $c_1(L) \equiv w_2(M) \,\ {\rm mod} \,\ 2$, the same is
obviously  true for $L\otimes L_{\alpha}^2$. In this way one can obtain all
the different 
$\sp^c$-structures on the manifold $M$ (although not all of them will be
different). 

In the case of an almost complex manifold $M$, the canonical complex spinor 
bundle associated  to the canonical $\sp^c$-structure has a simple
interpretation. From the structure  of the complex representation of the
Clifford algebra \cite{sg, morgan}, we have the  isomorphism:
\be S_{K^{-1}}(M)=\Omega^*_{\bf C} TM, 
\label{canospin}
\ee where the right hand side denotes the complex exterior powers of the
holomorphic  tangent bundle of $M$. Positive chirality spinors correspond to
the even powers, and  negative chirality spinors to the odd ones. The
determinant line bundle of the  canonical $\sp^c$-structure is precisely
$\Omega^n_{\bf C} TM = K^{-1}$, where $K$  is the canonical line bundle of
$M$. This is the reason of the notation in  (\ref{canospin}). The other
complex spinor bundles, associated to the  different
$\sp^c$-structures, are obtained by twisting this canonical  spinor bundle
with some line bundles over $M$. If $M$ is Spin, $w_2(M)=0$,  or
equivalently, 
$c_1(M)=-c_1(K) = 0 \,\ {\rm mod} \,\ 2$. This in turn implies that there
must be well  defined square roots of the canonical bundle $K$. Using
(\ref{spincfun}), together  with the fact that $L=K^{-1}$, we see that the
spinor bundle $S$ associated to  the square root $K^{1/2}$ of an almost
complex  manifold, is related to the canonical  spinor bundle
(\ref{canospin}) by
\be S=S_{K^{-1}} \otimes K^{1/2}.
\label{spinacopla}
\ee

Complex spinor bundles associated to a $\sp^c$-structure can be endowed with a
metric and  connection, obtaining in this way Dirac bundles. To do this, we
use the representation  (\ref{spincfun}), and we consider a Hermitian metric
and unitary connection on  the line bundle $L$. This induces  a connection on
$L^{1/2}$. Locally, both the  spinor bundle and $L^{1/2}$ exist and we can
consider the tensor product connection. The  connection is well defined
globally. Also, one can  define a Dirac operator $D_{L}$ according to
(\ref{dirope}) in the previous section.  The Weitzenb\"ock formula 
(\ref{wei}) reads in this case:
\be D_{L}^2=\nabla^{*} \nabla + {1\over 4}R + {i \over 2} \Omega, 
\label{weic}
\ee where $\Omega$ is the curvature of the connection on $L$, represented by
a real two-form. 

In the case of a K\"ahler manifold $M$, the Dirac operator for the canonical
 $\sp^c$-structure has a simple description in terms of Dolbeault  operators
\cite{sg, morgan}. First of all, we use the Hermitian metric to identify 
$S_{K^{-1}}(M)$ with the exterior algebra bundle of complex $(0,p)$-forms.
Again,  positive chirality spinors are identified with forms of even
antiholomorphic degree,  and negative chirality spinors with forms of odd
antiholomorphic degree. In this case,  the Dirac operator can be identified
with:
\be {\sqrt 2}\Big( {\overline \partial} + {\overline \partial}^*\Big) : 
\Omega^{0, {\rm even}} \too \Omega^{0, {\rm odd}}.
\label{diropcan}
\ee From (\ref{spinacopla}) follow that, if $M$ is also Spin, then the Dirac
operator is  the sum of the Dolbeault operators in (\ref{diropcan}) twisted
by the bundle $K^{1/2}$. 

$\sp_c$ structures are associated to cohomology classes in 
$H^2(M,{\bf Z})$ that are congruent to $w_2(X)$ ${\rm mod} \,\ 2$. In the
case of  four-dimensional manifolds, cohomology classes verifying this
condition are  characterized by a simple property. If we denote by $( \,\ ,
\,\ )$ the intersection  form of $M$, an element $c \in H^2(M,{\bf Z})$ is
called {\it characteristic} if 
\be (c,x) = (x,x) \,\,\ {\rm mod} \,\ 2, 
\label{caracter}
\ee for all $x \in H^2(M,{\bf Z})$. These are precisely the cohomology classes
defining 
$\sp^c$-structures. Once such a structure is defined, we can form the complex
spinor  bundles associated to the unique irreducible complex representation
of $\cl_4$. Many  constructions that we defined in the previous section for
usual spinors also hold  in the $\sp^c$ case. In particular, (\ref{casi})
also holds with $S^+$ the complex  spinor bundle associated to a
$\sp^c$-structure on $M$. This is because in this  isomorphism  we consider
the trace-free part of the endomorphisms of $S^+$, and the part of the 
complex spinor bundle associated to the determinant line bundle does not
appear.  Our final comment concerns the form of this isomorphism on a
K\"ahler manifold 
$M$, for the  canonical $\sp^c$-structure. A real SD form on a K\"ahler
manifold has  the decomposition  given in (\ref{desco}), and we can write it
as:
\be F^+=f \omega + F^{2,0}+F^{0,2}, 
\label{des}
\ee where $f$ is a real function on $M$ and $F^{2,0}={\overline F^{0,2}}$. If
we  denote an element in $S^+ =\Omega^0 \oplus 
\Omega^{0,2}$ as a column vector $(\alpha \,\ \beta)^t$, the trace-free,
skew-adjoint  endomorphism of $S^+$ is given by \cite{morgan}:
\be F^+ \too 2 \left(\begin{array}{cc}-if&-*F^{2,0}\wedge \\
                F^{0,2}\wedge &if \end{array} \right),
\label{endokal}
\ee where $*$ is the Hodge star operator and the $\wedge$ denotes the wedge
product with the  corresponding complex form. Notice that this endomorphism
is the same for any $\sp^c$- structure, and therefore it also holds for the
$\sp$ case. The norm of a self-dual form  is then given, according to
(\ref{mimetra}) and (\ref{hula}), by:
\be |F^+|^2=4\Big(|f|^2+|F^{2,0}|^2 \Big).
\label{endometrica}
\ee

\end{document}